## THE RECONSTRUCTION OF THE ANTIKYTHERA MECHANISM INSTRUCTION MANUAL AFTER A DILIGENT STUDY AND ANALYSIS OF THE BACK COVER INSCRIPTION <u>PART-1</u> (pages 01-60)

## THE RECONSTRUCTION OF THE ANTIKYTHERA MECHANISM INSTRUCTION MANUAL OF THE BACK DIAL PLATE, AFTER A DILIGENT STUDY AND ANALYSIS OF THE BACK COVER INSCRIPTION <u>PART-2</u> (pages 60-100)

**A. Voulgaris[1], C. Irakleous[2], C. Mouratidis[3], A. Vossinakis[4]**

[1]*City of Thessaloniki, Directorate Culture and Tourism, Thessaloniki, GR-54625, Greece*
[2]*Independent Researcher, Thessaloniki, GR-54248, Greece*
[3]*Merchant Marine Academy of Syros, GR-84100, Greece*
[4]*Thessaloniki Astronomy Club, Thessaloniki, GR-54646, Greece*

### Abstract

*The Inscriptions which were on the Back Cover of the Antikythera Mechanism were the User's Manual of the Antikythera Mechanism. In the Back Cover Inscription Part-1 text, the Presentation, Position, and Operation of the outer parts of the Mechanism's Front face was engraved by the ancient manufacturer. Although this technical text is partially/poorly preserved, useful information, results, and conclusions were retracted by studying the words/sentences adapted on their original geometrical positions, concerning the image of the Mechanism's front face. After the study and the analysis of the preserved text and its characteristics, the missing text completion is presented. A significant part of the reconstructed text was reproduced using the preserved words and phrases. In many cases the reconstructed text can be considered as definite. The use of the author's Antikythera Mechanism functional model was critical for the analysis and understanding the preserved text, contributing to the design and reconstruction of the Mechanism's Front Face parts, which are not preserved nowadays.*

### 1. Introduction

The Antikythera Mechanism was a Time Machine and an astronomical events calculator, constructed and (probably) used during the Hellenistic Era of 200-100 BC. It consists of a large number of gears, axes, shafts, pointers, scales, and plates. After 2000 years under the sea and its retraction in 1901, today the Mechanism is partially preserved in seven relatively large fragments (A-G) and 75 small pieces (minor fragments 1-75).[1] The fragments' corrosion is evident, as well as the strong deformation and displacement of the parts resulted from the material/density change.[2] The corrosion destroyed many parts and erased a large portion of the inscriptions.

On all the available bronze areas of the Mechanism the ancient manufacturer engraved important information related to the extruded results of the pointers.

---

[1] http://www.antikythera-mechanism.gr/data/fragments
[2] Voulgaris et al., 2019b and Voulgaris et al., 2021

Additionally, on the two bronze Cover plates, which were located on Front and Back side on the Mechanism's main body, a large text was engraved on each.[3]

Each of these thin bronze/copper plates was possibly stabilized by pins on a wooden plate: The bronze plates of the Mechanism's main body were also supported sideways by a wooden decorative case. The probable existence of two wooden bearing plates of the Back Cover (bronze) plates creates an intergraded wooden case.[4]

These two covers could be permanently adapted to the main body, acting as the "*doors*" of the Mechanism. On the plates, extensive information with a large number of sentences was engraved. In order to be read easily, these two covers could be totally independent, playing the role of a bronze page book. The two cover plates could be assembled and secured on the main body, by a leather strip or a bronze bordure, creating a small wooden chest.[5]

On the Back Cover the ancient Manufacturer - in this work we call him *The Engraver* - engraved a text that was the *User's Manual* of the Antikythera Mechanism.[6]

Claudius Ptolemaeus, in Almagest describes the construction and use of a *Dioptra* for angle measurements.[7] He refers to the words: ΚΥΚΛΟΝ ΧΑΛΚΕΟΝ, ΚΥΛΙΝΔΡΙΑ, ΤΕΤΟΡΝΕΥΜΕΝΑ, ΓΝΩΜΟΝΙΑ, ΚΥΚΛΙΣΚΟΣ, ΞΥΛΙΝΗΝ ΠΛΙΝΘΙΔΑ (circular copper plate, small cylinders, formed parts by the use of lathe, pointers, small circle, wooden base etc.) and giving information for the use of the instrument.[8] Heron Alexandreus (Schöne 1903) in *ΠΕΡΙ ΔΙΟΠΤΡΑΣ* describes the construction and use of a *Dioptra* for measuring angles, which are useful for the construction of buildings, walls, ports, and also for celestial sphere measurements, solar and lunar eclipses, stars etc. He refers to the words: ΣΤΥΛΙΣΚΟΣ, ΤΥΜΠΑΝΙΟΝ, ΤΥΜΠΑΝΙΟΝ ΟΔΟΝΤΩΜΕΝΟΝ, ΕΛΙΚΑ, ΧΑΛΚΑ ΣΤΗΜΑΤΙΑ, ΣΩΛΗΝ, ΥΑΛΙΝΑ ΚΥΛΙΝΔΡΙΑ (small pillar, base, base with teeth, helix, copper bearings, tube, glass small cylinders etc.).[9]

Today, a significant percentage of the Antikythera Mechanism User's Manual is missing. The first attempt for the letter recognition was started by I. Svoronos, A. Rehm, I. Theophanidis and then continued by D.S. Price.[10]

---

[3] Price 1959 and 1974; Freeth et al., 2006; Bitsakis and Jones 2016b
[4] Voulgaris et al., 2019b
[5] Voulgaris et al., 2019b
[6] Bitsakis and Jones 2016b
[7] Claudius Ptolemaeus,1898, Syntaxis mathematica, Book A, ΙΒ' Περὶ τῆς μεταξὺ τῶν τροπικῶν περιφερείας
[8] Voulgaris et al., 2019a
[9] Heron Alexandreus (Schöne 1903), ΠΕΡΙ ΔΙΟΠΤΡΑΣ (Heronis Alexandrini Opera quae Supersunt Vol. III).
[10] Svoronos 1903; Rehm 1905/1906; Theophanidis 1927-1930; Price 1974.

In 2004, during the implementation of the *Antikythera Mechanism Research Project*, an X-Ray tomograph 450keV was used for the study of the Mechanism's fragments: The pinpoint X-Ray focal spot around 50μm and the small voxels of the row data (in sub dimension of the engraved letters' depth) was the advantage of this instrument, allowing the recognition of non-visible, engraved letters with relative high contrast.[11] Using the AMRP X-Ray CTs, a large number of invisible letters was detected.[12] In Bitsakis and Jones 2016b, is published the updated Back Cover Inscription Part-1/2 text.

## 2. The preserved Back Cover plate

In the present, the remaining parts of the Back Cover are preserved on Fragment B (Part-1) and Fragment A (Part-2).[13] Many preserved letters on the fragments are easily visible to the naked eye. These letters appeared on their vertical - mirror symmetry, left/right – and are visible in the negative face, like a stamp (A. Rehm refers to them as "*patina-offsets*"). For this reason, many researchers suggested that these parts are not the original corroded copper/bronze plate, but they are formatted by sediment deposits on the original engraved letters, creating the "*negative*" ones.

The sediment deposits (usually material of $CaCO_3$ and sand) have usually beige or dark color shades in irregular distribution. As shown by **Fig. 1**, the colors of the preserved Back Cover plate are (relative regular) dark green shades, almost equal to Atacamite which is the main product of the bronze/copper corrosion under the sea.[14] In a short time on a bronze/copper plate under the sea, is created an amorphous crystal layer of Atacamite[15], which follows the shape and the mechanical characteristics of the plate **Fig. 1**. It seems that the area of the preserved Back Cover is the amorphous crystal layer of Atacamite, which was directly formatted on the Back Cover engraved plate surface, which today is mostly missing.

---

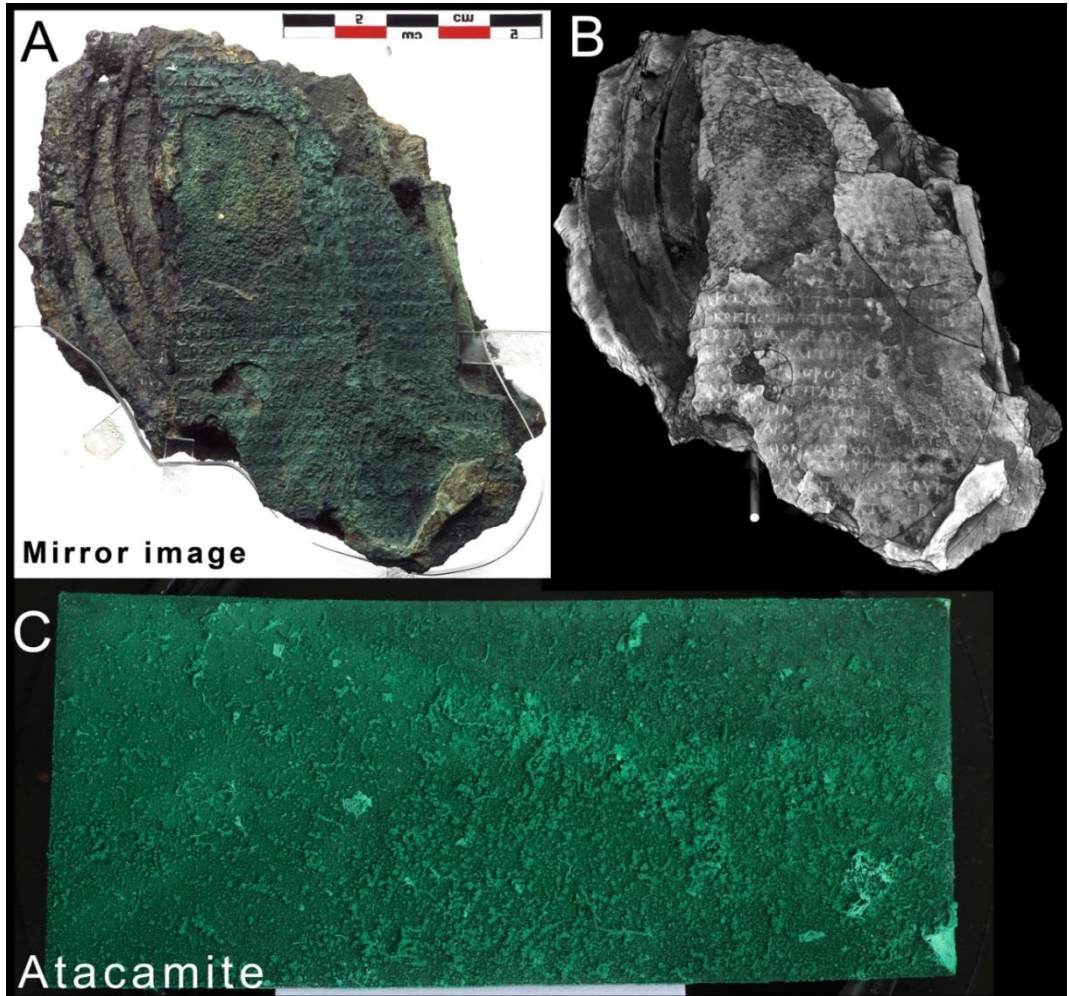

Fig. 1. A) Mirror image of Fragment B1. The dark green shades correspond to the preserved Back Cover copper/bronze plate Part-1. On the left, the preserved Metonic Helix turns are visible in darker shades. Credits: National Archaeological Museum, Athens, Greece, K. Xenikakis, Copyright Hellenic Ministry of Culture & Sports/Archaeological Receipts Fund. B) The reconstructed outer face of Fragment B, by using AMRP X-Ray CTs and the *Real3D VolViCon* software.[16] X-Ray image processed by the Authors.[17] C) Atacamite [$Cu_2(OH)_3Cl$] formation on a copper surface: Copper pieces were immersed in artificial sea water, in salinity around 38.5, equal to the Antikythera Sea.[18] For the artificial sea water unprocessed sea salt from Kythera was used. The total time for Atacamite formation under the artificial sea water was 33 months. Although the images A and C are captured by different cameras, the similarity in the shades of the Back Cover plate and Atacamite is evident. Material process and image by the Authors

By observing the preserved engraved letters on the Antikythera Mechanism fragments, their large engraving depth becomes evident. This depth is relative difficult to achieve with the ordinary engraving tools. In **Fig. 2** a corrosion experiment is presented, giving a probable answer for the large depth of the engraved letters.

---

[16] Real3D VolViCon (software) Version 4.30.1214.
[17] Voulgaris et al., 2018c.
[18] Voulgaris et al., 2019b.

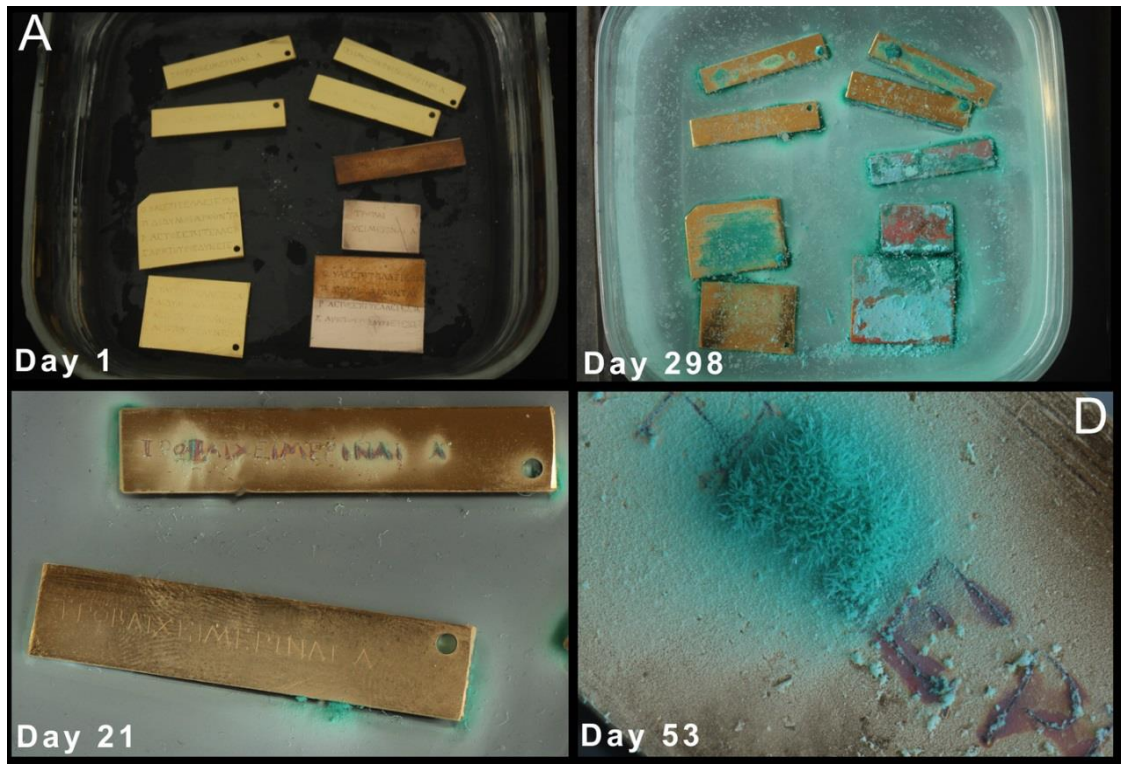

Fig. 2. A) Three pairs of gilded bronze pieces (two pairs on top and one on bottom left) were used for the following corrosion experiment: One piece of each pair was firstly gilded and afterwards engraved (therefore, each engraved letter is not gilded). The second piece was firstly engraved and then was gilded. Afterwards, the pieces were immersed in a solution of sea water with salinity of about 38.5, close to that of the Antikythera sea water, using raw sea salt from Kythera. The three pieces on bottom right are (totally) non-gilded bronze. B) After 298 days, each piece exhibits a different corrosion level. Visible products of bronze corrosion by sea water are: copper chlorides in turquoise shades, copper oxides in dark red/brown shades.[19] C, D) After about 2-3 weeks, non-gilded letters begin to show corrosion signs and as times passes, the corrosion increases the letter depth (for the chemical procedure see Voulgaris et al., 2019b). On the other hand, totally gilded pieces don't show any corrosion signs, since gold is highly resistant to sea water corrosion.[20] Since the Mechanism's bronze material is prone to short term oxidization, the manufacturer could have gilded them before the engraving. C) The corrosion of the simple bronze engraved piece is evident. D) The concentration of crystal copper chlorides is visible. Material process and corrosion experiment by the Authors. The images were taken after the use of an optomechanical system with several focal lengths, in macro, with micro x-y movements and differential mechanical micro focus, designed/constructed/used by the first Author

## 3. The Back Cover text's metrology

As aforementioned, the Mechanism was formed in a small wooden chest. The dimensional specifications of a box cover are dependent on the dimension of the box's remaining parts. According to measurements of the Mechanism's dimensions in Voulgaris et al., 2019b; Allen et al., 2016, the dimension of the Mechanism's (bronze) main body is around 325mmX174mm each of the two covers would have a similar dimension.

---

[19] Voulgaris et al., 2019b.
[20] Scott, 1990.

Based on our measurements of **Table I**, it seems that each of *The Engraver's* letter (left-right) width dimension (including the left/right "semi-spaces"), varies between 2.0mm and 2.7mm, with a statistically mean of 2.4mm/letter, **Fig. 3**. At the same time, some areas of the Back Cover plate's left boundary (text and also the plate) are preserved. For Symmetry reasons, we believe *The Engraver* kept the same space/gap of the recessed page at the end/right boundary of the Back Cover plate, which is not preserved today. The left recessed space dimension (boundary copper/bronze plate up to the boundary of the first letter of line) is around 2mm. Therefore, 174mm–(2mm left+2mm right)= 170mm. According to **Table I**, each line varies between 63-82 letters/line (mean 72).[21]

| Text | Letters number | Dimension | mm/ Letter | Letters/ cm | Letters/ Line |
|---|---|---|---|---|---|
| **16 ΠΡΟΕΧΟΝΑΥΤΟΥΓΝΩΜΟΝΙΟΝΣ** | 22 | 55mm | 2.5 | 4 | 68 |
| **17 ΦΕΡΕΙΩΝΗΜΕΝΕΧΟΜΕΝ** | 17 | 39mm | 2.3 | 4.3 | 73 |
| **18 ΤΟΣΤΟΔΕΔΙΑΥΤΟΥΦΕΡΟΜΕΝ** | 21 | 55mm | 2.6 | 3.8 | 65 |
| **19 ΤΗΣΑΦΡΟΔΙΤΗΦΩΣΦΟΡΟΥ** | 19 | 50mm | 2.6 | 3.8 | 65 |
| **20 ΤΟΥ[ΦΩ]ΣΦΟΡΟΥΠΕΡΙΦΕΡΕΙΑΝ** | 22 | <54mm | 2.4 | 4.1 | 70 |
| **21 ΓΝΩΜΩ[.]ΚΕΙΤΑΙΧΡΥΣΟΥΝΣΦΑΙΡΙΟΝ** | 27 | 57mm | 2.1 | 4.7 | 80 |
| **22 ΑΚΤΙΝΥΠΕΡΔΕΤΟΝΗΛΙΟΝΕΣΤΙΝ** | 24 | 50mm | 2.1 | 4.8 | max 82 |
| **24 ΕΘΟΝΤΟΣΤΟΔΕΔΙΑΠΟΡΕΥΟΜΕΝ** | 23 | 62mm | 2.7 | 3.7 | min 63 |
| | | Mean: | **2.4** | **4.15** | **71 (-8/+11)** |

Table I. A selection of well-preserved texts of the Back Cover Inscription Part-1, located on Fragment B. In order to achieve the best approach statistically (letters/cm), as many visible to the naked eye letters were chosen. Then, the dimension of each selected text was measured using calibrated visual photographs (Xenikakis 2004 and Voulgaris 2016/2019). As detailed below, the ratios mm/letter and letters/cm were calculated. These ratios were extrapolated in order to calculate the total number of letters for each line-text (see also Bitsakis and Jones 2016b). These ratios revealed *The Engraver's* handwriting. For the letters' recognition on the visual photographs by Xenikakis 2004; Voulgaris 2016/2018/2021, the X-Ray computed tomographies of AMRP were also contributed

The height of 21 successive preserved lines (from the 1[st] line/letter base to the 21[th] line/letter base), was measured around 75mm. Therefore, the mean height line is 75/21= 3.57mm/line (letter height + inter-line space). Setting the height of the Mechanism 325mm (Voulgaris et al., 2019b), the estimated maximum number of the Back Cover plate lines is around 91 (but it could be 2-4 lines less, 89 or 87, if *The Engraver* left some spaces on the top and the bottom edges) (also equal measurements in Bitsakis and Jones 2016b). Today, the total number of (partially/poorly preserved) lines of Part-1 and Part-2, is 30+25= 55 lines. According to these calculations, the estimated maximum number of letters for the Back Cover plate is around 7350.

The Back cover inscription presents information regarding the design, positioning and the role/operation of the outer parts of the Mechanism, which were visible to the naked eye. On BCI Part-1 (upper plate's area-Fragment B1), the text describes the Front face's outer parts of the Mechanism, visible to the naked eye, such as the Lunar Disc, its lunar phases sphere and pointer, the Sun's pointer. Also, the names of the seven planets (observable to the naked eye, according to the Hellenistic Astronomy), are referred.

---

[21] Also, equal measurements in Bitsakis and Jones 2016b.

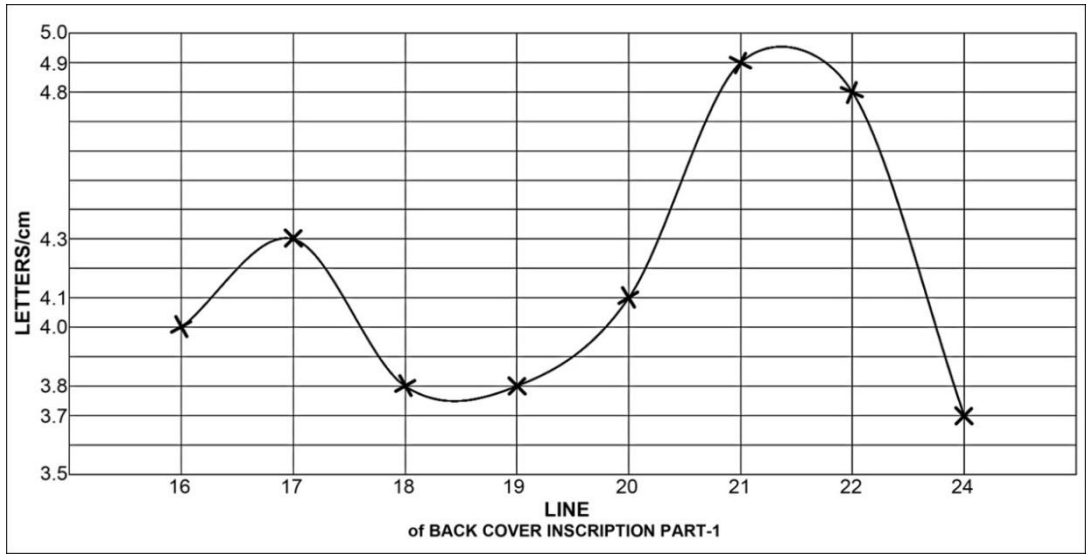

Fig. 3. Letters per centimeter vs line-text (based on 8 partially preserved lines in total). The difference is due to the hand writing, the total number of the "thin" letters such as "I" and "Z" (the presence of a large number of these letters, especially "I", creates space for additional letters), and also by the material shrinkage and deformation after its corrosion

On BCI Part-2 (bottom plate's area-Fragments A2, E, 19, 67), the text describes the design and operation of the Back face's outer parts, such as the two Helices and their subdivisions and also their pointers' operation.[22]

The Antikythera Mechanism consists of a large number of moving parts, gears, axes, pointers, and also measuring scales. As with any other machine, simple or complex, the presentation and the operation of the mechanical parts on a technical text/manual is crucial in order to explain "*how this devise works*".

The text of the Back Cover plate is considered the *User's Manual* (see section 3.1) of this time calculator-events predictor machine. This specific machine was too complicated. For this reason, the Antikythera Mechanism's users should possess a deep understanding of Astronomy, Celestial Mechanics, the four lunar cycles[23], and time measuring techniques, in order to understand and comprehend the Mechanism's extruded information/results, which were visible on the Front and Back plate pointers.

## 4. What is the User's manual

---

[22] Bitsakis and Jones 2016b.
[23] The four lunar cycles: Lunar month of 29.53[d], Sidereal month of 27.321[d], Anomalistic month of 27.55[d] and Draconic/Draconitic month of 27.21[d].

**4.a. The *User's Manual* for machines handling**

Some products or constructions are complicated and sophisticated. They could consist of a large number of parts that could be difficult to handle by a person that is not familiar with these complicated constructions. In most cases, a user of a product is interested in using it for a procedure, but it is not necessary to know how this construction works. E.g. it's not necessary for the user of an electric drill to be an electrical engineer. For this reason, a technical text with instructions on *how the product must be used* is necessary.

This text is called the *Instruction Manual*, named also *the User's manual*, *Reference manual*, *Product manual*, *Working manual*, *Owner's manual* etc. Such technical documents provide any information necessary to help someone to use a product. The manual is a one-way communication that is provided by a person who knows the operation of the product in depth and gives information and instructions to someone who is non-expert and needs to know how to use this product.

The manual presents the parts of the product, their function, the procedure of the start/close and its operation. The presentation of the parts is not random, but follows a specific scenario, mostly reproducing the procedure of the start/use/end.[24]

In an explanatory technical text, as the User's Manual, after a general description (i.e. *Introduction and Purpose*, presenting general information about the product and the purpose it was constructed for), follow the steps for the use of the product:

1) **Definition/Description** of the parts (*This is…., There is…, Is named…., Is consisted by…etc.*).

2) **Positioning** of the parts (*The part is located on…., Is stabilized on…, Is visible on…etc.*).

3) **Operation/Function** of the parts (*This part starts…, Opens/Closes the…, Is rotated CCW…etc.*). These steps (which we now call *DDPO*) explain how to start, use, operate, assemble, a device.

A shorter text document that includes only the basic features and procedures offering information on *How to use it*, but without the information on "*How to assemble it, How to repair it*", and it is useful for a quick start of the device, called *The User's Guide*.

**4.b. Technical information, Observations and Comments on the Back Cover Inscription Part-1 text**

On the Antikythera Mechanism's Back cover plate 55 lines are partially preserved, in which *The Engraver* presents and describes the outer parts located on the Front face of the

---

Mechanism (BCI Part-1, 30 lines), see **Table II**, as well as the parts of the Back face (BCI Part-2 (25 lines).[25]

On the Antikythera Mechanism, two main kinds of parts exist:

The *operational* and the *supportive parts*.

The *operational parts* are considered the parts located on the outer faces of the Mechanism and are visible to the naked eye. The *operational parts* have a prominent role in the use of the Mechanism. These parts are the Lunar Disc, the pointers and the measuring scales with the subdivisions. Via these parts, the extracted results are presented, after the calculations were made by the *supportive parts*.

The *supportive parts* are gears, axes and shafts, stabilizing pins, spacers etc. and they are not visible, because they are located inside the Mechanism's closed wooden box. In addition, there are *decorative parts* that do not have an important function in the Mechanism. An example of this can be seen in the decorative wooden case.[26]

The Front face/plate of the Mechanism presents a characteristic *central symmetry*[27] (Lunar Disc adapted on the Front face's center, which coincides with the Zodiac and Egyptian rings' centers) and also *horizontal axis symmetry*[28] (top and bottom Parapegma plates PP-1 and PP-2, Bitsakis and Jones 2016a). The Front plate consisted of three independent pieces (Bitsakis and Jones 2016a) creating an oblong plate in dimensions around 174mmX345mm (as also the Back plate) (Voulgaris et al., 2019b). On the Front plate center defined by the cross point of the two diagonals, there is a central large hole around 135mm in diameter (see also Allen et al., 2016). In the center of the hole, the axis $b_{out}$ is located which is stabilized on the Middle plate (Voulgaris et al., 2018b). On axis $b_{out}$, the large gear b1 is adapted. In the center of the inner $b_{in}$ axis the Earth is located (the Mechanism is geocentric). On the upper edge of $b_{in}$ axis the Lunar Disc is adapted, which is the proper and ideal input/driving of the Mechanism. The Lunar Disc's body is a cylinder in diameter around 68mm-70mm and height about 10mm-13mm. The authors drive their functional model of the Mechanism rotating the Lunar Disc/Input by hand.

After the analysis of the original preserved text, and although the text is poorly preserved (without any fully preserved line), a reconstruction of the missing text is not impossible. This reconstructed text could be ideal and representative if it is based on specific criteria and parameters observing the text's original characteristics. In most cases, the text reconstruction can be considered as definite, especially in the area of the planets' orbit presentation.

---

[25] Bitsakis and Jones 2016b.
[26] Voulgaris et al., 2019b.
[27] https://www.mathsisfun.com/geometry/symmetry-point.html
[28] https://www.mathsisfun.com/geometry/symmetry-point.html

Below, critical observations, results, relative information and comments, are presented, after the study of the preserved text Part-1:

1) The preserved text is a technical text/the *User's Manual of the Antikythera Mechanism*. *The Engraver Defines/Presents* the parts of its creation, their *Position* and *Operation*, in order to teach the user "*what parts you see/what the parts represent and how they work*".

2) The lost text is not random, but it has a specific structure: *The Engraver* presents the Front face's outer/*operational parts*, starting from the geometrical center and then, following a radial distribution, continues the presentation of the parts.

3) In the text, phrases or equal meaning phrases are preserved, especially in the text of the planet orbit presentation. The syntax of these phrases presents a characteristic *Symmetry*, using the same words in (about) the same positions.

4) The critical left edge of the text, i.e. the beginning some of the Lines is preserved (Bitsakis and Jones 2016b). Therefore, in some cases what was written is known at the (non-preserved) end of the previous line (or can be predicted): E.g. on the beginning of left boundary text in Line 25 the phrase 25 [-5-]**ΙΝΟΝΤΟΣ ΚΥΚΛΟΣ**[…… is preserved, i.e.

25 [ΝΟΥ ΦΑ]**ΙΝΟΝΤΟΣ ΚΥΚΛΟΣ**[… (-*nos Phainon*).[29] Without doubt, at the end of the previous line (Line 24), the last three letters should be ΚΡΟ], i.e. the three first letters of the word ΚΡΟΝΟΥ,

24 …………………………………ΚΡΟ]

25 ΝΟΥ ΦΑ]**ΙΝΟΝΤΟΣ ΚΥΚΛΟΣ**[……. (*Kro-nos Phainon*).

At the beginning of Line 17 the phrase **ΦΕΡΕΙΩΝ Η ΜΕΝ ΕΧΟΜΕΝΗ ΤΩΙ ΤΗΣ** is preserved, and at the end of the previous line (Line 16) the last four letters are ΠΕΡΙ, i.e. [ΠΕΡΙ]**ΦΕΡΕΙΩΝ**, (*circu-mferences*). In this case, using the maximum number of letters per line, the remaining missing letters are easily assumed.

5) For the lost words of planet Y description, the words and the same meaning pattern were reproduced as the preserved words of another planet X. E.g. based on the phrase ΚΡΟΝΟΥ ΦΑ]**ΙΝΟΝΤΟΣ ΚΥΚΛΟΣ** (*Kronos Phainon's circle/orbit*) the phrases of the other planets can be reconstructed: [ΔΙΟΣ ΦΑ]**ΕΘΟΝΤΟΣ ΚΥΚΛΟΣ** (*Circle/orbit of Zeus Phaethon*), ΚΥΚΛΟΣ ΤΟ]**Υ ΑΡΕΩΣ ΠΥΡΟΕΝΤΟΣ** (*Circle of Ares Pyroes*) (for Mercury's and Venus' orbits see further below).

---

```
       ...........................................................................................................................
       ...........................................................................................................................
01)  . .[.........................................................................................................................
02)  ΤΑΥΤΗΝΔ[......................................................................................................................
03)  ΔΕΙΔΥΠΟΛΑΒ[ΕΙΝ................................................................................................................
04)  ΥΠΟΔΕΤΟΝΤΩ .[..................................................................................................................
05)  Δ .[- - - - - -]ΟΙΚΑ[..........................................................................................................
06)  Ε[- - - - - - - - -]ΗΙΣΠ[......................................................................................................
07)  [- - - - - - - - - -]ΠΡΟΣ[.....................................................................................................
08)  Ω [- - - - - - - - -]ΜΘΕ .[.....................................................................................................
09)  . [- - - - - - - - - -]ΝΗΡΜΟΣ[.................................................................................................
10)  [- - - - - - - - - -]ΕΠΑΚΡΟΥΔ[ ................................................................................................
11)  [- - - - - - - - - -].ΩΣΜΕΝΩΝ .[...............................................................................................
12)  [- - - - - - - - - - -]ΕΜΕΛΑΝΟΤ .[.............................................................................................
13)  [- - - - - - - - - -]. . . . . .ΛΩΝΓΕΓ[.........................................................................................
14)  [- - - - - - - - - -]. Ε . ΔΥΠΟΛΑΒΕΙ[Ν..........................................................................................
15)  [. .]ΟΥΈ . Ι* ΤΟΣΦΑΙΡΙΟΝΦΕΡΕ .[...............................................................................................
16)  ΠΡΟΕΧΟΝΑΥΤΟΥΓΝΩΜΟΝΙΟΝΣ[.......................................................................................................
17)  ΦΕΡΕΙΩΝΗΜΕΝΕΧΟΜΕΝΗΤΩΙΤΗΣ[......................................................................................................
18)  ΤΟΣΤΟΔΕΔΙΑΥΤΟΥΦΕΡΟΜΕΝ[ΟΝ.......................................................................................................
19)  ΤΗΣΑΦΡΟΔΙΤΗ<Σ>ΦΩΣΦΟΡΟΥ . . .[.................................................................................................
20)  ΤΟΥ [ΦΩ]ΣΦΟΡΟΥΠΕΡΙΦΕΡΕΙΑΝ .[..................................................................................................
21)  ΓΝΩΜΩ[.]ΚΕΙΤΑΙΧΡΥΣΟΥΝΣΦΑΙΡΙΟΝ . .[.............................................................................................
22)  ΗΛΙ[ΟΥ]ΑΚΤΙΝΥΠΕΡΔΕΤΟΝΗΛΙΟΝΕΣΤΙΝΚΥ[............................................................................................
23)  [- - -ΤΟ]ΥΑΡΕΩΣΠΥΡΟΕΝΤΟΣΤΟΔΕΔΙΑΠΟΡΕ[ΥΟΜΕΝΟΝ .................................................................................
24)  [ΔΙΟΣΦΑ]ΕΘΟΝΤΟΣΤΟΔΕΔΙΑΠΟΡΕΥΟΜΕΝΟΝ[............................................................................................
25)  [ΝΟΥΦΑ]ΙΝΟΝΤΟΣΚΥΚΛΟΣΤΟΔΕΣΦΑΙΡΙΟΝΦΛ[...........................................................................................
26)  [- - - - - - -]ΕΡΑΔΕΤΟΥΚΟΣΜΟΥΚΕΙΤΑΙ . . .[.....................................................................................
27)  [- - - - - - - - -]ΜΕΝ[.]ΣΤΟΙΧΕΙΑ ΠΑΡΑΚΕΙΜ[ΕΝΑ................................................................................
28)  [- - - - - - - - - - -] . ΑΥΤΑΤΑΙΣΑΣΠΙΔ[ΙΣΚΑΙΣ.................................................................................
29)  [- - - - - - - - - -]ΠΡΟΕΙΡΗΜΕΝΑ[..............................................................................................
30)  [- - - - - - - - - - - - - -]ΑΣΠ[. .].[........................................................................................
31)  ...........................................................................................................................
```

Table II. The preserved words/letters of the Back Cover plate by Bitsakis and Jones 2016b. The preserved letters in bold and the possible ones in regular. The letters/words are presented in their about original position and form, without gaps (see also **Fig. 1A**). The background color of the letters corresponds to a specific part description/meaning, starting from the Lunar Disc in yellow. An additional letter (resulting the re-position of the previous letters) was recognized by the authors of the present work: in Line 15 (‡) the letter "Υ" instead of letter Ǫ. Also in Line 15 (*) the letter "Ι" (iota) was recognized before the phrase "ΤΟ ΣΦΑΙΡΙΟΝ". In the AMRP tomography of the Fragment B, the letter "Ι" is relative well preserved with its two serifs. Using the position (numbering) of the well preserved letters in the next Line 16, the letter "Ι" should be the 8[th] letter from the line beginning

6) The language used in BCI text by *The Engraver* seems to be the Koine Hellenistic language. The ancient Greek language follows specific rules for language syntax.

7) In order to avoid repeating the same words, *The Engraver* slightly modifies/alters the words and their position, especially about the planets' orbit description:

| |
|---|
| 23 (ΚΥΚΛΟΣ) ΤΟ]**Υ ΑΡΕΩΣ ΠΥΡΟΕΝΤΟΣ ΤΟ ΔΕ ΔΙΑΠΟΡΕ**[ΥΟΜΕΝΟΝ ΣΦΑΙΡΙΟΝ................... |
| 25 (ΚΡΟ)- ΝΟΥ ΦΑ]**ΙΝΟΝΤΟΣ ΚΥΚΛΟ ΤΟ ΔΕ ΣΦΑΙΡΙΟΝ Φ**[ΕΡΟΜΕΝΟΝ[30]........................... |
| *23 circle of Ares Pyroes and the traveling around sphere…...................................................* |
| *25 (Kro)-nos Phainon circle and the sphere travels around…..................................................* |

---

[30] **ΤΟ ΔΕ ΣΦΑΙΡΙΟΝ Φ**[ΕΡΟΜΕΝΟΝ instead of **ΤΟ ΔΕ ΣΦΑΙΡΙΟΝ ΦΛ**[

*The Engraver* uses two words for the same meaning ΔΙΑΠΟΡΕΥΟΜΕΝΟΝ and ΦΕΡΟΜΕΝΟΝ (travelling along/rotating through). In these two - same meaning - sentences, he changed the position/syntax of words ΚΥΚΛΟΣ, ΔΙΑΠΟΡΕΥΟΜΕΝΟΝ and ΣΦΑΙΡΙΟΝ.

8) The Ancient Greek language has a large number of words, with multifactorial meanings. In many cases, a word appears to have different and usually, non-related meanings (see §6.e. regarding the word ΓΝΩΜΟΝΙΟΝ - perpendicular pillar/pointer). E.g. today, a mechanical part of a lathe is named in the Greek language ΚΟΥΚΟΥΒΑΓΙΑ "*the Owl*" (*bird*). If a researcher studies the Greek User's manual of the lathe after 2000 years from today, then, reading the word "*Owl*" in the lathe's manual: either he will try to find "*how the flight of an owl affects the operation of the mechanical lathe*", or he will try to find which characteristic part of the lathe corresponds to the word "*owl*".

Hence, the authors did not depend precisely on the standard translation of the ancient Greek words. In order to translate the meaning of the preserved words on the BCI text, authors tried to correlate these words with the Mechanism's parts.

9) The authors have also studied the preserved text by simultaneous cross-examination/use of their Antikythera Mechanism functional model, in order to correlate the text with the real operation of the Mechanism. This procedure offers the advantage of the *interactive communication* between the instrument and the corresponding text, allowing the comparison of the text references and parts' operation from *the Manufacturer's perspective*. The authors based the design and construction of the mechanical parts in their model, mostly on the visual photographs of the Fragments, on the AMRP X-Ray tomographies and partially on the published bibliography, and assembled the Mechanism parts. By observing and operating the outer parts which were located/visible on the Front face, the authors corresponded the bronze reconstructed parts to the referred parts on the text.

10) Furthermore, the study of the User's Manual parts description, could be useful for the reconstruction of a part which is not preserved on the Mechanism's fragments (such as the Lunar Disc pointer, the Golden Sphere, the solar pointer etc.).

11) The User's Manual of the Antikythera Mechanism is a technical text, presenting the Mechanism's parts and their operation. For the parts' presentation, it appears that the text follows the (necessary) pattern *Definition/Description, Position and Operation* (if this exists) of a part.

E.g. for the *Definition* of a part, *The Engraver* uses the characteristic ancient Greek word ΕΣΤΙΝ (*there is*/exists):

22 **…ΥΠΕΡ ΔΕ ΤΟΝ ΗΛΙΟΝ ΕΣΤΙΝ ΚΥ**[ΚΛΟΣ… (*beyond Sun, there is the circle…*), i.e. *The Engraver Defines/Places* a circle which is located, beyond the Sun.

*Definition*: there is a circle,

*Position* (where is it located?): the circle is beyond Sun's (circle).

In Line (20)-21 [ΕΙΣ] **ΓΝΩΜΩ**[Ν] **ΚΕΙΤΑΙ.**[31] (*A gnomon is adapted on…*),

- *Definition:* a gnomon is adapted (somewhere). The *Position* and the *Operation* of **ΓΝΩΜΩ**[Ν] is missing.

In Line 21 **ΧΡΥΣΟΥΝ ΣΦΑΙΡΙΟΝ** [ΕΠΙ ΓΝΩΜΟΝΑ ΕΣΤΙΝ… (The Golden Sphere is adapted on the gnomon).

- *Definition:* **ΧΡΥΣΟΥΝ ΣΦΑΙΡΙΟΝ** (*there is a sphere which has a golden color-Sun*).

- *Position (where is it located?):* [ΕΠΙ ΓΝΩΜΩΝΑ (*the Golden Sphere is located/stabilized on the gnomon*).

- *Operation* (*what is the operation of the golden sphere? Is it traveling along? Is it measuring? …*).

At the same time the *Operation* of **ΓΝΩΜΩ**[Ν] is revealed: The **ΓΝΩΜΩ**[Ν] exists in order the ΧΡΥΣΟΥΝ ΣΦΑΙΡΙΟΝ-Sun to be stabilized on the **ΓΝΩΜΩ**[Ν].

The operation of the Golden Sphere-Sun is missing. However, during Hellenistic Astronomy it was claimed that the *Sun* rotated around the Earth in one Tropical year, named ΕΝΙΑΥΤΟΣ.[32] Geminus (Spandagos 2002) writes *"Ὁ δὲ ἥλιος ἐνιαυτῷ διαπορεύεται τὸν ζωδιακὸν κύκλον"* (*The Sun travels along the Zodiac Circle in Eniautos/one Tropical year*). Therefore, the operation of the Golden Sphere-Sun could be its rotation around the Center/Earth in one Tropical year. A probable reconstructed text could be … ΕΙΣ] **ΓΝΩΜΩ**[Ν] **ΚΕΙΤΑΙ. ΧΡΥΣΟΥΝ ΣΦΑΙΡΙΟΝ** [ΕΠΙ ΓΝΩΜΟΝΑ ΕΣΤΙΝ, ΕΝ ΕΝΙΑΥΤΩ ΔΙΑΠΟΡΕΥΟΜΕΝΟΝ (*…a pillar is stabilized. The Golden Sphere is adapted on the pillar and rotated in Eniautos/one tropical year*).

## 5. Analyzing the preserved text of the BCI Part-1
### 5.a. Stylistic information for the text presentation

The BCI Part-1 text analysis was based on the revised published text, "*The Back Cover Inscription*", by Bitsakis and Jones 2016b.

In the present work, the Leiden system symbols are used and additionally:

a) All of the preserved original words/letters are written in capital Greek letters, as the original ancient text and are noted in bold.

---

[31] For the line 21 see also comments in Bitsakis and Jones 2016b, page 243. If their suggestion is correct, then the phrase could be ΕΠΙ ΤΩ] **ΓΝΩΜΩ**[Ν]<Ι> **ΚΕΙΤΑΙ ΧΡΥΣΟΥΝ ΣΦΑΙΡΙΟΝ**. See page 33 of the present work.

[32] Ptolemy in Heiberg 1898; Danezis and Theodosiou 1994; Geminus in Spandagos 2002.

b) The preserved original words/letters are located on their corresponding geometrical position (as much as possible).

c) By the first step analysis, the maximum possible number of letters per line was calculated into 82 letters/line and by adding +4 letters the final maximum possible number of letters per line is 86. Hence, the beginning-left boundary of the text is preserved in several lines, the maximum number of letters can (theoretically) be defined as the right boundary of the text. Naturally, when *The Engraver* engraved/wrote the text on the Back cover plate, he was not defining a constant number of letters/line. His text was "free" of such restrictions (after the analysis of the BCI Part-1 and during the final adaptation of the reconstructed text, the arbitrary constant number of letters per line will be not valid).

d) For the text analysis of a specific phrase and meaning, the corresponding text of interest is presented in the usual font size (11) and the rest preserved/reconstructed/non-related text is presented by a shorter font size (9) or it is replaced by dots.

**5.b. The preserved Back Cover Inscription-Part 1, presentation**

    *The Engraver* uses a specific way to write the User's Manual of his creation. He presents the Front face parts starting with the geometrical center of the Front plate and then, radially presents the other parts. For each *operational part's* presentation he defines and presents a part, afterward refers to its position, and finally describes its operation, following the pattern **Definition/Description, Positioning, Operation**. Generally, he follows the same procedure as someone would follow today to write the User's manual for a product.

    *The Engraver* probably starts the *operational parts' presentation* from the Earth-*Center of Cosmos* and continues the presentation in radial distribution from the center outwards: he *Describes* and *Places* the Lunar Disc-Moon and its parts. After that, he presents the planets, starting from the two inferior planets-after Earth (between Lunar and Solar orbit), and also refers to their theophoric names ΕΡΜΟΥ ΣΤΙΛΒΟΝ]**ΤΟΣ** (*Hermes Stilbon*, Mercury), **ΑΦΡΟΔΙΤΗ**<Σ> **ΦΩΣΦΟΡΟΥ** (*Aphrodites Phosphoros*, Venus). After planet Sun **ΗΛΙΟΣ** and its parts, he presents the three superior planets **ΑΡΕΩΣ ΠΥΡΟΕΝΤΟΣ** (*Ares Pyroeis*, Mars), [ΔΙΟΣ ΦΑ]**ΕΘΟΝΤΟΣ** (*Dias-Zeus Phaethon*, Jupiter) and ΚΡΟ[ΝΟΥ ΦΑ]**ΙΝΟΝΤΟΣ** (*Kronos Phainon*, Saturn).[33]

---

[33] The theophoric names of the planets are also referred by Pseudo-Aristotle 1831 (On the Cosmos); Geminus (in Manitius 1880/Spandagos 2002); Dorotheus of Sidon in Pingree 1976, in Papyrus Oxyrhynchus POxy. 307 (Neugebauer and Van Hoesen 1987); Cleomedes in Spandagos 2002; Bowen and Todd 2004.

After planet Saturn, *The Engraver* refers to the word ΚΟΣΜΟΥ (*Cosmos*, Line 26) and the words **ΣΤΟΙΧΕΙΑ** (*Letters*) and **ΑΣΠΙΔ[**ΙΣΚΑΙΣ (*Small Shields*) follow.

More specifically,

In Line 7 the letters [-10-]**ΠΡΟΣ[**…. are preserved, which could be the phrase **ΠΡΟΣ**ΘΙΑ ΟΨΙΣ (*Front face*).

The preserved word in Line 9 **ΗΡΜΟΣ**[ΘΑΙ] (*fitted on*…, see Bitsakis and Jones 2016b) and Line 10 **ΕΠ ΑΚΡΟΥ** (*on the edge of*…) related to the Lunar Disc positioning (which is located at the geometrical center of the Front face): its adaptation and stabilization on the edge of axis b$_{in}$.[34] The letters in Line 11, **ΩΣΜΕΝΩΝ** could be part of one word or two/three separated words: **ΩΣ ΜΕΝ ΩΝ** (*as he also is*) or **ΩΣ ΜΕΝΩΝ** ("*As he is stable in a state for a long time*" or "*As he is stable in a situation*" or "*As he is constant*" or "*As he stays on*" or "*As he lives*" or "*He continuously…*" or [.….]**ΩΣ ΜΕΝΩΝ**.[35]

The Lunar Disc and its parts are presented in Lines 9-16. The words in Line 12 **ΜΕΛΑΝ** (*black*) and Line 15 **ΤΟ ΣΦΑΙΡΙΟΝ ΦΕΡΕ**[ΤΑΙ] (*the little sphere is rotated/traveling through*) were correlated to an important part of the Lunar Disc, the Lunar phases sphere. Some parts of the Lunar Phases Sphere, the crown gear-z and its axis are visible on AMRP tomographies of Fragment C. Line 16 … **ΠΡΟΕΧΟΝ ΑΥΤΟΥ ΓΝΩΜΟΝΙΟΝ Σ**[… (…*little pointer protrudes from it*…) refers to the Lunar Disc pointer, which protrudes from the Lunar Disc and travels around during the Lunar Disc rotation.

Afterwards, in Line 17 the phrase ΠΕΡΙ**ΦΕΡΕΙΩΝ** H MEN **ΕΧΟΜΕΝΗ** (*Circumferences, the one is close to*…) is preserved. Usually, in the Ancient Greek language, when the words **H ΜΕΝ…** (*the first*…) exist in a phrase, a second phrase follows including the words **H ΔΕ…** (*the next, the following*…). Therefore, the word ΠΕΡΙ**ΦΕΡΕΙΩΝ** (in plural i.e. at least two) concerns the H ΜΕΝ ΠΕΡΙΦΕΡΕΙΑ 1 ΚΑΙ H ΔΕ ΠΕΡΙΦΕΡΕΙΑ 2 (*the first circumference and then the following circumference*).

In Line 20 the words [ΦΩ]**ΣΦΟΡΟΥ ΠΕΡΙΦΕΡΕΙΑΝ** are preserved. Therefore, the word ΠΕΡΙ**ΦΕΡΕΙΩΝ** (Line 17) is correlated to the H ΜΕΝ ΠΕΡΙΦΕΡΕΙΑ ΤΟΥ ΕΡΜΟΥ ΚΑΙ H ΔΕ ΠΕΡΙΦΕΡΕΙΑ ΤΗΣ ΑΦΡΟΔΙΤΗΣ (*firstly the orbit of Hermes/Mercury and follows the orbit of Aphrodite/Venus*). For the inferior planets' orbits, *The Engraver* uses the word **ΠΕΡΙΦΕΡΕΙΑ** (*circumference*) and for the superior planets' orbits the word **ΚΥΚΛΟΣ** (circle, see below). Euclides in *Elements of Geometry*[36] writes "*Κύκλος ἐστί σχῆμα ἐπίπεδον ὑπό μιᾶς γραμμῆς περιεχόμενον [ἡ καλεῖται περιφέρεια], πρός ἥν ἀφ'ένός σημείου τῶν*

---

[34] Wright 2006; Freeth et al., 2006; Carman and Di Cocco, 2016; Voulgaris et al., 2018b.
[35] Liddell and Scott's lexicon 2007.
[36] Euclidis in Heiberg 1883–1885.

*ἐντὸς τοῦ σχήματος κειμένων πᾶσαι αἱ προσπίπτουσαι εὐθεῖαι [πρὸς τὴν τοῦ κύκλου περιφέρειαν] ἴσαι ἀλλήλαις εἰσίν."* [The circle is a plane scheme which is defined by a line (which is called circumference), and from a point located inside the scheme, all of the incident lines (directed towards to the circumference) are equal to each other].

The planet Venus is referred to twice: in Line 19 (planet + theophoric name) and in Line 20 (only by its theophoric name).

After Venus, follows ΗΛΙΟΣ (*Helios-Sun*). *The Engraver* writes that Sun has a **ΓΝΩΜΩ**[Ν] (*perpendicular small pillar* according to the present work, see §6.e.)[37], **ΧΡΥΣΟΥΝ ΣΦΑΙΡΙΟΝ** (*golden little sphere*) and **ΗΛΙΟΥ ΑΚΤΙΝ** (*ray of Sun-the real pointer*, see the analysis on §6.f.).

After the Sun, follows the reference of the three superior planets, with their corresponding theophoric names and their orbits **ΑΡΕΩΣ ΠΥΡΟΕΝΤΟΣ** (ΚΥΚΛΟΣ) (*Ares Pyroes-Mars' orbit*), [ΔΙΟΣ ΦΑ]**ΕΘΟΝΤΟΣ** (ΚΥΚΛΟΣ) (*Zeus Phaethon-Jupiter's orbit*) and ΚΡΟ[ΝΟΥ ΦΑ]**ΙΝΟΝΤΟΣ ΚΥΚΛΟΣ** (*Kronos-Saturn Phainon's orbit*).

In Line 26 …]**ΕΡΑ ΔΕ ΤΟΥ ΚΟΣΜΟΥ, ΚΕΙΤΑΙ** probably Π]**ΕΡΑ ΔΕ ΤΟΥ ΚΟΣΜΟΥ, ΚΕΙΤΑΙ** (*beyond Cosmos, there is…*).[38] This phrase is correlated with the Cosmos which consists of the Earth-Center of Cosmos, Moon, and the other six planets. At this point, *The Writer* describes what exists beyond Cosmos:

According to Hellenistic Astronomy, after the planets follows Η ΣΦΑΙΡΑ ΤΩΝ ΑΠΛΑΝΩΝ ΑΣΤΕΡΩΝ (*the Celestial Sphere with the fixed stars*).[39] Moon, Sun and planets travel around the Earth-*The Center of Cosmos,* and are projected on the Zodiac constellations-the Ecliptic zone/Zodiac cycle.

On the Mechanism, after the presentation of Cosmos, follows the "*Sky*", the Zodiac month ring, divided into 12 (unequal) parts-Dodecatemoria, the Zodiac constellations/Zodiac months.[40] Right after follows the Egyptian calendar ring, which is an additional calendar in the Mechanism, which was also in use during the Hellenistic era.[41]

| Part description on the preserved BCI Part-1 text | Corresponding (rounded per line) Number of lines | Percentage of 17 selected Lines 9-25 | Percentage of 30 preserved Lines 1-30 |
|---|---|---|---|
| *Lunar Disc* | Lines 9-16 (total 5-8) | 29%-47% | 16%-26% |
| *Mercury* | Line 18 (total 1) | 6% | 3% |

---

[37] In Bitsakis and Jones 2016b the word ΓΝΩΜΟΝΙΟΝ is translated as a pointer.
[38] For the words ΠΕΡΑ and ΠΕΡΑΝ see LSJ lexicon.
[39] Geminus (in Manitius 1880/Spandagos 2002); Cleomedes in Spandagos 2002.
[40] Voulgaris et al., 2018a.
[41] Ptolemy in Heiberg 1898; Toomer 1984

| | | | |
|---|---|---|---|
| *Venus* | Lines 19-20 (total 2) | 12% | 6% |
| *Sun* | Lines 21-22 (total 2) | 12% | 6% |
| *Mars* | Line 23 (total 1) | 6% | 3% |
| *Jupiter* | Line 24 (total 1) | 6% | 3% |
| *Saturn* | Line 25 (total 1) | 6% | 3% |

Table III. The total number of lines, correlated to the Mechanism's Front Face parts, measured on the BCI Part-1 preserved text and their corresponding percentages. For the lines correlated to Mars, Jupiter and Saturn, their percentages are definite (one line per planet)

At the end of the BCI Part-1, the words **ΣΤΟΙΧΕΙΑ** (*letters*) in Line 27 and the phrase **ΤΑΙΣ ΑΣΠΙΔ**[ΙΣΚΑΙΣ (on the small shields, see §6.h.) in Lines 28 and 30 are preserved, correlated to the Parapegma events and its position presentation.

Preliminary statistics regarding the lines per referred parts are presented on **Table III**.

# 6. RECONSTRUCTING THE BACK COVER INSCRIPTION PART-1
## 6.a.1. The ΚΥΚΛΟΙ, Circles/orbits of the superior planets (Lines 22-25)

*The Engraver* refers the word ΚΥΚΛΟΣ on the text area for the planets presentation:

| |
|---|
| 22 **ΗΛΙΟΥ ΑΚΤΙΝ ΥΠΕΡ ΔΕ ΤΟΝ ΗΛΙΟΝ ΕΣΤΙΝ ΚΥ**[ΚΛΟΣ - - - - - - - - - - 51 - - - - - - - - - - - - - - - - - - -] |
| 23 …………………………………………………………………………………………………………………………………………………………. |
| 24 …………………………………………………………………………………………………………………………………………… ΚΡΟ] |
| 25 **ΝΟΥ ΦΑ**]**ΙΝΟΝΤΟΣ ΚΥΚΛΟΣ ΤΟ ΔΕ ΣΦΑΙΡΙΟΝ ΦΛ** [ - - - - - - - - - - - - - 51 - - - - - - - - - - - - - - - -] |
| 22 *Beyond Sun, there is (the) circle*…………………………………………………………………………………… |
| 25 *Saturn Phainon circle*……………………………………………………………………………………………… |

The word ΚΥΚΛΟΣ concerns the orbit of planet Saturn (Line 25) and also the orbit of a planet which is located after/beyond the Sun's orbit, i.e. the orbit of Mars. Therefore, the word ΚΥΚΛΟΣ should refer to each of the superior planets (for the inferior planets see next paragraph).

## 6.a.2. The ΠΕΡΙΦΕΡΕΙΑ, Circumferences/orbits of the inferior planets (Lines 16, 17 and 20)

For the inferior planet Venus's orbit on Line 20, *The Engraver* refers the word ΠΕΡΙΦΕΡΕΙΑ (*circumference*):

| |
|---|
| 16 **ΠΡΟΕΧΟΝ ΑΥΤΟΥ ΓΝΩΜΟΝΙΟΝ Σ**[- - - - - - - - - - - - - - - - - - 60 - - - - - - - - - - - - - - - - - - - ΠΕΡΙ |
| 17 **ΦΕΡΕΙΩΝ Η ΜΕΝ ΕΧΟΜΕΝΗ ΤΩΙ ΤΗΣ**[- - - - - - - - - - - - - - - - - -63- - - - - - - - - - - - - - - - - - - - ] |
| 18 …………………………………………………………………………………………………………………………………………………. |
| 19 …………………………………………………………………………………………………………………………………………………… |
| 20 **ΤΟΥ** [**ΦΩ**]**ΣΦΟΡΟΥ ΠΕΡΙΦΕΡΕΙΑΝ** [- - - - - - - - - - - - - - - - - - - - -62- - - - - - - - - - - - - - - - - - - - ] |
| 16 *pointer projecting from it* ………………………………………………………………………………………*circu-* |
| 17 *mferences, the one* …………………………………………………………………………………………………… |
| 18 …………………………………………………………………………………………………………………………………………… |
| 19…………………………………………………………………………………………………………………………………………… |
| 20 *Phosphoros' circumference/orbit*………………………………………………………………………………… |

In the previous lines, Line (16)/17 the word [ΠΕΡΙ]**ΦΕΡΕΙΩΝ** is referred to in the plural (*circumferences*), i.e. at least two circumferences. This text's area is correlated to the two inferior planets, Mercury and Venus. *The Engraver* uses the word ΠΕΡΙΦΕΡΕΙΑ (*circumference*) when he refers to the orbits of the two inferior planets, which were located between the orbit of the Moon and the Sun (according to Hellenistic Astronomy).

Thus, there should be an equal *Definition* about Mercury's orbit in Lines 17-18:

| |
|---|
| 16 **ΠΡΟΕΧΟΝ ΑΥΤΟΥ ΓΝΩΜΟΝΙΟΝ Σ**[- - - - - - - - - - - - - - - - - - - -60- - - - - - - - - - - - - - - - - - - ΠΕΡΙ- |
| 17 **ΦΕΡΕΙΩΝ Η ΜΕΝ ΕΧΟΜΕΝΗ ΤΩΙ ΤΗΣ**[- - - - - 36- - - - - - ΕΣΤΙΝ ΠΕΡΙΦΕΡΕΙΑ ΕΡΜΟΥ ΣΤΙΛΒΟΝ- |
| 18 **ΤΟΣ ΤΟ ΔΕ ΔΙ ΑΥΤΟΥ ΦΕΡΟΜΕΝ**[- - - - - - - - - - - - - - - 65- - - - - - - - - - - - - - - - - - - - - - - - - - ] |
| 19 ..................................................................................................................................................... |
| 20 **ΤΟΥ** [ΦΩ]**ΣΦΟΡΟΥ ΠΕΡΙΦΕΡΕΙΑΝ** [- - - - - - - - - - - - - - - - - - -64- - - - - - - - - - - - - - - - - - - - - - - ] |
| *16 projecting from it ...........................................................................................................circu-* |
| *17 mferences, the one ...........................................................................circumference of Mercury Stil-* |
| *18 bon .........................................................................................................................................* |
| *19 ...........................................................................................................................................................* |
| *20 Phosphoros' circumference/orbit.......................................................................................................* |

Heron Alexandrinus in *Definitiones geometricae* (Hasenbalg 1826) refers for the word ΠΕΡΙΦΕΡΕΙΑ: Ὡς γὰρ κύκλου ὅρος ἐστίν ἡ περιφέρεια. *(The circumference is the boundary of the circle)*.[42]

### 6.a.3 The word ΥΠΕΡ (beyond/next/after) and the Repetition for the planet names

The word ΥΠΕΡ (beyond), is used by *The Writer* in order to define the position of a part B which is located after the position of the part A:

| |
|---|
| 22 **ΗΛΙ**[ΟΥ] **ΑΚΤΙΝ ΥΠΕΡ ΔΕ ΤΟΝ ΗΛΙΟΝ ΕΣΤΙΝ ΚΥ**[ΚΛΟΣ - - - - - - - - - - - - -51- - - - - - - - - - - - - - - - - ] |
| 22 *...beyond/after the Sun there is a circle/orbit*, (i.e. the orbit of Mars) |

Therefore, the word **ΥΠΕΡ** (*After/Beyond*) should appear for each of the planets, following the phrase pattern:

*Beyond the planet A .........................................*

*Beyond the planet B .........................................*

*Beyond the planet C..................................... etc.*

Two references of Venus are well-preserved:

| |
|---|
| 19 **ΤΗΣ ΑΦΡΟΔΙΤΗ<Σ> ΦΩΣΦΟΡΟΥ** [- - - - - - - - - - - - - - - - - - 67- - - - - - - - - - - - - - - ·- - - - - - - - - - - -] |

---

| 20 **ΤΟΥ [ΦΩ]ΣΦΟΡΟΥ ΠΕΡΙΦΕΡΕΙΑΝ** [- - - - - - - - - - - - - - - - - - -64- - - - - - - - - - - - - - - - - - - -] |
| :--- |
| *19 Aprodites' Phosphoros…………………………………………………………………………………………………………………* |
| *20 Phosphoros' circumference/orbit………………………………………………………………………………………………* |

- (*The Engraver* omitted by mistake the last letter of **ΑΦΡΟΔΙΤΗ**<Σ>).[43]

The second time Venus is referred to, *The Engraver* uses only its theophoric name *Phosphoros*. Generally, *The Engraver* avoids repeating the same or non-necessary words.

The double reference to Venus can be well justified by the following interpretation:

Firstly, Venus (planet's name + theophoric name) is referred to as the *Definition* of the planet and the *Position* of its orbit:

| 19 **ΤΗΣ ΑΦΡΟΔΙΤΗ**<Σ> **ΦΩΣΦΟΡΟΥ** ΠΕΡΙΦΕΡΕΙΑ ΕΣΤΙΝ[- - - - - - - - - - - - - 51- - - - - - - - - - - - - - -] |
| :--- |
| 20 **ΤΟΥ [ΦΩ]ΣΦΟΡΟΥ ΠΕΡΙΦΕΡΕΙΑΝ** [- - - - - - - - - - - - - - - - - - - 64 - - - - - - - - - - - - - - - - - - - - -] |
| *19 There is Aphrodites' Phosphoros circumference…………………………………………………………………………..* |
| *20 Phosphoros' circumference………………………………………………………………………………………………………* |

Secondly, the theophoric name of Venus is used to *Present/Define* the next planet and its *Position* referring the word ΥΠΕΡ ΔΕ ΤΟΥ ΦΩΣΦΟΡΟΥ (*beyond Phosphoros there is the Sun's Circumference/Circle*), as the word ΥΠΕΡ (*beyond*) exists in Line 22 **ΥΠΕΡ ΔΕ ΤΟΝ ΗΛΙΟΝ ΕΣΤΙΝ ΚΥ**[ΚΛΟΣ:

Correlating the Lines 19, 20 and 22:

| 19 **ΤΗΣ ΑΦΡΟΔΙΤΗ**<Σ> **ΦΩΣΦΟΡΟΥ** ΠΕΡΙΦΕΡΕΙΑ ΕΣΤΙΝ [- - - - - - - - - - - -45- - - - - - - - - -  ΥΠΕΡ ΔΕ] |
| :--- |
| 20 **ΤΟΥ [ΦΩ]ΣΦΟΡΟΥ, ΠΕΡΙΦΕΡΕΙΑΝ** ΗΛΙΟΥ ΕΣΤΙΝ [- - - - - - - - - - - - - - - - -54- - - - - - - - - - - - - - -] |
| *19 There is Aphrodites' Phosphoros circumference…………………………………………………………………………..* |
| *20 beyond Phosphoros' circumference/orbit, there is the Sun's orbit…………………………………………………..* |

## 6.a.4. A preliminary text reconstruction of Lines 17-25

Based on the observations and analysis of 6.a.1.-6.a.3., in Lines 17-25, the names of the planets should be referred twice:

1) *planet + its theophoric name*
2) only by its *theophoric name*.

The words ΥΠΕΡ and ΠΕΡΙΦΕΡΕΙΑ/ΚΥΚΛΟΣ should also appear according to the following pattern:

- *There is (ΕΣΤΙΝ) the orbit of A (planet+theophoric)* ……………………………………….

- *Beyond (ΥΠΕΡ) the A (theophoric),*
*There is (ΕΣΤΙΝ) the orbit of B (planet+theophoric)* …………………………………………..

- *Beyond the B (theophoric),*

---

*There is the orbit of C (planet+theophoric) …………………………………..*

*- Beyond the C (theophoric),*
*There is the orbit of D (planet+theophoric) ………………………………………………*

*i.e.:*

- ΥΠΕΡ ΔΕ ΤΗΣ ΣΕΛΗΝΗΣ,
ΕΣΤΙΝ ΠΕΡΙΦΕΡΕΙΑ ΤΟΥ ΕΡΜΟΥ ΣΤΙΛΒΟΝ]**ΤΟΣ**………………………..
(*beyond the Moon there is the circumference of Hermes Stilbon…*)

- ΥΠΕΡ ΔΕ ΤΟΥ ΣΤΙΛΒΟΝΤΟΣ,
ΕΣΤΙΝ ΠΕΡΙΦΕΡΕΙΑ] **ΤΗΣ ΑΦΡΟΔΙΤΗΣ ΦΩΣΦΟΡΟΥ**………………
(*beyond Stilbon there is the circumference of Aphrodites Phosphoros…*)

- ΥΠΕΡ **ΤΟΥ [ΦΩ]ΣΦΟΡΟΥ,**
**ΠΕΡΙΦΕΡΕΙΑΝ** [ΗΛΙΟΥ ΕΣΤΙΝ…………………………………..
(*beyond the circumference of Phosphoros, there is the circle of Sun…*)
(see comments in footnotes 44)

- **ΥΠΕΡ ΔΕ ΤΟΝ ΗΛΙΟΝ,**
**ΕΣΤΙΝ ΚΥ**[ΚΛΟΣ ΑΡΕΩΣ ΠΥΡΟΕΝΤΟΣ……………………
(*beyond the Sun there is the circle of Ares Pyroes…*)

- ΥΠΕΡ ΔΕ ΤΟΥ ΠΥΡΟΕΝΤΟΣ,
ΕΣΤΙΝ ΚΥΚΛΟΣ ΔΙΟΣ ΦΑ]**ΕΘΟΝΤΟΣ**………………
(*beyond Pyroes there is the circle of Jupiter Phaethon…*)

- ΥΠΕΡ ΔΕ ΤΟΥ ΦΑΕΘΟΝΤΟΣ,
ΕΣΤΙΝ ΚΥΚΛΟΣ ΤΟΥ ΚΡΟ[ΝΟΥ ΦΑ]**ΙΝΟΝΤΟΣ**………………
(*beyond Phaethon there is the circle of Saturn Phainon…*).[44]

Applying the reconstructed text and taking into account the original positions of the preserved words/phrases as well as the geometrical text limits (the preserved text left boundary and the maximum amount of 86 letters/line), the reconstructed text located on its original position should be:

| |
|---|
| 17 **ΦΕΡΕΙΩΝ Η ΜΕΝ ΕΧΟΜΕΝΗ ΤΩΙ ΤΗΣ** [- - - - - - - 33 - - - - - - ΕΣΤΙΝ ΠΕΡΙΦΕΡΕΙΑ ΤΟΥ ΕΡΜΟΥ ΣΤΙΛΒΟΝ] |
| 18 **ΤΟΣ ΤΟ ΔΕ ΔΙ ΑΥΤΟΥ ΦΕΡΟΜΕΝ**[- - - - - - - - - - - - 46 - - - - - - --- - ΥΠΕΡ ΔΕ ΤΟΥ ΣΤΙΛΒΟΝΤΟΣ, Η] |
| 19 **ΤΗΣ ΑΦΡΟΔΙΤΗ<Σ> ΦΩΣΦΟΡΟΥ[** ΠΕΡΙΦΕΡΕΙΑ ΕΣΤΙΝ [- - - - - - - - - - - - - 44 - - - - - - - - - ΥΠΕΡ ΔΕ] |

---

[44] Keeping this specific pattern, the phrase ΥΠΕΡ **ΤΟΥ ΦΩΣΦΟΡΟΥ ΠΕΡΙΦΕΡΕΙΑΝ**… should be ΥΠΕΡ **ΤΟΥ** ΦΩΣΦΟΡΟΥ (only theophoric), **ΠΕΡΙΦΕΡΕΙΑΝ** ΗΛΙΟΥ ΕΣΤΙΝ…
(*beyond Phosphoros, there is the Circumference of Sun*), but the correct word should be **ΠΕΡΙΦΕΡΕΙΑ** (without "Ν"). If the letter Ν was engraved by mistake, then the Sun's orbit is named ΠΕΡΙΦΕΡΕΙΑ. Otherwise, the above phrase should be ΥΠΕΡ ΤΗΝ **ΤΟΥ** ΦΩΣΦΟΡΟΥ **ΠΕΡΙΦΕΡΕΙΑΝ**, ΕΣΤΙΝ ΚΥΚΛΟΣ/ΠΕΡΙΦΕΡΕΙΑ ΤΟΥ ΗΛΙΟΥ… (*beyond Phosphoros' circumference, there is the circumference/circle of Sun*). The authors believe that the letter "Ν" was mistakenly written by the Engraver. Thus, after ΥΠΕΡ ΤΟΥ ΦΩΣΦΟΡΟΥ a comma is considered.

| |
|---|
| 20 **ΤΟΥ ΦΩΣΦΟΡΟΥ, ΠΕΡΙΦΕΡΕΙΑΝ** [ΗΛΙΟΥ ΕΣΤΙΝ - - - - - - - - - - - - - - - - - - - 53- - - - - - - - - - - - - -] |
| 21 …………………………………………………………………………………………………………………………………… |
| 22 ΗΛΙΟΥ ΑΚΤΙΝ **ΥΠΕΡ ΔΕ ΤΟΝ ΗΛΙΟΝ, ΕΣΤΙΝ ΚΥ**]- - - - - - - - 29 - - - - - - ΥΠΕΡ ΔΕ ΤΟΥ ΗΛΙΟΥ, ΚΥΚΛΟΣ ΕΣ] |
| 23 ΤΙΝ ΤΟ]**Υ ΑΡΕΩΣ ΠΥΡΟΕΝΤΟΣ ΤΟ ΔΕ ΔΙΑΠΟΡΕ**[- -25 - - ΥΠΕΡ ΔΕ ΤΟΥ ΠΥΡΟΕΝΤΟΣ, ΚΥΚΛΟΣ ΕΣΤΙΝ ΤΟΥ] |
| 24 ΔΙΟΣ ΦΑ]**ΕΘΟΝΤΟΣ ΤΟ ΔΕ ΔΙΑΠΟΡΕΥΟΜΕΝΟΝ**[-36- ΥΠΕΡ ΔΕ ΤΟΥ ΦΑΕΘΟΝΤΟΣ ΕΣΤΙΝ ΤΟΥ ΚΡΟ] |
| 25 ΝΟΥ ΦΑ]**ΙΝΟΝΤΟΣ ΚΥΚΛΟΣ ΤΟ ΔΕ ΣΦΑΙΡΙΟΝ ΦΛ** [- - - - - - - - - - - - 54- - - - - - - - - - - - - - - - - - ] |
| *17 ………………………………………………………………………there is the circumference of Hermes Stil-* <br> *18 bon ………………………………………………………………………………beyond Stilbon* <br> *19 there is the circumference of Aprodites Phosphoros…………………………………………… beyond* <br> *20 Phosphoros circumference there is the circle of Sun………………………………………………………….* <br> *21 ……………………………………………………………………………………………………………………………………….* <br> *22 …………………………………………………………………………beyond Sun there is the circle* <br> *23 of Ares Pyroes…………………………………………………………beyond Ares there is the circle of* <br> *24 Zeus Phaethon…………………………………………………………beyond Phaethon there is* <br> *25 the circle of Kronos Phainon…………………………………………………………………………………* |

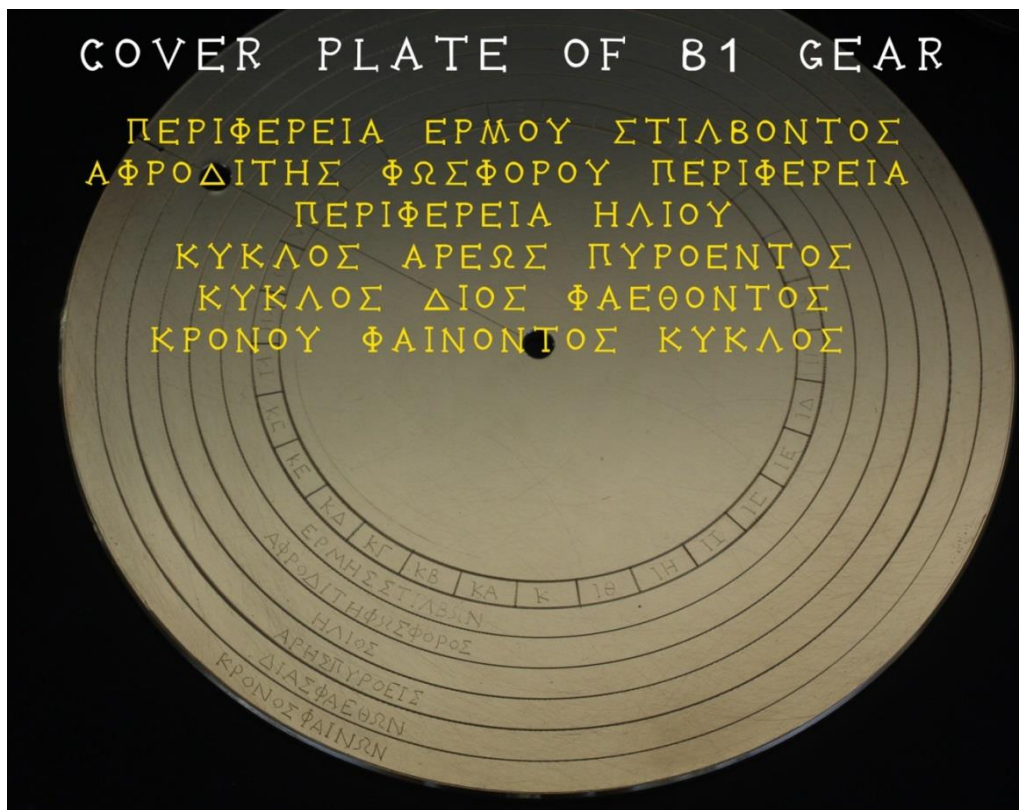

Fig. 4. The ΠΕΡΙΦΕΡΕΙΕΣ (circumferences/orbits) of Mercury, Venus, Sun and the ΚΥΚΛΟΙ (circles/orbits) of Mars, Jupiter and Saturn engraved as homocentric circles on the bronze plate which covers the Front central area of the Mechanism. The plate is stabilized on the central annual gear b1, through the four pillars (one totally preserved and the other three partially preserved, Freeth and Jones 2012). At the same time, this plate hides the internal mechanical system of the Mechanism (gears and axes). Bronze front central plate constructed by the authors (here is presented before the final material process and assembly)

For Lines 22, and 23, see also §6.f.

In the two preserved sentences concerning the presentation of the inferior planet Mercury (Lines 17-18) and the superior planet Jupiter (Lines 23-24), a Symmetry in the syntax and the meaning is revealed:

| |
|---|
| 17-18 ΕΣΤΙΝ ΠΕΡΙΦΕΡΕΙΑ ΕΡΜΟΥ ΣΤΙΛΒΟΝ]**ΤΟΣ ΤΟ ΔΕ ΔΙ ΑΥΤΟΥ ΦΕΡΟΜΕΝ**[ΟΝ... |
| 23-24 ΕΣΤΙΝ ΚΥΚΛΟΣ ΔΙΟΣ ΦΑ]**ΕΘΟΝΤΟΣ ΤΟ ΔΕ ΔΙΑΠΟΡΕΥΟΜΕΝΟΝ**... |
| *17-18 There is the circumference of Mercury Stilbon and the traveling through it.......* |
| *23-24 There is the circle of Zeus Phaethon and the traveling through...* |

The bronze reconstruction of ΠΕΡΙΦΕΡΕΙΕΣ and ΚΥΚΛΟΙ according to the Lines 17-25, is presented in **Fig. 4**.

### 6.b. The missing text of the Introduction (Lines 1-6)

*The Engraver* has probably written the *Introduction*, a *General Description*, *Presenting*, *Naming* and *Reasoning* (Purpose) of his construction and probably his name, in (or before) Lines 1-6. The text is difficult to reconstruct because it is a general description and not a clear technical text. The real name of the Antikythera Mechanism is not known.

A reference to the Back Cover copper plate in which the parts of the Mechanism are presented, could exist.

In Line 1, no letter is preserved. In Line 2, there could be a possible reference to the copper/bronze plate:[45]

| |
|---|
| 01 - - - - - - - - - - - - - - - - - - - - - - - - - - - - - 83 - - - - - - - - - - - - - -··- - - - - - - - - - - - - - - - - - - ΕΙΣ |
| 02 **ΤΑΥΤΗΝ Δ**[Ε ΤΗΝ ΕΝΕΠΙΓΡΑΦΟΝ ΧΑΛΚΗΝ ΠΛΑΚΑ ΕΙΣΙΝ ΑΝΑΓΡΑΦΟΜΕΝΑΙ ΑΙ - - - - - - 33 - - - - - ] |
| *01 ..........................................................................................................................................on* |
| *02 this copper engraved plate, there are written.........................................................................* |

In Line 3 the phrase **ΔΕΙ Δ ΥΠΟΛΑΒΕΙΝ** (*one should understand*) is preserved, which is well-correlated to the phrase "*this text was written by the manufacturer (his name…..) in order for someone to understand the use of this creation, named…. etc.*". (No text reconstruction).

### 6.c.1. Reconstructing the presentation text for Lines 7-11

Lines 7-11 should be correlated to the *Presentation/Description* and *Position* of the Mechanism's Front face very central parts. In order to reconstruct the missing text the following methodology was adopted by the authors:

1) While using the functional model of Antikythera Mechanism, a text without limitation on letter number per line, was written in the ancient Greek language. The text was concerned with the *Presentation* and the *Description* of the very central area parts and their operation

---

[45] The word ΤΑΥΤΗΝ (*on this*) is feminine in Greek language, as well as the word ΠΛΑΞ/ΠΛΑΚΑ (*plate*).

(central axis, Lunar Disc, central hole of Lunar Disc, in which the edge of the central axis is adapted etc.).

2) For the missing text specific and relative keywords/phrases, such as: ΓΕΩΜΕΤΡΙΚΟ ΚΕΝΤΡΟ (*geometrical center*), Η ΓΑΙΑ ΕΣΤΙΝ ΤΟ ΚΕΝΤΡΟ ΚΟΣΜΟΥ (*Earth is the Center of Cosmos*), ΣΕΛΗΝΗΣ ΚΥΛΙΝΔΡΟΝ] **ΗΡΜΟΣ**[ΘΑΙ (*The Lunar Disc/Cylinder is fitted*), ΤΟ ΑΞΟΝΙΟΝ ΚΑΙ **ΕΠ ΑΚΡΟΥ Δ**[Ι ΑΥΤΟΥ (*the axis and its edge*), (ΦΕΡΕΤΑΙ) ΔΕΞΙΟΣΤΡΟΦ]**ΩΣ** (*is continuously rotated clockwise*, were used.

3) Then, they applied the text according to the text's limitations and the preserved letter position.

### 6.c.2. The Front face geometrical center and the Earth locating at the center of Cosmos
### Lines 6-7

In Line 7 the letters **ΠΡΟΣ**[.... are preserved, which could be part of the phrase **ΠΡΟΣ**[ΘΙΑ ΟΨΙΣ (*Front face*). The phrase ΠΡΟΣΘΙΑ ΟΨΙΣ *Defines/Refers to* the position from which *The Engraver* starts the presentation of the Front face *operational parts*. At the geometrical center of the Front face is located the ΓΗ ΤΟ ΚΕΝΤΡΟ ΤΟΥ ΚΟΣΜΟΥ (*Earth, the Center of Cosmos*).[46] The fixed axis $b_{out}$ and the rotatable axis $b_{in}$ are located on the same center.[47]

| |
|---|
| 06.................................................................................................................................................. ΕΝ Τ |
| 07 Ω ΚΕΝΤΡΟ ΤΗΣ **ΠΡΟΣ**[ΘΙΑΣ ΟΨΕΩΣ ΤΩ ΓΕΩΜΕΤΡΙΚΩ Η ΓΑΙΑ ΙΣΤΑΤΑΙ Η ΕΣΤΙΝ ΣΗΜΕΡΟΝ ΤΟ ΚΕΝΤΡΟΝ ΤΟΥ ΚΟΣΜ |
| 08 **Ο**[Υ - - - - - - - - - - ] **ΘΕ**[.................................................................................................................. |
| 06/07/08 *at the geometrical center of the Front Face, the Earth is located at the Center of Cosmos* |

### 6.d.1. Reconstructing the presentation text for the Lunar Disc *Definition* and *Position* in
### Lines 8-11

For the moon-Lunar Disc presentation and function, *The Engraver* uses a large part of the BCI text. Lines 8-16 should refer to the *Definition/Description, Position, and Operation* of the Lunar Disc and its parts.

---

[46] In order to cover the Lunar Disc's stabilizing pin and the crown gear z (Voulgaris et al., 2018b), the adaptation of a colored semi sphere on the Lunar Disc central area depicting the colored Earth is possible. The material for the Earth's sphere could be a (blue color?) painted wood or a colored rock/stone of Azurite (Pliny the Elder 1847, in *Natural History* named ΚΥΑΝΟΣ 35, 47) or Chalcanthon "*the flower of copper*" (Χάλκανθον, Pliny, Natural History in Bostock and Riley 1855, 34, 134) or Crysocolla (cyan/blue-green color) see Theophrastus, 1956; Katsaros et al., 2010. The blue colored Earth's semi sphere could be named in Greek as ΚΥΑΝΗ ΓΑΙΑ.

[47] Voulgaris et al., 2018b and 2019b.

The Lunar Disc is stabilized at the upper edge of the axis $b_{in}$. The preserved words in Line 9, **ΗΡΜΟΣ**[ΘΑΙ (*fitted on*) and in Line 10, **ΕΠ ΑΚΡΟΥ** (*at the edge*) are perfectly matched with the procedure of the Lunar Disc adaptation on the edge of the axis $b_{in}$.[48]

A possible name for the Lunar Disc could be ΚΥΛΙΝΔΡΟΣ ΣΕΛΗΝΗΣ (*Lunar Cylinder*) or less probable ΔΙΣΚΟΣ ΣΕΛΗΝΗΣ (*Lunar Disc*) **Fig. 5**. For critical mechanical and handling reasons, the Lunar Disc is the ideal and proper Input of the Mechanism.[49] Rotating the Lunar Disc continuously clockwise[50], the gears and pointers, start rotating as well.[51]

| |
|---|
| 08 **Ο**[Υ ΕΠΙ ΚΕΝΤΡΟΥ] **ΘΕ**[ΣΙΝ ΚΥΛΙΝΔΡΟΣ ΕΣΤΙΝ ΔΙΑ ΧΕΙΡΟΣ ΦΕΡΟΜΕΝΟΣ, ΟΣ ΣΕΛΗΝΗΣ ΚΥΛΙΝΔΡΟΣ ΚΑΛΕΙΤΑΙ. ΤΟΝ |
| 08 ………. on the position of the center there is *a cylinder rotated by hand, which is called Lunar Cylinder (or Lunar Disc)* |

After the *Definition* of the Lunar Cylinder, follows the Lunar Disc *Positioning*. The words ΤΡΗΜΑΤΩΝ (*holes on a part*) and ΤΡΗΜΑ[…. , appear on BCI Part-2, Lines 6-7.[52] On the Lunar Cylinder center (AMRP CTs, also visually observed) there is a square hole[53] in which the axis $b_{in}$ enters and the Lunar Disc is stabilized.

| |
|---|
| 09 ΚΥΛΙΝΔΡΟΝ] **ΗΡΜΟΣ**[ΘΑΙ ΔΙΑ ΤΡΗΜΑΤΟΣ ΚΕΝΤΡΟΥ ΑΥΤΟΥ ΕΝ ΤΩ ΑΞΟΝΙΩ, Ο ΕΙΣ ΚΕΝΤΡΟΝ ΓΕΩΜΕΤΡΙΚΟΝ ΕΣΤΙΝ. ΕΙΣ Τ |
| 09 *stabilizing the cylinder on the center via its central hole, on the axis located on the geometrical* |

---

[48] See the Antikythera Mechanism gearing scheme Freeth and Jones 2012 and Voulgaris et al., 2019b

[49] Voulgaris et al., 2018b; Roumeliotis 2018.

[50] The Zodiac constellations preserved in the Zodiac month ring of Fragment C, are located in clockwise direction. Therefore, the Sun pointer (also the gear b1) rotates clockwise. From the gearing scheme presented in Freeth and Jones, 2012 and Voulgaris et al., 2019b, it results that the Lunar Cylinder also rotates in clockwise direction. As time always moves from the Past to the Future (i.e. from left to the right in the axis-x, Voulgaris et al., 2018a), therefore the Mechanism's pointers rotate from the Past to the Future. In order to achieve this procedure, the Lunar Cylinder must be continuously rotated in CW direction. A direction change of the Lunar Cylinder rotation (CCW) could create mechanical problems or bad engagement and immobilization of gears.
*The Manufacturer* writes an equal instruction regarding the operation of the Metonic and Saros pointers in Bitsakis and Jones 2016b, pages 235/245 BCI Part-2:
15 […] ΠΕΡΟΝΗΝ ΟΘΕΝ ΕΞΗΛΚΥΣ[ΘΗ (…pin from whence it was pulled out…)
16 […] ΤΗΣ ΠΡΩΤΗΣ ΧΩΡΑΣ Ν Μ[ (… the first space…). *The Manufacturer* informs the reader of the manual that the user must pull out the pointers and reposition them at the spiral beginning when the two spiral pointers reach at the end of their spiral. (Otherwise the pointers will be immobilized and several gears will be destroyed).

[51] Text for the Lunar Disc (Cylinder) Definition/Presentation/Operation. Regarding the preserved letters ΜΘΕ:
1) it is too difficult to detect a word in Greek literature beginning/ending with these three letters.
2) The pattern of two separated words: …**Μ ΘΕ**… i.e. the last letter of the first word is the letter Μ or …**ΜΘ Ε**, is also difficult to exist (the number ΜΘ = 49 is difficult to be correlated with the text).
On the CTs are clearly visible the letter **Θ** and the letter **Ε**, while the letter Μ is not visible. We consider as definite the letters …]. **ΘΕ** .[ ….. ]. The missing letter could also be the letter Υ, we adopt the pattern …Υ]**ΘΕ**[…

[52] Bitsakis and Jones 2016b.

[53] Carman and DiCocco 2016; Voulgaris et al., 2018b.

| center. On… |
|---|

Afterwards, in Line 10 should follow a text for the Lunar Disc Positioning. The phrase **ΕΠ ΑΚΡΟΥ** refers to the edge of axis $b_{in}$, in which the Lunar Disc is stabilized via its central square hole.

| 09 ………………………………………………………………………………………………………………………………………… ΕΙΣ Τ] |
|---|
| 10 [Ο ΑΞΟΝΙΟΝ ΚΑΙ] **ΕΠ ΑΚΡΟΥ Δ**[Ι ΑΥΤΟΥ, Ο ΤΗΣ ΣΕΛΗΝΗΣ ΚΥΛΙΝΔΡΟΣ ΦΕΡΟΜΕΝΟΣ ΕΣΤΙΝ ΚΑΙ ΕΙΣ ΤΟ ΔΙΗΝΕΚ |
| 11 [ΕΣ ΔΕΞΙΟΣΤΡΟΦ]**ΩΣ ΜΕΝΩΝ**[….. |
| 10 (*on*) *the edge of the small axis the Lunar Cylinder is rotated and continuously (rotating)* |
| 11 *clockwise* |

The phrase ΚΑΙ ΕΙΣ ΤΟ ΔΙΗΝΕΚΕΣ ΔΕΞΙΟΣΤΡΟΦ]**ΩΣ ΜΕΝΩΝ** implies an instruction by *The Writer*: "*Do not turn the Lunar Cylinder in counter clockwise direction*" (see footnotes 50).

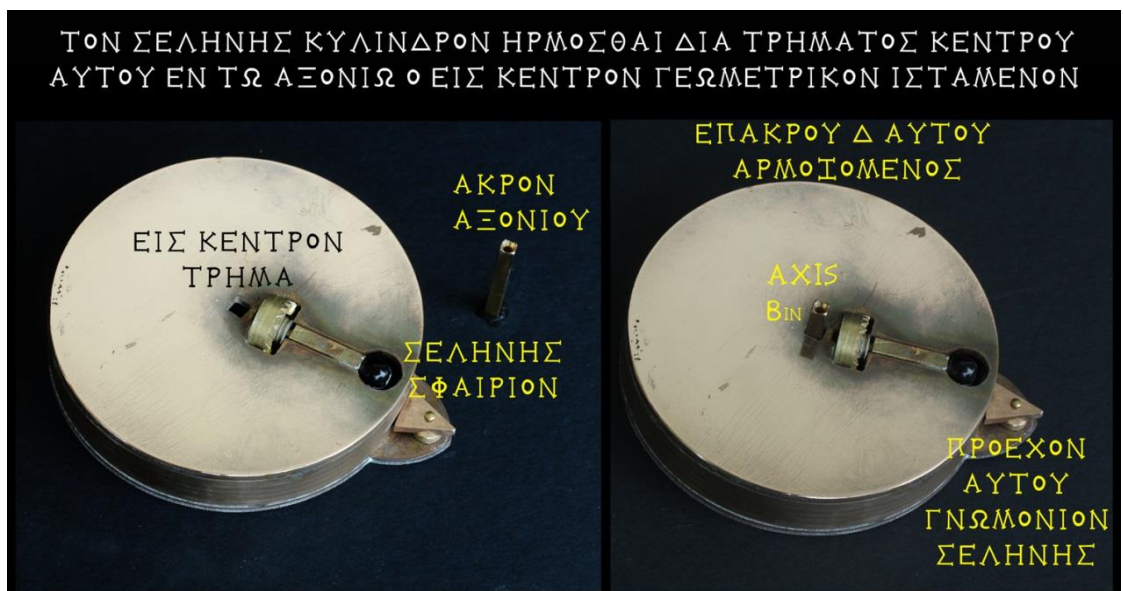

Fig. 5. The Lunar Disc/Cylinder adaptation on the αxis-$b_{in}$, which is located on the geometrical center of the Mechanism front face. The parts of the Lunar Disc/Cylinder are visible: the Lunar phases sphere and the Lunar pointer. At the center of the disc, the square central hole (TPHMA) is also visible.
Bronze Lunar Cylinder constructed by the authors

### 6.d.2. The parts of the Lunar Disc in Lines 11-16

In Lines 11-16 *The Engraver Defines/Describes, Places*, and Presents the operation of the Lunar Disc parts, the Lunar phases sphere and the Lunar pointer. On the Lunar Disc perimeter, a small sphere half white/half blackened is located, which presents the lunar phases during its rotation.[54] In Line 12, the word **ΜΕΛΑΝ** (*black*) is preserved, which is

---

related to the half blackened (and half white[55]) hemisphere of the Lunar sphere[56] **Fig. 6**. Therefore, before the color description, the lunar sphere must be *Defined* at the end of the previous line (Line 11). Additionally, Lunar Disc *Operation* (rotating clockwise) and then, the Lunar Phases sphere *Definition* and *Position* should also be referred. (The word ΠΕΡΙΦΕΡΕΙΑ also refers to Mercury and Venus orbits and it is the possible name for the lunar orbit).

| |
|---|
| 11 [ΕΣ ΔΕΞΙΟΣΤΡΟΦ]**ΩΣ ΜΕΝΩΝ** [ΕΙΣ ΤΗΝ ΠΕΡΙΦΕΡΕΙΑΝ ΤΗΣ ΣΕΛΗΝΗΣ ΚΥΛΙΝΔΡΟΥ, ΣΦΑΙΡΙΟΝ ΣΕΛΗΝΗΣ ΦΕΡΟΜΕΝΟΝ ΕΣΤΙΝ] |
| 12 [ΤΟ ΣΦΑΙΡΙΟΝ ΟΤ]**Ε ΜΕΛΑΝ ΟΤ**[Ε ΛΕΥΚΟΧΡΟΥΝ ΤΗΝ ΝΟΥΜΗΝΙΑΝ ΚΑΙ ΤΗΝ ΠΑΝΣΕΛΗΝΟΝ ΣΥΝ ΤΩ ΧΡΟΝΩ ΔΕΙΚ] |
| 13 [ΝΥΣΙ |
| 11 (continuously) *rotated clockwise. On the circumference of the Cylinder, there is the rotated lunar little sphere* |
| 12 *the little sphere, either in black or in white color appears, and through time depicting the New moon and the Full moon* |

In Line 13, *The Engraver* probably mentions that during the Lunar Disc rotation a large number of lunar and solar events i.e. the solar and lunar eclipses[57] can be presented based on the synodic cycle (and half-cycle) of the Lunar Disc:

| |
|---|
| 13 [ΤΑ ΜΕΓΙΣΤΑ ΤΩΝ ΟΛ]**ΩΝ ΓΕΓ**[ΟΝΟΤΩΝ ΚΑΙ Η ΣΕΛΗΝΗΣ ΚΑΙ Η ΗΛΙΟΥ ΕΚΛΕΙΨΙΣ ΕΝ ΤΗ ΔΙΧΟΜΗΝΙΑ ΚΑΙ ΕΝ ΤΗ ΤΡΙΑΚΑΔ] |
| 14 [Ι ΓΙΓΝΟΝΤΑΙ |
| 13 *the most important events, the Lunar and Solar eclipses, during mid-month and the last day of the month,* |
| 14 *are occurred* |

In Line 14, *The Engraver* could probably explain how *The User* can understand the mid-month/ΔΙΧΟΜΗΝΙΣ and the last day of month/ΤΡΙΑΚΑΣ on the Mechanism by observing the lunar days. Since the existence of the Tropical Year circular scale of the Sun (the Zodiac month ring is preserved in Fragment C), a circular Synodic month scale of the Moon, divided

---

[55] ΛΕΥΚΟΧΡΟΥΝ in Eur. Ph. 322; in Ptol. 7,2; Dorotheus of Sidon in Pingree 1976. Or ΛΕΥΚΟΝ referred by Hesiod in Girgenis 2001, Aristotle 1831; Theophrastus 1956.

[56] The *Description* and *Operation* of the Lunar Phases sphere. The critical positions of the Lunar sphere are in total black color (New moon-the last day of each synodic lunar month) and in total white (Full moon-the mid-month).

[57] A lunar eclipse occurred during Full moon. In ancient Greek calendars every Full moon occurred at 15th day of each month, Mid-month named ΔΙΧΟΜΗΝΙΣ *(Dichominis)*. A solar eclipse occurred during New moon. In ancient Greek calendars every New moon occurred at 29th/30th day of each month, the last day of month, named ΤΡΙΑΚΑΣ *(Triakas)*, Geminus (in Manitius 1880/Spandagos 2002); Danezis and Theodosiou 1992; Jones 2017; Voulgaris et al., 2021.

into 29.5 cells/lunar days (numbering 1-30, A, B, Γ,… KH and Λ) is quite possible and necessary.[58]

| 14 [I ΓΙΓΝΟΝΤΑΙ Δ]Ε[Ι] **Δ ΥΠΟΛΑΒΕΙ**[N[59] ΚΑΙ ΤΟ ΧΡΩΜΑ ΤΟΥ ΣΦΑΙΡΙΟΥ ΚΑΙ ΤΑΣ ΤΗΣ ΣΕΛΗΝΗΣ ΗΜΕΡΑΣ, ΑΙ ΠΕΡΙΞ ΤΟΥ ΚΥΛΙΝ] <br> 15 [ΔΡΟΥ <br> *14 someone should understand the dates of these events (Dichominis and Triakas) by observing the engraved days around the Lunar Cylin-* <br> *15 der….* |
| --- |

On the Antikythera Mechanism ΔΙΧΟΜΗΝΙΣ occurs when the Lunar pointer aims just before the number ΙΕ (about to 14.75 cell) and ΤΡΙΑΚΑΣ when the pointer aims at the first line of A, **Fig. 7**. The number ΙΕ is about in opposite position to the Golden Sphere.

In order to understand the synodic month and the operation of the lunar sphere, a reference to the text, regarding the time duration of one full rotation of the lunar phases sphere (synodic month) is necessary.[60]

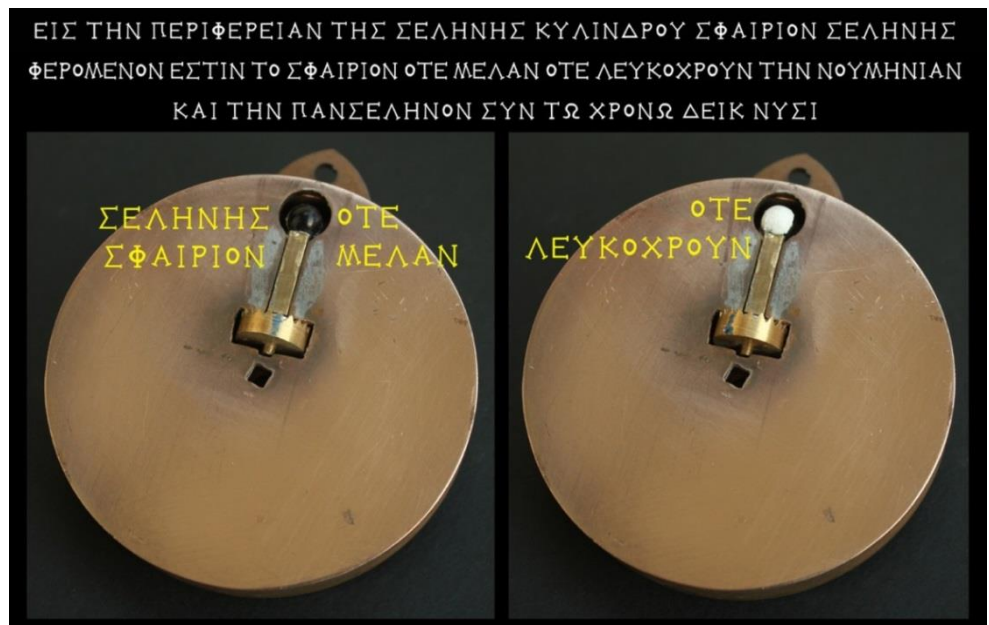

Fig. 6. The Lunar phases sphere which is half white/half blackened. Black and white color alternate during the time of rotation of the small crown gear. Bronze Lunar Cylinder constructed by the authors

In Line 15 [. .]**ΟΥΕ**. .**ΤΟ ΣΦΑΙΡΙΟΝ ΦΕΡΕ .**[……….

---

1) By studying the tomographies on the area of letter "Y", the arms and the two top serifs of letter "Y"are visible. Bitsakis and Jones 2016b write the letter **Θ** but in page 233 refer: "*a deformed Y cannot be ruled out*".

2) Using AMRP CTs and the *Real3D VolViCon* software[61] the new letter "Ι" before the phrase ΤΟ ΣΦΑΙΡΙΟΝ, was detected by the authors. For the positional numbering of the new letter, the well preserved letters of the next Line 16 **ΠΡΟΕΧΟΝ ΑΥΤΟΥ ΓΝΩΜΟΝΙΟΝ Σ**[… , were used. The new letter "Ι" seems to be the 8th letter of the Line 15, as is located on the same position with the letter "Α"of Line 16.

In Lines 15-16 a text concerning the Lunar Disc's pointer *Definition* and *Position* should exist. Today, the Lunar Disc pointer is not preserved, but it is totally necessary for the Mechanism to function. According to the preserved text, the lunar pointer protrudes from the Lunar Cylinder.

| |
|---|
| 14 [Ι ΓΙΓΝΟΝΤΑΙ Δ]Ε[Ι] **Δ ΥΠΟΛΑΒΕΙΝ** [ΚΑΙ ΤΟ ΧΡΩΜΑ ΤΟΥ ΣΦΑΙΡΙΟΥ ΚΑΙ ΤΑΣ ΤΗΣ ΣΕΛΗΝΗΣ ΗΜΕΡΑΣ, ΑΙ ΠΕΡΙΞ ΤΟΥ ΚΥΛΙΝ] |
| 15 [ΔΡ]**ΟΥ Ε**[ΙΣ]**Ι ΤΟ ΣΦΑΙΡΙΟΝ ΦΕΡΕ**[ΤΑΙ ΕΝ ΗΜΕΡΑΙΣ ΚΘ Β ΟΥΤΟΣ Ο ΜΗΝ ΣΥΝΟΔΙΚΟΣ ΚΑΛΕΙΤΑΙ ΑΠΟ ΤΟΥ ΚΥΛΙΝΔΡΟΥ] |
| 17 **ΠΡΟΕΧΟΝ ΑΥΤΟΥ ΓΝΩΜΟΝΙΟΝ Σ**[ΕΛΗΝΗΣ - - - - - - - - - - - - - - 58 - - - - - - - - - - - - - - - - - - -] |
| *14 occurred. One should take into account the little sphere's color and the lunar days, which around the cylin-* |
| *15 der exist. The little (phases) sphere is rotated in (time span) of 29.5 days, which is called synodic month. From the (Lunar) Cylinder* |
| *16 it protrudes the lunar pointer…………..* |

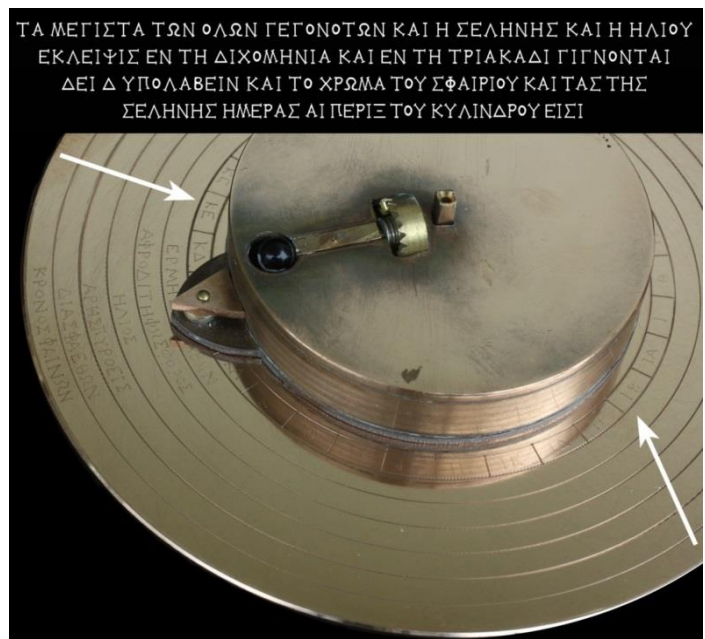

Fig. 7. The Synodic month scale is located right after the circumference of the Lunar Disc (white arrows). The scale is divided into 29.5 cells/days, equal to the duration of the Lunar Synodic month.

---

Dichominis and Triakas located in opposite positions (see also **Fig. 5**). Bronze parts, before their final assembly

| |
|---|
| 14 [Ι ΓΙΓΝΟΝΤΑΙ ΔΕΙ Δ ΥΠΟΛΑΒΕΙΝ ΚΑΙ ΤΟ ΧΡΩΜΑ ΤΟΥ ΣΦΑΙΡΙΟΥ ΚΑΙ ΤΑΣ ΤΗΣ ΣΕΛΗΝΗΣ ΗΜΕΡΑΣ, ΑΙ ΠΕΡΙΞ ΤΟΥ ΚΥΛΙΝ<br>15 [ΔΡ]**ΟΥ Ε**[ΙΣ]**Ι ΤΟ ΣΦΑΙΡΙΟΝ ΦΕΡΕ**[ΤΑΙ ΕΝ ΗΜΕΡΑΙΣ ΚΘ Β ΟΥΤΟΣ Ο ΜΗΝ ΣΥΝΟΔΙΚΟΣ ΚΑΛΕΙΤΑΙ -23-] |
| 14 occurred. One should take into account the little sphere's color and the lunar days, which around the cylin-<br>15 der exist. *The little (phases) sphere is rotated in (time span) of 29.5 days, which is called synodic month………* |

### 6e. The operation and the mechanical characteristics of the Lunar pointer (Lines 16-17)

In Lines 16-17, a text for the Lunar pointer Operation, should be referred to. The lunar pointer protrudes from the Lunar Cylinder. It has a specific length and during its rotation (via the Lunar Cylinder), travels through the two inferior planet's circumferences.

Afterwards, follows a text in which *The Engraver Describes* and *Places* the position of the two *circumferences*: the first circumference is located close (ΕΧΟΜΕΝΗ) to the Lunar Cylinder and the second one follows (ΕΠΟΜΕΝΗ).

| |
|---|
| 16 **ΠΡΟΕΧΟΝ ΑΥΤΟΥ ΓΝΩΜΟΝΙΟΝ Σ**[ΕΛΗΝΗΣ ΕΣΤΙΝ  -25- ΤΟ ΓΝΩΜΟΝΙΟΝ ΦΕΡΕΤΑΙ ΕΠΙ ΤΩΝ ΠΕΡΙ]<br>17 **ΦΕΡΕΙΩΝ Η ΜΕΝ ΕΧΟΜΕΝΗ ΤΩΙ ΤΗΣ** ΣΕΛΗΝΗΣ ΚΥΛΙΝΔΡΩ (ΕΣΤΙΝ), Η ΔΕ ΕΠΟΜΕΝΗ [- - - 33 - - -] |
| 16 *it protruded the lunar pointer…………… The lunar pointer travels along the two circu-*<br>17 *mferences, the one that is located close to the Lunar Cylinder and the next follows ……….* |

### 6.e.1. The Golden sphere-Sun, its pillar and the solar ray-pointer (Lines 20-22), The word ΓΝΩΜΟΝΙΟΝ

In Lines 20-22 *The Engraver Defines*, *Describes*, and *Places* three parts which were related to ΠΕΡΙΦΕΡΕΙΑ/ΚΥΚΛΟΣ ΗΛΙΟΥ (*orbit of the Sun*): **ΓΝΩΜΩ**[Ν] (*Pillar*, analyzed below), **ΧΡΥΣΟΥΝ ΣΦΑΙΡΙΟΝ** (*Golden Sphere*) and **ΗΛΙΟΥ ΑΚΤΙΝ** (*Solar Ray*).

| |
|---|
| 20 **ΤΟΥ ΦΩΣΦΟΡΟΥ, ΠΕΡΙΦΕΡΕΙΑΝ** ΗΛΙΟΥ ΕΣΤΙΝ [- - - - - - - - - - - - - - - - - 53 - - - - - - - - - - - - - - - ] |
| 21 **ΓΝΩΜΩ**[Ν] **ΚΕΙΤΑΙ ΧΡΥΣΟΥΝ ΣΦΑΙΡΙΟΝ** [- - - - - - - - - - - - - - -56- - - - - - - - - - - - - - - - - - ΤΟΥ] |
| 22 **ΗΛΙΟΥ ΑΚΤΙΝ ΥΠΕΡ ΔΕ ΤΟΝ ΗΛΙΟΝ ΕΣΤΙΝ ΚΥ**[- - - - - - - - - - - - - - - - - 55 - - - - - - - - - - - - - - - - - - ] |
| *20 beyond Phosphoros' circumference, there is the Sun's circle…………………………………………………on the*<br>*21 Pillar is stabilized the Golden Sphere…………………………………………………………………………………the*<br>*22 Sun's Ray………………………………………………………………………………………………………………………….* |

### 6.e.2. The word ΓΝΩΜΩΝ

During antiquity, the Greek word ΓΝΩΜΩΝ (gnomon) had many different meanings. Below, an analysis of the word ΓΝΩΜΩΝ is presented, and also the mechanical correlation to the Golden Sphere and the Solar Ray is discussed.

The word ΓΝΩΜΩΝ/ΓΝΩΜΟΝΙΟΝ comes from the words *ΓΝΩΜΗ + ΩΝ* (meaning the person who expresses an opinion with a significant meaning, one that knows, or the decision-maker, Liddell and Scott's lexicon 2007) or by the word *ΓΩΝΙΩΜΩΝ* (one that measures/calculates the angle).[62]

The ΓΝΩΜΟΝΙΚΗ (Gnomonics) was one of the oldest sections of the Practical/Observational Astronomy of Babylonians, Egyptians, and Ancient Greeks.[63]

In Ancient Greece, the word ΓΝΩΜΩΝ/ΓΝΩΜΟΝΙΟΝ was used in order to describe many different meanings:

- ΓΝΩΜΩΝ was a thin tall pillar, stabilized on a base, making a shadow, acting as a pointer/indicator for the sundial.[64] The inclination of this ΓΝΩΜΩΝ could be vertical (ΟΡΘΩΣ/ΟΡΘΟΝ) or at an angle (ΚΛΙΜΑ-ΚΛΙΣΙΣ) equal to the latitude of the place of observation:

-Proclus in Manitius 1906, also writes *"Ἡ δὲ μεσημβρινὴ γραμμὴ λαμβάνεται γνώμονος ὀρθοῦ στάντος ἐπὶ τῆς πλακὸς ταύτης καὶ κύκλου γραφέντος περὶ τὴν ῥίζαν τοῦ γνώμονος ὡς περὶ κέντρον"* (*the Meridian is defined by a perpendicular/vertical gnomon fixed on the base plate and a sketched circle with center, the gnomon's base*). Here, the word ΓΝΩΜΩΝ represents a simple pillar.

- Heron Alexandrinus in Dioptra (Schöne 1903) *Γνωμονίων τινῶν περιαγωμένων (the rotating pointers of the odometer).*

- ΓΝΩΜΩΝΕΣ were named the teeth that indicated the age of a horse, Xenophon (Marchant 1966), On the Art of Horsemanship Chap. 3.1.

- Apollodorus from Damascus in Poliorcitica 149.4 (in Wescher 1867), refers ΓΝΩΜΩΝ as the front point of a drill.

- Euclides, in Elements of Geometry (Heiberg 1883/2008) uses extendedly the word ΓΝΩΜΩΝ in Geometry (*Theorem of the Gnomon*).

- Proclus in Friedlein 2011, *in Primum Euclidis Elementorum librum commentarii* writes: *Τοῦτο τὸ πρόβλημα πρῶτον Οἰνοπίδης ἐζήτησεν χρήσιμον αὐτὸ πρὸς ἀστρολογίαν οἰόμενος. ὀνομάζει δὲ τὴν κάθετον ἀρχαϊκῶς κατὰ γνώμονα, διότι καὶ ὁ γνώμων πρὸς ὀρθάς ἐστι τῷ ὁρίζοντι.* (*This problem, which was firstly discussed by Oenopides, is useful for the Astrology (Astronomy). For the perpendicular angle direction he uses the archaic phrase KATA ΓΝΩΜΟΝΑ, because the Gnomon's direction is vertical to the horizon*).

### 6.e.3. ΓΝΩΜΟΝΙΑ in the Antikythera Mechanism

On the Antikythera Mechanism the ΓΝΩΜΟΝΙΟΝ of the Lunar Disc should be a part which protruded (ΠΡΟΕΧΟΝ) from the Lunar Disc with a pointy edge. This ΓΝΩΜΟΝΙΟΝ/Lunar pointer was adapted by *The Manufacturer* in order to aim (during its rotation) at the Golden Sphere-Sun (New moon) or in the opposite direction (Full moon). These two relative positions were critical for the measurement of Time, and also during New moon/Full moon a solar/lunar eclipse occurred.[65] The Lunar pointer/ΓΝΩΜΟΝΙΟΝ acts as a "*real*" pointer.

---

[62] Danezis and Theodosiou 1998.
[63] Danezis and Theodosiou 1998; Thibodeau 2017.
[64] Hipparchus in Manitius 1894, 1.3.6 and 1.4.8; Plutarch in Bernadotte 1916 in *The Life of Pericles* 6.3; *Suidas Lexicon* Bekkeri 1854; Danezis and Theodosiou 1994.

It seems that the Sun indicator/pointer is not a simple part. *The Engraver* refers the words **ΓΝΩΜΩ[Ν] ΚΕΙΤΑΙ**

**ΧΡΥΣΟΥΝ ΣΦΑΙΡΙΟΝ** and then

**ΗΛΙΟΥ ΑΚΤΙΝ**

Therefore, the Sun's indicator is a complex design, consisted by three parts.

At the end of Line 20 should be existed the *Definition and the Position* of **ΓΝΩΜΩ[Ν]**.

| |
|---|
| 19 ………………………………………………………………………………………………………………………ΥΠΕΡ ΔΕ<br>20 **ΤΟΥ [ΦΩ]ΣΦΟΡΟΥ, ΠΕΡΙΦΕΡΕΙΑΝ** [ΗΛΙΟΥ ΕΣΤΙΝ. ΕΙΣ ΤΗΝ ΤΟΥ ΗΛΙΟΥ ΠΕΡΙΦΕΡΕΙΑΝ ΕΙΣ ΣΤΑΘΕΡΟΣ ΚΑΙ ΟΡΘΟΣ]<br>21 **ΓΝΩΜΩ[Ν] ΚΕΙΤΑΙ. ΧΡΥΣΟΥΝ ΣΦΑΙΡΙΟΝ**] ΕΠΙ ΓΝΩΜΟΝΑ ΕΣΤΙΝ<br>*19…………………………………………………………………………………………………………………………beyond*<br>*20 Phosphoros, there is the Sun's circumference. On the (Sun's) circumference is located one fixed and perpendicular*<br>*21 pillar. The Golden Sphere is adapted on the pillar…* |

The **ΓΝΩΜΩ[Ν]** is a fixed and perpendicular pillar[66], adapted on the Front plate of b1 gear, just on the Sun's engraved circumference. This phrase is the *Definition, Description* and *Position* of the **ΓΝΩΜΩ[Ν]**.

On the **ΓΝΩΜΩ[Ν]** is adapted the **ΧΡΥΣΟΥΝ ΣΦΑΙΡΙΟΝ**, the Golden Sphere-Sun. Therefore, the fixed and perpendicular pillar (ΣΤΑΘΕΡΟΣ ΚΑΙ ΟΡΘΟΣ ΓΝΩΜΩΝ) acts as a supporting base for the **ΧΡΥΣΟΥΝ ΣΦΑΙΡΙΟΝ, Fig. 8**. In this way, the *Position* of the ΧΡΥΣΟΥΝ ΣΦΑΙΡΙΟΝ is described (for the Operation of the ΧΡΥΣΟΥΝ ΣΦΑΙΡΙΟΝ see §6.f.). The word ΣΤΑΘΕΡΟΣ (fixed) could also act as a Manufacturer's instruction: "*Do not remove the fixed pillar*".

The **ΧΡΥΣΟΥΝ ΣΦΑΙΡΙΟΝ** in order to be adapted on the **ΓΝΩΜΩ[Ν]** it needs to be perforated on its center. Hence, this pillar acts as a stabilizer for the Golden sphere and not as an operational solar pointer, **Fig. 9**.

The real Sun's pointer/indicator should be a construction which was stabilized on the ΧΡΥΣΟΥΝ ΣΦΑΙΡΙΟΝ and protruded from it. Characteristic formations that protrude from the Sun are the solar rays. In the text the phrase ΗΛΙΟΥ ΑΚΤΙΝ (*solar ray*) is preserved.[67] *The*

---

[65] Freeth et al., 2006; Voulgaris et al., 2021
[66] See §6.e.2.
[67] The word ΑΚΤΙΝ (in the nominative): The word "ΑΚΤΙΝ" (instead of "ΑΚΤΙΣ" in nom?) is written in the BCI text by the ancient engraver. The word "ἡ ἀκτίν" (nom.) is referred in Aelius Herodianus in Lentz 1867, 2.511./ 3.1. In Bubeník, 1989 referred the words Χάλκιν, Φράγγελιν, Πάννιν). Additionally, for equal words: Δελφίς (Geminus in Spandagos 2002, p. 63); Aratus 1821, p. 24)/ Δελφίν Mosch. 3, 37 ; Man. 5, 157. Πελκίς/Πελκίν/Πέλκιν in Vassilakis 2000. Γόρτυν/Γόρτις in Gillis 2018. In Eustathii archiepiscopi Thessalonicensis (Stallbaum 1970) in Commentarii ad Homeri Odysseam, is referred: "Ἰστέον δὲ καὶ ὡς Ὅμηρος μὲν Γόρτυν Γόρτυνος κλίνει". Herodianus in Lentz 1867,

*Manufacturer* probably constructed the ΧΡΥΣΟΥΝ ΣΦΑΙΡΙΟΝ, on which a small disc or a ring with some short solar rays was adapted. One of the solar rays was larger in length and extended/protruded (as also the Lunar Pointer protrudes from the Lunar Cylinder, see §6.d.2.). During the rotation of the ΧΡΥΣΟΥΝ ΣΦΑΙΡΙΟΝ-Sun, the pointer of the Sun-ΗΛΙΟΥ ΑΚΤΙΝ traveled through (scan/transit) the outer circles/orbits of the three superior planets. It seems that the reference ΗΛΙΟΥ ΑΚΤΙΝ was the real/operational Solar pointer. The length of ΗΛΙΟΥ ΑΚΤΙΝ-pointer reaches Saturn's circle (the last superior planet) **Fig. 10** and at the same time it should also be aiming at the subdivisions of the Zodiac Month ring, depicting the position of the Sun on the Ecliptic.

Through this analysis it is inferred that the pointer of the Sun is a composition of three different parts: a fixed perpendicular pillar, a sphere, and a cosmetic component which also acts as the real pointer/indicator of the Sun **Fig. 11**.

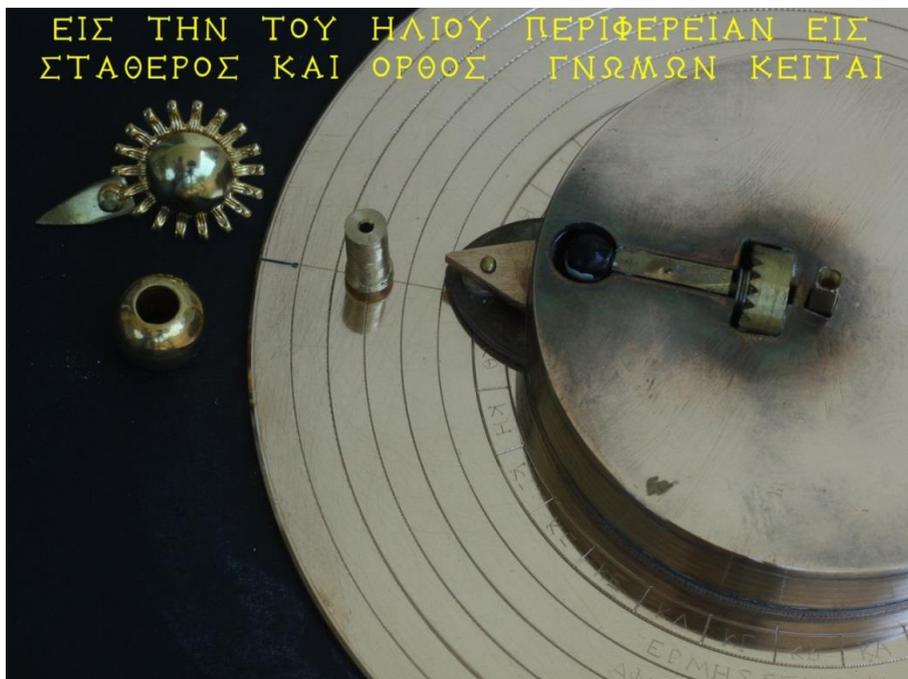

Fig. 8. A perpendicular pillar is placed on the circumference/orbit of the Sun. Material and image by the Authors

In Line 22, *The Engraver* refers ΗΛΙΟΥ ΑΚΤΙΝ, which is directly correlated to the ΧΡΥΣΟΥΝ ΣΦΑΙΡΙΟΝ. After the missing *Definition* and the *Position* of ΗΛΙΟΥ ΑΚΤΙΝ (Line 21), the *Operation* of the ΗΛΙΟΥ ΑΚΤΙΝ which should be referred to in Line 22 is also missing.

---

Grammatici Graeci, vol. 3.1, p. 18, refers: "κίνδυν· οὕτω δὲ ἔφη Σαπφὼ τὸν κίνδυνον". Also, Φόρκυν, Πόλτυν, Κότυν, Κάπυν, μόσυν (same writer). Arcadius (Schmidt 1860) in Gramm., De accentibus [Sp.] (2116: 001) "Ἐπιτομὴ τῆς καθολικῆς προσῳδίας Ἡρωδιανοῦ", p.8 writes: δελφίν καὶ δελφίς, Τελχίν καὶ Τελχίς, Σαλαμίν καὶ Σαλαμίς, ἀκτίν καὶ ἀκτίς.

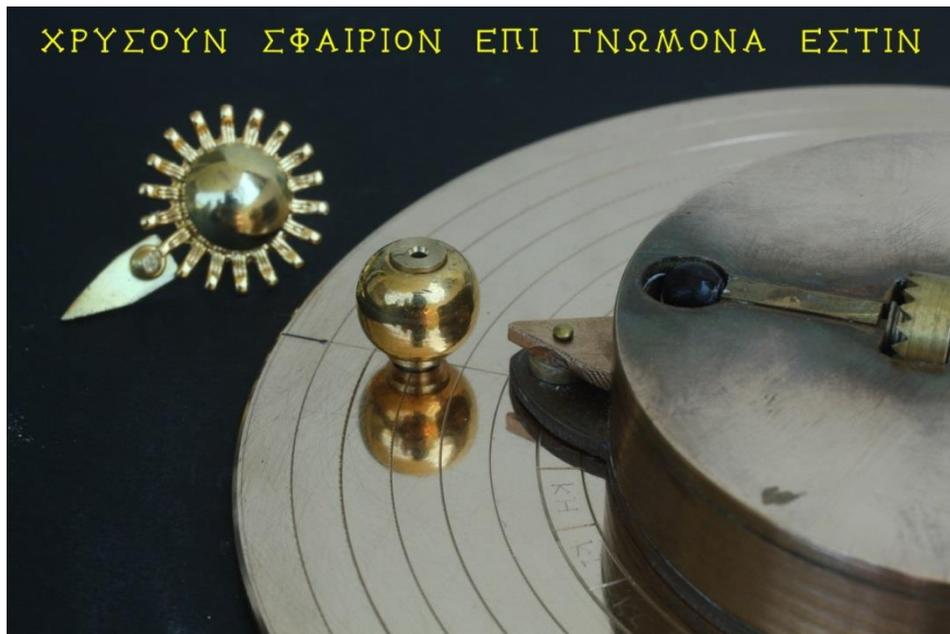

Fig. 9. On the perpendicular pillar is adapted the Golden Sphere-Sun, which rotates in one tropical year. Material and image by the Authors

| |
|---|
| 20 **ΤΟΥ** [ΦΩ]**ΣΦΟΡΟΥ, ΠΕΡΙΦΕΡΕΙΑΝ** ΗΛΙΟΥ ΕΣΤΙΝ. ΕΙΣ ΤΗΝ ΤΟΥ ΗΛΙΟΥ ΠΕΡΙΦΕΡΕΙΑΝ ΕΙΣ ΟΡΘΟΣ ΚΑΙ ΣΤΑΘΕΡΟΣ] 21 **ΓΝΩΜΩ**[Ν] **ΚΕΙΤΑΙ. ΧΡΥΣΟΥΝ ΣΦΑΙΡΙΟΝ** [ΕΠΙ ΓΝΩΜΟΝΑ ΕΣΤΙΝ - -  17 - - ΑΠΟ ΤΟΥ ΣΦΑΙΡΙΟΥ ΠΡΟΕΧΕΙ ΑΥΤΟΥ Η ΤΟΥ] 22 **ΗΛΙΟΥ ΑΚΤΙΝ** |
| 20 *…there is the Sun's circumference. On the Sun's circumference, one fixed and perpendicular* 21 *pillar is located. The Golden Sphere is adapted on the pillar ……………………. from the Golden Sphere is protruded* 22 *the Sun's Ray …………………………………………………………………………………….* |

In Line 21, if we based on the suggestion by Bitsakis and Jones 2016b[68], **ΓΝΩΜΩ**[Ν]**<Ι>**

**ΚΕΙΤΑΙ ΧΡΥΣΟΥΝ ΣΦΑΙΡΙΟΝ**[… ,  an alternative reconstruction for the Lines 20, 21 and 22 is:

| |
|---|
| 20 **ΤΟΥ ΦΩΣΦΟΡΟΥ, ΠΕΡΙΦΕΡΕΙΑΝ** [ΗΛΙΟΥ ΕΣΤΙΝ. ΕΙΣ ΤΗΝ ΗΛΙΟΥ ΠΕΡΙΦΕΡΕΙΑΝ ΙΣΤΑΤΑΙ ΓΝΩΜΟΝΙΟΝ ΟΡΘΟΝ. ΕΠΙ ΤΩ] 21 **ΓΝΩΜΩ**[Ν]**<Ι> ΚΕΙΤΑΙ ΧΡΥΣΟΥΝ ΣΦΑΙΡΙΟΝ.** . . . . . . . . . [ΑΠΟ ΤΟΥ ΣΦΑΙΡΙΟΥ ΑΥΤΟΥ ΠΡΟΕΧΕΙ Η ΤΟΥ 22 **ΗΛΙΟΥ ΑΚΤΙΝ**[ |
| 20 *… there is the Sun's circumference. On the Sun's circumference is located a perpendicular pillar. On the* 21 *pillar is adapted the Golden Sphere ……………………. From the Golden Sphere is protruded* 22 *the Sun's Ray …………………………………………………………………………………….* |

Following the same design reference pattern, at the end of Line 22 **ΗΛΙΟΥ ΑΚΤΙΝ ΥΠΕΡ**

**ΔΕ ΤΟΝ ΗΛΙΟΝ ΕΣΤΙΝ ΚΥ**[-52-] should be the phrase ΥΠΕΡ ΤΟΥ ΗΛΙΟΥ, ΚΥΚΛΟΣ ΕΣΤΙΝ Ο ΤΟ]**Υ**

**ΑΡΕΩΣ ΠΥΡΟΕΝΤΟΣ** follows on next Line 23) (*beyond Sun there is the circle of Ares Pyroes*),

i.e.:

---

| |
|---|
| 22 **ΗΛΙΟΥ ΑΚΤΙΝ ΥΠΕΡ ΔΕ ΤΟΝ ΗΛΙΟΝ ΕΣΤΙΝ ΚΥ**[- - - - - - -34- - - - - -] ΥΠΕΡ ΤΟΥ ΗΛΙΟΥ, ΚΥΚΛΟΣ ΕΣ] |
| 23 ΙΝ Ο ΤΟ]**Υ ΑΡΕΩΣ ΠΥΡΟΕΝΤΟΣ ΤΟ ΔΕ ΔΙΑΠΟΡΕ**[- - - - - - - - - - - - - - 55 - - - - - - - - - - - - - - - - - -] |
| *22 Sun's ray. Beyond Sun there is circle………………………… Beyond Sun there is circle* |
| *23 of Ares Pyroes……………………………………………………………………………………………………………* |

However, the same phrase fits in the middle of line 22:

| |
|---|
| 22 **ΗΛΙΟΥ ΑΚΤΙΝ ΥΠΕΡ ΔΕ ΤΟΝ ΗΛΙΟΝ ΕΣΤΙΝ ΚΥ**[ΚΛΟΣ ΑΡΕΩΣ ΠΥΡΟΕΝΤΟΣ[- - - - 17 - - - - - ] ΥΠΕΡ ΗΛΙΟΥ ΚΥΚΛΟΣ ΕΣΤ] |
| *23* ΙΝ Ο ΤΟ]**Υ ΑΡΕΩΣ ΠΥΡΟΕΝΤΟΣ ΤΟ ΔΕ ΔΙΑΠΟΡΕ**[- - - - - - - - - - - - - - 52 - - - - - - - - - - - - - - - - - -] |
| *22 ……….Beyond Sun there is the circle of Mars Pyroes. ………..Beyond Sun there is the circle* |
| *23 of Mars Pyroes……………………………………………………………………………………………………………..* |

The full name of Mars Pyroes is preserved in Line 23. Therefore, the *Definition/Description* ΕΣΤΙΝ ΚΥΚΛΟΣ ΑΡΕΩΣ ΠΥΡΟΕΝΤΟΣ could not be in the middle of Line 22 (if this phrase was added, the planet Mars will be referred to three times, instead of the other planets which were referred to twice).

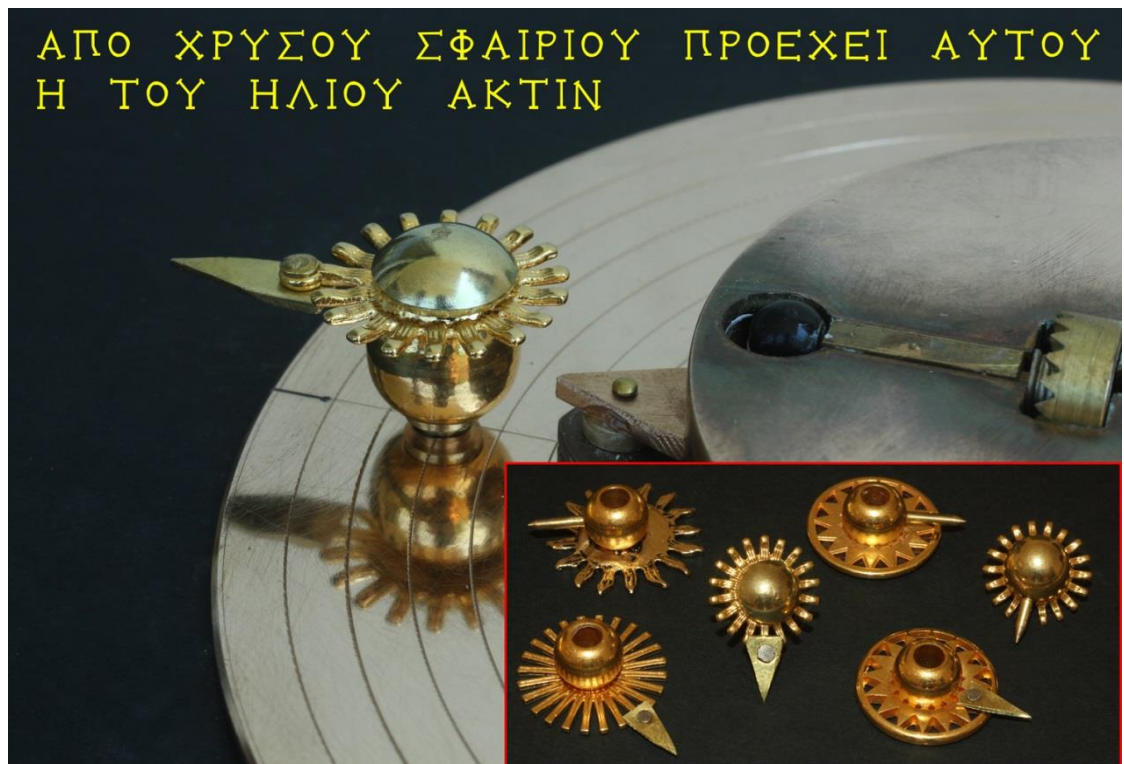

Fig. 10. From the Golden Sphere the Solar Ray (pointer of the Sun) it protrudes. This travels through the three orbits of the superior planets Mars, Jupiter and Saturn. The Lunar pointer travels through the circumferences of Mercury and Venus (see §6.a.). Insert, different designs of the part ΧΡΥΣΟΥΝ ΣΦΑΙΡΙΟΝ with ΗΛΙΟΥ ΑΚΤΙΝ, were constructed by the first Author. Material and image by the Authors

An ideal explanation is that in the middle of Line 22, *The Engraver* refers to the missing *Operation* of the Sun indicator-ΗΛΙΟΥ ΑΚΤΙΝ, which travels through/around the three next planet orbits (ΚΥΚΛΟΙ). The same procedure is written for the Lunar pointer, which travels through/around the two circumferences of the inferior planets, Mercury and Venus:

| |
|---|
| 16 ……………………………………………………… ΤΟ (ΣΕΛΗΝΗΣ) ΓΝΩΜΟΝΙΟΝ ΦΕΡΕΤΑΙ ΕΠΙ ΤΩΝ ΠΕΡΙ] |
| 17 **ΦΕΡΕΙΩΝ, Η ΜΕΝ ΕΧΟΜΕΝΗ ΤΩΙ ΤΗΣ** [ΣΕΛΗΝΗΣ ΚΥΛΙΝΔΡΩ (ΕΣΤΙΝ), Η ΔΕ ΕΠΟΜΕΝΗ …………… |
| *16……………………………………………….the Lunar Disc pointer travels through the two circum-* |
| *17 –ferences, the first close to the lunar circumference and the next follows* |

Applying the same description, the Sun's Ray-pointer travels through the three superior planets' orbit/ΚΥΚΛΟΣ (ΚΥΚΛΟΣ ΠΡΩΤΟΣ-ΜΕΣΟΣ-ΕΣΧΑΤΟΣ, first-middle-last[69] or ΚΥΚΛΟΣ ΠΡΩΤΟΣ, ΔΕΥΤΕΡΟΣ, ΤΡΙΤΟΣ, first, second, third[70]):

| |
|---|
| 22 **ΗΛΙ**[ΟΥ] **ΑΚΤΙΝ ΥΠΕΡ ΔΕ ΤΟΝ ΗΛΙΟΝ ΕΣΤΙΝ ΚΥ**[ΛΟΣ ΠΡΩΤΟΣ, ΜΕΣΟΣ, ΕΣΧΑΤΟΣ, ΟΥΣ Η ΑΚΤΙΝ ΦΕΡΕΤΑΙ. ΥΠΕΡ ΗΛΙΟΥ ΚΥΚΛΟΣ ΕΣΤ] |
| 23 ΙΝ Ο ΤΟΥ **ΑΡΕΩΣ ΠΥΡΟΕΝΤΟΣ**…………………………………………………………………………….. |
| *22….beyond Sun's orbit there is the first, the middle, the last circle, in which the sun's ray travels.* |
| *Beyond Sun there is the circle* |
| *23 of Mars Pyroes* |

In order to avoid the second repetition of the words ΥΠΕΡ ΔΕ ΤΟΝ ΗΛΙΟΝ ΕΣΤΙΝ, the ideal reference could be ΠΡΩΤΟΣ ΚΥΚΛΟΣ (*The first circle*):

| |
|---|
| 22 **ΗΛΙ**[ΟΥ] **ΑΚΤΙΝ ΥΠΕΡ ΔΕ ΤΟΝ ΗΛΙΟΝ ΕΣΤΙΝ ΚΥ**[ΛΟΣ ΠΡΩΤΟΣ, ΜΕΣΟΣ, ΕΣΧΑΤΟΣ, ΟΥΣ Η ΑΚΤΙΝ ΦΕΡΕΤΑΙ. ΠΡΩΤΟΣ ΚΥΚΛΟΣ ΕΣΤ] |
| 23 [ΙΝ Ο ΤΟ]**Υ ΑΡΕΩΣ ΠΥΡΟΕΝΤΟΣ** |
| *22….Beyond Sun's orbit there is the first, middle, last circles in which the sun's ray travels. The first* |
| *is the circle* |
| *23 of Mars Pyroes* |

A relative symmetry appears on the texts of the Lunar Pointer and the Sun's pointer:

On the one hand, the pointer of the Lunar Disc ΓΝΩΜΟΝΙΟΝ ΣΕΛΗΝΗΣ travels/transits through the orbits of inferior planets Mercury and Venus.

On the other hand, the Sun's pointer/indicator-ΗΛΙΟΥ ΑΚΤΙΝ travels/transits through the orbits of the three superior planets. Therefore, *The Engraver* describes the *Operation* of the two pointers and also the specific length of each pointer (analyzed in §6.f.).

---

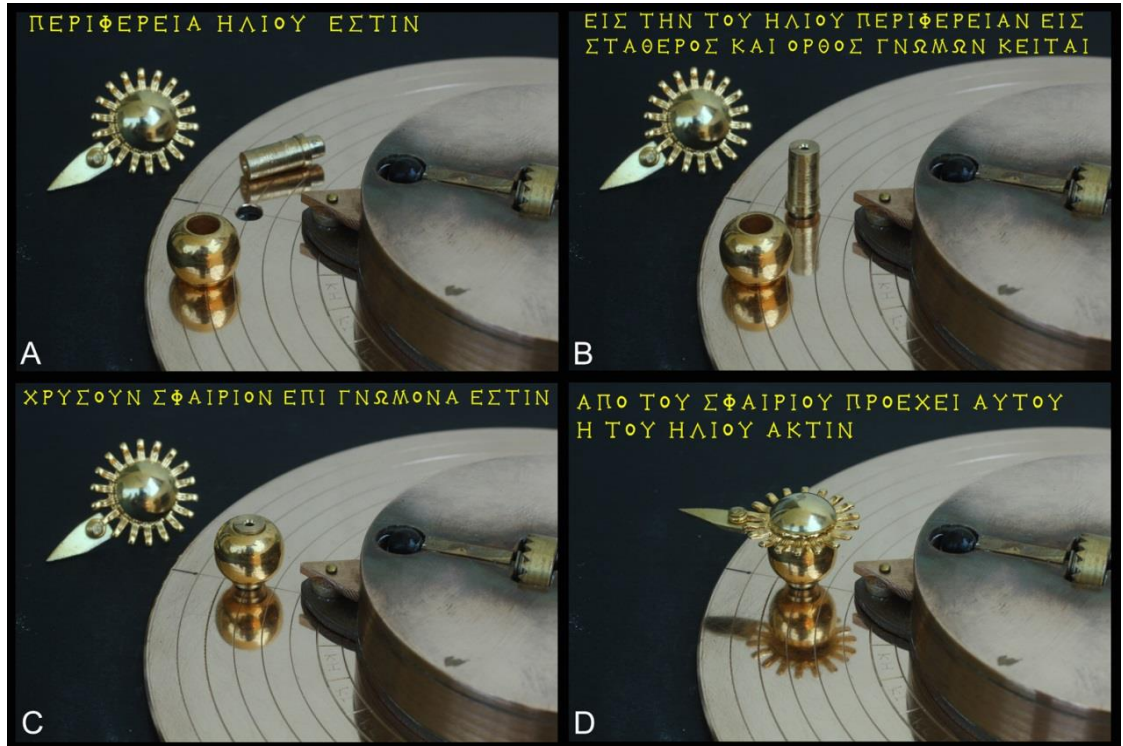

Fig. 11. Assembling the three mechanical parts of the Sun pointer. Material and images by the Authors

Based on the reference of Lunar/Solar sphere and pointer, and the preserved detailed reference for each planet orbit on the Back Cover Inscription Part-1, an equal *Definition/Presentation/Position* of the planets ΓΝΩΜΟΝΙΟΝ and ΣΦΑΙΡΙΟΝ (*pointer and sphere*) should exist for each planet of the Antikythera Mechanism, if the existence of the planet indication gearing was realistic and not hypothetical.

### 6.e.4. A significant question arises: Could the hypothetical/suggested planets, including their rotating spheres adapted on their corresponding pointers, be described on the Back Cover Inscription Part-1 text?

Many researchers maintain that a planet indication gearing existed on the Antikythera Mechanism Front central area: they suggest that, despite the rotating lunar and solar pointers/spheres, there were planet rotating pointers and colored spheres, presenting the exact planets' position on the Ecliptic.

If the assumption of the rotating planets' existence on the Antikythera Mechanism was realistic, a relative *Definition/Description/Position* and *Operation* for the planet pointers and their spheres should exist in the User's Manual of the Mechanism. The planets parts Spheres/Pointers are very important, as well as the Golden Sphere and its Pointer. The position of these descriptions should be on the text's planet orbits presentation.

Assuming there is a planet indication gearing presenting the rotating planets with their corresponding pointers and colored spheres, Freeth and Jones 2012 suggest for all of the planets the phrase:

…**ΤΟ ΔΕ ΔΙΑΠΟΡΕΥΟΜΕΝΟΝ** <u>ΑΥΤΟΥ ΣΦΑΙΡΙΟΝ</u>, e.g.

23 [-3- ΤΟ]**Υ ΑΡΕΩΣ ΠΥΡΟΕΝΤΟΣ ΤΟ ΔΕ ΔΙΑΠΟΡΕ**[ΥΟΜΕΝΟΝ <u>*ΑΥΤΟΥ ΣΦΑΙΡΙΟΝ*</u>

24 [ΔΙΟΣ ΦΑ]**ΕΘΟΝΤΟΣ ΤΟ ΔΕ ΔΙΑΠΟΡΕΥΟΜΕΝΟΝ** <u>*ΑΥΤΟΥ ΣΦΑΙΡΙΟΝ*</u>

(23 *Ares Pyroes, and its traveling sphere*
24 *Zeus Phaethon and its traveling sphere*)

Bitsakis and Jones also suggest

23 [-3- ΤΟ]**Υ ΑΡΕΩΣ ΠΥΡΟΕΝΤΟΣ ΤΟ ΔΕ ΔΙΑΠΟΡΕ**ΥΟΜΕΝΟΝ ΑΥΤΟΥ ΣΦΑΙΡΙΟΝ ΠΥΡΡΟΝ

(*23 And the little sphere making its way through it is fire-red…*)

Thus, before the reference of "its traveling sphere", *The Engraver* should have *Defined/Described and Placed* the planet pointers with their spheres as he has done with *Definition/Description/Places* of the Lunar pointer and the Sun's pointer with its golden sphere, see §6.e.).

Hypothetical phrases for the *Definition/Operation* of the other planet pointers could be:

ΑΡΕΩΣ ΠΥΡΟΕΝΤΟΣ ΦΕΡΟΜΕΝΟΝ ΓΝΩΜΟΝΙΟΝ ΕΣΤΙΝ
(*there is the traveling/rotating pointer of Mars*)

ΕΠΙ ΤΩ (ΑΡΕΩΣ) ΓΝΩΜΟΝΙ ΚΕΙΤΑΙ ΠΥΡΟΧΡΟΥΝ ΣΦΑΙΡΙΟΝ
(*on Mars' pointer is adapted a red-colored little sphere*) or

ΕΣΤΙΝ ΣΦΑΙΡΙΟΝ ΔΙΟΣ ΙΣΤΑΜΕΝΟΝ ΕΙΣ ΤΟΝ ΑΥΤΟΥ ΓΝΩΜΟΝΑΝ
(there is Zeus's sphere, adapted on its pointer)

Also a reference regarding the *Operation* of each planet's sphere should exists:

e.g. *Jupiter's sphere transits the Zodiac circle in … or rotates around in …*

Reconstructing the relative text based on Freeth and Jones 2012 assumption:

23 [-3-ΤΟ]**Υ ΑΡΕΩΣ ΠΥΡΟΕΝΤΟΣ ΤΟ ΔΕ ΔΙΑΠΟΡΕ**[ΥΟΜΕΝΟΝ <u>*ΑΥΤΟΥ ΣΦΑΙΡΙΟΝ*</u> ΠΥΡΡΟΝ -29-]

24 [ΔΙΟΣ ΦΑ]**ΕΘΟΝΤΟΣ ΤΟ ΔΕ ΔΙΑΠΟΡΕΥΟΜΕΝΟΝ** [*ΑΥΤΟΥ ΣΦΑΙΡΙΟΝ {COLOR*, Minimum 5 Letters} -37-]

(*23 Ares Pyroes and its traveling sphere in reddish color* …………………………………………………….
*25 Zeus Phaethon and its traveling sphere {color}*……………………………………………………….)

Assuming Freeth and Jones 2012 and Bitsakis and Jones 2016b words suggestion, the missing maximum number of letters, in Line 23, is 29. At the end of Line 23 the phrase ΥΠΕΡ (ΔΕ) ΤΟΝ ΠΥΡΟΕΝΤΑ ΕΣΤΙΝ Ο ΚΥΚΛΟΣ (*beyond Pyroes, there is the circle of…*) is necessary, as well as at the end of Line 24, the phrase ΥΠΕΡ (ΔΕ) ΤΟΝ ΦΑΕΘΟΝΤΑ ΕΣΤΙΝ Ο ΚΥΚΛΟΣ ΤΟΥ

ΚΡΟ (*beyond Phaethon, there is the circle of Kro-…*) (see §6.a.). Therefore, the maximum missing letters of Line 23 are 0 or -1 and the missing letters of Line 24 are 8.

23 [-3- ΤΟ]**Υ ΑΡΕΩΣ ΠΥΡΟΕΝΤΟΣ ΤΟ ΔΕ ΔΙΑΠΟΡΕ**[ΥΟΜΕΝΟΝ *ΑΥΤΟΥ ΣΦΑΙΡΙΟΝ ΠΥΡΡΟΝ* ΥΠΕΡ ΤΟΝ ΠΥΡΟΕΝΤΑ ΕΣΤΙΝ Ο ΚΥΚΛΟΣ ΤΟΥ] {*total number of letters 87*}
24 [ΔΙΟΣ ΦΑ]**ΕΘΟΝΤΟΣ ΤΟ ΔΕ ΔΙΑΠΟΡΕΥΟΜΕΝΟΝ** [*ΑΥΤΟΥ ΣΦΑΙΡΙΟΝ* - - - - - {Min 5 Letters for the color} ΥΠΕΡ ΔΕ ΤΟΥ ΦΑΕΘΟΝΤΟΣ ΕΣΤΙΝ Ο ΤΟΥ ΚΡΟ] {*total number of letters 79*}

25 [ΝΟΥ ΦΑ]**ΙΝΟΝΤΟΣ ΚΥΚΛΟΣ ΤΟ ΔΕ ΣΦΑΙΡΙΟΝ Φ**[ΕΡΟΜΕΝΟΝ - 47 -]

In the remaining letters (0/-1 in Line 23 and 7 letters in Line 24, *The Engraver* should present the *Definition/Description, Position* and *Operation*, of the rotating spheres of Mars/Jupiter, which are adapted to their corresponding pointers, as he presents the *Definition/Description, Position* and *Operation* for the Sun and Moon sphere and their pointers.

*The Engraver* dedicated (at least) two lines from the original text in order to *Define/Describe/Place* in detail, the Golden Sphere/pointer of the Sun, which was considered as one of the seven planets of the Hellenistic Astronomy:

- [ΠΕΡΙΦΕΡΕΙΑΝ] **ΗΛΙΟΥ**

- [ΕΙΣ ΟΡΘΟΣ] **ΓΝΩΜΩ**[Ν]

- **ΧΡΥΣΟΥΝ ΣΦΑΙΡΙΟΝ**

- **ΗΛΙΟΥ ΑΚΤΙΝ**

*The Engraver* also *Defines* in great detail the position of the Lunar Disc pointer:

| |
|---|
| 15 …………………………………………………………………………………………… ΑΠΟ ΤΟΥ ΚΥΛΙΝΔΡΟΥ |
| 16 **ΠΡΟΕΧΟΝ ΑΥΤΟΥ ΓΝΩΜΟΝΙΟΝ Σ**[ΕΛΗΝΗΣ - - - - - - - - - - - - - - - - - - 58- - - - - - - - - - - - - - - - - - -] |
| *15 ………………………………………………………………………………From the Lunar Disc* |
| *16 it protrudes the Lunar Pointer………………………………………………………………………* |

For the (*Definition*) and the *Operation* of the Lunar phases sphere, *The Engraver* also refers:

| |
|---|
| 12 ………ΟΤ]**Ε ΜΕΛΑΝ ΟΤ**[Ε ΛΕΥΚΟΧΡΟΥΝ |
| 13 ………………………………. |
| 14 ……………………………. |
| 15 ……… **ΤΟ ΣΦΑΙΡΙΟΝ ΦΕΡΕ**[ΤΑΙ |
| *12 …………………either black, or white colored* |
| *13 …………………………* |
| *14 …………………………* |
| *15 ……………the sphere is rotated* |

Although *The Engraver Defines/Describes/Places* the rotating parts of the Moon and Sun, a similar well-referred to presentation for the pointers and spheres of Mars and Jupiter seems difficult to exist. In the missing text there is no sufficient space in order to describe the hypothetical/suggested planet pointers and spheres as a *Definition/Position/Operation* text.

It is difficult for someone to justify "*Why The Engraver does not present the other five planet pointers and their spheres, as he presents in detail the pointers/spheres of the Sun and Moon*". *The Engraver* uses about 195 letters/ ≈2.5 text lines for the *Definition/Description, Position and Operation of* Suns' sphere and its pointer (see the full text reconstruction in **Table V**).

As this text is the User's Manual of the Antikythera Mechanism, *The Engraver* should be clear and refer to the (also important) other planet pointers and spheres (if they existed as rotating planets on the Mechanism).

The absence of a satisfying *Description/Position/Operation* of the planet pointers/spheres unjustifiably downgrades the role and the existence of the rotating planet pointers and spheres, relative to the well-presented Solar sphere and its pointer, as well as the Lunar parts.

Someone could assume that all of the other planet pointers and spheres were presented by a general reference, as a common description for all the other planet parts in one sentence. In this way, the role of the other planet pointers/spheres is also downgraded. It is clear that *The Engraver* presents each planet separately. Moreover, there is no space for this hypothetical general description about the other planet pointers, spheres, and their operation.

Although the names of the planets are referred to on the BCI Part-1, it seems that the assumption for a text regarding the *Definition Position* and *Operation* of the planet pointers/spheres is difficult to support.

On the preserved text, clear descriptions regarding the ΧΡΥΣΟΥΝ ΣΦΑΙΡΙΟΝ (Sun), ΗΛΙΟΥ ΑΚΤΙΝ (pointer), also the Lunar phases sphere and the pointer of the Lunar Cylinder, exist. A simple reference to the planet names/orbits could not be considered as proof of the existence of the planet indication gearing on the Antikythera Mechanism.

Moreover, the adaptation of about 35 hypothetical gears, plus 50 additional non-existing parts that are needed for the hypothetical planet indication gearing, raises questions and further doubts regarding the real existence of the suggested rotating planets with pointers and spheres on the Antikythera Mechanism.

**6.f. Lunar pointer traveling through and Sun's ray traveling through (Lines 17-19, 23-25)**

After the *Definition* of Mars and Jupiter's orbits follows the phrase **ΔΙΑΠΟΡΕΥΟΜΕΝΟΝ** ΣΦΑΙΡΙΟΝ (after Saturn's orbit should follow the phrase **ΤΟ ΔΕ ΣΦΑΙΡΙΟΝ ΦΕ**[ΡΟΜΕΝΟΝ (*traveling sphere*) (last preserved letter "Ε", instead of letter "Λ").

Before lines 23-25, the only clear correlation for a traveling sphere is the Golden Sphere-Sun. The ancient Greek astronomers, in order to describe the Sun's travel along the Zodiac/Ecliptic used the word ΔΙΑΠΟΡΕΥΕΤΑΙ: Geminus (in Manitius 1880 and Spandagos 2002) in Chap. 1 refers: *"Ὁ δὲ ἥλιος ἐνιαυτῷ διαπορεύεται τὸν ζῳδιακὸν κύκλον"*; in Théon d'Alexandrie 1821 in Hypomnema (page 48): *"την περιφέρειαν διαπορευόμενος ὁ ἥλιος τάς τῶν γνωμόνων σκιάς"*, also in Theon's Scholia of Aratus Soleus (Halma 1821, scholion on page 18): *"ὁ δὲ ζωδιακός, ὅτι διαπορευόμενος τοῦτον ὁ ἥλιος"*; in Autolycus of Pitane (in Mogenet 1950, Lib. 1): *"ὁ ἥλιος τὴν ΕΗ περιφέρειαν διαπορεύεται"*).

| |
|---|
| 23 [-3-ΤΟ]**Υ ΑΡΕΩΣ ΠΥΡΟΕΝΤΟΣ ΤΟ ΔΕ ΔΙΑΠΟΡΕ**[ΥΟΜΕΝΟΝ ΣΦΑΙΡΙΟΝ  - - - - -7 - - - - ΥΠΕΡ ΔΕ ΤΟΥ ΠΥΡΟΕΝΤΟΣ, ΚΥΚΛΟΣ ΕΣΤΙΝ Ο ΤΟΥ] |
| 24 [ΔΙΟΣ ΦΑ]**ΕΘΟΝΤΟΣ ΤΟ ΔΕ ΔΙΑΠΟΡΕΥΟΜΕΝΟΝ** [ΣΦΑΙΡΙΟΝ  - - - - - -20 - - - - - ΥΠΕΡ ΔΕ ΤΟΥ ΦΑΕΘΟΝΤΟΣ ΕΣΤΙΝ Ο ΤΟΥ ΚΡΟ] |
| 25 [ΝΟΥ ΦΑ]**ΙΝΟΝΤΟΣ ΚΥΚΛΟΣ ΤΟ ΔΕ ΣΦΑΙΡΙΟΝ Φ**<Ε>[ΡΟΜΕΝΟΝ - - - - - - - -41 - - - - - - - - - - - - - -] |
| *23 circle of Mars Pyroes and the traveling sphere……………….beyond Pyroes there is circle of* |
| *24 Zeus Phaethon and the traveling sphere……………………. beyond Phaethon, there is Kro-* |
| *25 –nos Phainon circle and the sphere travels…………………………..…………………………………………* |

It is logical to directly correlate the preserved words ΔΙΑΠΟΡΕΥΟΜΕΝΟΝ ΣΦΑΙΡΙΟΝ and ΣΦΑΙΡΙΟΝ ΦΕΡΟΜΕΝΟΝ (traveling/rotating sphere) with the ΔΙΑΠΟΡΕΥΟΜΕΝΟΝ ΧΡΥΣΟΥΝ ΣΦΑΙΡΙΟΝ-Golden sphere, which was aforementioned before the description of the three superior planet orbits.

In Lines 17 and 18, in which the inferior planets Mercury and Venus are referred to, the phrase (Line 18) **ΤΟ ΔΕ ΔΙ ΑΥΤΟΥ ΦΕΡΟΜΕΝΟΝ** (*and the traveling …..through it……….*) is preserved. Applying the same meaning to the traveling Golden Sphere on the superior planet orbits, the only two parts which are traveling around the inferior planet's orbits, are the Lunar phases sphere and the Lunar Disc pointer. Of these two parts, only the Lunar Disc pointer travels on the circumferences of Mercury and Venus. Therefore, the phrase **ΤΟ ΔΕ ΔΙ ΑΥΤΟΥ ΦΕΡΟΜΕΝΟΝ** is well-correlated to the Lunar Disc pointer:

| |
|---|
| 16 **ΠΡΟΕΧΟΝ ΑΥΤΟΥ ΓΝΩΜΟΝΙΟΝ Σ**[ΕΛΗΝΗΣ - - - -33- - - ΤΟ ΓΝΩΜΟΝΙΟΝ ΦΕΡΕΤΑΙ ΕΠΙ ΤΩΝ ΠΕΡΙ] |
| 17 **ΦΕΡΕΙΩΝ, Η ΜΕΝ ΕΧΟΜΕΝΗ ΤΩΙ ΤΗΣ** [ΣΕΛΗΝΗΣ ΚΥΛΙΝΔΡΩ, Η ΔΕ ΕΠΟΜΕΝΗ. ΠΡΩΤΗ ΕΣΤΙΝ Η ΠΕΡΙΦΕΡΕΙΑ ΤΟΥ ΕΡΜΟΥ ΣΤΙΛΒΟΝ] |
| 18 **ΤΟΣ. ΤΟ ΔΕ ΔΙ ΑΥΤΟΥ ΦΕΡΟΜΕΝ**[ΟΝ ΤΗΣ ΣΕΛΗΝΗΣ ΓΝΩΜΟΝΙΟΝ - -25- - ΥΠΕΡ ΔΕ ΤΟΝ ΣΤΙΛΒΟΝΤΑ], |
| 19 **ΤΗΣ ΑΦΡΟΔΙΤΗ**<Σ> **ΦΩΣΦΟΡΟΥ** [ΠΕΡΙΦΕΡΕΙΑΝ ΕΣΤΙΝ. ΤΟ ΔΕ ΔΙ ΑΥΤΗΣ ΦΕΡΟΜΕΝΟΝ ΣΕΛΗΝΗΣ ΓΝΩΜΟΝΙΟΝ - - 8 - - ΥΠΕΡ ΔΕ] |

*16 projecting from it the lunar Disc pointer ……. The pointer travels through the two circu-*
*17 mferences, the one is located close to the Moon and the next follows. Beyond Moon, there is the circumference of Mercury Sti-*
*18 bon the traveling lunar pointer through this circle ………………… Beyond Stilbon there is*
*19 Aphrodites Phosphoros circumference, and the traveling pointer ………………… Beyond…*

From the analysis of §6.e.4., we can conclude that the existence of the planet pointers/spheres is difficult to support. On the other hand, the existence of the Lunar and Solar pointers is real and definite. Both of the pointers travel along the planet (inferior/superior) orbits:

The Lunar Disc pointer travels along the orbits of Mercury and Venus (Lines 16-17).

The Sun's sphere with its ΗΛΙΟΥ ΑΚΤΙΝ/indicator/pointer travels along the orbits of Mars, Jupiter and Saturn (Line 22).

During the Sun and the Moon independent rotation, at some point they meet/cross with a planet. This event is called conjunction, ΣΥΝΟΔΟΣ.

A conjunction (ΣΥΝΟΔΟΣ) of the Moon and Mercury occurs approximately every 30 days = Λ.[71]

A conjunction of the Moon and Venus also occurs about every 30 days.[72]

A conjunction of the Sun and Mars occurs about every 780 days=$2^y$ $1^{mon}$ $20^d \approx 2 + ^1/_7$ years = BL Z (L= year),[73]

The Sun and Jupiter about every 399 days=$1^y$ $1^{mo}$ $3.5^d \approx 1 + ^1/_{10}$ = AL I,

The Sun and Saturn about every 378.09 days= $1^y$ $13^d \approx 1 + ^1/_{28}$ = AL KH.[74]

Using this information, *The User* of the Mechanism could also calculate the position (constellation) of a planet which will be in conjunction with the Sun/Moon, if he knows the initial position of each planet in the Zodiac during the Initial Calibration Date of the Antikythera Mechanism (i.e. the date which was defined by the ancient manufacturer as the initial date for the pointers' position).

---

[71] Varies between 27 and 31 days. The equation for the calculation of the inferior planets synodic rotation was used. http://astronomyonline.org/science/siderealsynodicperiod.asp.

[72] Calculations based on the data from the description of Geminus in Manitius 1880 and Spandagos 2002, also in Pseudo-Plutarch 892b-c in Plutarch (in Bernardakis 1893). The equation for the calculation of the inferior planets synodic rotation was used.
http://astronomyonline.org/science/siderealsynodicperiod.asp.

[73] Calculated data, based on the Antikythera Mechanism Front cover Inscription, Anastasiou et al., 2016a. The equation for the calculation of the superior planets synodic rotation was used.
http://astronomyonline.org/science/siderealsynodicperiod.asp.

[74] See Fragment 19; Bitsakis and Jones 2016b. From the preserved text of the AM, it seems that he ancient engraver does not use the usual punctuation for the numbering system (ι′, κ′, λ′,….. for 10, 20, 30, …). In Bitsakis and Jones 2016 (Back cover) page 235 and 246: …ΤΗΝ ΤΗΣ ΟCL ΙΘL ΤΟΥ… (…της 19ετούς περιόδου των 76 ετών…). In Anastasiou et al., 2016b (Front Cover) page 266, in about ten preserved (large) numbers there is no punctuation for the numbers. Generally the engraver does not use any punctuation when he refers to numbers. Fractions arise from the way numbers are been written: First, the number of full years is referred, then, the year symbol "L" and logically any number after the symbol "L" should be the fraction of the year (before "L" full years, after "L" fraction of year): BL Z: 2 ΕΤΗ (years) + 1/7, AL KH: 1 ΕΤΟΣ (year) + 1/28.

We suppose that during the initial calibration date of the Mechanism[75] the Sun was on the 1st Zodiac day of Capricorn, and Jupiter was in conjunction with the Sun. *The User* can estimate the position of Jupiter after 5 conjunctions (i.e. in which constellation it will be located), starting from the initial calibration date:

$5 \times 399^d = 1995^d$, $1995^d/365.25^d = 5.462^y = 5$ full years $+ 0.462^y$. Then, $0.462^y = 168.7^d = 166.3°$ western to Sun, $166.3/30° = 5.544$ dodecatemoria $\approx 5$ full $+ 0.5$ dodecatemoria after 1st day of Capricorn i.e. around the middle area of the constellation of Gemini.

*The Engraver* would also refer to the time of the Lunar pointer's conjunction with each of the inferior planets and the time for the Solar pointer's conjunction with each of the superior planets.

On the corresponding circumference/circle (orbit) was sketched on the central front cover of gear b1, he could also mark the position of each planet during the specific starting date of the Mechanism pointers.

Taking into account the aforementioned discussion, the text reconstruction of the lunar conjunction (ΕΝ ΣΥΝΟΔΩ) with Mercury and Venus is:

| |
|---|
| 17 …………………………………………………… ΠΡΩΤΗ ΕΣΤΙΝ Η ΠΕΡΙΦΕΡΕΙΑ ΤΟΥ ΕΡΜΟΥ ΣΤΙΛΒΟΝ] |
| 18 **ΤΟΣ. ΤΟ ΔΕ ΔΙ ΑΥΤΟΥ ΦΕΡΟΜΕΝ**[ΟΝ ΤΗΣ ΣΕΛΗΝΗΣ ΓΝΩΜΟΝΙΟΝ, ΕΝ ΣΥΝΟΔΩ ΗΜΕΡΑΙΣ Λ. ΥΠΕΡ ΔΕ ΤΟΝ ΣΤΙΛΒΟΝΤΑ] |
| 19 **ΤΗΣ ΑΦΡΟΔΙΤΗ<Σ> ΦΩΣΦΟΡΟΥ** [Η ΠΕΡΙΦΕΡΕΙΑ ΕΣΤΙΝ. ΤΟ ΔΕ ΦΕΡΟΜΕΝΟΝ (ΣΕΛΗΝΗΣ) ΓΝΩΜΟΝΙΟΝ ΕΝ ΣΥΝΟΔΩ ΗΜΕΡΑΙΣ Λ. ………………… |
| 17 …………………………………………beyond Moon, there is the circumference of Mercury Sti 18 von the traveling lunar pointer in conjunction at 28 days. Beyond Stilbon 19 there is Aphrodites Phosphoros circumference, and the traveling Lunar pointer in conjunction at 30 days |

The text reconstruction of the Sun's conjunction with the three superior planets is:

| |
|---|
| 20 …………………………………………… ΕΙΣ ΤΗΝ ΤΟΥ ΗΛΙΟΥ ΠΕΡΙΦΕΡΕΙΑΝ ΕΙΣ ΣΤΑΘΕΡΟΣ ΚΑΙ ΟΡΘΟΣ |
| 21 **ΓΝΩΜΩ**[Ν] **ΚΕΙΤΑΙ. ΧΡΥΣΟΥΝ ΣΦΑΙΡΙΟΝ** [ΕΠΙ ΓΝΩΜΟΝΑ ΕΣΤΙΝ, ΕΝ ΕΝΙΑΥΤΩ ΔΙΑΠΟΡΕΥΟΜΕΝΟΝ - - - - - 21 - - - - - -] |
| 22 …………………………………………………………………… ΠΡΩΤΟΣ ΚΥΚΛΟΣ ΕΣΤ |
| 23 [ΙΝ Ο ΤΟ]**Υ ΑΡΕΩΣ ΠΥΡΟΕΝΤΟΣ ΤΟ ΔΕ ΔΙΑΠΟΡΕ**[ΥΟΜΕΝΟΝ (ΧΡΥΣΟΥΝ) ΣΦΑΙΡΙΟΝ ΕΝ ΣΥΝΟΔΩ ΒL Z. ΥΠΕΡ ΔΕ ΤΟΝ ΠΥΡΟΕΝΤΑ, ΚΥΚΛΟΣ ΕΣΤΙΝ] |
| 24 [ΔΙΟΣ ΦΑ]**ΕΘΟΝΤΟΣ, ΤΟ ΔΕ ΔΙΑΠΟΡΕΥΟΜΕΝΟΝ** [(ΧΡΥΣΟΥΝ) ΣΦΑΙΡΙΟΝ ΕΝ ΣΥΝΟΔΩ ΑL Ι. |

---

[75] The initial calibration date of the Antikythera Mechanism defines the initial specific position for each of the seven pointers of the Mechanism.

> ΥΠΕΡ ΔΕ ΤΟΝ ΦΑΕΘΟΝΤΑ ΕΣΤΙΝ Ο ΤΟΥ ΚΡΟ]
>
> 25 [ΝΟΥ ΦΑ]**ΙΝΟΝΤΟΣ ΚΥΚΛΟΣ, ΤΟ ΔΕ** (ΧΡΥΣΟΥΝ) **ΣΦΑΙΡΙΟΝ** Φ[ΕΡΟΜΕΝΟΝ ΕΝ ΣΥΝΟΔΩ ΑL ΚΗ.
>
> *20 ................................................On the Sun's circle is located a perpendicular pillar. On the*
> *21 pillar is adapted the Golden Sphere, rotated in one full turn per year...............................*
> *22 ........................................................................................................Firstly, is the circle of*
> *23 Ares Pyroes. The traveling sun's sphere in conjunction at 2 plus 1/7 years. Beyond Pyroes there is the circle of*
> *24 Zeus Phaethon. The traveling (sun's golden) sphere in conjunction to 1 year plus 1/10. Beyond Phaethon, there is*
> *25 Saturn Phainon circle. The traveling (sun's golden) sphere in conjunction at 1 year plus 1/28*

As the pointer of the Sun, i.e. the Golden Sphere (ΧΡΥΣΟΥΝ ΣΦΑΙΡΙΟΝ) with the Solar Ray (ΗΛΙΟΥ ΑΚΤΙΝ), rotates once per tropical year (ΕΝ ΕΝΙΑΥΤΩ ΔΙΑΠΟΡΕΥΟΜΕΝΟΝ or ΕΝ ΕΝΙΑΥΤΩ ΦΕΡΟΜΕΝΟΝ), a relative reference for the Lunar Cylinder's pointer (ΓΝΩΜΟΝΙΟΝ ΣΕΛΗΝΗΣ), should exist.

> 16 **ΠΡΟΕΧΟΝ ΑΥΤΟΥ ΓΝΩΜΟΝΙΟΝ** Σ[ΕΛΗΝΗΣ. ΕΣΤΙΝ ΕΝ ΣΥΝΟΔΩ ΑΣΤΡΑΣΙ ΗΜΕΡΑΙΣ ΚΖ Γ. ΤΟ ΓΝΩΜΟΝΙΟΝ ΦΕΡΕΤΑΙ ΕΠΙ ΤΩΝ ΠΕΡΙ]
>
> 17 **ΦΕΡΕΙΩΝ Η ΜΕΝ ΕΧΟΜΕΝΗ ΤΩΙ ΤΗΣ** [====================63====================]
>
> *16 it protruded the lunar pointer, which comes in conjunction to the stars in 27.3 days. The lunar pointer travels along the two circu-*
> *17 -mferences ........................................................................................................*

Therefore, a text for the Lunar pointer *Operation*, is presented. The lunar pointer protrudes from the Lunar Cylinder and rotates with it. One full turn of the Lunar pointer represents the Sidereal lunar cycle.[76] The Sidereal cycle of the moon is the time needed for the moon to return to the same position in the Sky/stars, i.e. to the same zodiac constellation and position. The Moon turns around the celestial sphere (as observed from the Earth) in about 27.3 days (ΚΖ Γ).[77]

In this way, the two basic lunar cycles, the Synodic and Sidereal (235 Synodic cycles equal to 254 Sidereal cycles), were extensively used since Meton's era (around 430BC) and during the Mechanism's era (200BC-100BC). This can be referred to on the User's Manual of the Mechanism (and also represented via the Mechanism's pointers): the Synodic cycle, via the half black/half white lunar phases sphere, the lunar days scale (see § 6.d.) and the Sidereal cycle via the rotation of the Lunar pointer around the Mechanism's Ecliptic Sky/Zodiac ring (see §6.h.).

**6.g. The ΚΟΣΜΟΣ of the Antikythera Mechanism (Lines 25-26)**

---

[76] Voulgaris et al., 2018b.
[77] Voulgaris et al., 2018b.

After presenting the planet orbits KYKΛOI in Lines 17-25, *The Engraver* refers to the word KOΣMOY (*Cosmos-World*) in the next line:

| |
|---|
| 26 [- - - - - - Π]ΕΡΑ ΔΕ ΤΟΥ ΚΟΣΜΟΥ ΚΕΙΤΑΙ [- - - - - - - - - - - - - - - - - -59- - - - - - - - - - - - - - - - - - - -] |
| *26 - - - - - - beyond the Cosmos there is…………………………………………………………………………………………* |

The word Π]**ΕΡΑ**[78] implies that something other follows after the KOΣMOΣ of the Mechanism, but the *Definition* of KOΣMOΣ is missing and should be referred to between the end of the previous line (Line 25) and the beginning of Line 26. The Hellenistic KOΣMOΣ starts with the Earth, *The Center of Cosmos* and the last boundary of Cosmos is defined by the orbit of Saturn KYKΛOΣ KPONOY, **Fig. 12**.

In Lines 7-25 *The Writer Presents/Defines* the parts of KOΣMOΣ. At the end of Line 25, he summarizes these parts, *Defining* the existing Cosmos:

| |
|---|
| 25 ………………………………………………………………………………………………… OYTOΣ MEN EΣTIN O THN ΣHMEPON] |
| 26 [ΚΟΣΜΟΣ. Π]**ΕΡΑ ΔΕ ΤΟΥ ΚΟΣΜΟΥ ΚΕΙΤΑΙ** [- - - - - - - - - - - - - - - -59- - - - - - - - - - - - - - - - - - - ] |
| *25 …………………………………………………………………………………………………This is today the Cosmos* |
| *26 out of this Cosmos, there is…………………………………………………………………………………………………………………* |

In Line 26 the *Definition/Position* of an additional *operational part*, which is located "*beyond/out of Cosmos*", is presented (see next paragraph).

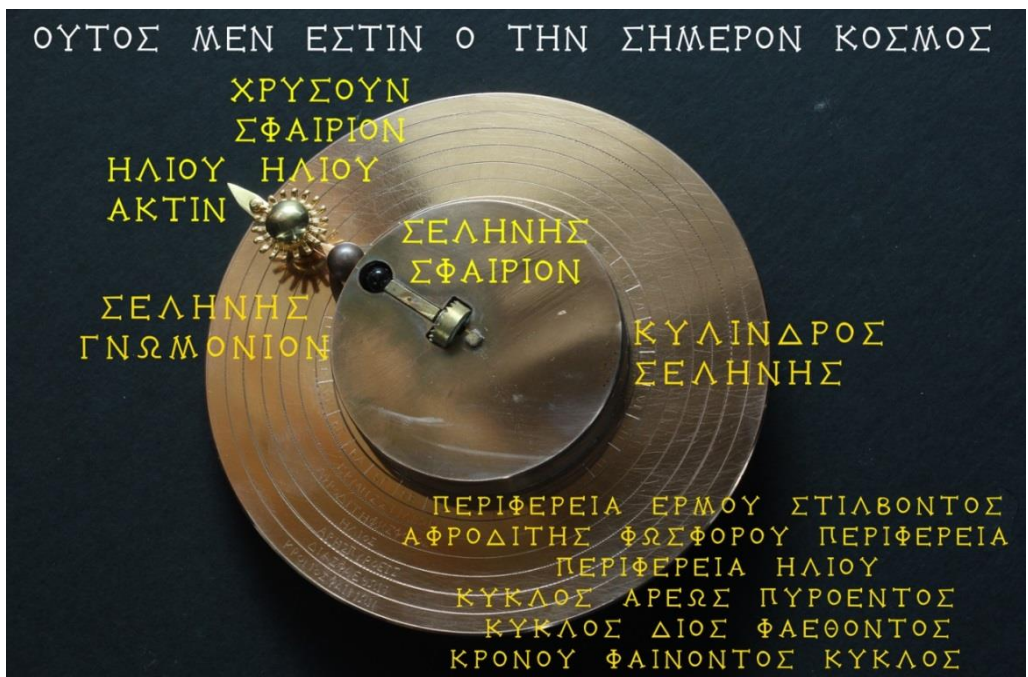

Fig. 12. The Presentation of the Hellenistic KOΣMOΣ (Cosmos) of the Antikythera Mechanism: The Earth is located in the geometrical center, the Lunar Disc/Cylinder, the Lunar Phases Sphere, the orbits of the inferior planets, the Golden Sphere with the Solar Ray-pointer and the orbits of the superior planets. Material and image by the Authors

---

[78] For the word ΠΕΡΑ/ΠΕΡΑΝ see LSJ lexicon

**6.h. The *Sky* and the Parapegma presentation in the Mechanism User's Manual (Lines 26-30)**

After the reference of the Lunar Disc, the seven planets and the Definition of ΚΟΣΜΟΣ, the text of the Antikythera Mechanism User's Manual continues in Lines 26-30.

During Hellenistic astronomy, after the last orbit of planet Saturn follows the Celestial Sphere - the Sky with the ΑΠΛΑΝΕΙΣ ΑΣΤΕΡΕΣ (stars with a fixed position in the sky).

In contrast, ΠΛΑΝΗΤΕΣ (*planets*) were considered as stars which changed their position relative to the "fixed" background" of the ΑΠΛΑΝΕΙΣ ΑΣΤΕΡΕΣ (*stars*).[79]

On the Antikythera Mechanism, the Sky is represented by the Zodiac month ring-the Ecliptic sky, **Fig. 13**. The Zodiac month ring is divided into twelve Zodiac months and into 365 equal subdivisions/Zodiac days.[80] Each Zodiac month has a different number of days, therefore the Zodiac ring is divided into twelve unequal parts/months. On some of the Zodiac days/subdivisions, *The Engraver* noted the index letters of the Parapegma events, probably corresponding to the word ΣΤΟΙΧΕΙΑ.[81] Beyond the Zodiac ring, the Egyptian calendar ring follows, divided into 12X30 + 5 Epagomenai days.[82]

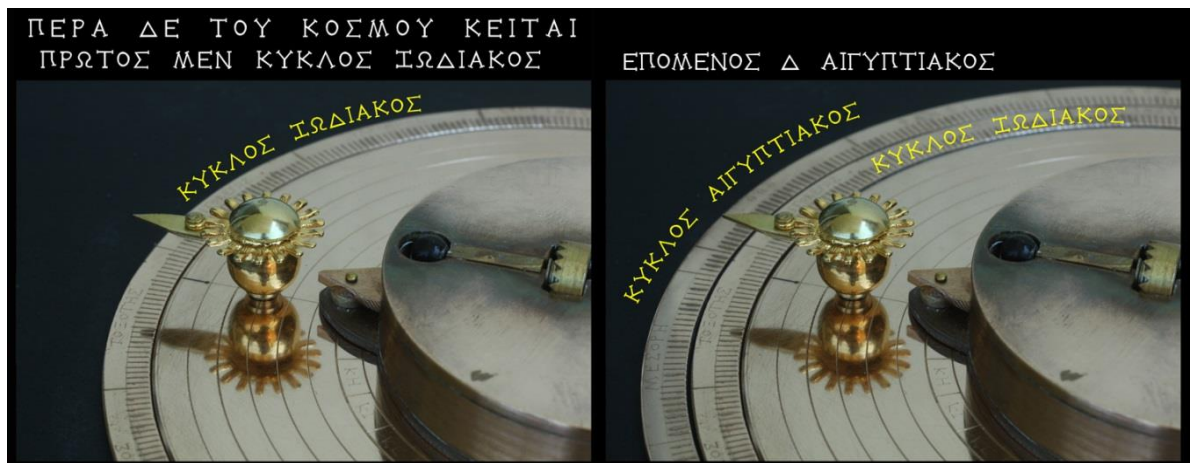

Fig. 13. Beyond Cosmos, there is the Zodiac month ring which is the Ecliptic Sky of the Mechanism. The Egyptian calendar ring follows. Material and image by the Authors

The positions of ΣΤΟΙΧΕΙΑ correspond to the dates of the Parapegma events (morning/evening, rising/setting). A possible phrase for the Parapegma events reference could be the phrase ΑΠΛΑΝΩΝ (ΑΣΤΕΡΩΝ) ΓΕΓΟΝΟΤΑ (star events). The star events were engraved on the two oblong Parapegma plates. On the top plate are engraved the Winter

---

and the Vernal months and on the bottom plate are engraved the Summer and the Autumn months.[83]

In lines 28 and 30 the word **ΑΣΠΙΔ**ΙΣΚΑΙ (*small shields*), which has many different meanings, is referred to. The ΑΣΠΙΣ (*shield*) was a defensive weapon, extensively used by the ancient Greek hoplites and Roman soldiers. During the Greco-Persian Wars (507BC-449BC), Great Alexander's kingdom (up to 323 BC) and the Hellenistic Era (323BC-31 BC), the Greek shields had a circular shape.[84] After the changes in the Roman Military System, the roman soldiers-*legionaries* used shields in (curved) oblong/rectangular shape, named *Scutum*.[85] Additionally, the word ΑΣΠΙΔΙΣΚΗ (small shield) was used to name the decorative pin heads (ΕΦΗΛΙΔΕΣ), with a circular shape and design, as a shield. These pins with heads were placed on the outer surface of wooden doors, on windows or furniture.[86] ΑΣΠΙΔΙΣΚΑΙ were also named the decorative parts (pin button/stick pin) used on the ancient Greek garments and metal utensils.[87]

Bitsakis and Jones 2016a suggest that the ΑΣΠΙΔΙΣΚΑΙ referred on BCI text are the four circular buttons on the corners of the Front dial plate, which were adapted on the sliding catches, stabilizing the Front Dial plate. These buttons seem more as *supportive (decorative)/parts*, than *operational parts* of the Mechanism and do not affect the operation of the Mechanism.

The word ΑΣΠΙΔΙΣΚΑΙ, as it is referred to on the Mechanism User's Manual right after the ΚΟΣΜΟΣ and the two rings (Zodiac and Egyptian), can be well-correlated to the two independent oblong plates, located on the top and the bottom Front face, **Fig. 14**, in which the Parapegma star events were engraved.[88]

26 [ΚΟΣΜΟΣ Π]**ΕΡΑ ΔΕ ΤΟΥ ΚΟΣΜΟΥ ΚΕΙΤΑΙ**, [ΠΡΩΤΟΣ ΜΕΝ ΚΥΚΛΟΣ ΖΩΔΙΑΚΟΣ, ΕΠΟΜΕΝΟΣ Δ ΑΙΓΥΠΤΙΑΚΟΣ. ΤΑ ΤΩΝ ΑΠΛΑΝ]

27 [ΩΝ ΑΝΑΓΡΑΦΟ]**ΜΕΝ**Α **ΣΤΟΙΧΕΙΑ ΠΑΡΑΚΕΙΜ**[ΕΝΑ ΕΝ ΤΑΙΣ ΗΜΕΡΑΙΣ ΤΟΥ ΖΩΔΙΑΚΟΥ ΚΥΚΛΟΥ ΕΣΤΙ. ΠΑΡΑΤΙΘΕΝΤΑΙ]

28 [ΔΕ ΤΑ ΓΕΓΟΝΟΤΑ Τ]**ΑΥΤΑ ΤΑΙΣ ΑΣΠΙΔ**[ΙΣΚΑΙΣ ΑΙ ΕΙΣΙΝ ΑΝΩΤΕΡΩ ΚΑΙ ΚΑΤΩΤΕΡΩ ΤΩΝ ΕΙΣ ΚΕΝΤΡΟΝ ΚΥΚΛΩΝ.]

29 [ΤΑ ΤΩΝ ΑΠΛΑΝΩΝ] **ΠΡΟΕΙΡΗΜΕΝΑ** [ΓΕΓΟΝΟΤΑ ΕΩΙΩΝ ΤΕ ΕΣΠΕΡΙΩΝ ΚΑΙ ΕΠΙΤΟΛΩΝ ΤΕ ΔΥΣΕΩΝ ΑΝΑΓΡΑΦΟΝΤΑΙ ΑΥΤΟΘΙ]

30 [ΕΝ ΤΗ ΑΝΩΤΕΡΑ ΕΚ ΤΩΝ] **ΑΣΠ**[ΙΔΙΣΚΩΝ, ΟΙ ΧΕΙΜΕΡΙΝΟΙ ΚΑΙ ΕΑΡΙΝΟΙ ΜΗΝΕΣ ΖΩΔΙΑΚΟΙ ΚΑΙ ΕΝ

---

| |
|---|
| ΤΗ ΚΑΤΩΤΕΡΑ ΟΙ ΘΕΡΙΝΟΙ ΚΑΙ] |
| 31 [ΦΘΙΝΟΠΩΡΙΝΟΙ ΜΗΝΑΙ ΖΩΔΙΑΚΟΙ ΠΑΡΑΤΙΘΕΝΤΑΙ ============49=============] |
| *26 beyond Cosmos, firstly there is the Ecliptic ring and follows the Egyptian ring. The stars'* <br> *27 index letters are presented on the Zodiac days (subdivisions) of the Zodiac ring* <br> *28 the star events are engraved on the two oblong plates on top and bottom from the central cycles* <br> *29 the aforementioned morning/evening rising/setting star events are engraved* <br> *30 on the top oblong plate the winter and the vernal zodiac months and on the bottom oblong plate the summer and* <br> *31 the autumnal zodiac months are presented……………………………………..* |

After Line 30(31,…), *The Engraver* continues the text for the *Definition/Description/Position/Operation* of the *operational* outer parts, visible to the naked eye on the Back face of his creation (Metonic and Saros Helices).[89] The Back Cover Inscription Part-2 is also partially preserved.[90]

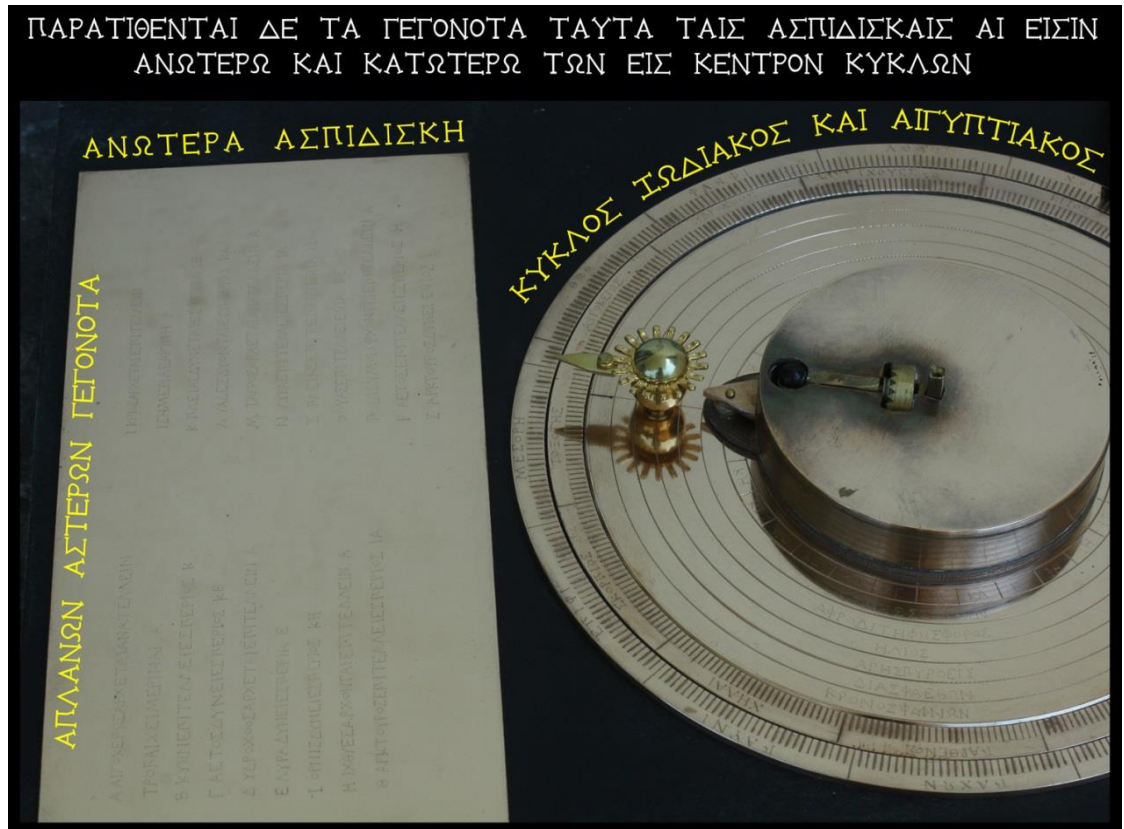

Fig. 14. On the top and bottom of the central circles there are two rectangular plates on which the Parapegma events are engraved. Material and image by the Authors

### 7.a. Summarizing the Antikythera Mechanism Front Face parts

In the previous chapters, the text which was located on the Back Cover of the Mechanism was analyzed and reconstructed. Based on the preserved and the reconstructed text, the Front face outer parts were reconstructed in bronze material by the authors.

---

[89] Freeth et al., 2006; Voulgaris et al., 2021.
[90] Bitsakis and Jones 2016b.

On **Table IV** the outer parts of the Front face are presented, starting from the Earth, the Center of Cosmos up to the last parts, the two ΑΣΠΙΔΙΣΚΑΙ the Parapegma oblong parts, **Fig. 15**. On this table, the *Definition/Naming*, *Presentation/Description*, *Position* and *Operation* of each part is presented.

| | Front face Part with Ancient Greek name | OPERATIONAL PART/ DEFINITION | *DESCRIPTION* | *POSITION* | *OPERATION* |
|---|---|---|---|---|---|
| 1 | ΓΑΙΑ ΤΟ ΚΕΝΤΡΟΝ ΤΟΥ ΚΟΣΜΟΥ | Central geometrical point, b*in* axis | ΓΑΙΑ (Earth) | On the Front face geometrical center | The Center of Cosmos |
| 2 | ΚΥΛΙΝΔΡΟΣ/ ΔΙΣΚΟΣ ΤΗΣ ΣΕΛΗΝΗΣ | Lunar Cylinder or Lunar Disc | A rotating Cylinder | Stabilized on the Front face's geometrical center | The driving/Input of Mechanism |
| 3 | ΣΦΑΙΡΙΟΝ ΦΑΣΕΩΝ ΤΗΣ ΣΕΛΗΝΗΣ | Lunar phases little sphere | A colored sphere half black, half white | At the circumference of the Lunar Disc | Represents New moon and Full moon during its rotation |
| 4 | ΣΕΛΗΝΗΣ ΓΝΩΜΟΝΙΟΝ | Lunar Pointer | A pointer is protruded from the Lunar Disc | Is stabilized on the Lunar Disc | Rotating and traveling through Mercury and Venus circumferences |
| 5 | ΣΕΛΗΝΗΣ ΗΜΕΡΑΙ (ΣΥΝΟΔΙΚΟΣ ΚΥΚΛΟΣ ΣΕΛΗΝΗΣ) | Days of the Lunar Synodic month | A circular scale divided in 29.5 sectors-days | Engraved around the Lunar Cylinder | A measuring scale with subdivisions/days |
| 6 | ΠΕΡΙΦΕΡΕΙΑ ΕΡΜΟΥ ΣΤΙΛΒΟΝΤΟΣ | Circumference of Mercury | A circumference around the Earth | After Lunar Disc's circumference | An engraved circle (no part) |
| 7 | ΠΕΡΙΦΕΡΕΙΑ ΑΦΡΟΔΙΤΗΣ ΦΩΣΦΟΡΟΥ | Circumference of Venus | A circumference around the Earth | Beyond Mercury's circumference | An engraved circle (no part) |
| 8 | ΠΕΡΙΦΕΡΕΙΑ ΗΛΙΟΥ | Circumference of Sun | A circle around the Earth | Beyond Venus' circumference | An engraved circle (no part) |
| 9 | ΓΝΩΜΩΝ ΗΛΙΟΥ | A pillar of Sun | Pillar | Perpendicular stabilized on Sun's circumference | Mechanical part for other parts stabilization |
| 10 | ΧΡΥΣΟΥΝ ΣΦΑΙΡΙΟΝ | Golden Sphere | Sphere with Golden color | Stabilized on the Sun's pillar | Rotating around the Earth in one Tropical year |
| 11 | ΗΛΙΟΥ ΑΚΤΙΝ | Sun Ray-pointer | Sun's indicator/pointer | Stabilized on the Golden Sphere | Rotates and travels through the three outer planet circles |
| 12 | ΚΥΚΛΟΣ ΑΡΕΩΣ ΠΥΡΟΕΝΤΟΣ | Circle of Mars | A circle around the Earth | Beyond Sun's circumference | An engraved circle (no part) |
| 13 | ΚΥΚΛΟΣ ΔΙΟΣ ΦΑΕΟΟΝΤΟΣ | Circle of Jupiter | A circle around the Earth | Beyond Mars' circle | An engraved circle (no part) |
| 14 | ΚΥΚΛΟΣ ΚΡΟΝΟΥ ΦΑΙΝΟΝΤΟΣ | Circle of Saturn | A circle around the Earth | Beyond Jupiter's circle | An engraved circle (no part) |
| 15 | ΚΟΣΜΟΣ | Cosmos | The present Cosmos | From the Earth up to Saturn's circle | Definition |
| 16 | ΚΥΚΛΟΣ ΖΩΔΙΑΚΟΣ | Zodiac Month ring | Ring divided in twelve Zodiac constellations and subdivisions | Beyond Cosmos | The Celestial *Sky* of the Mechanism (Ecliptic) |
| 17 | ΚΥΚΛΟΣ ΑΙΓΥΠΤΙΑΚΟΣ | Egyptian Month ring | The ring with the Egyptian Months | After the Zodiac month ring | Additional Calendar |
| 18 | ΑΣΠΙΔΙΣΚΑΙ | Small shields | Oblong engraved bronze plates | On top and bottom the central rings | Plates for Parapegma events engraving |

| 19 | ΣΤΟΙΧΕΙΑ ΑΠΛΑΝΩΝ ΑΣΤΕΡΩΝ | Star's letters | Index numbers of the Parapegma events | Engraved on the Zodiac month subdivisions | Index number |
|----|--------------------------|----------------|----------------------------------------|-------------------------------------------|--------------|
| 20 | ΓΕΓΟΝΟΤΑ ΑΠΛΑΝΩΝ ΑΣΤΕΡΩΝ | Star events text | Parapegma events | Engraved on top/bottom oblong plate shields | Star events calendar |

Table IV. Synoptic Catalogue of the Antikythera Mechanism Front face parts
Defined/Described/Positioned/Operation were presented on the BCI-Part 1

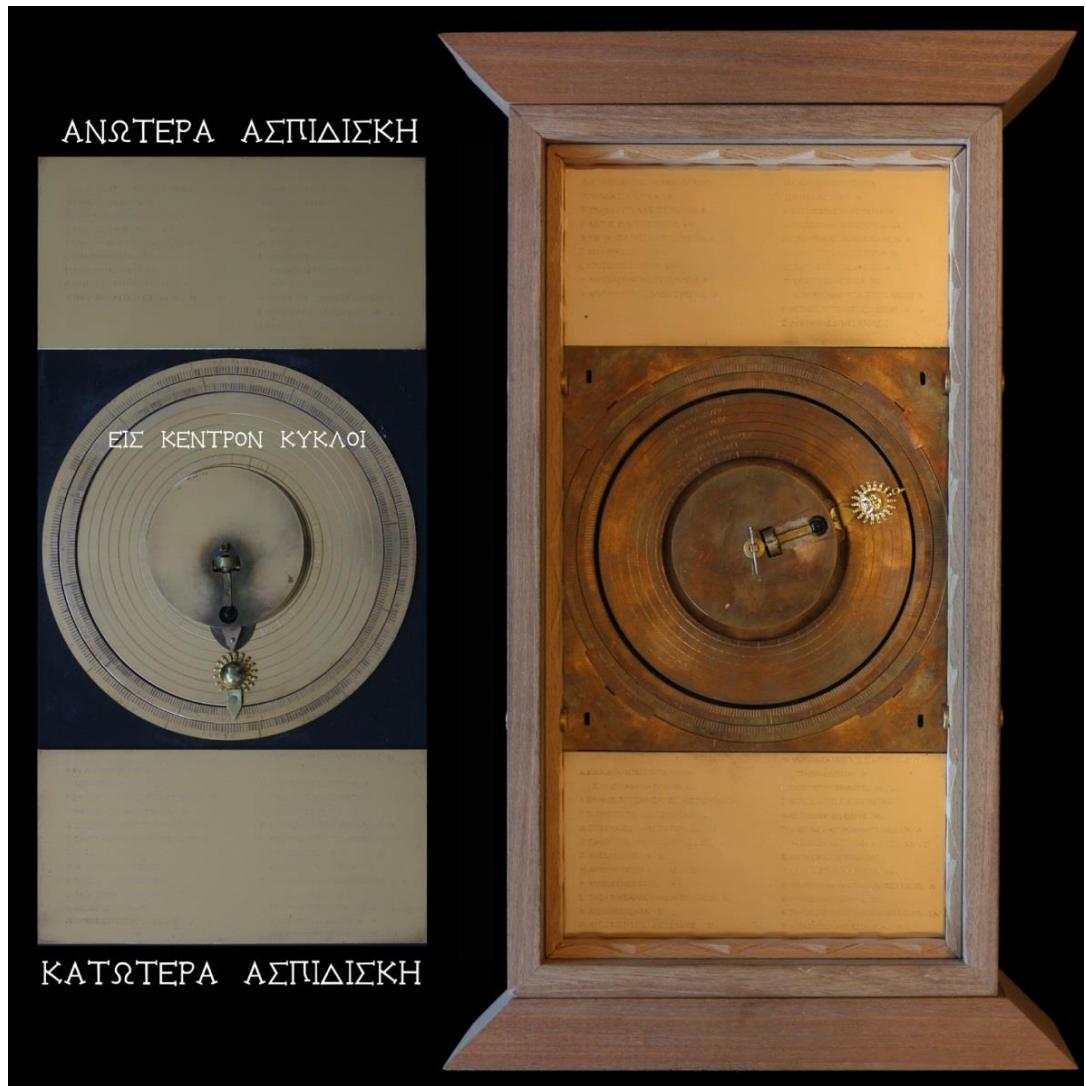

Fig. 15. Left, assembling the parts which are *Defined/Described/Referred* on the BCI Part-1 text. Right, one of the Antikythera Mechanism functional models designed/constructed by authors (after the final assembling of the parts presented on BCI Part-1). The oxidized central front area was deliberately not polished. Parts constructed in bronze material (94%Cu, 6%Tin). The wooden design of this model is a design view of the first Author. Images by the Authors

### 7.b. Symmetry on the Mechanism's Front face two pointers' operation and Symmetry on the text

1) From the analysis of the preserved text of BCI Part-1 we concluded that the rotating

pointers of the Mechanism's Front face are two: the Lunar Pointer and the Solar

Pointer/Solar Ray. As presented in **Table IV**, both pointers present symmetry in their operation:

| Rotating Parts around Earth ($b_{in}$ axis) | The two rotating spheres of the Mechanism | The location of the spheres | Rotation of the two spheres | The two pointers of the Mechanism's Front face | Travelling pointer through the planet's orbits | Synodic duration of moon and inferior/Sun and superior |
|---|---|---|---|---|---|---|
| *LUNAR DISC/ CYLINDER* | *Lunar Phases sphere* | *Located on the Lunar orbit (circumference)* | *Lunar Sphere: One rotation/ synodic month MHN ΣΥΝΟΔΙΚΟΣ* | *Lunar pointer, protruded from the Lunar Disc* | *The Lunar pointer travels through the orbits of the two inferior planets* | *The Lunar pointer in conjunction to Mercury and Venus* |
| *GEAR b1/ SOLAR CIRCUMFE RENCE* | *Golden Sphere* | *Located on the Sun's orbit (circumference)* | *Golden Sphere: One rotation per tropical year (ΕΝΙΑΥΤΟΣ)* | *Solar Ray protruded from the Golden Sphere* | *The Solar Ray travels through the orbits of the three superior planets* | *The Solar Ray in conjunction to Mars, Jupiter and Saturn* |

Table IV: A parallel correlation/Symmetry between the Lunar Disc and the solar circle, including their operational parts

2) A Symmetry is revealed in the text for the parts' presentation:

I) ΑΠΟ (ΣΕΛΗΝΗΣ) ΚΥΛΙΝΔΡΟΥ **ΠΡΟΕΧΟΝ ΑΥΤΟΥ ΓΝΩΜΟΝΙΟΝ** ΣΕΛΗΝΗΣ
II) ΑΠΟ (ΧΡΥΣΟΥ) ΣΦΑΙΡΙΟΥ ΠΡΟΕΧΕΙ ΑΥΤΟΥ Η ΤΟΥ **ΗΛΙΟΥ ΑΚΤΙΝ**

i) *From the Lunar cylinder is protruded the Lunar pointer*
ii) *From the (Golden) sphere is protruded the Solar Ray-pointer*

I) ΤΟ ΓΝΩΜΟΝΙΟΝ ΦΕΡΕΤΑΙ ΤΩΝ ΠΕΡΙ**ΦΕΡΕΙΩΝ** Η ΜΕΝ ΕΧΟΜΕΝΗ… ΚΑΙ ΔΕ ΕΠΟΜΕΝΗ
II) **ΚΥ**ΚΛΟΣ ΠΡΩΤΟΣ ΜΕΣΟΣ ΕΣΧΑΤΟΣ ΟΥΣ Η (ΗΛΙΟΥ) ΑΚΤΙΝ ΦΕΡΕΤΑΙ

i) *The Lunar pointer travels through the circumferences (inferior planets orbits) on the first and on the next*
ii) *The solar ray-pointer travels through the first, middle, last circles (superior planets orbits)*

I) **ΤΟ** (ΣΕΛΗΝΗΣ) **ΣΦΑΙΡΙΟΝ ΦΕΡΕ**ΤΑΙ ΕΝ ΗΜΕΡΑΙΣ ΚΘ Β ΟΥΤΟΣ ΜΗΝ ΣΥΝΟΔΙΚΟΣ ΚΑΛΕΙΤΑΙ
II) **ΧΡΥΣΟΥΝ ΣΦΑΙΡΙΟΝ** (ΗΛΙΟΥ) ΕΙΣ ΕΝΙΑΥΤΟΝ ΦΕΡΟΜΕΝΟΝ

i) *The Lunar sphere is rotated in 29.5 days, named synodic month*
ii) *The Golden sphere is rotated in one tropical year/Eniautos.*

I) ΠΡΩΤΗ ΕΣΤΙΝ Η ΠΕΡΙΦΕΡΕΙΑ ΤΟΥ ΕΡΜΟΥ ΣΤΙΛΒΟΝ**ΤΟΣ**
II) ΠΡΩΤΟΣ ΚΥΚΛΟΣ ΕΣΤΙΝ Ο ΤΟΥ **ΑΡΕΩΣ ΠΥΡΟΕΝΤΟΣ**

i) *The first is the circumference of Hermes Stilbon (inferior planet)*
ii) *The first is the circle of Ares Pyroes (superior planet)*

I) ΕΡΜΟΥ ΣΤΙΛΒΟΝ**ΤΟΣ ΤΟ ΔΕ ΔΙ ΑΥΤΟΥ ΦΕΡΟΜΕΝ**ΟΝ ΤΗΣ ΣΕΛΗΝΗΣ ΓΝΩΜΟΝΙΟΝ ΕΝ ΣΥΝΟΔΩ ΗΜΕΡΑΙΣ Λ
II) **ΑΡΕΩΣ ΠΥΡΟΕΝΤΟΣ ΤΟ ΔΕ ΔΙΑΠΟΡΕ**ΥΟΜΕΝΟΝ (ΧΡΥΣΟΥΝ) ΣΦΑΙΡΙΟΝ ΕΝ ΣΥΝΟΔΩ BL Z

i) *The travelling Lunar pointer in conjunction to Hermes Stilbon, into 30 days*
ii) *The travelling Golden Sphere in conjunction to Ares Pyroes, into 2 years plus 1/7.*

I) ΥΠΕΡ ΔΕ ΤΟΥ ΣΤΙΛΒΟΝΤΟΣ, **ΤΗΣ ΑΦΡΟΔΙΤΗΣ ΦΩΣΦΟΡΟΥ** ΠΕΡΙΦΕΡΕΙΑΝ ΕΣΤΙΝ

| II) ΥΠΕΡ ΔΕ ΤΟΥ ΦΑΕΘΟΝΤΟΣ ΕΣΤΙΝ Ο ΤΟΥ ΚΡΟΝΟΥ ΦΑ**ΙΝΟΝΤΟΣ ΚΥΚΛΟΣ** |
|---|

i) *Out of Stilbon there is Aphrodite's Phosphoros circumference*
ii) *Out of Phaethon there is Kronos' Phainon circle.*

## 8. The final arrangement/refinition of the BCI Part-1 reconstructed text

Based on the previous analysis and the text reconstruction line by line, the reconstruction of the text parts were firstly connected in an intergraded preliminary text, keeping the specific position of the preserved words and the maximum limit of 86 letters per line.

Afterwards, following the style of *The Writer*, the intergraded text was further processed by subtracting, adding, changing a short number of words, to avoid repetition.

In **Table V**, the final reconstructed text is presented (also in **Table VI** the text translation in English and in **Table VII** the text translation in Modern Greek). Each of the colored background sentences/phrases corresponds to the *Definition/Description, Position, Operation* of a specific part.

**01)** ......................................................................................................................................... ΕΙΣ

**02) ΤΑΥΤΗΝ** ΔΕ ΤΗΝ ΕΝΕΠΙΓΡΑΦΟΝ ΧΑΛΚΗΝ ΠΛΑΚΑ ΕΙΣΙΝ ΑΝΑΓΡΑΦΟΜΕΝΑΙ ΑΙ .............................................

**03) ΔΕΙ Δ ΥΠΟΛΑΒΕ**ΙΝ...........................................................................................................................

**04) ΥΠΟ ΔΕ ΤΟΝ ΤΩ**......................................................................................................................

**05) Δ .[- - - - - -]ΟΙΚΑ**...............................................................................................................

**06) Ε - - - - - - - - - ΗΙΣΠ**........................................................................... ΕΝ Τ

**07)** Ω ΚΕΝΤΡΩ ΤΗΣ **ΠΡΟΣ**ΘΙΑΣ ΟΨΕΩΣ ΤΩ ΓΕΩΜΕΤΡΙΚΩ, Η ΓΑΙΑ ΙΣΤΑΤΑΙ Η ΕΣΤΙ ΣΗΜΕΡΟΝ ΤΟ ΚΕΝΤΡΟΝ ΤΟΥ ΚΟΣΜ

**08) ΟΥ**. ΕΠΙ ΚΕΝΤΡΟΥ **ΘΕ**ΣΙΝ ΚΥΛΙΝΔΡΟΣ ΕΣΤΙΝ ΔΙΑ ΧΕΙΡΟΣ ΦΕΡΟΜΕΝΟΣ, ΟΣ ΣΕΛΗΝΗΣ ΚΥΛΙΝΔΡΟΣ ΚΑΛΕΙΤΑΙ.

**09)** ΤΟΝ ΚΥΛΙΝΔΡΟΝ **ΗΡΜΟΣ**ΘΑΙ ΔΙΑ ΤΡΗΜΑΤΟΣ ΚΕΝΤΡΟΥ ΑΥΤΟΥ ΕΝ ΤΩ ΑΞΟΝΙΩ, Ο ΕΙΣ ΚΕΝΤΡΟΝ ΓΕΩΜΕΤΡΙΚΟΝ ΕΣΤΙΝ. ΕΙΣ Τ

**10)** Ο ΑΞΟΝΙΟΝ ΚΑΙ **ΕΠ ΑΚΡΟΥ ΔΙ** ΑΥΤΟΥ, Ο ΤΗΣ ΣΕΛΗΝΗΣ ΚΥΛΙΝΔΡΟΣ ΦΕΡΟΜΕΝΟΣ ΕΣΤΙΝ ΚΑΙ ΕΙΣ ΤΟ ΔΙΗΝΕΚ

**11)** ΕΣ ΔΕΞΙΟΣΤΡΟΦ**ΩΣ ΜΕΝΩΝ**. ΕΙΣ ΤΗΝ ΠΕΡΙΦΕΡΕΙΑΝ ΤΗΣ ΣΕΛΗΝΗΣ ΚΥΛΙΝΔΡΟΥ, ΣΦΑΙΡΙΟΝ ΣΕΛΗΝΗΣ ΦΕΡΟΜΕΝΟΝ ΕΣΤΙΝ.

**12)** ΤΟ ΣΦΑΙΡΙΟΝ, ΟΤ**Ε ΜΕΛΑΝ ΟΤ**Ε ΛΕΥΚΟΧΡΟΥΝ, ΤΗΝ ΝΟΥΜΗΝΙΑΝ ΚΑΙ ΤΗΝ ΠΑΝΣΕΛΗΝΟΝ ΣΥΝ ΤΩ ΧΡΟΝΩ ΔΕΙΚ

**13)** ΝΥΣΙ. ΤΑ ΜΕΓΙΣΤΑ ΤΩΝ **ΟΛΩΝ** ΓΕΓΟΝΟΤΩΝ ΚΑΙ Η ΣΕΛΗΝΗΣ ΚΑΙ Η ΗΛΙΟΥ ΕΚΛΕΙΨΙΣ ΕΝ ΤΗ ΔΙΧΟΜΗΝΙΑ ΚΑΙ ΕΝ ΤΗ ΤΡΙΑΚΑΔ

**14)** Ι ΓΙΓΝΟΝΤΑΙ. **ΔΕΙ Δ ΥΠΟΛΑΒΕΙΝ** ΚΑΙ ΤΟ ΧΡΩΜΑ ΤΟΥ ΣΦΑΙΡΙΟΥ ΚΑΙ ΤΑΣ ΤΗΣ ΣΕΛΗΝΗΣ ΗΜΕΡΑΣ, ΑΙ ΠΕΡΙΞ ΤΟΥ ΚΥΛΙΝ

**15)** ΔΡΟΥ ΕΙΣΙ. **ΤΟ ΣΦΑΙΡΙΟΝ ΦΕΡΕΤ**ΑΙ ΕΝ ΗΜΕΡΑΙΣ ΚΘ Β, ΟΥΤΟΣ Ο ΜΗΝ ΣΥΝΟΔΙΚΟΣ ΚΑΛΕΙΤΑΙ. ΑΠΟ ΤΟΥ ΚΥΛΙΝΔΡΟΥ

**16) ΠΡΟΕΧΟΝ** ΑΥΤΟΥ ΓΝΩΜΟΝΙΟΝ Σ**ΕΛΗΝΗΣ ΕΣΤΙΝ ΕΝ ΣΥΝΟΔΩ ΑΣΤΡΑΣΙ ΗΜΕΡΑΙΣ ΚΖ Γ. ΤΟ ΓΝΩΜΟΝΙΟΝ ΦΕΡΕΤΑΙ ΕΠΙ ΤΩΝ ΠΕΡΙ

**17) ΦΕΡΕΙΩΝ, Η ΜΕΝ ΕΧΟΜΕΝΗ ΤΩΙ ΤΗΣ** ΣΕΛΗΝΗΣ ΚΥΛΙΝΔΡΩ, Η ΔΕ ΕΠΟΜΕΝΗ. ΠΡΩΤΗ ΕΣΤΙΝ Η ΠΕΡΙΦΕΡΕΙΑ ΤΟΥ ΕΡΜΟΥ ΣΤΙΛΒΟΝ

**18)** ΤΟΣ, ΤΟ ΔΕ ΔΙ ΑΥΤΟΥ ΦΕΡΟΜΕΝΟΝ ΤΗΣ ΣΕΛΗΝΗΣ ΓΝΩΜΟΝΙΟΝ ΕΝ ΣΥΝΟΔΩ ΗΜΕΡΑΙΣ Λ. ΥΠΕΡ ΔΕ ΤΟΝ ΣΤΙΛΒΟΝΤΑ Η

**19) ΤΗΣ ΑΦΡΟΔΙΤΗΣ ΦΩΣΦΟΡΟΥ** ΠΕΡΙΦΕΡΕΙΑ ΕΣΤΙΝ. ΤΟ ΔΕ ΦΕΡΟΜΕΝΟΝ ΓΝΩΜΟΝΙΟΝ ΕΝ ΣΥΝΟΔΩ ΗΜΕΡΑΙΣ Λ. ΥΠΕΡ ΔΕ

**20) ΤΟΥ** ΦΩΣ**ΦΟΡΟΥ, ΠΕΡΙΦΕΡΕΙΑΝ** ΗΛΙΟΥ ΕΣΤΙΝ. ΕΙΣ ΤΗΝ ΤΟΥ ΗΛΙΟΥ ΠΕΡΙΦΕΡΕΙΑΝ ΕΙΣ ΣΤΑΘΕΡΟΣ ΚΑΙ ΟΡΘΟΣ

**21) ΓΝΩΜΩΝ ΚΕΙΤΑΙ. ΧΡΥΣΟΥΝ ΣΦΑΙΡΙΟΝ** ΕΠΙ ΓΝΩΜΟΝΑ ΕΣΤΙΝ, ΕΝ ΕΝΙΑΥΤΩ ΔΙΑΠΟΡΕΥΟΜΕΝΟΝ. ΑΠΟ ΣΦΑΙΡΙΟΥ ΠΡΟΕΧΕΙ ΑΥΤΟΥ Η ΤΟΥ

**22)** ΗΛΙΟΥ **ΑΚΤΙΝ. ΥΠΕΡ ΔΕ ΤΟΝ ΗΛΙΟΝ ΕΣΤΙΝ ΚΥΚΛ**ΟΣ ΠΡΩΤΟΣ, ΜΕΣΟΣ, ΕΣΧΑΤΟΣ, ΟΥΣ Η ΑΚΤΙΝ ΦΕΡΕΤΑΙ. ΠΡΩΤΟΣ ΚΥΚΛΟΣ ΕΣΤ

**23)** ΙΝ ΤΟ **ΤΟΥ ΑΡΕΩΣ ΠΥΡΟΕΝΤΟΣ**. ΤΟ ΔΕ ΔΙΑΠΟΡΕΥΟΜΕΝΟΝ (ΧΡΥΣΟΥΝ) ΣΦΑΙΡΙΟΝ ΕΝ ΣΥΝΟΔΩ BL Z. ΥΠΕΡ ΔΕ ΤΟΝ ΠΥΡΟΕΝΤΑ, ΚΥΚΛΟΣ ΕΣΤΙΝ

**24)** ΔΙΟΣ ΦΑΕΘΟΝΤΟΣ. ΤΟ ΔΕΔΙΑΠΟΡΕΥΟΜΕΝΟΝ (ΧΡΥΣΟΥΝ) ΣΦΑΙΡΙΟΝ ΕΝ ΣΥΝΟΔΩ AL R. ΥΠΕΡ ΔΕ ΤΟΝ ΦΑΕΘΟΝΤΑ ΕΣΤΙΝ Ο ΤΟΥ ΚΡΟ

**25)** ΝΟΥ ΦΑΙΝΟΝΤΟΣ ΚΥΚΛΟΣ. ΤΟ ΔΕ (ΧΡΥΣΟΥΝ) **ΣΦΑΙΡΙΟΝ Φ**ΕΡΟΜΕΝΟΝ ΕΝ ΣΥΝΟΔΩ ΑΙ ΚΗ. ΟΥΤΟΣ ΜΕΝ ΕΣΤΙΝ Ο ΤΗΝ ΣΗΜΕΡΟΝ

**26)** ΚΟΣΜΟΣ. **ΠΕΡΑ ΔΕ ΤΟΥ ΚΟΣΜΟΥ ΚΕΙΤΑΙ,** ΠΡΩΤΟΣ ΜΕΝ ΚΥΚΛΟΣ ΖΩΔΙΑΚΟΣ, ΕΠΟΜΕΝΟΣ Δ ΑΙΓΥΠΤΙΑΚΟΣ. ΤΑ ΤΩΝ ΑΠΛΑΝ

**27)** ΩΝ ΑΝΑΓΡΑΦΟΜΕΝΑ **ΣΤΟΙΧΕΙΑ ΠΑΡΑΚΕΙΜΕΝΑ** ΕΝ ΤΑΙΣ ΗΜΕΡΑΙΣ ΤΟΥ ΖΩΔΙΑΚΟΥ ΚΥΚΛΟΥ ΕΣΤΙ. ΠΑΡΑΤΙΘΕΝΤΑΙ

**28)** ΔΕ ΤΑ ΓΕΓΟΝΟΤΑ **ΤΑΥΤΑ ΤΑΙΣ ΑΣΠΙ**ΔΙΣΚΑΙΣ ΑΙ ΕΙΣΙΝ ΑΝΩΤΕΡΩ ΚΑΙ ΚΑΤΩΤΕΡΩ ΤΩΝ ΕΙΣ ΚΕΝΤΡΟΝ ΚΥΚΛΩΝ.

**29)** ΤΑ ΤΩΝ ΑΠΛΑΝΩΝ **ΠΡΟΕΙΡΗΜΕΝΑ** ΓΕΓΟΝΟΤΑ, ΕΩΙΩΝ ΤΕ ΕΣΠΕΡΙΩΝ ΚΑΙ ΕΠΙΤΟΛΩΝ ΤΕ ΔΥΣΕΩΝ, ΑΝΑΓΡΑΦΟΝΤΑΙ ΑΥΤΟΘΙ.

**30)** ΕΝ ΤΗ ΑΝΩΤΕΡΑ ΕΚ ΤΩΝ **ΑΣΠΙ**ΔΙΣΚΩΝ, ΟΙ ΧΕΙΜΕΡΙΝΟΙ ΚΑΙ ΕΑΡΙΝΟΙ ΜΗΝΕΣ ΖΩΔΙΑΚΟΙ ΚΑΙ ΕΝ ΤΗ ΚΑΤΩΤΕΡΑ, ΟΙ ΘΕΡΙΝΟΙ ΚΑΙ

**31)** ΦΘΙΝΟΠΩΡΙΝΟΙ ΜΗΝΑΙ ΖΩΔΙΑΚΟΙ ΠΑΡΑΤΙΘΕΝΤΑΙ …………………………………………………………………………………………………

Table V: Reconstructed text of the Back Cover Inscription Part-1 based on the analysis of Section 5

1…………………………………………………………………………………………………………………………… *on*

2 *this coppered engraved plate, there is*…………………………………………………………………………..

3 *should be understood*………………………………………………………………………………………………

4 *below the*……………………………………………………………………………………………………………….

5 …………………………………………………………………………………………………………………………….

6 ……………………………………………………………………………………………………………… *at the*

7 *geometrical center of the Front Face, the Earth is located, which is the Center of Cosmos.*

8 *At the center there is a cylinder, rotated by hand and is called Lunar Cylinder.*

9 *The cylinder, via its central hole, is adapted on the axis, which is located on the geometrical center.*

10 *On the axis's edge the Lunar Cylinder, is continuously rotating*

11 *clockwise. On the circumference of the Lunar Cylinder there is the rotating Lunar little sphere.*

12 *The little sphere, either in black or in white color, depicts the New moon and the Full moon in course of time.*

13 *The most important Lunar and Solar events, the Lunar and Solar eclipses occurring during Dichominis and Triakas,*

14 *these (events) should be understood by observing the color of the Lunar little sphere and the Lunar days were engraved around the Lunar Cylinder.*

15 *The little sphere is rotated continuously completing 29.5 days, called Synodic Month. From the Cylinder*

16 *a pointer protrudes and it is in conjunction with stars every 27.3 days. The pointer travels on the two circu-*

17 *mferences, the one is located close to the Lunar Cylinder and the next one follows. The first is the circumference of Mercury Stil-*

18 *bon. Through this circumference, the traveling Lunar pointer comes in conjunction every 30 days. Beyond Stilbon*

19 *there is Aphrodites Phosphoros' circumference and the traveling Lunar pointer comes in conjunction every 30 days. Beyond*

20 *Phosphoros, there is the circumference of Sun. On the circumference a vertical pillar is stabilized.*

> *On the*
>
> *21 pillar, the Golden sphere is adapted, rotated in Eniautos. From the Golden little sphere, the Sun's ray protrudes.*
>
> *22 Beyond Sun, there is the first circle, the middle and the last one, on which the Sun's ray travels. The first is the circle of*
>
> *23 Ares Pyroes. The traveling Sun's sphere comes in conjunction every 2+1/7 years. Beyond Pyroes, there is the circle of*
>
> *24 Zeus Phaethon. The traveling Sun's sphere comes in conjunction every 1+1/10 years. Beyond Phaethon, there is*
>
> *25 Saturn Phainon circle. The traveling Sun's sphere comes in conjunction every 1+1/28 years. This is the today's*
>
> *26 Cosmos. Beyond Cosmos, there is firstly the Zodiac circle and the Egyptian circle follows.*
>
> *27 The index letters of the stars are engraved on the days located on the Zodiac ring.*
>
> *28 The star events are engraved on the two small shields, located above and below the central rings.*
>
> *29 The aforementioned star events morning/evening, rising/setting are presented.*
>
> *30 On the top small shield, the winter and the vernal Zodiac months, and on the bottom (shield) the summer and*
>
> *31 the autumnal Zodiac months are engraved.……………………………………………………………….…………*
>
> *………………………………………………………………………………………………………………………………………*

Table VI: BCI PART-1 text translation in the English language

01) ………………………………………………………………………………………………………………………… ΣΕ

02) ΑΥΤΗ ΤΗΝ ΧΑΛΚΙΝΗ ΕΓΧΑΡΑΚΤΗ ΠΛΑΚΑ ΑΝΑΓΡΑΦΟΝΤΑΙ …………………………………………………………

03) ΚΑΠΟΙΟΣ ΜΠΟΡΕΙ ΝΑ ΚΑΤΑΛΑΒΕΙ ……………………………………………………………………………………..

04) ΥΠΟ ΔΕ ΤΟΝ ΤΩ …………………………………………………………………………………………………………..

05) Δ . - - - - - - ΟΙΚΑ…………………………………………………………………………………………………………..

06) Ε - - - - - - - - - ΗΙΣΠ……………………………………………………………………………………………… ΣΤΟ

07) ΓΕΩΜΕΤΡΙΚΟ ΚΕΝΤΡΟ ΤΗΣ ΠΡΟΣΘΙΑΣ ΟΨΗΣ  ΒΡΙΣΚΕΤΑΙ Η ΓΗ, Η ΟΠΟΙΑ ΣΗΜΕΡΑ ΕΙΝΑΙ ΤΟ ΚΕΝΤΡΟ ΤΟΥ ΚΟΣΜ-

08) ΟΥ. ΣΤΗ ΘΕΣΗ ΤΟΥ ΚΕΝΤΡΟΥ ΒΡΙΣΚΕΤΑΙ ΕΝΑΣ ΚΥΛΙΝΔΡΟΣ ΠΟΥ ΠΕΡΙΣΤΡΕΦΕΤΑΙ ΜΕ ΤΟ ΧΕΡΙ ΚΑΙ ΟΝΟΜΑΖΕΤΑΙ ΚΥΛΙΝΔΡΟΣ ΤΗΣ ΣΕΛΗΝΗΣ.

09) Ο ΚΥΛΙΝΔΡΟΣ ΣΤΕΡΩΝΕΤΑΙ ΜΕΣΩ ΤΗΣ ΚΕΝΤΡΙΚΗΣ ΟΠΗΣ ΤΟΥ ΣΤΟΝ ΑΞΟΝΑ ΠΟΥ ΒΡΙΣΚΕΤΑΙ ΤΟΠΟΘΕΤΗΜΕΝΟΣ ΣΤΟ ΓΕΩΜΕΤΡΙΚΟ ΚΕΝΤΡΟ. ΣΤΟΝ

10) ΑΞΟΝΑ ΚΑΙ ΣΤΟ ΑΚΡΟ ΑΥΤΟΥ, Ο ΚΥΛΙΝΔΡΟΣ ΤΗΣ ΣΕΛΗΝΗΣ ΠΕΡΙΣΤΡΕΦΕΤΑΙ ΣΥΝΕΧΩΣ

11) ΔΕΞΙΟΣΤΡΟΦΑ. ΣΤΗΝ ΠΕΡΙΦΕΡΕΙΑ ΤΟΥ ΚΥΛΙΝΔΡΟΥ ΤΗΣ ΣΕΛΗΝΗΣ ΥΠΑΡΧΕΙ ΤΟ ΠΕΡΙΣΤΡΕΦΟΜΕΝΟ ΣΦΑΙΡΙΟ ΤΗΣ ΣΕΛΗΝΗΣ.

12) ΤΟ ΣΦΑΙΡΙΟ ΑΛΛΟΤΕ (ΕΙΝΑΙ) ΜΑΥΡΟΥ ΚΑΙ ΑΛΛΟΤΕ ΛΕΥΚΟΥ ΧΡΩΜΑΤΟΣ ΔΕΙΧΝΟΝΤΑΣ ΣΤΑΔΙΑΚΑ ΤΗΝ ΝΟΥΜΗΝΙΑ ΚΑΙ ΤΗΝ ΠΑΝΣΕΛΗΝΟ.

13) ΤΑ ΣΠΟΥΔΑΙΟΤΕΡΑ ΑΠ' ΟΛΑ ΤΑ ΓΕΓΟΝΟΤΑ ΚΑΙ Η ΕΚΛΕΙΨΗ ΤΗΣ ΣΕΛΗΝΗΣ ΚΑΙ ΤΟΥ ΗΛΙΟΥ, ΣΥΜΒΑΙΝΟΥΝ ΣΤΗ ΜΕΣΗ ΤΟΥ ΜΗΝΑ ΚΑΙ ΣΤΗΝ ΤΡΙΑΚΟΣΤΗ ΗΜΕΡΑ ΤΟΥ.

14) ΚΑΠΟΙΟΣ ΜΠΟΡΕΙ ΝΑ ΑΝΤΙΛΗΦΘΕΙ (ΤΗΝ ΝΟΥΜΗΝΙΑ ΚΑΙ ΤΗΝ ΔΙΧΟΜΗΝΙΑ) ΠΑΡΑΤΗΡΩΝΤΑΣ ΚΑΙ

ΤΟ ΧΡΩΜΑ ΤΟΥ ΣΦΑΙΡΙΟΥ ΚΑΙ ΤΙΣ ΗΜΕΡΕΣ ΤΗΣ ΣΕΛΗΝΗΣ ΠΟΥ ΑΝΑΓΡΑΦΟΝΤΑΙ ΓΥΡΩ ΑΠΟ ΤΟΝ ΚΥΛΙΝΔΡΟ ΤΗΣ.

15) ΤΟ ΣΦΑΙΡΙΟ ΠΕΡΙΣΤΡΕΦΕΤΑΙ ΣΕ 29.5 ΗΜΕΡΕΣ. ΑΥΤΟΣ Ο ΜΗΝΑΣ ΟΝΟΜΑΖΕΤΑΙ ΣΥΝΟΔΙΚΟΣ ΜΗΝΑΣ. ΑΠΟ ΤΟΝ ΚΥΛΙΝΔΡΟ ΑΥΤΟΝ

16) ΤΟ ΓΝΩΜΟΝΙΟ ΤΗΣ ΣΕΛΗΝΗΣ ΠΟΥ ΠΡΟΕΞΕΧΕΙ ΕΡΧΕΤΑΙ ΣΕ ΣΥΝΟΔΟ ΜΕ ΤΑ ΑΣΤΕΡΙΑ ΣΕ 27.3 ΗΜΕΡΕΣ. ΤΟ ΓΝΩΜΟΝΙΟ ΔΙΑΒΑΙΝΕΙ ΔΥΟ ΠΕΡΙ

17) ΦΕΡΕΙΕΣ. Η ΜΙΑ ΒΡΙΣΚΕΤΑΙ ΚΟΝΤΑ ΣΤΗ ΣΕΛΗΝΗ ΚΑΙ Η ΑΛΛΗ ΕΠΕΤΑΙ. ΠΡΩΤΗ ΕΙΝΑΙ Η ΠΕΡΙΦΕΡΕΙΑ ΤΟΥ ΕΡΜΟΥ ΣΤΙΛΒΟΝ

18) ΤΟΣ.  ΤΟ ΓΝΩΜΟΝΙΟ ΤΗΣ ΣΕΛΗΝΗΣ ΠΟΥ ΠΕΡΙΦΕΡΕΤΑΙ ΣΕ ΑΥΤΟΝ ΒΡΙΣΚΕΤΑΙ ΣΕ ΣΥΝΟΔΟ ΚΑΘΕ 30 ΗΜΕΡΕΣ. ΠΕΡΑ ΑΠΟ ΤΟΝ ΣΤΙΛΒΟΝΤΑ

19) ΒΡΙΣΚΕΤΑΙ Η ΠΕΡΙΦΕΡΕΙΑ ΤΗΣ ΑΦΡΟΔΙΤΗΣ ΦΩΣΦΟΡΟΥ. ΤΟ ΠΕΡΙΦΕΡΟΜΕΝΟ ΓΝΩΜΟΝΙΟ ΕΡΧΕΤΑΙ ΣΕ ΣΥΝΟΔΟ ΚΑΘΕ 30 ΗΜΕΡΕΣ. ΠΕΡΑ ΑΠΟ ΤΟΝ

20) ΦΩΣΦΟΡΟ ΒΡΙΣΚΕΤΑΙ Η ΠΕΡΙΦΕΡΕΙΑ ΤΟΥ ΗΛΙΟΥ. ΣΤΗΝ ΠΕΡΙΦΕΡΕΙΑ ΤΟΥ ΗΛΙΟΥ ΕΙΝΑΙ ΣΤΕΡΕΩΜΕΝΟΣ ΕΝΑΣ ΚΑΘΕΤΟΣ ΚΑΙ ΑΚΙΝΗΤΟΣ ΣΤΥΛΙΣΚΟΣ. ΣΤΟΝ

21) ΣΤΥΛΙΣΚΟ ΤΟΠΟΘΕΤΕΙΤΑΙ ΤΟ ΧΡΥΣΟ ΣΦΑΙΡΙΟ ΤΟΥ ΗΛΙΟΥ, ΠΕΡΙΦΕΡΟΜΕΝΟ ΣΕ ΕΝΑ ΤΡΟΠΙΚΟ ΕΤΟΣ. ΑΠΟ ΤΟ ΣΦΑΙΡΙΟ ΠΡΟΕΚΤΕΙΝΕΤΑΙ Η

22) ΑΚΤΙΝΑ ΤΟΥ ΗΛΙΟΥ. ΠΕΡΑ ΑΠΟ ΤΟΝ ΗΛΙΟ ΥΠΑΡΧΕΙ Ο ΠΡΩΤΟΣ, Ο ΜΕΣΟΣ ΚΑΙ Ο ΕΣΧΑΤΟΣ ΚΥΚΛΟΣ, ΤΟΥΣ ΟΠΟΙΟΥΣ ΔΙΑΒΑΙΝΕΙ Η ΑΚΤΙΝΑ ΤΟΥ ΗΛΙΟΥ. Ο ΠΡΩΤΟΣ ΚΥΚΛΟΣ ΕΙΝΑΙ

23) ΤΟΥ ΑΡΕΩΣ ΠΥΡΟΕΝΤΟΣ ΚΑΙ ΤΟ ΔΙΑΠΟΡΕΥΟΜΕΝΟ ΣΦΑΙΡΙΟ (ΤΟΥ ΗΛΙΟΥ) ΕΡΧΕΤΑΙ ΣΕ ΣΥΝΟΔΟ ΚΑΘΕ 2 ΕΤΗ ΚΑΙ 1/7. ΠΕΡΑ ΑΠΟ ΤΟΝ ΠΥΡΟΕΝΤΑ ΒΡΙΣΚΕΤΑΙ Ο ΚΥΚΛΟΣ ΤΟΥ

24) ΔΙΟΣ ΦΑΕΘΟΝΤΟΣ ΚΑΙ ΤΟ ΔΙΑΠΟΡΕΥΟΜΕΝΟ ΣΦΑΙΡΙΟ (ΤΟΥ ΗΛΙΟΥ) ΕΡΧΕΤΑΙ ΣΕ ΣΥΝΟΔΟ ΚΑΘΕ 1 ΕΤΟΣ ΚΑΙ 1/10. ΠΕΡΑ ΑΠΟ ΤΟΝ ΦΑΕΘΟΝΤΑ ΒΡΙΣΚΕΤΑΙ Ο

25) ΚΥΚΛΟΣ ΤΟΥ ΚΡΟΝΟΥ ΦΑΙΝΟΝΤΟΣ ΚΑΙ ΤΟ ΠΕΡΙΦΕΡΟΜΕΝΟ ΣΦΑΙΡΙΟ (ΤΟΥ ΗΛΙΟΥ) ΕΡΧΕΤΑΙ ΣΕ ΣΥΝΟΔΟ ΚΑΘΕ 1 ΕΤΟΣ ΚΑΙ 1/28. ΑΥΤΟΣ ΛΟΙΠΟΝ ΕΙΝΑΙ Ο ΚΟΣΜΟΣ ΣΗΜΕΡΑ.

26) ΠΕΡΑ ΑΠΟ ΑΥΤΟΝ ΤΟΝ ΚΟΣΜΟ ΒΡΙΣΚΕΤΑΙ ΠΡΩΤΑ Ο ΖΩΔΙΑΚΟΣ ΔΑΚΤΥΛΙΟΣ ΚΑΙ ΕΠΕΙΤΑ Ο ΑΙΓΥΠΤΙΑΚΟΣ (ΔΑΚΤΥΛΙΟΣ).

27) ΤΑ ΑΝΑΓΡΑΦΟΜΕΝΑ ΣΤΟΙΧΕΙΑ (ΟΙ ΠΑΡΑΠΟΜΠΕΣ) ΤΩΝ ΑΠΛΑΝΩΝ ΑΣΤΕΡΩΝ ΕΙΝΑΙ ΣΗΜΕΙΩΜΕΝΑ ΣΤΙΣ ΗΜΕΡΕΣ ΤΟΥ ΖΩΔΙΑΚΟΥ ΔΑΚΤΥΛΙΟΥ.

28) ΤΑ ΓΕΓΟΝΟΤΑ ΤΩΝ ΑΣΤΕΡΩΝ ΕΙΝΑΙ ΓΡΑΜΜΕΝΑ ΣΤΙΣ ΔΥΟ ΜΑΚΡΟΣΤΕΝΕΣ ΠΛΑΚΕΣ, ΟΙ ΟΠΟΙΕΣ ΒΡΙΣΚΟΝΤΑΙ ΕΠΑΝΩ ΚΑΙ ΚΑΤΩ ΑΠΟ ΤΟΥΣ ΚΕΝΤΡΙΚΟΥΣ ΔΑΚΤΥΛΙΟΥΣ.

29) ΤΑ ΠΡΟΑΝΑΦΕΡΘΕΝΤΑ ΓΕΓΟΝΟΤΑ ΤΩΝ ΑΠΛΑΝΩΝ ΑΣΤΕΡΩΝ ΚΑΙ ΠΡΩΙΝΩΝ ΚΑΙ ΑΠΟΓΕΥΜΑΤΙΝΩΝ ΚΑΙ ΑΝΑΤΟΛΩΝ ΚΑΙ ΔΥΣΕΩΝ ΤΟΥΣ, ΠΑΡΑΤΙΘΕΝΤΑΙ ΕΔΩ (ΣΤΙΣ ΠΛΑΚΕΣ).

30) ΣΤΗΝ ΕΠΑΝΩ ΠΛΑΚΑ ΟΙ ΧΕΙΜΕΡΙΝΟΙ ΚΑΙ ΟΙ ΕΑΡΙΝΟΙ ΖΩΔΙΑΚΟΙ ΜΗΝΕΣ ΚΑΙ ΣΤΗΝ ΚΑΤΩ ΠΛΑΚΑ

31) ΟΙ ΘΕΡΙΝΟΙ ΚΑΙ ΟΙ ΦΘΙΝΟΠΩΡΙΝΟΙ ΖΩΔΙΑΚΟΙ ΜΗΝΕΣ ΕΙΝΑΙ ΧΑΡΑΓΜΕΝΟΙ ……………………………..

Table VII: BCI PART-1 text translation in the Modern Greek language

As can be deduced from the distribution of the colors on the reconstructed text in **Table V**, the largest percentage of the text, about 30% is dedicated to the *Definition/Description, Position, Operation* of the Lunar Disc and its parts, and the Presentation of the Sun and its parts follows with about 9.4%. Regarding the presentation of the planets, the percentage varies between 2.8% and 3.6% (can be considered as definite for Mars, Jupiter and Saturn), significantly smaller than the Lunar Disc's and Suns' percentage.

The difference in the percentages between the Moon and Sun which have pointers and spheres and the simple reference to the planet circles reveals that the Lunar Disc and the Golden Sphere-Sun are more important planets than the other five, which were only presented by their orbits.

| Part description on the preserved BCI Part-1 text | Corresponding number of lines | Total letters | Percentage % of total 2475 letters in 31 lines |
|---|---|---|---|
| *Introduction* | Lines 1-6 | 516 max | 20% |
| *Lunar Disc* | Lines 8-17 | 735 | 29.7% |
| *Mercury* | Lines 17-18 | 84 | 3.3% |
| *Venus* | Lines 18-19 | 90 | 3.6% |
| *Sun* | Lines 19-22 | 235 | 9.5% |
| *Mars* | Line 22-23 | 71 | 2.8% |
| *Jupiter* | Line 23-24 | 78 | 3.1% |
| *Saturn* | Line 24-25 | 80 | 3.2% |
| *Cosmos* | Line 25 | 30 | 1.2% |
| *Zodiac and Egyptian rings* | Line 26 | 64 | 2.5% |
| *Index letters* | Line 27 | 80 | 3.2% |
| *Star events, Parapegma plates (small shields)* | Lines 28-31 | (238+59)=297 | 12% |

Table V. The final letter percentage per each operational part on the BCI Part-1 reconstructed text

To be more exact, the corresponding text for the Lunar Disc has the most extensive reference, indicating an important role in the Antikythera Mechanism's operation. This gives the impression that all of the Mechanism's procedures are dependent, based and operated by the Lunar Disc. For mechanical and handling reasons the most proper and ideal input/driving of the Mechanism is the Lunar Disc.[91] Hence, *The Engraver/The Manufacturer* of the Mechanism logically dedicates a large–section of the text to this part. From our perspective, this adds another significant argument: the Antikythera Mechanism was a Luni-Solar Time calculating machine, based on the synodic Lunar cycle.

## 9. Discussion and Summary

In this work, the reconstruction of the BCI Part-1 text was presented after the analysis of the preserved text.

Although a notable part of the original text is missing, this absence did not prevent the text reconstruction, because the preserved text can be correlated to the visual photographs and the AMRP tomographies. This information was not only the engraved words, but also the mechanical operational parts of the Mechanism. In many cases, the text, and the parts are both preserved. Moreover, the preserved original text leads to the bronze parts reconstruction. On the other hand, the adaptation and the use of these parts on the functional model of the Mechanism lead to the reconstruction of the text.

---

[91] Voulgaris et al., 2018b; Roumeliotis 2018.

Reading the partially preserved original text, and keeping the Symmetry on specific sentences, it results that the ancient writer presents the circles-planet orbits of Mercury, Venus, the Sun, Mars, Jupiter, and Saturn. The ancient manufacturer designed six homocentric circles on the cover of the b1 gear, with the $b_{in}$ axis-Earth in the center. The Lunar pointer travels through the first two circumferences of Mercury and Venus, and the pointer of the Sun through the other three planet circles (Mars, Jupiter and Saturn).

The analysis of the original text concludes there is no space for a reference to the rotating planet pointers with spheres. A prominent question arises from the absence of planet pointers and their spheres: if the rotating planets existed on the Antikythera Mechanism, why did *The Engraver* not extensively *Describe, Present* and *Place* these parts on the User's Manual of his creation? Moreover, the absence of any reference (or poor description) to these parts, downgrades their importance and it is too difficult to justify.

The construction, the assembly of the Mechanism parts and the large number of "*flight hours*" using the authors' Antikythera Mechanism functional models, was critical and decisive for the reconstruction of the Back Cover Inscription Part-1 text.

**Acknowledgments**

\*\*\*\*\*\*\*\*\*\*\*\*\*\*\*\*\*\*\*\*\*\*\*\*\*\*\*\*\*\*\*\*\*\*\*\*\*\*\*\*\*\*\*\*\*\*\*\*\*\*\*\*\*\*\*\*\*\*\*\*\*\*\*\*\*\*\*\*\*\*\*\*\*

## THE RECONSTRUCTION OF THE ANTIKYTHERA MECHANISM INSTRUCTION MANUAL OF THE BACK DIAL PLATE, AFTER A DILIGENT STUDY AND ANALYSIS OF THE BACK COVER INSCRIPTION PART-2

**A. Voulgaris[1], C. Mouratidis[2], A. Vossinakis[3]**

[1]*City of Thessaloniki, Directorate Culture and Tourism, Thessaloniki, GR-54625, Greece*
[2]*Merchant Marine Academy of Syros, GR-84100, Greece*
[3]*Thessaloniki Astronomy Club, Thessaloniki, GR-54646, Greece*

*Abstract*

*We present the Part-2 text reconstruction of the Back Cover Inscription (BCI). In the BCI Part-2 text, the ancient manufacturer of the Antikythera Mechanism has engraved the information related to the Presentation, the Position, and the Operation of the outer operational mechanical parts, which were located at the Back plate. Although this technical text is also partially/poorly preserved as is the Part-1, the words and the phrases present a characteristic repeatability, especially when the Engraver presents and describes the two helical (spiral) dials, their pointers and their parts. A significant percentage of the missing text was completed using the existed words and phrases. Some of the spiral parts' were preserved on the text, and their geometrical characteristics were studied and named based on Archimedes' work Περί Ελίκων (De Lineis Spiralibus). In order to describe the geometry and the position of some of the mechanical parts mentioned in the text, words from the* technical text *of Heron's of Alexandria work Περί Διόπτρας (Dioptra), as well as from works written by Ptolemy, Euclid and Geminus were used. The style of the reconstructed text was based on the preserved text's style and is presented at the simplest grammar and syntax form of the koine Hellenist Greek language. In many cases the reconstructed text can be considered as definite.*

## 1. Introduction

The Antikythera Mechanism was an ancient Greek time measuring machine, constructed between 200-100 BC. The Mechanism has a complex mechanical design with gears, axes, scales, pointers, in order to calculate via a mechanical procedure the position of the Moon and the Sun on the zodiac sky, the lunar phases, the dates of the solar and lunar eclipses, even their times of occurrence (Freeth et al., 2006; Seiradakis and Edmunds 2018, Antikythera Mechanism Research Project, http://www.antikythera-mechanism.gr). It could also calculate the year of the athletic Stephanites Games (Freeth et al., 2008).

For such a complex machine, an instruction manual, giving information about its use, was necessary. For this reason, *The Manufacturer* of the Mechanism has engraved the *Instruction Manual* of his creation, on the Back Cover plate. The Front and Back Cover plates were adapted on the main body of the Mechanism, creating an integrated wooden case (Voulgaris et al., 2019b). Today, the *Instruction Manual* of the Antikythera Mechanism is partially preserved on some of the fragments. The text is preserved in two parts (1 and 2, Bitsakis and Jones 2016b). On Part-1, the *Description*, *Definition*, *Position* and the *Operation* of the external *operational parts*, located on the Front Dial plate are presented (Authors' BCI Part-1, submitted). On Part-2 the *Description*, *Definition*, *Position* and the *Operation* of the *operational parts*, located on the Back Dial plate are presented (Anastasiou et al., 2014; Bitsakis and Jones 2016b).

## 2. The Metonic calendar and the Saros cycle

A famous cycle for time measuring at the ancient times was the 19 (solar tropical) years cycle. equal to 235 lunar synodic months. This cycle could easily be detected, by making observations and recordings of the position of the Sun and the Moon for 2-3 decades, and was used by the ancient Babylonian, Egyptian, Greek and Chinese astronomers (Fotheringham 1924; Britton 2007). The period of 19 years-offered a relation of integer number of solar tropical years and lunar the synodic months, with an accepted error for the era of 400BC.

Meton of Athens, a Greek famous mathematician, architect, astronomer, geometer, and engineer who lived in Athens, based on this cycle presented the Metonic calendar with a number of specific arrangements: As 12 synodic months equals to 354.36 days, and 13 synodic months correspond to 383.89, they both deviate a great deal from the tropical solar year of 365.245 days. Meton made the proper arrangements in order to match as well as possible the tropical solar year with alternately - 12 or 13 lunar synodic months, so that 4-5 Lunar years would approach 4-5 tropical years, as Geminus mentions (Spandagos 2002 and Manitius 1898). Meton's mathematical algorithm also included an arrangement in order to avoid the time span of 29.53 days per synodic month: he arranged his calculations, so that some synodic months have 30 days (full months) and the rest 29 days (hollow months). Meton's calculations resulted to the pattern that every 64 days one day was subtracted; this day as named ΕΞΑΙΡΕΣΙΜΟΣ ΗΜΕΡΑ (exeresimos-omitted day, in singular).

The first Metonic cycle (ΚΥΚΛΟΣ ΤΟΥ ΜΕΤΩΝΟΣ or ΜΕΤΩΝΟΣ ΕΝΙΑΥΤΟΣ or ΕΝΝΕΑΚΑΙΔΕΚΑΕΤΗΡΙΣ eneakaidekateris 9+10 years) of the attic calendar started on the Summer Solstice, right after the New Moon of 27/28 June 432BC (Danezis and Theodosiou 1995). After 19 years, the next Metonic cycle started right after the Summer solstice of 413BC.

About 100 years later, Callippus having the advantage of the recorded observations of the Metonic cycle has detected timing errors in the Metonic cycle and he has presented a better corrected cycle, the Callippic cycle, which consisted of four Metonic cycles minus 1 day. The first Callippic cycle of the attic calendar started with the Summer solstice on 330BC, right after the New Moon of 28 June.

The Saros cycle is a time span of 18 solar tropical years plus 10 or 11 days (18.03 years) equal to 223 lunar synodic months. The Saros cycle related to the eclipse sequence and their times of occurrence: In every Saros cycle, the eclipse sequence is repeated with the same kind of eclipse, but with 8 hours delay on the time of all the events (≈8h/Saros). After an additional Saros cycle, the eclipses were repeated with a delay of 16 hours relatively to the initial cycle. In order for the eclipse events hour correction to be clear, a specific naming for each of the three Saros cycles must have existed e.g. 1st, 2nd and 3rd (ΠΡΩΤΟΣ-ΔΕΥΤΕΡΟΣ-ΤΡΙΤΟΣ) Saros cycle (or ΠΡΩΤΟΣ-ΜΕΣΟΣ-ΕΣΧΑΤΟΣ, *First-Middle-Last* or such). After three successive Saros (3X8h/Saros =24h =0 hours correction) the eclipse events are repeated at (about) the same hour. The time span of 3 Saros is equal to 54 years plus 1 synodic month and is named Exeligmos as Geminus refers (Spandagos 2002) and also Ptolemy in Almagest (Heiberg 1898; Toomer 1984). Each Saros cycle started when the three lunar cycles where at their beginning (reset position), with the New Moon at Apogee and on the Node (i.e. on the Ecliptic). The beginning of Saros is marked with a large duration annular solar eclipse as results from Geminus' description for Exeligmos/(Saros).

## 3. The Metonic and the Saros cycle introduced on the Antikythera Mechanism

The Metonic cycle calendar and the eclipse information-Saros cycle, of the Antikythera Mechanism are presented on the Back plate, which is divided in two dial scales. Usually, a measuring scale consists of a circle whose perimeter is divided in a number of subdivisions. A rotating pointer, whose one edge is pinned to the center of the circle, gradually transits all of the scale's subdivisions. Such a design is the Egyptian dial ring and the Zodiac month ring, located on the Front plate of the Mechanism, and the Golden Sphere-The Sun of the Mechanism (Voulgaris et al., 2018b) transits their subdivisions/days (Voulgaris et al., 2018a).

On the contrary, the two dials of the Back plate were designed in spiral form rather than circular. The ancient Manufacturer applied this remarkable measuring scale design for reason of space necessity: For the representation of both the cycles, the Metonic with 235 words/names of synodic months and the Saros with 223 synodic months, with engraved information on some of these, a lot of space is needed, in order for the engraved information to be readable. The width of the Mechanism is around 175mm (Allen et al., 2016; Voulgaris et al., 2019b), therefore a circle of 160mm diameter offers 160Χπ/235 words = 2.14mm on the perimeter per each word. In this dimension, it is impossible (or practically useless) the month e.g.ΔΩΔΕΚΑΤΕΥΣ to be engraved. The 235 words could be in radial distribution, but the resolution of the pointer would be very low and highly doubtful (Voulgaris et al., 2022). Also, additional gears are needed (and the mechanical errors would be increased). For this reason *The Manufacturer* adopted the unique idea of the measuring scale in a spiral shape distribution. The Metonic 5 full turns spiral offers satisfactory space (total spiral length ≈1726mm, instead of 502mm corresponding to a circular perimeter) for the month names to be engraved on their subdivisions/cells, with a mean width of about 7.3mm/synodic month cell (about equal dimension/cell for the 4 turn Saros spiral). Even the 7.3 mm is a short dimension for a word and finally *The Engraver* distributes each word in two or three lines.

The pattern of the omitted - ΕΞΑΙΡΕΣΙΜΟΙ days, which was calculated by Meton, is also (partially) preserved on the Metonic first spiral turn and their specific position indicates the month from which a day will be subtracted, i.e. these months will have 29 days (hollow month). Each of the omitted days is engraved by an ancient Greek number, signifying the numbered day which will be omitted by the specific month.

The Mechanism could also calculate the year the Athletic Games would take place. This was very useful, as the ΣΠΟΝΔΟΦΟΡΟΙ ΑΓΓΕΛΙΟΦΟΡΟΙ (*spondophoroi messengers*) needed to know the Games' starting date well ahead, in order to have time to travel from city to city to announce it.

In AMRP tomographies of Fragment B, at the internal area of the Metonic spiral a small circular scale is visible, divided in four sectors/quadrants. On each part one of the symbols LA, LB, LΓ, LΔ is engraved, corresponding to the year 1, 2, 3, and 4 (Freeth et al., 2008). The names of the ΣΤΕΦΑΝΙΤΕΣ ΑΓΩΝΕΣ (Stephanitic-Crown athletic Games) are engraved at the outer part of the four quadrants. These Games were named ΣΤΕΦΑΝΙΤΕΣ, because the prize for the victors was only a twig curved like a crown and named ΣΤΕΦΑΝΙ; no money prize was awarded. In the Olympic Games the prize was a branch from a wild olive, pine in Isthmia, celery in Nemea and laurel in Pythia. The Stephanitic Games were also named ΙΕΡΟΙ ΑΓΩΝΕΣ (Sacred Games). In contradiction, the price for the winners in ΧΡΗΜΑΤΙΤΕΣ ΑΓΩΝΕΣ (Chrematitic-Money Games) was money. The pointer of the Games scale (today lost) was rotated by one full turn/4 years.

On each of the spirals, the ancient Manufacturer made a furrow (perforating the Back plate) following the spiral pattern design. At the "*center*" of each of the spiral dial, there is a rotating pointer (Archimedes names the "center" of a Spiral-Helix *Beginning-Origin*, see Section 6). The pointers-ΓΝΩΜΟΝΙΑ (today only one is preserved on Fragment B) is not a simple design, a bronze oblong straight bar: rather, at the edge each of the two pointers continues in perpendicular direction. The edge of this "bended" pointer is pointed and travels inside the furrow during the pointer's rotation (Anastasiou 2014). This procedure seems to be the same to the operation of the stylus (needle) of a vinyl record player. In this way, the pointer gradually scans all of the spiral turns (and their subdivisions), like the stylus of the record player scans all the grooves of a vinyl record (see **Figure 7**). As the spiral furrow increases its distance from the center of spiral, the pointer must be capable to increasing its length. The ancient Manufacturer supported the two pointers with a very special design construction, in order to achieve variable length of the pointers during their rotation (Anastasiou et al., 2014).

The 4 turn Saros spiral was divided in 223 cells-synodic months. The eclipse information, symbol *H* (ΗΛΙΟΣ-Helios-Sun) for solar eclipse and *Σ* (ΣΕΛΗΝΗ-Selene-Moon) for a lunar eclipse and also their corresponding times of the events were engraved on some of the cells (Freeth et al., 2006 and 2008). On the internal area of the Saros spiral on Fragment A2 a small circular scale of Exeligmos, which is divided in three equal sectors is also preserved. On the second sector ($2^{nd}$ Saros cycle) the ancient Greek number *H* (8) is engraved and on the third ($3^{rd}$ Saros cycle) the number *IC* (16). When the pointer of Exeligmos was aimed to the number *H* ($2^{nd}$ Saros cycle), then the user of the Mechanism should add on each hour of the eclipse event, +8 additional hours, while +16 hours when the pointer aimed to *IC* number ($3^{rd}$ Saros cycle).

## 4. The preserved Back Cover plate

The BCI Part-2 text is located at the lower area of the Back cover plate and resulted after the combination/stapling of the Fragments A2, E, 19 and 67 (Bitsakis and Jones 2016b). Today, 25 lines with partially/poorly preserved phrases or words or parts of words and letters comprise the Part-2 text (see **Table I**), as the Mechanism was about 2000 years on the Antikythera seabed, and suffered from corrosion material alteration (Scott 1990; Ingo et al., 2006; Voulgaris et al., 2019b). The text presents the *Description* (design), the *Position* and the *Operation* of the outer *operational* parts located on the Back plate and it is the Instruction Manual for the *operational* parts located on the Back Dial plate.
The use of the authors' functional models of the Antikythera Mechanism revealed that the specific way of the use/adjustment of some of the parts was totally necessary to be engraved on the Instruction Manual of the Mechanism (especially the pointers' Reset procedure, presented further below).

**Table I**. The preserved words/letters of the Back Cover plate Part-2, by Bitsakis and Jones 2016b. The preserved letters are in bold and the presumed ones in regular. The letters/words are presented in their original position and form, without gaps. The gaps between words left by *The Engraver* are noted with "_" for one letter dimension gap and in "-" for a half letter dimension.

```
           ...................................................................................................................................
           ...................................................................................................................................
01)  . . . . ΛΟΣ[ ..........................................................................................................................
02)  [. . . .]ΑΠΟΤΩΝΔΙΑΙΡΕΣΕ[ΩΝ.............................................................................................
03)  [. . Ε]ΝΟΛΗ<Ι>ΤΗΙΕΛΙΚΙΤΜΗΜΑΤΑ_ΣΛΕ[-.............................................................................
04)  ΤΑΙΔΕΚΑΙΑΙΕΞΑΙΡΕΣΙΜΟΙΗΜΕΡΑΙΚΑ[...................................................................................
05)  [Ε]ΧΟΝΣΤΗΜΑΤΙΑΔΥΟ-ΠΕΡΙΤΥΜΠΑΝΙ[ΟΥ............................................................................
06)  [Τ]ΑΠΡΟΕΙΡΗΜΕΝΑΣΤΗΜΑΤΙΑΤΡΗΜΑ[.............................................................................
07)  [ΔΙ]ΑΤΩΝΤΡΗΜΑΤΩΝΔΙΕΛΚΕΣΘΑΙ . .[......................................................................
08)  ΟΜΟΙΩΣΤΟΙΣΙΠΡΩ . [ ]. . .[..................................................................................
09)  ΦΥΕΣΠΟΙΗΣ[. .]ΤΥΜΠ[.......................................................................................
10)  ΚΑΙΣΥΜΦΥΕ[Σ.....................................................................................................
11)  [.]ΑΣΤΗΜΑΤΙΑ[....................................................................................................
12)  [. . . . . . .]Ξ[. .]ΑΓΕΣ[.........................................................................................
13)  [. .]ΡΟΥ . . ΘΟΔΟΥ[.]Η[........................................................................................
14)  [. . . .]ΤΗΝΕΝΑΝΤΙΑΝ_Ε[.....................................................................................
15)  [. . .]ΠΕΡΟΝΗΝΟΘΕΝΕΞΥΛΚΗΣ[............................................................................
16)  [. . .]ΤΗΣΠΡΩΤΗΣΧΩΡΑΣ-Μ[................................................................................
17)  [ΓΝΩ]ΜΟΝΙΑΔΥΟ-ΩΝΤΑΑΚΡΑΦΕ[ΡΟΝΤΑΙ............................................................
18)  [. .]ΤΕΣΣΑΡΑΔΗΛΟΙΔΟΜΕΝΤΗ[Ν............................................................................
19)  . ΣΤΗΝΤΗΣ-_ΟCL_ΙΘΙ_ΤΟΥ[...............................................................................
20)  ΜΟΣΕΙΣΙΣΑ_ΣΚΓ-ΣΥΝΤΕΣ[.................................................................................
21)  ΤΕ . Α . . ΟΣΔΙΑΙΡΕΘΗ<Ι> -Η-ΟΛΗ[....................................................................
22)  ΜΩΝ[. . . .]ΟΙΕΓΛΕΙΠΤΙΚΟΙΧΡ[ΟΝΟΙ.....................................................................
23)  ΟΜΟ[ΙΩ]ΣΤΟΙΣΕΠΙΤΗΣΕ[..................................................................................
24)  ΑΚΡΟΝΦΕΡΕΤΑΙΚ[.]. .[......................................................................................
25)  . . . ΜΕΝΤ . ΥΠ[..............................................................................................
           ...................................................................................................................................
           ...................................................................................................................................
```

## 5.1 Back Dial Parts presentation on the Instruction Manual of the Mechanism

- On the BCI Part-2 there is preserved evidence for the two spirals-helices and their subdivisions:

03) . .Ε]Ν ΟΛΗ<Ι> ΤΗΙ ΕΛΙΚΙ ΤΜΗΜΑΤΑ_ΣΛΕ (*along the full length of the helix, 235 parts of the Metonic helix-spiral*)

20) ΜΟΣ ΕΙΣ ΙΣΑ_ΣΚΓ-ΣΥΝ ΤΕΣ[ΣΑΡΣΙ ΠΕΡΙΦΟΡΑΙΣ (*in 223 equal parts in four turns, divide the Saros helix-spiral*). Considering the design of the 223 helix parts-subdivisions it can be assumed that they are not in same linear dimension. The Engraver means that the parts have equal epicenter angles-same angular dimension per each cell.

- The preserved phrase:

04) ΤΑΙ ΔΕ ΚΑΙ ΑΙ ΕΞΑΙΡΕΣΙΜΟΙ ΗΜΕΡΑΙ ΚΑ… (*Omitted-Exeresimoi days*) concerns information regarding the Metonic cycle days arrangement.

- Very characteristic preserved words and phrases concerning a terminology for the description of mechanical parts, the spiral pointers:

05) Ε]ΧΟΝ ΣΤΗΜΑΤΙΑ ΔΥΟ (*having two bearings*)

05) ΠΕΡΙ ΤΥΜΠΑΝΙ[ΟΥ (*about the base*)

06) ΤΑ ΠΡΟΕΙΡΗΜΕΝΑ ΣΤΗΜΑΤΙΑ ΤΡΗΜΑ [ΤΑ ΕΧΟΥΣΙΝ (*the aforementioned bearing, have holes*)

07) ΔΙ]Α ΤΩΝ ΤΡΗΜΑΤΩΝ ΔΙΕΛΚΕΣΘΑΙ (*it passes/slides via the holes*)

08) …………………………………………………………………………….. ΣΥΜ-
09) **ΦΥΕΣ ΠΟΙΗΣ**[ΑΙ] **ΤΥΜΠ**ΑΝΙΟΝ (*made the same Base*, see Section 8).

   - And also the mechanical procedure for the pointers' reset positioning is referred:
15) ΤΗΝ **ΠΕΡΟΝΗΝ ΟΘΕΝ ΕΞΗΛΚΥΣ**ΘΑΙ (*pull out the needle*).

   - The description of the two small circular scales is found in the phrases:
17) ΓΝΩ]**ΜΟΝΙΑ ΔΥΟ-ΩΝ ΤΑ ΑΚΡΑ ΦΕ**[ΡΟΝΤΑΙ ΕΠ' ΕΛΑΣΣΟΝΩΝ ΚΥΚΛΩΝ, (*two pointers, which are rotating on their corresponding small circular scales*)
18) . . **ΤΕΣΣΑΡΑ ΔΗΛΟΙ Δ' Ο ΜΕΝ** (ΚΥΚΛΟΣ) **ΤΗ**[ …….. ΚΑΙ Ο ΔΕ ΚΥΚΛΟΣ… (*the one cycle… and the other cycle… ,* Ο ΜΕΝ ΚΥΚΛΟΣ ΚΑΙ Ο ΔΕ ΚΥΚΛΟΣ).
   (After further analysis and the text reconstruction, it seems that on the Back Dial plate of the Mechanism there are exist two small circles: the first circle depicts the athletic Stephanites Games scale, located on the internal area of the Metonic spiral, while the second is the Exeligmos scale which depicts each of the Saros cycles and the hour correction of the eclipse events, located on the Saros spiral internal area, see Section 8, for Lines-17-20).

   - Additionally, two preserved words correspond to the information engraved on some of the Saros cells, in which a solar or lunar eclipse occurred:
**ΕΓΛΕΙΠΤΙΚΟΙ ΧΡ**ΟΝΟΙ (*ecliptic times/ecliptic hours*, see Section 8, Line 22).

   - Also, the words:
02) . . . . **ΑΠΟ ΤΩΝ ΔΙΑΙΡΕΣΕ**[ΩΝ (*from the divisions*)
03) . . Ε]**Ν ΟΛΗ**<Ι> **ΤΗΙ ΕΛΙΚΙ** (*along the full length of the helix-spiral*)
16) . . . **ΤΗΣ ΠΡΩΤΗΣ ΧΩΡΑΣ** (*1$^{st}$ Area*),
20_ ………**ΣΥΝ ΤΕΣ**[ΣΑΡΣΙ ΠΕΡΙΦΟΡΑΙΣ (*in four turns*)
correspond to the geometrical shape of a helix (spiral) and its geometrical features (see Section 6).

   Same/similar words exist on the work of Heron Alexandreus (Schöne 1903) in *ΠΕΡΙ ΔΙΟΠΤΡΑΣ (Dioptra)*. Heron describes some devices, their mechanical parts and also the mechanical procedure of handling these devices. Heron well describes the construction and use of the *Dioptra*, a device for angle measuring, which is necessary for the construction of buildings, walls, ports, and also for celestial sphere measurements, solar and lunar eclipses, angular distances between stars etc. He mentions the words: ΓΝΩΜΩΝ (*pointer*), ΤΜΗΜΑΤΑ (*divisions*), ΤΥΜΠΑΝΙΟΝ (*small circular plate-disc*, though in the BCI Part-2 it means *a Base*, not-necessarily circular), ΣΤΗΜΑΤΙΑ (*bearings/supporting parts*), ΤΡΗΜΑ (*hole*), ΣΥΜΦΥΕΣ (*stabilized*, though in the BCI Part-2 it means *same part* or *an additional part coexists with the initial part*, see Section 5.4), ΠΕΡΟΝΗ (*needle pin*), ΧΩΡΑΝ and ΧΩΡΙΟΝ (*area*), ΑΚΡΟΝ (*edge*), ΦΕΡΕΤΑΙ (*rotated-travels through*).

   Archimedes invented the geometrical scheme of ΕΛΙΞ/ΕΛΙΚΑ (see Section 6) and in his work *ΠΕΡΙ ΕΛΙΚΩΝ* (*On Spirals*) extensively describes the ΕΛΙΞ, its parts and the geometrical characteristics. Archimedes work was very important and critical for the BCI Part-2 text reconstruction, because the names for the specific parts of the helices were necessary for the *Description*, the *Position* and the *Operation* of the mechanical operational parts of the Back Dial plate.

### 5.2 The Back Cover text's metrology

We repeated the same dimensional measurements in order to detect the number of letters per each line text as in Authors' submitted BCI Part-1. According to **Table II**, the letter number per text line of the BCI Part-2, varies between 66-87 letters per text line, equal to the BCI Part-1 text (equal measurements in Bitsakis and Jones 2016b).

**Table II**. A selection of well-preserved texts of the Back Cover Inscription Part-2 is located on Fragments A and 19. The dimension of each selected text was measured using calibrated visual photographs by the first author 2016/2019 and AMRP PTM. The ratios mm/letter, letters/cm and the corresponding extrapolation for the Back cover width dimension of 170mm (Voulgaris et al., 2019b and Authors' submitted BCI Part-1), in order to estimate the total number of letters for each line-text (see also Bitsakis and Jones 2016b) are presented. The Letters per line width of 170mm is equal to the Part-1 text, reaches between 66-87 letters per line (as is *The Engraver's* handwriting). For the text reconstruction a maximum number of letters per line was set at 88 letters/line.

| Selected Text | Letters number | Dimension | mm/ Letter | Letters/ cm | Letters/ Line width of 170mm |
|---|---|---|---|---|---|
| 05) ΣΤΗΜΑΤΙΑΔΥΟ-ΠΕΡΙΤΥΜΠΑΝ | 21.5 | 42mm | 1.95 | 5.12 | 87 |
| 06) ΠΡΟΕΙΡΗΜΕΝΑΣΤΗΜΑΤΙΑΤ | 20 | 42mm | 2.10 | 4.76 | 81 |
| 16) ΣΠΡΩΤΗΣΧΩΡΑΣ | 12 | 28mm | 2.33 | 4.9 | 73 |
| 17) ΜΟΝΙΑΔΥΟ-ΩΝΤΑΑΚΡΑΦ | 17.5 | 37mm | 2.11 | 4.74 | 80.5 |
| 18) ΡΗΤΕΣΣΑΡΑΔΗΛΟΙΔΟΜΕΝΤΗ | 17 | 38mm | 2.23 | 4.48 | 76 |
| 19) ΝΤΗΣ-_OCL_IΘL_ΤΟΥ | 16.5 | 38.5mm | 2.33 | 4.30 | 73 |
| 20) ΣΙΣΑ_ΣΚΓ-ΣΥΝΤΕΣ | 14.5mm | 37.5mm | 2.58 | 3.87 | 66 |
| 21) ΟΣΔΙΑΙΡΕΘΗ-Η-ΟΛΗ | 15 | 32mm | 2.13 | 4.70 | 80 |
| | | Mean: | **2.22** | **4.5** | **77 (-11/+10)** |

### 5.3 Instruction Manual characteristics

As the Antikythera Mechanism is a complex mechanical device, an explanatory text which defines, presents, describes the *operational mechanical parts*, and also their position and their operation, (i.e. *how this device works*) is needed. The Instruction Manual is totally necessary if the user of the device is not the Manufacturer of the device. (The first author also constructs and uses optomechanical constructions/instruments and in some cases some notes are needed in order for these instruments to be properly operated even by him).

For the BCI Part-2 text reconstruction we applied the same procedure as we did for the reconstruction of the Part-1: The Instruction Manual of the Mechanism for the Back plate operational parts must give comprehensible information regarding the:

1) ***Definition/Description*** of the parts (*This is…., There is…, Is named…., Is consisted by…etc.*).

2) ***Position*** of the parts (*The part is located on…., Is stabilized on…, Is visible on…etc.*).

3) ***Operation/Function*** of the parts (*This part starts…, Opens/Closes…, Is rotated CCW…etc.*),
https://blog.bit.ai/write-instruction-manual/
https://www.douglaskrantz.com/SCManual.html
https://grammar.yourdictionary.com/grammar-rules-and-tips/tips-on-writing-user-manuals.html
https://makeitclear.com/insight/how-is-a-quick-start-guide-different-to-an-instruction-manual.

These steps (which we call *DDPO*) explain how to start, use, operate, and assemble any device.

### 5.4 Parameters for the BCI Part-2 text reconstruction

1) The text presents the parts which affect/adjust the operation of the device i.e. the *operational parts,* (as opposed to the supportive parts which are not related to the device's operation, e.g. the wooden decorative case, see Authors' submitted BCI Part-1).

2) The preserved text is a technical text which describes the Definition/Description, the Position and the Operation of the parts.

3) A large percentage of the text appears Symmetrical, especially when *The Engraver* describes the two helices-spirals division and the pointers of the two helices-spirals.

4) For the syntax of the reconstructed text, the syntax style retracted by the preserved text was applied, e.g. the following phrase is preserved: 23) **ΟΜΟ**ΙΩΣ **ΤΟΙΣ ΕΠΙ ΤΗΣ Ε**[… and this syntax was maintained: ΤΗΙ ΕΠΙ ΤΗΣ ΕΛΙΚΟΣ ΑΡΧΗΙ and ΤΟΙΣ ΕΠΙ ΤΟΥ ΕΞΕΛΙΓΜΟΥ ΜΕΡΕΣΙ.

5) For phrases of the reconstructed text with the same meaning, the preserved words were used.

6) In order to avoid repeating the same words on the same sentence e.g. ΤΗΣ ΤΟΥ ΜΕΤΩΝΟΣ ΧΩΡΑΣ ΚΑΙ ΤΗΣ ΤΟΥ ΠΕΡΙΟΔΙΚΟΥ ΧΩΡΑΣ, a text collapse was applied, e.g. ΤΗΣ ΤΟΥ ΜΕΤΩΝΟΣ ΚΑΙ ΤΗΣ ΤΟΥ ΠΕΡΙΟΔΙΚΟΥ ΧΩΡΑΣ or ΤΗΣ ΤΟΥ ΜΕΤΩΝΟΣ ΧΩΡΑΣ ΚΑΙ ΤΗΣ ΤΟΥ ΠΕΡΙΟΔΙΚΟΥ.

7) For the meaning of some of the preserved words on the BCI text, the authors did not choose the standard translation of the ancient Greek words instead they translated those words considering the mechanical characteristics of the Back plate parts, which were detected on the AMRP CTs. For example, in BCI Part-2, the word ΣΥΜΦΥΕΣ (ΣΥΝ+ΦΥΩ) is used by *The Engraver*. The word ΣΥΜΦΥΕΣ is mentioned in Heron's *Dioptra* and means "*is stabilized/attached/fixed*". In the BCI Part-2, Line 9(8) the phrase ΣΥΜ**ΦΥΕΣ ΠΟΙΗΣ**ΑΙ **ΤΥΜΠ**ΑΝΙΟΝ is preserved. As the ΤΥΜΠΑΝΙΟΝ (which is the pointer's base, see below) is continuously rotated, cannot be fixed/stabilized. Therefore, the aforementioned translation does not fit for the pointer's part description/operation, inferred from the AMRP Computed Tomographies. The word ΦΥΩ is also means "*growing*" or "*produce*" or "*generate*" (see Mousouros Etymologicum Magnum in Kalliergis 1499; Hofmann 1974) and therefore ΣΥΜΦΥΕΣ means  "*born with one*" or "*grown together*" or "*produced together with something else*" or "*exist with another same part*" or "*component of the same kind*" or "*same part*" (see also Liddell and Scott dictionary 2007).

The word ΤΥΜΠΑΝΟΝ in Heron's engineering means a circular plate. The word ΤΥΜΠΑΝΙΟΝ (small ΤΥΜΠΑΝΟΝ) is mentioned two times in BCI Part-2 text and it indicates the base of the Metonic (and the missing Saros) pointer. But on the AMRP CT's it is clearly visible that this base has parallelogram shape design rather-than circular. Therefore the word ΤΥΜΠΑΝΙΟΝ is translated on the Instruction Manual of the Mechanism as a small Base (with no indication of shape) and not as a circular small plate.

8) For the definition of the geometrical features defining the spiral, the terminology presented in Archimedes work ΠΕΡΙ ΕΛΙΚΩΝ was adopted. E.g. Archimedes called the "*center*" of the helix ΕΛΙΚΟΣ ΑΡΧΑ (ΑΡΧΗ, *beginning/origin of the helix*).

9) The preserved mechanical parts were visible on the AMRP CTs and their digital reconstruction/completion (Anastasiou et al., 2014) was used as the original information in order to describe the parts in the ancient Greek language, e.g. the bearings on the Metonic pointer are located fixed on the pointer's base, opposite to each other and perpendicular to their base.

10) For the geometrical description of the parts, the words found in Heron's work were adopted, since many words on the BCI part-2 text are already mentioned in Heron's work (Dioptra in Schöne 1903) and also in Euclid's work ΣΤΟΙΧΕΙΑ-Elements, (Heiberg 1883–1885; Fitzpatrick 2008), e.g. the holes of the bearings have oblong shape ΕΤΕΡΟΜΗΚΗ ΤΩΙ ΣΧΗΜΑΤΙ (also mentioned by Ptolemy (Heiberg 1898).

We believe that the Instruction Manual of the Antikythera Mechanism was written by *The Engraver*, who was the same person as *The Manufacturer* and not by some expert in the Greek language and grammar during the Hellenistic era (since also the preserved phrases are not samples of advanced grammar). Thus, we present the reconstructed text in its simplest grammar form, leaning towards the colloquial Greek language rather than the formal/literary one. The modern punctuation symbols of *comma* and *dot* are included on the reconstructed text, for better understanding of the meanings.

## 6. Archimedes' work ΠΕΡΙ ΕΛΙΚΩΝ

The work of *Archimedes of Syracuse*, *ΠΕΡΙ ΕΛΙΚΩΝ* (*On Spirals*, Heath 2009; see also Netz 2017) describes the formation of a helix, its geometrical parts, its characteristics and a number of results. The ancient title of *ΠΕΡΙ ΕΛΙΚΩΝ* (*About Helices*) was translated (around the 13[th] century, Israel 2015) in Latin as *De Spiral Libus* where by a mistake the word ΕΛΙΞ (Helix) was faulty translated as spiral (ΣΠΕΙΡΑ). As the word ΕΛΙΚΙ (*dat* of the word ΕΛΙΞ) is preserved on the BCI Part-2, the name for the two spirals is definitely ΕΛΙΚΑ and not ΣΠΕΙΡΑ.

*Archimedes*, in his work ΠΕΡΙ ΕΛΙΚΩΝ, defines the creation of the Helix, which is a geometrical curve and has specific characteristics, parts and terminology, **Figure 1**.

Below we quote the listed Definitions in Archimedes' work, translated in English and edited by T. Heath 2009. The English translation is in italics and our comments in regular:

1) *If a straight line drawn in a plane revolve at a uniform rate about one extremity which remains fixed and return to the position from which it started, and if, at the same time as the line revolves, a point move at a uniform rate along the straight line beginning from the extremity which remains fixed, the point will describe a spiral (pug) in the plane*.
(i.e. the shape formation of an Archimedean helix demands a continuously change position of a point, in two different motions, linear and rotation. According to this definition, it results that the Archimedean helix presents same distances between successive turns).

The two helices/spirals of the Antikythera Mechanism are not shaped according to the Archimedean design pattern, but are a combination of semicircles with gradually increased radius. The end of a semicircle coincides with the start of the next semicircle, which has a larger radius (Anastasiou et al., 2014). This design pattern is easier construct, as only one motion (rotation) is needed for the formation of each semicircle. The final shape of this helix design quite simulates the Archimedean helix and has not any mechanical malfunction or impact to the measurements.

*2. Let the extremity of the straight line which remains fixed while the straight line revolves be called the origin (beginning, ΑΡΧΑ/ΑΡΧΗ) of the spiral.*

The helix has not a center, but a beginning/origin, which is the point of the rotating straight line that is not changing its position.

*3. And let the position of the line from which the straight line began to revolve be called the initial line in the revolution (ΑΡΧΑ ΤΑΣ ΠΕΡΙΦΟΡΑΣ).*

*4. Let the length which the point that moves along the straight line describes in one revolution be called the first distance (ΠΡΩΤΑ ΕΥΘΕΙΑ), that which the same point describes in the second revolution the second distance (ΔΕΥΤΕΡΑ ΕΥΘΕΙΑ), and similarly let the distances described in further revolutions be called after the number of the particular revolution.*

*5. Let the area bounded by the spiral described in the first revolution and the first distance be called the first area (1ˢᵗ Area), that bounded by the spiral described in the second revolution and the second distance the second area, and similarly for the rest in order.*

The integrated area which is defined by the 1ˢᵗ distance through the 1ˢᵗ revolution created the 1ˢᵗ Area (also the same for 2ⁿᵈ, 3ʳᵈ, 4ᵗʰ, etc.), **Figure 1**.

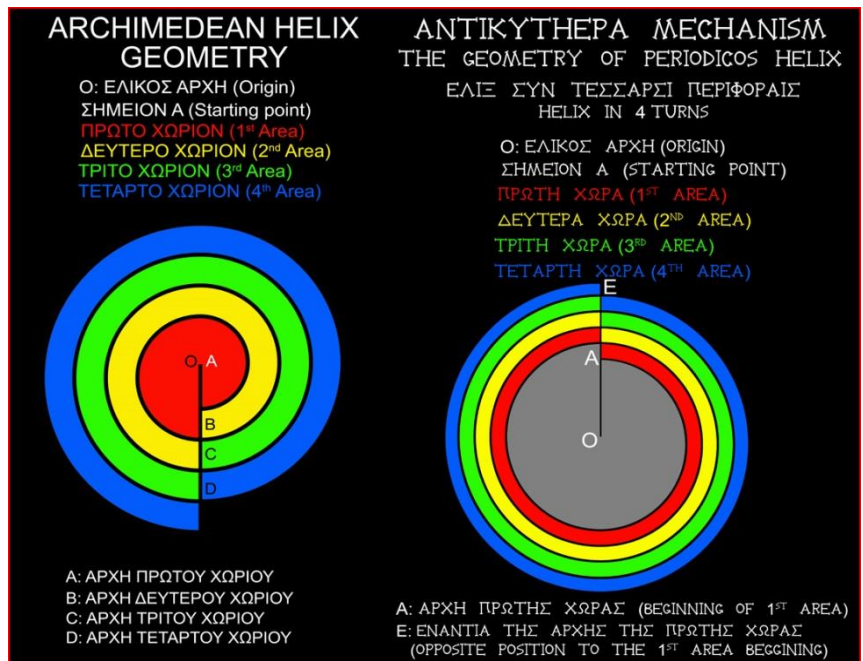

**Figure 1:** Left, The Archimedean helix and its geometrical parts. Each colored strip turn corresponds to an area (1ˢᵗ, 2ⁿᵈ, 3ʳᵈ and 4ᵗʰ). The 1ˢᵗ area is defined by the point A, which travels in constant velocity through the line OE (in this design point A starts traveling at point O). Right: The Saros (Periodicos) Helix also in four turns. In the same manner, point A (which starts from a position which does not coincide with point O-helix origin, and travels through the line O(A)E, (i.e. starts from point A and ends at point E) defines the 1ˢᵗ area of this Helix (red colored strip). Point A is the beginning of the 1ˢᵗ area (ΑΡΧΗ ΠΡΩΤΗΣ ΧΩΡΑΣ). Point E is the end of the 4ᵗʰ area (i.e. the opposite position of the 1ˢᵗ area beginning). Equal geometry and naming apply to the 5 turn Metonic helix (with 5 areas).

The two helices of the Antikythera Mechanism present an additional peculiarity: the point that moves along the (rotated) straight line does not start at the Helix origin point, but at cell-0 (the 1ˢᵗ boundary line of cell-01). Therefore, by applying the aforementioned definitions for the helix parts naming and numbering, the specific position of the point/Cell-0 was taken into account. So, the starting point for the turns numbering is located on Cell-0 and not on the helix origin, **Figure 1**. This consideration is supported by the preserved phrase ΤΗΣ ΠΡΩΤΗΣ ΧΩΡΑΣ (1ˢᵗ Area) corresponding to the reset positon of Metonic pointer on cell-01 (see Line 16).

### 7. The word ΕΝΙΑΥΤΟΣ and the Naming for the Back plate two Helices

The word ΕΝΙΑΥΤΟΣ was used in order to describe a repeated cycle of time: Geminus in Section 8 writes ΚΑΘ ΗΛΙΟΝ ΕΝΙΑΥΤΟΣ (solar tropical cycle-one year = 1 ΕΝΙΑΥΤΟΣ/year i.e. 365.25 days) or ΚΑΤΑ ΣΕΛΗΝΗΝ ΕΝΙΑΥΤΟΣ (Lunar cycle = 1 ΕΝΙΑΥΤΟΣ/29.53 days), or ΚΑΤ' ΑΙΓΥΠΤΙΟΥΣ ΕΝΙΑΥΤΟΣ (the Egyptian calendar cycle ΑΙΓΥΠΤΙΩΝ 1 ΕΝΙΑΥΤΟΣ/365 days. Also ΜΕΤΩΝΟΣ ΕΝΙΑΥΤΟΣ, the Metonic cycle of 19 years = 1 ΕΝΙΑΥΤΟΣ/19 years, Diodorus Siculus 12,36 in Bekker 1903-1906; Suidas lexicon in Bekker 1854).

The upper helix presents the Metonic calendar, the 13/12 months per year, and also the omitted-Exeresimoi days. A very probable ancient Greek naming for this helix (Voulgaris et al., 2021), is Η ΤΟΥ ΜΕΤΩΝΟΣ ΕΝΙΑΥΤΟΣ ΕΛΙΞ (*the Helix of the Metonic eniautos-cycle*) or Η ΤΟΥ ΜΕΤΩΝΟΣ ΕΛΙΞ (*the Helix of Meton*) or Η ΕΛΙΞ ΤΗΣ ΕΝΝΕΑΚΑΙΔΕΚΑΕΤΗΡΙΔΟΣ (*the Helix of Enneakedekateris*). The last option includes a word of 20 letters and it seems to be less probable to be continuously engraved.

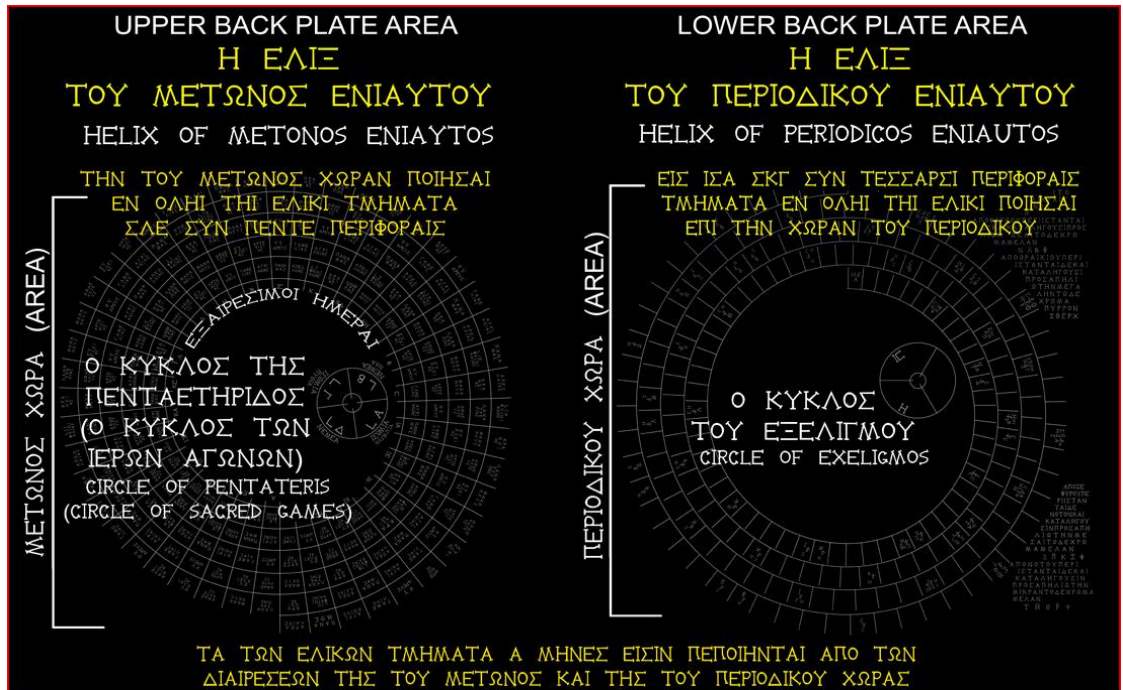

**Figure 2:** The nomenclature of the *operational* parts located at the Back Dial plate of the Antikythera Mechanism. The names for the specific parts are presented, analyzed and discussed in the following Sections. The ΕΞΑΙΡΕΣΙΜΟΙ ΗΜΕΡΑΙ (omitted days), are engraved along the 1[st] turn of the Metonic helix.

At the first presentation of the Metonic helix (*Definition* of helix), which should be located 2-3 lines before the currently preserved Line 1, *The Engraver* could probably use the extended naming Η ΤΟΥ ΜΕΤΩΝΟΣ ΕΝΙΑΥΤΟΥ ΕΛΙΞ but in the following text he probably used the shorter name Η ΤΟΥ ΜΕΤΩΝΟΣ ΕΛΙΞ (or Η ΕΛΙΞ ΤΟΥ ΜΕΤΩΝΟΣ), **Figure 2**.

The Saros cycle is the minimum time span (223 synodic months), in which the sequence pattern of the solar and lunar eclipses is repeated. Pliny the Elder (Naturalis Historia II.10,56, Bostock and Riley 1855) mentions the 223-month eclipse cycle without mentioning the name of this period. Ptolemy in Almagest (Book V,II) writes "Οἱ παλαιοὶ μαθηματικοί... Ἐκάλεσαν δέ τόν χρόνον τοῦτον περιοδικόν (*the ancient mathematicians/astronomers called this time span Periodicos-ΠΕΡΙΟΔΙΚΟΣ ΧΡΟΝΟΣ*). He also mentions that Hipparchus re-measured and corrected the duration of Periodicos. The word Saros is mentioned in the lexicon by the Byzantine Suidas (Bekker 1854), who described a different cycle; and E. Halley who mistakenly connected the name of Saros to the Periodicos cycle.

As the word ΕΝΙΑΥΤΟΣ describes any cycle of time, a very probable name for the Saros helix is Η ΕΛΙΞ ΤΟΥ ΠΕΡΙΟΔΙΚΟΥ ΕΝΙΑΥΤΟΥ (*the Helix of Periodikos Eniautos*) or Η ΕΛΙΞ ΤΟΥ ΠΕΡΙΟΔΙΚΟΥ (*the Helix of Periodicos*).

In the present work the word *Helix* instead of the *Spiral* and the word *Periodicos* instead of *Saros* are used.

## 8. Reconstructing the BCI Part-2 text

For the reconstruction of the Instruction Manual of the Antikythera Mechanism concerning the Back plate operational parts, the text is presented in the simplest grammar and syntax. For the description of equal mechanical procedures the existing words were used as much as possible.

An introductory text-general description (in Line i) regarding *what is presented on the Back plate* should have existed before Line 1 of the preserved text. Also, a reference for the specific *Naming*, *Definition* and *Position* of the two helices is necessary (Lines i, ii).

**- Lines i, ii, iii** (not preserved, hypothetical but necessary):

i) ΕΙΣ ΤΗΝ ΟΠΙΣΘΙΑΝ ΟΨΙΝ, ΤΟ ΗΜΕΡΟΛΟΓΙΟΝ ΚΑΙ ΤΩΝ ΕΓΛΕΙΨΕΩΝ ΤΑ ΓΕΓΟΝΟΤΑ ΚΑΙ ΣΕΛΗΝΗΣ ΚΑΙ ΗΛΙΟΥ, ΑΝΑΓΡΑΦΟΝΤΑΙ

ii) ΑΥΤΟΘΙ. ΔΙΗΡΗΜΕΝΗ ΕΣΤΙΝ Η ΠΛΑΞ ΕΙΣ ΕΛΙΚΑΣ ΔΥΟ ΚΑΙ ΜΕΤ' ΑΥΛΑΚΟΣ ΕΚΑΣΤΗ ΕΣΤΙΝ. ΑΝΩΘΕΝ ΚΕΙΤΑΙ Η ΤΟΥ ΜΕΤΩΝΟΣ

iii) ΕΝΙΑΥΤΟΥ ΕΛΙΞ ΚΑΙ ΚΑΤΩΘΕΝ Η ΕΛΙΞ ΤΟΥ ΠΕΡΙΟΔΙΚΟΥ ΕΝΙΑΥΤΟΥ. ...................................................................

i) On the Back face, the calendar and the lunar and solar eclipse events are engraved.

ii) The Back plate is divided in two helices and each of them has a furrow. At the top there is the Helix of Metonic

iii) eniautos and at the bottom there is the Helix of Periodicos Eniautos.

The word ΚΕΙΤΑΙ *(is located, lies)* is preserved in Line 21 of the BCI Part-1**.** For the word ΕΝΙΑΥΤΟΣ see Section 7. The word ΕΓΛΕΙΨΙΣ (eclipse) is presented instead of ΕΚΛΕΙΨΙΣ, as *The Engraver* writes in Line 22 **ΟΙ ΕΓΛΕΙΠΤΙΚΟΙ ΧΡ**ΟΝΟΙ instead of **ΟΙ ΕΚΛΕΙΠΤΙΚΟΙ ΧΡ**ΟΝΟΙ. The word ΕΓΛΕΙΨΙΣ is also repeated in Lines 11 and 21 and ΕΓΛΕΙΠΤΙΚΟΙ in Line 23.

**- Line 1**: the specific three preserved letters …**ΛΟΣ**… (or …**ΛΟΣ** or …**ΛΟ** *vac* Σ) do not give any hint to help us guess the actual word or meaning.

**- Line 2:** The phrase **ΑΠΟ ΤΩΝ ΔΙΑΙΡΕΣΕ**ΩΝ (…*resulted from the divisions*, in plural, i.e. at least two divisions) is related to the procedure of dividing the two helices in segments (cells). Each of these segments corresponds to a lunar synodic month (ΣΥΝΟΔΙΚΟΣ ΜΗΝΑΣ). Therefore, in Line 1 there should be a general *Definition/Description* for the subdivisions of the two helices, which correspond to ΜΗΝΕΣ (ΣΥΝΟΔΙΚΟΙ) (*lunar synodic months*). Afterwards, *The Engraver* explains the way for the first helix division (continuing in Line 3).

**- Line 3:** The words Ε**Ν ΟΛΗ<Ι> ΤΗΙ ΕΛΙΚΙ** (*along the full path of the helix*), **ΤΜΗΜΑΤΑ** (*segments*) and **ΣΛΕ** (235) are directly related to the Metonic cycle and the division of the Metonic helix area (ΧΩΡΑ see Section 6) in 235 segments.

The phrase in Line 20: **ΜΟΣ. ΕΙΣ ΙΣΑ_ΣΚΓ-ΣΥΝ ΤΕΣ**ΣΑΡΣΙ ΠΕΡΙΦΟΡΑΙΣ (*in equal 223 parts in four turns*) describes the four turn helix of Periodicos Eniautos. A similar reference for the Metonic helix in 5 turns – ΣΥΝ ΠΕΝΤΕ ΠΕΡΙΦΟΡΑΙΣ was reconstructed for Line 3.

| |
|---|
| 01) . . . . . ΛΟΣ………………………………………   ΤΑ ΤΩΝ ΕΛΙΚΩΝ ΤΜΗΜΑΤΑ, Α ΜΗΝΕΣ ΕΙΣΙΝ, ΠΕΠΟΙΗ- |
| 02) ΝΤΑΙ **ΑΠΟ ΤΩΝ ΔΙΑΙΡΕΣΕ**ΩΝ ΤΗΣ ΤΟΥ ΜΕΤΩΝΟΣ ΚΑΙ ΤΗΣ ΤΟΥ ΠΕΡΙΟΔΙΚΟΥ ΧΩΡΑΣ. ΕΠΙ ΤΗΝ ΤΟΥ ΜΕΤΩΝΟΣ ΧΩΡΑΝ ΕΙΣ- |
| 03) ΙΝ ΕΝ **ΟΛΗ**<I> **ΤΗΙ ΕΛΙΚΙ, ΤΜΗΜΑΤΑ_ΣΛΕ_**ΣΥΝ ΠΕΝΤΕ ΠΕΡΙΦΟΡΑΙΣ……. |
| *01) …………………………………………………… The parts of helix, which are synodic months, have been* |
| *02) created by the division of the Metonic and Periodicos helix area. The Metonic helix area is made (divided)* |
| *03) along the full path of the helix, in  235 parts  in five turns…….* |

**- Lines 3-4**: The omitted days **ΕΞΑΙΡΕΣΙΜΟΙ ΗΜΕΡΑΙ ΚΑ**[….. , are engraved along the first turn of the helix (*Description*). The specific sequence pattern of the omitted days was calculated by Meton (*reference*), so the description and the reference were added to the text.

| |
|---|
| 03)  ΙΝ ΕΝ **ΟΛΗ**<I>  **ΤΗΙ  ΕΛΙΚΙ, ΤΜΗΜΑΤΑ_ΣΛΕ_**ΣΥΝ ΠΕΝΤΕ ΠΕΡΙΦΟΡΑΙΣ. ΚΑΤΑ ΤΗΝ ΤΗΣ ΕΛΙΚΟΣ ΠΡΩΤΗΝ ΠΕΡΙΦΟΡΑΝ ΑΝΑΓΡΑΦΟΝ- |
| 04) **ΤΑΙ ΔΕ ΚΑΙ ΑΙ ΕΞΑΙΡΕΣΙΜΟΙ ΗΜΕΡΑΙ ΚΑ**ΤΑ ΤΟΝ ΜΕΤΩΝΑΝ ΛΕΛΟΓΙΣΜΕΝΑΙ. ΤΗΙ ΕΠΙ ΤΗΣ ΕΛΙΚΟΣ ΑΡΧΗΙ ΤΥΜΠΑΝΙΟΝ ΚΕΙΤΑΙ |
| *03) along the full path of the helix, in  235 parts  in five turns. Along the first turn there are also engraved* |
| *04) the Exeresimoi (omitted) days, calculated by Meton. At the beginning (origin) of the (Metonic) helix there is a base,* |

**- Lines 4-6**: The ***Definition/Description***, the ***Position*** of the Metonic pointer parts is presented.

**- Line 4:** The *Definition* and the *Position* of **ΤΥΜΠΑΝΙΟΝ** (*Base*) are missing.

- *Definition*: ΤΥΜΠΑΝΙΟΝ ΕΣΤΙΝ (there is a Base) or ΤΥΜΠΑΝΙΟΝ ΚΕΙΤΑΙ (a Base is located)

- *Position* (where is it located?): ΤΗΙ ΕΠΙ ΤΗΣ ΕΛΙΚΟΣ ΑΡΧΗΙ (*on the Helix origin-beginning*), **Figure 3**.

- *Operation*: ΤΥΜΠΑΝΙΟΝ is a part of the Metonic pointer.

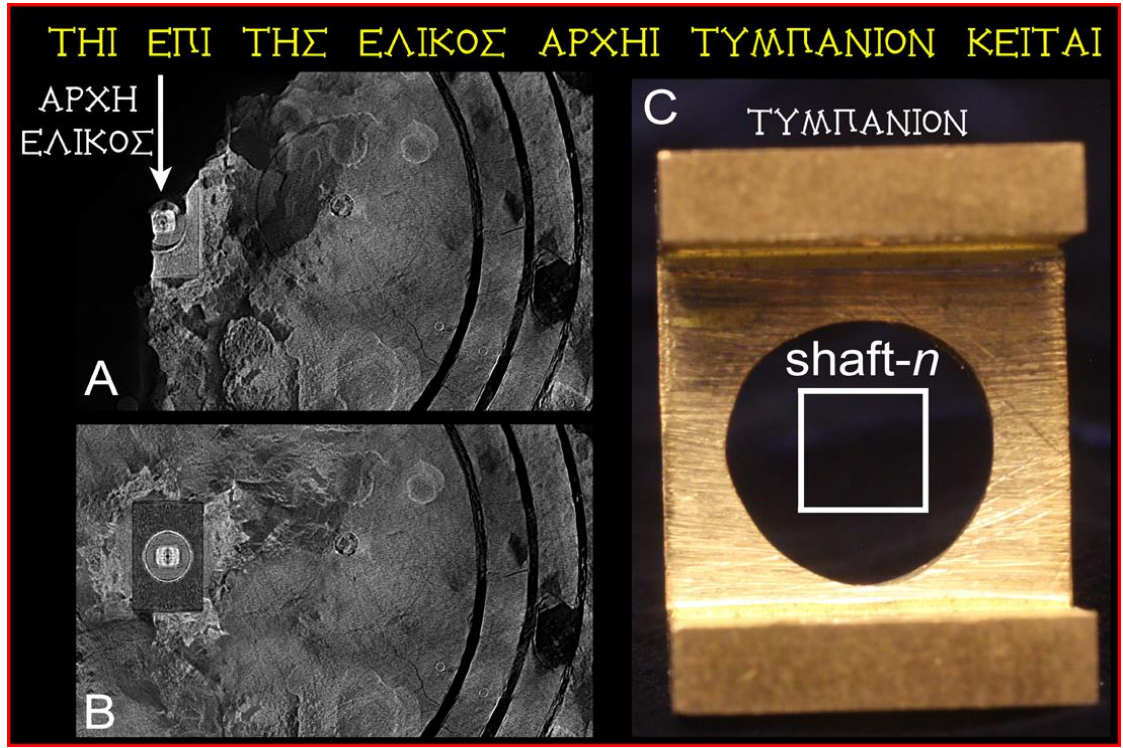

**Figure 3:** A) AMRP tomography (axial view) of Fragment B central part, presenting the position, and the shape of the Metonic pointer's Base (ΤΥΜΠΑΝΙΟΝ). Shaft-n is located in the central hole. The Volume Data was aligned to the Base direction. B) For the digital reconstruction of the lost parts, the central/axial Symmetry was applied and the oblong shape of the Base is revealed (see also Anastasiou et al., 2014). The AMRP Computed Tomography was processed by the authors and was taken from the Raw Data Volume, using the software VolviCon. C) The reconstructed Base of the pointer in bronze (spare part), by first author.

- **Line 5:** The *Definition* and the *Position* of **ΣΤΗΜΑΤΙΑ** (*bearings/supporting arms*) is preserved. The *Description* is missing.

- *Definition*: **ΣΤΗΜΑΤΙΑ ΔΥΟ** (*two bearings/supporting arms*).

- *Description*: ΤΑ ΣΤΗΜΑΤΙΑ ΚΑΙ ΟΜΟΙΑ ΚΑΙ ΠΑΡΑΛΛΗΛΑ ΕΙΣΙΝ (*The bearings are similar and parallel to each other*), (*Ptolemy* in Heiberg 1898, p. 403, 17-18 *ὀρθά πρισμάτια ἴσα τε καί παράλληλα, perpendicular plates of similar size and parallel to each other,* describes the parts for a parallactic instrument construction).

- *Position*: (where are they located?): ΟΡΘΑ ΙΣΤΑΝΤΑΙ ΕΠΙ ΤΟΥ ΤΥΜΠΑΝΙΟΥ (*they are perpendicularly stabilized/fixed on the Base*) (*Heron* in Schoene 1903, p. 204, 12 *Οἱ δε κανόνες ὀρθοί σταθήσονται, the plates are perpendicularly located*) **Figure 4**. Instead of the word ΙΣΤΑΝΤΑΙ there could be the word ΣΥΜΠΑΓΗ or ΤΕΤΑΓΜΕΝΑ (*joined together, compacted, see LSG*).

The phrase **ΠΕΡΙ ΤΥΜΠΑΝΙ**ΟΥ is translated as a reference, meaning *About the Base*, as in Archimedes' work ΠΕΡΙ ΕΛΙΚΩΝ (*On Spirals*) and not as positioning detail, like *Around the Base*.

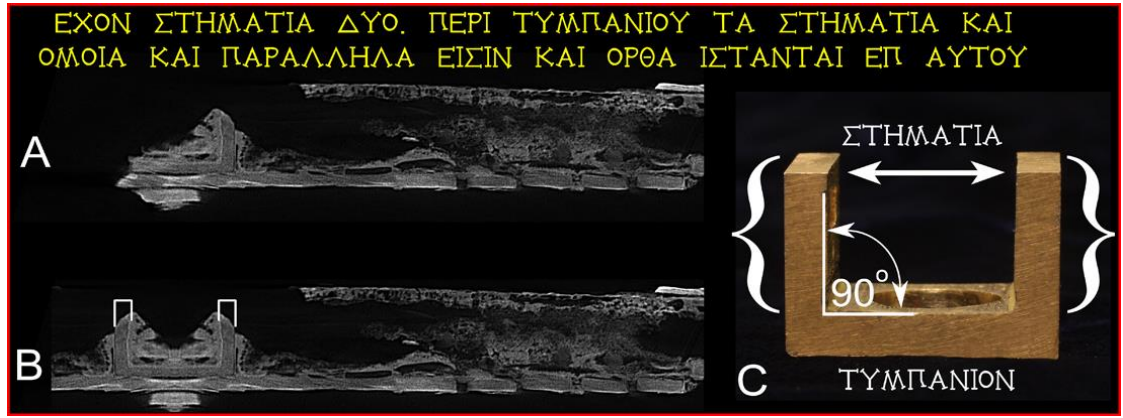

**Figure 4:** A) AMRP tomography of a part of Fragment B (sagittal view), presenting the profile of the Metonic pointer bearings (ΣΤΗΜΑΤΙΑ), which are parts of the Base (ΤΥΜΠΑΝΙΟΝ). The Volume Data was aligned to the Base direction. B) For the digital reconstruction of the lost parts, the central/axial Symmetry was applied and the profile of the bearings is revealed (see also Anastasiou et al., 2014). The white lines show the probable height of the bearings, as today these parts are missing. The AMRP Computed Tomography was processed by the authors and was taken from the Raw Data Volume, using the software VolviCon. C) The reconstructed Base of the pointer with its bearings in bronze (spare part), by first author.

- **Line 6:** The *Definition* and the *Position* of **ΤΡΗΜΑ**ΤΑ (*holes*) is preserved. The *Description* is missing.

- *Definition*: Τ**Α ΠΡΟΕΙΡΗΜΕΝΑ ΣΤΗΜΑΤΙΑ ΤΡΗΜΑ**ΤΑ ΕΧΟΝ (*The aforementioned bearings have holes*).

- *Description*: ΚΑΙ ΕΤΕΡΟΜΗΚΗ ΤΩ ΣΧΗΜΑΤΙ ΚΑΙ ΟΜΟΙΑ ΚΑΙ ΕΝΑΝΤΙΑ ΚΑΙ ΙΣΟΥΨΗ ΕΣΤΙΝ (correct ΙΣΟΫΨΗ) (*the holes are similar, in oblong shape, opposite to each other and of the same height*). The word ΕΤΕΡΟΜΗΚΗ is mentioned by *Heron* in Schöne 1903, p. 6 *Ἔστω χωρίον ετερόμηκες*) and *Euclid* in Heiberg 1883–1885, p. 6 *τέ ἔστι καὶ ὀρθογώνιον, ἑτερόμηκες δέ* ) and ΙΣΟΫΨΗ by *Heron* in Schöne 1903, p. 212 and by Euclid in Heiberg 1883–1885, XI p. 134 *Ἐὰν ᾖ δύο πρίσματα ἰσοϋψῆ.*

- *Position*: The holes located/created on the bearings, **Figure 5**.

---

04) **ΤΑΙ ΔΕ ΚΑΙ ΑΙ ΕΞΑΙΡΕΣΙΜΟΙ ΗΜΕΡΑΙ ΚΑ**ΤΑ ΤΟΝ ΜΕΤΩΝΑ ΛΕΛΟΓΙΣΜΕΝΑΙ. ΤΗΙ ΕΠΙ ΤΗΣ ΕΛΙΚΟΣ ΑΡΧΗΙ ΤΥΜΠΑΝΙΟΝ ΚΕΙΤΑΙ
05) Ε**ΧΟΝ ΣΤΗΜΑΤΙΑ ΔΥΟ. -ΠΕΡΙ ΤΥΜΠΑΝΙ**ΟΥ ΤΑ ΣΤΗΜΑΤΙΑ ΚΑΙ ΟΜΟΙΑ ΚΑΙ ΠΑΡΑΛΛΗΛΑ ΕΙΣΙΝ ΚΑΙ ΟΡΘΑ ΙΣΤΑΝΤΑΙ ΕΠ ΑΥΤΟΥ.
06) Τ**Α ΠΡΟΕΙΡΗΜΕΝΑ ΣΤΗΜΑΤΙΑ ΤΡΗΜΑ**ΤΑ ΟΜΟΙΑ ΕΧΟΥΣΙΝ Α ΚΑΙ ΕΤΕΡΟΜΗΚΗ ΤΩΙ ΣΧΗΜΑΤΙ ΚΑΙ ΕΝΑΝΤΙΑ ΚΑΙ ΙΣΟΫΨΗ ΕΙΣΙΝ.

*04) the Exeresimoi (omitted) days, calculated by Meton. At the beginning (origin) of the (Metonic) helix there is a base*
*05) which has two bearings. The bearings of the base are similar, parallel to each other and perpendicular to the base and fixed on it.*
*06) The aforementioned bearings have holes, which are similar, in oblong shape, opposite to each other and of the same height.*

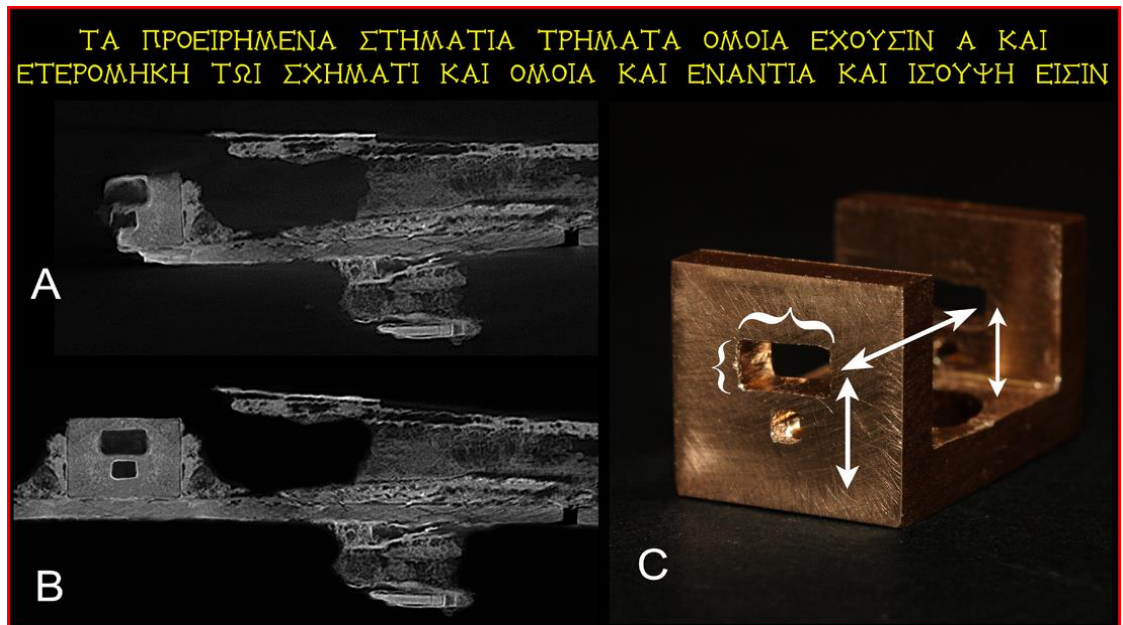

**Figure 5:** A) AMRP tomography of a part of Fragment B (coronal view), presenting the front face of the Metonic pointer bearings (ΣΤΗΜΑΤΙΑ), which have two holes (the Metonic pointer was adapted on the upper hole, while on the lower hole a pin for stabilization was adapted, missing now, see Anastasiou et al., 2014). Note that in the lower hole, even at a size as small as this, the angle of 90° on the preserved shape is quite sharp and precise, as a result of the high quality tools –a delicate file– used (and probably constructed) by the ancient Manufacturer. We can confidently say that *the ancient Manufacturer was "painting" in 3D on bronze material*. B) For the digital reconstruction of the lost parts, the central/axial Symmetry was applied and the profile of the bearings is revealed (see also Anastasiou et al., 2014). The AMRP Computed Tomography was processed by the authors and was taken from the Raw Data Volume, using the software VolviCon. (The Volume Data was aligned to the pointer's Base direction). C) The reconstructed Base of the pointer with its bearings and their holes in bronze (spare part), by first author.

    **- Line 7**: The *Definition/Description* of ΓΝΩΜΟΝΙΟΝ and the *Definition* of ΠΕΡΟΝΗ (pointer of the Metonic helix) is missing.

- *Definition:* ΓΝΩΜΟΝΙΟΝ ΜΕΤΑ ΠΕΡΟΝΗΣ (pointer with a pin).

- *Operation* of the pointer is partially preserved: Δl**A ΤΩΝ ΤΡΗΜΑΤΩΝ ΔΙΕΛΚΕΣΘΑΙ** (*the pointer slides via the holes*), (instead of ΔΙΕΛΚΕΣΘΑΙ-infinitive, it might have been written ΔΙΕΛΚΕΤΑΙ-verb), also ΔΕΞΙΟΣΤΡΟΦΩΣ ΦΕΡΕΤΑΙ, ΤΟΥΣ ΜΗΝΑΣ ΔΕΙΚΝΥΣΙ (and is rotated in CW direction indicating the synodic months).

The phrase ΔΕΞΙΟΣΤΡΟΦΩΣ ΦΕΡΕΤΑΙ was also adopted on the BCI Part-1 (Authors' submitted BCI Part-1), in order to explain that the Lunar Cylinder (the proper and ideal Input of the Antikythera Mechanism see Voulgaris et al., 2018b and 2022), must be rotated in CW direction (for mechanical and operational reasons).

- *Position:* The pointer is adapted on the Base (through the holes) and it is a part of the Base, **Figure 6**.

| 07) ΔΙ**Α ΤΩΝ ΤΡΗΜΑΤΩΝ ΔΙΕΛΚΕΣΘΑΙ** ΓΝΩΜΟΝΙΟΝ ΜΕΤΑ ΠΕΡΟΝΗΣ, ΔΕΞΙΟΣΤΡΟΦΩΣ ΦΕΡΕΤΑΙ ΚΑΙ ΤΟΥΣ ΜΗΝΑΣ ΔΕΙΚΝΥΣΙ. |
|---|
| *07) Through the holes slides a pointer which has a perpendicular pin and is rotated CW indicating the (lunar synodic) months.* |

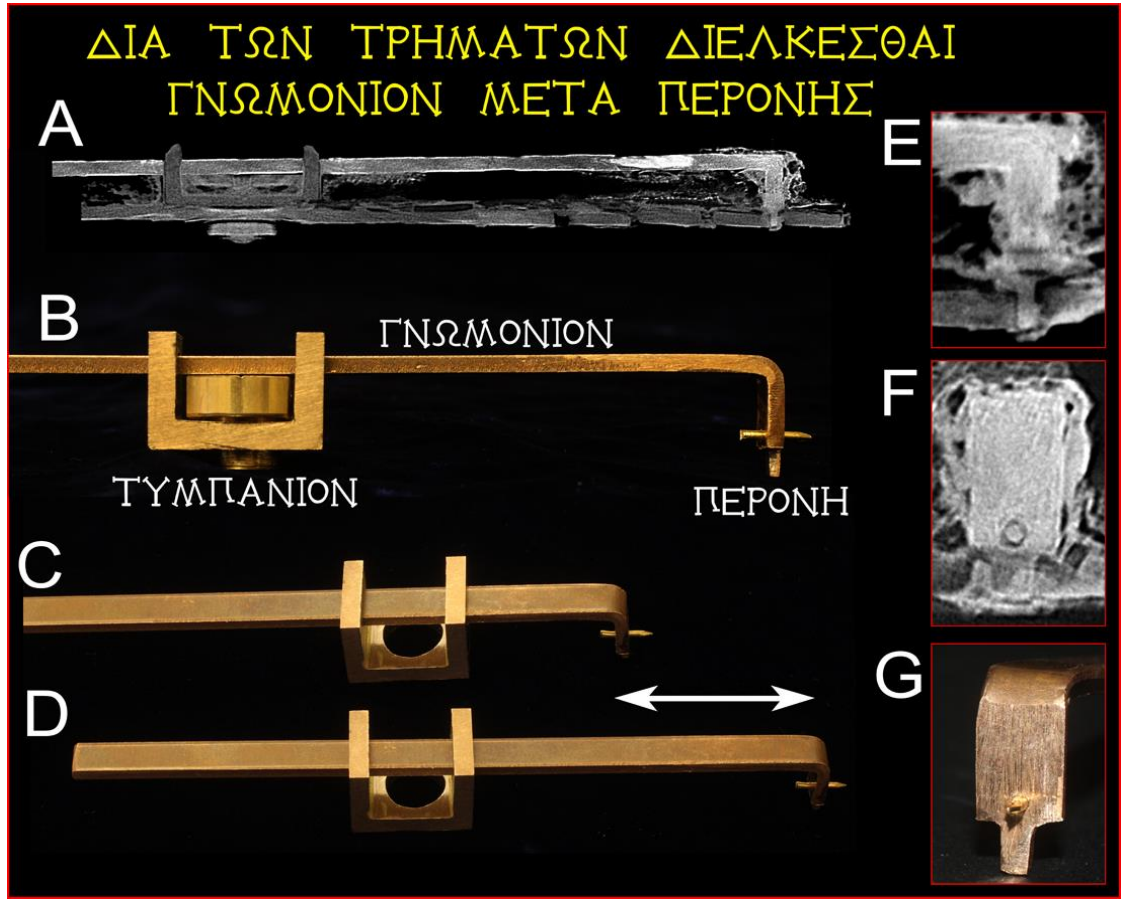

**Figure 6:** A) Digital reconstruction of the Metonic pointer and its Base using AMRP multi-combined processed tomographies of a part of Fragment B (sagittal view), using the software VolviCon. The pointer's perpendicular needle pin is visible inside the furrow of the Metonic spiral. The pointer is adapted on its Base through the holes of the bearings (see also Anastasiou et al., 2014). The Volume Data was aligned to the Base/pointer direction. B, C, D) The pointer slides via the holes and its length can be varied (spare parts reconstructed in bronze by first author). E) The AMRP tomography was processed by the authors, presenting the front face of ΠΕΡΟΝΗ-needle pin and the corresponding reconstruction in bronze by the first author. The small perpendicular pin on ΠΕΡΟΝΗ acts as the pointer which indicates the synodic month/cell and at the same time it does not allow the ΠΕΡΟΝΗ to sink into the furrow, and immobilize the pointer (Figure 7), which would be a bad case for the gears of the Mechanism, (see the analysis for Lines 14-15).

**- Line 8**: Line 7 completes the DDPO for the first pointer, located on the Metonic helix. In Line 8 *The Engraver* writes a comment: **ΟΜΟΙΩΣ ΤΟΙΣ ΠΡΩ**ΤΟΙΣ ΛΟΓΟΙΣ ΚΑΙ ΤΟΙΟΥΤΟΙ ΛΟΓΟΙ ΕΠΟΝΤΑΙ (*Similar to the previous sentences are the sentences that follow*). He informs *the User* that the following presentation is similar to the aforementioned presentation: The following presentation concerns the *Definition*, *Position*, *and Operation* of the second pointer which is located on the Periodicos helix.

| 08) **ΟΜΟΙΩΣ ΤΟΙΣ ΠΡΩ**ΤΟΙΣ ΛΟΓΟΙΣ ΚΑΙ ΤΟΙΟΥΤΟΙ ΛΟΓΟΙ ΕΠΟΝΤΑΙ: …… |
| --- |
| *08) Similar to the previous sentences (/description) are the sentences (/description) that follow: ………* |

**- Lines 8-12**: In this area of the text there is the presentation of the Periodicos pointer. - *Description*: Since the Periodicos pointer has the same mechanical design and function as the Metonic pointer, the *Description*, *Position* and *Operation* for the Periodicos pointer is presented by using the same words and syntax. The word **ΣΥΜΦΥΕΣ** (equal, exists as the first, same kind) is repeated two times in order to emphasize that the two pointers are exactly the same.

Instead of the word ΠΟΙΗΣΑΙ (infinitive) it should be written ΣΥΜΦΥΕΣ ΠΕΠΟΙΕΙΤΑΙ (verb). The same style adopting the infinitive ΠΟΙΗΣΑΙ was introduced in Line 10: **ΚΑΙ ΣΥΜΦΥΕ**Σ ΠΟΙΗΣΑΙ ΓΝΩΜΟΝΙΟΝ ΜΕΤΑ ΠΕΡΟΝΗΣ (here is also the *Definition* of ΠΕΡΟΝΗ).

- *Operation*: The Periodicos pointer is also rotated ΔΕΞΙΟΣΤΡΟΦΩΣ, *in CW direction* (see comments for Line 7) and indicates the eclipse events, ΤΩΝ ΕΓΛΕΙΨΕΩΝ ΤΑ ΓΕΓΟΝΟΤΑ ΔΕΙΚΝΥΣΙ (a similar reference exists in BCI Part-1, Authors' submitted BCI Part-1, *Line 13: ΤΑ ΜΕΓΙΣΤΑ ΤΩΝ ΟΛΩΝ ΓΕΓΟΝΟΤΩΝ ΚΑΙ Η ΣΕΛΗΝΗΣ ΚΑΙ Η ΗΛΙΟΥ ΕΓΛΕΙΨΙΣ, The most important events are the Lunar and Solar eclipses*).

---

08) **ΟΜΟΙΩΣ ΤΟΙΣ ΠΡΩ**ΤΟΙΣ ΛΟΓΟΙΣ ΚΑΙ ΤΟΙΟΥΤΟΙ ΛΟΓΟΙ ΕΠΟΝΤΑΙ. ΤΗΙ ΕΠΙ ΤΗΣ ΤΟΥ ΠΕΡΙΟΔΙΚΟΥ ΕΛΙΚΟΣ ΑΡΧΗΙ ΣΥΜ-
09) **ΦΥΕΣ ΠΟΙΗΣ**ΑΙ **ΤΥΜΠ**ΑΝΙΟΝ ΕΧΟΝ ΣΤΗΜΑΤΙΑ ΔΥΟ Α ΚΑΙ ΟΜΟΙΑ ΚΑΙ ΠΑΡΑΛΛΗΛΑ ΕΙΣΙΝ, ΚΑΙ ΟΡΘΑ ΙΣΤΑΝΤΑΙ ΕΠ ΑΥΤΟΥ.
10) **ΚΑΙ ΣΥΜΦΥΕ**Σ ΠΟΙΗΣΑΙ ΓΝΩΜΟΝΙΟΝ ΜΕΤΑ ΠΕΡΟΝΗΣ. ΤΡΗΜΑΤΑ ΟΜΟΙΑ ΚΑΙ ΕΤΕΡΟΜΗΚΗ ΚΑΙ ΕΝΑΝΤΙΑ ΚΑΙ ΙΣΟΫΨΗ ΕΧΟΥΣΙΝ
11) **ΤΑ ΣΤΗΜΑΤΙΑ**. ΔΙΑ ΤΩΝ ΤΡΗΜΑΤΩΝ ΔΙΕΛΚΕΤΑΙ ΤΟ ΓΝΩΜΟΝΙΟΝ, ΔΕΞΙΟΣΤΡΟΦΩΣ ΦΕΡΕΤΑΙ ΚΑΙ ΤΩΝ ΕΓΛΕΙΨΕΩΝ ΤΑ ΓΕΓΟΝΟΤ-
12) Α ΔΕΙΚΝΥΣΙ. .................................................................................................................................

*08) Similar to the previous sentences (/description) are the sentences (/description) that follow: On the beginning (origin) of the Periodicos helix*
*09) a same kind of base is made (/there is). On the base there are two bearings, which are similar and parallel to each other, perpendicular and fixed on the base.*
*10) And a same kind of pointer is made, which has a perpendicular pin. Similar holes, in oblong shape, opposite each other and of the same height, have*
*11) the bearings. Through the holes the pointer slides and is rotated in CW direction, and the eclipse events*
*12) indicates. ....................................*

---

     **- Line 12**: In this sentence, the "units" of the Periodicos scale are presented: the relation between the CW rotating Periodicos pointer, the eclipse events which were indicated by the pointer and the instructions for inferring what eclipse event the Periodicos pointer is aiming at, are presented (see **Figures 12, 13**). The engraved symbol *Σ* (ΣΕΛΗΝΗ) corresponds to a lunar eclipse and the symbol *Η* (ΗΛΙΟΣ) to a solar eclipse (Freeth et al., 2008). The word ΣΤΟΙΧΕΙΟ (letter) corresponds to the word *symbol*, as also preserved in Authors' submitted BCI Part-1, see also Bitsakis and Jones 2016a:

*Line 27: ΑΝΑΓΡΑΦΟΜΕΝΑ ΣΤΟΙΧΕΙΑ ΠΑΡΑΚΕΙΜΕΝΑ (engraved letters, the index letters of the Parapegma star events).*

---

12) Α ΔΕΙΚΝΥΣΙ. **ΤΑ ΓΕ** ΣΕΛΗΝΗΣ ΓΕΓΟΝΟΤΑ ΜΕΤΑ ΤΟΥ ΣΤΟΙΧΕΙΟΥ -Σ- ΕΙΣΙΝ, ΤΑ Δ' ΗΛΙΟΥ ΜΕΤΑ ΤΟΥ ΣΤΟΙΧΕΙΟΥ -Η-. ........................
*12) indicates. The lunar (eclipse) events are marked with the letter (symbol) Σ and the solar (eclipse) events are marked with the letter Η. ............*

---

     **- Line 13**: Most of the letters at the line beginning [. .]**ΡΟΥ . . ΘΟΔΟΥ**[.]**Η**[… are indistinct and do not allow a safe reconstruction. The probable word ΜΕ**ΘΟΔΟΥ** (*method*, it could be also ΚΑΘΟΔΟΥ-*descent*, but no relation was found) is difficult to relate with to the previous letters. We didn't reconstruct this text area.

After the description of the pointers' parts, we thought it was critical to include a reference describing the procedure of the two pointers' operation: *in which way, the pointer travels along the helix path*, **Figure 7**. As each helix has a furrow (mentioned in Line ii, ΜΕΤ' ΑΥΛΑΚΟΣ ΕΚΑΣΤΗ), a description that the pin of the pointer travels along the furrow is necessary in order for *the User* to understand how the pointer works and the procedure of the pointers' reset position, which is described in the following Lines 14-15.

---

12) Α ΔΕΙΚΝΥΣΙ. **ΤΑ ΓΕ Σ**ΕΛΗΝΗΣ ΓΕΓΟΝΟΤΑ ΜΕΤΑ ΤΟΥ ΣΤΟΙΧΕΙΟΥ -Σ- ΕΣΤΙΝ, ΤΑ Δ' ΗΛΙΟΥ ΜΕΤΑ ΤΟΥ ΣΤΟΙΧΕΙΟΥ -Η-. [.......

13) [. .]**ΡΟΥ** . . **ΘΟΔΟΥ ΤΗ**Ν ΕΚΑΣΤΗΝ ΤΗΣ ΕΛΙΚΟΣ ΑΥΛΑΚΑ ΑΓΕΣΘΑΙ ΚΑΙ ΔΙΑΠΟΡΕΥΕΣΘΑΙ ΑΠΟ ΤΗΝ ΤΟΥ ΓΝΩΜΟΝΙΟΥ ΠΕΡΟΝΗΝ.

*12) indicates. The lunar (eclipse) events are marked with the letter (symbol) Σ and the solar (eclipse) events are marked with the letter H. ............*

*13) .................................. Each of the pointers is driven and travels along the helix's furrow.*

---

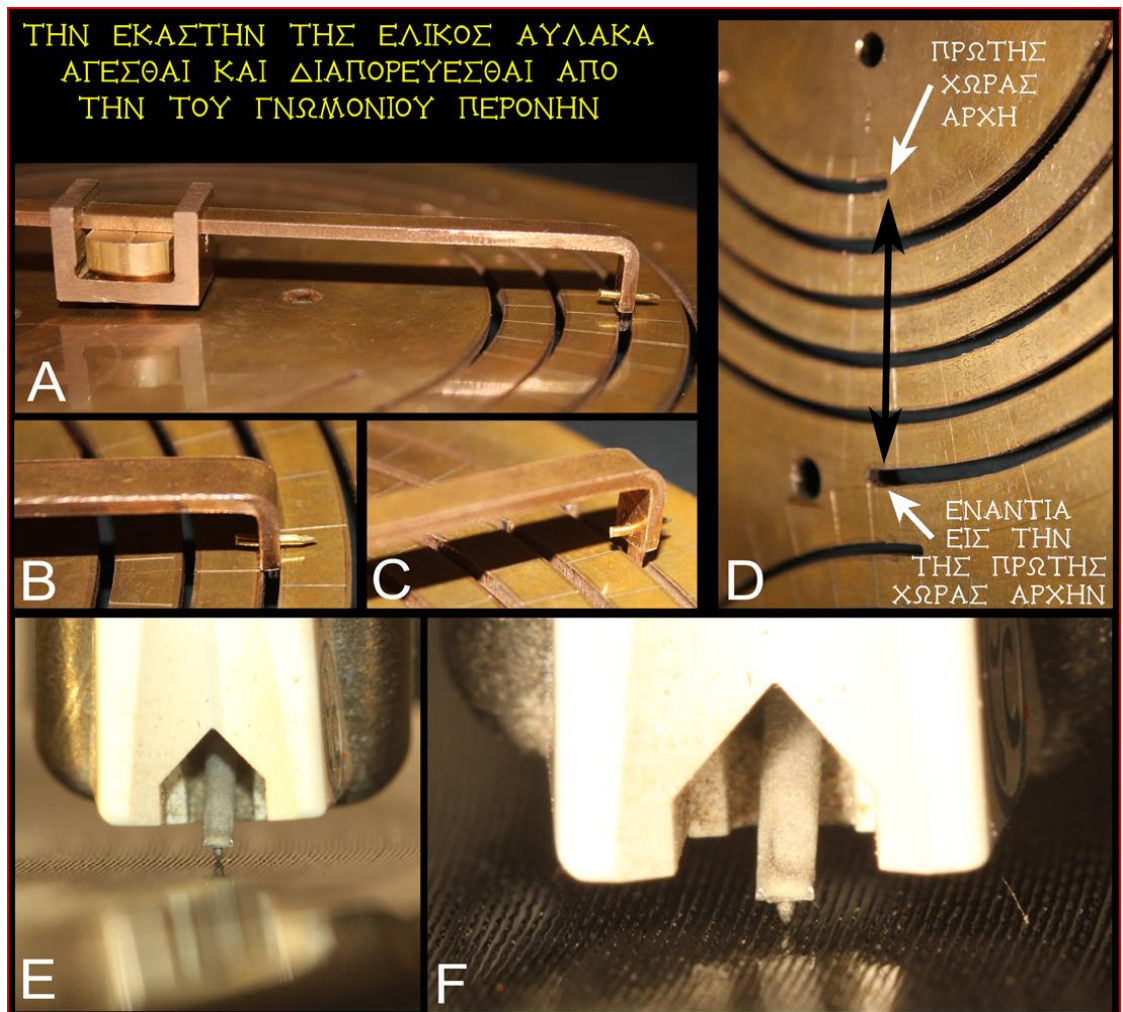

**Figure 7:** A) The needle pin is located on the furrow of the Periodicos helix and travels along it (see also Anastasiou et al., 2014). B, C) The small perpendicular pin is the tip of the pointer and is the real pointer which indicates the Periodicos and Metonic ΤΜΗΜΑΤΑ-subdivisions (cell-synodic months). D) The end of the helix is located in the opposite positon relative to the 1st area beginning. The ends of the two furrows are not coincided according to Voulgaris et al., 2021. Close up of the spare/testing parts designed/constructed by the first author. E, F) Close-up image of the stylus (needle) of a vinyl record player, which travels along the vinyl furrows. Images by the first author.

Then follow the reference that the pointer's pin travels along the furrow of helix as well as a description regarding *what should be done when the pointer's pin reaches the end of the furrow*:

　　**- Lines 14-15**: The relation of the words **ΕΝΑΝΤΙΑΝ** (*opposite*), **ΠΕΡΟΝΗΝ ΟΘΕΝ ΕΞΗΛΚΥΣ**ΘΑΙ (*pull out the pointer's pin*) and **ΤΗΣ ΠΡΩΤΗΣ ΧΩΡΑΣ** (1ˢᵗ Area of the helix), lead to the ***Pointers' Reset Position Procedure***: The Manufacturer/Engraver informs the reader of the manual that when the pointer arrives to the end of the Periodicos helix, the user must pull out the pointer's pin from the furrow and reposition it at the beginning of the Periodicos helix, which is the beginning of the 1ˢᵗ area (ΑΡΧΗΝ **ΤΗΣ ΠΡΩΤΗΣ ΧΩΡΑΣ**).

　　**The Antikythera Mechanism Helices *Pointer's Reset Position*** is achieved by the following procedure: when the pointer reaches the end of the helix, the user must pull out the pointer's pin, then pull the pointer so that it slides back through the holes of the bearings (**ΔΙΑ ΤΩΝ ΤΡΗΜΑΤΩΝ ΔΙΕΛΚΕΣΘΑΙ**, in Line 7) and reposition the pointer's pin in the furrow at the place of the 1ˢᵗ Area Beginning, **Figure 8**. During this procedure, the User must not rotate the pointer to any direction.

　　The word **ΕΝΑΝΤΙΑΝ** is related to the opposite position to the 1ˢᵗ Area Beginning. The opposite position of *the 1ˢᵗ Area Beginning* is the end of the last area, i.e. the end of the helix fourth area (end of cell-223 or a bit after).

　　As the Saros cycle/Periodicos helix lasts 223 months/cells and the Metonic cycle/helix lasts 235 months/cells, the Periodicos pointer will come first to the end of helix, thus the reset position procedure for the Periodicos pointer is firstly mentioned in the text, followed by the reset position procedure for the Metonic pointer.

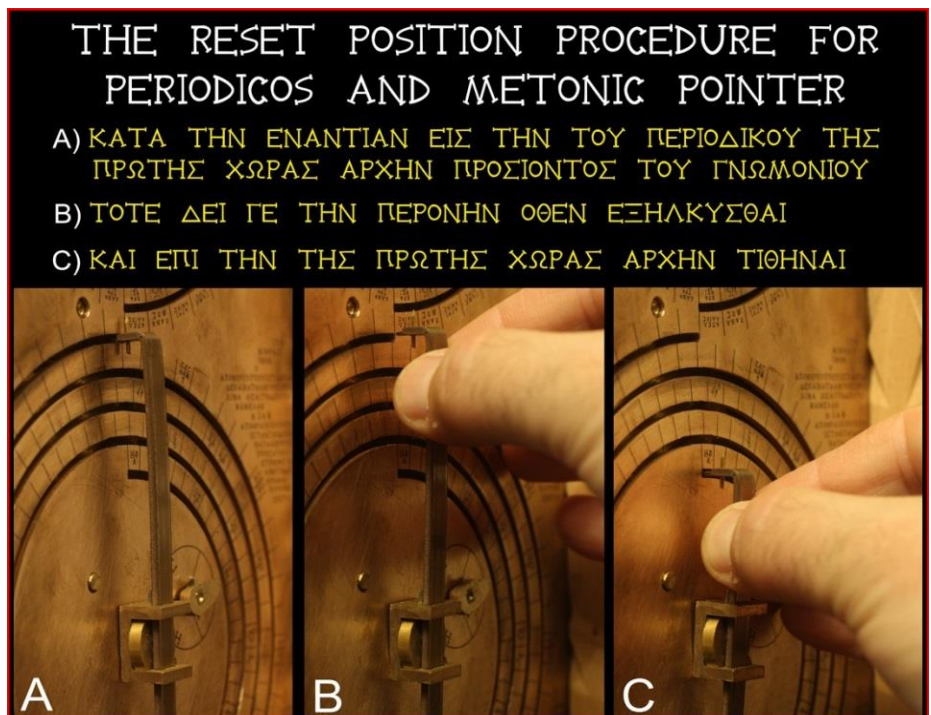

**Figure 8:** The reset position procedure for Periodicos pointer. *The User* pulls out the pointer, so that the needle-pin comes out of the furrow, then to pulls the pointer back in order to decrease its length by sliding it through the holes of the bearings. Finally, he places the needle-pin on the beginning of the furrow (1ˢᵗ area beginning), see Anastasiou et al., 2014. Close up image of the functional model of Antikythera Mechanism designed and constructed by the authors.

| The Procedure of the Pointers' reset position - A Caution/Notice in the Antikythera Mechanism Instruction Manual |
|---|
| 14) ΚΑΤΑ **ΤΗΝ ΕΝΑΝΤΙΑΝ_Ε**ΙΣ ΤΗΝ ΤΟΥ ΠΕΡΙΟΔΙΚΟΥ ΤΗΣ ΠΡΩΤΗΣ ΧΩΡΑΣ ΑΡΧΗΝ ΠΡΟΣΙΟΝΤΟΣ ΤΟΥ ΓΝΩΜΟΝΙΟΥ, ΤΟΤΕ ΔΕΙ ΓΕ<br>15) ΤΗΝ **ΠΕΡΟΝΗΝ ΟΘΕΝ ΕΞΗΛΚΥΣ**ΘΑΙ ΚΑΙ ΕΠΙ ΤΗΝ ΤΗΣ ΠΡΩΤΗΣ ΧΩΡΑΣ ΑΡΧΗΝ ΤΙΘΗΝΑΙ .................<br>*14) When the pointer reaches the opposite position to the beginning of the Periodicos' 1st Area,*<br>*15) pull out the pointer's pin end set it at the beginning of the 1st Area. ...................................................* |

In this area of the text, the ancient Manufacturer gives comprehensive instructions in order for the reset position of the pointer to be achieved and at the same time for a damage of the Mechanism parts to be avoided. Most of the modern Instruction Manuals contain important notes, including a special reference to the way of use of the device in order to avoid the user's injury or possible damage to the device:

**IMPORTANT**

**WARNING/CAUTION/NOTE**
Please read this manual and follow its instructions carefully. To emphasize special information, the symbol ⚠ and the words **WARNING**, **CAUTION** and **NOTE** have special meanings. Pay special attention to the messages highlighted by these signal words:
- ⚠ **WARNING:** Indicates a potential hazard that could result in death or injury.
- ⚠ **CAUTION:** Indicates a potential hazard that could result in device damage.
- ⚠ **NOTE:** Indicates special in formation to make maintenance easier or instructions clearer.
- The circle with a slash ⊘ means "***Don't do this***" or "***Don't let this happen***".

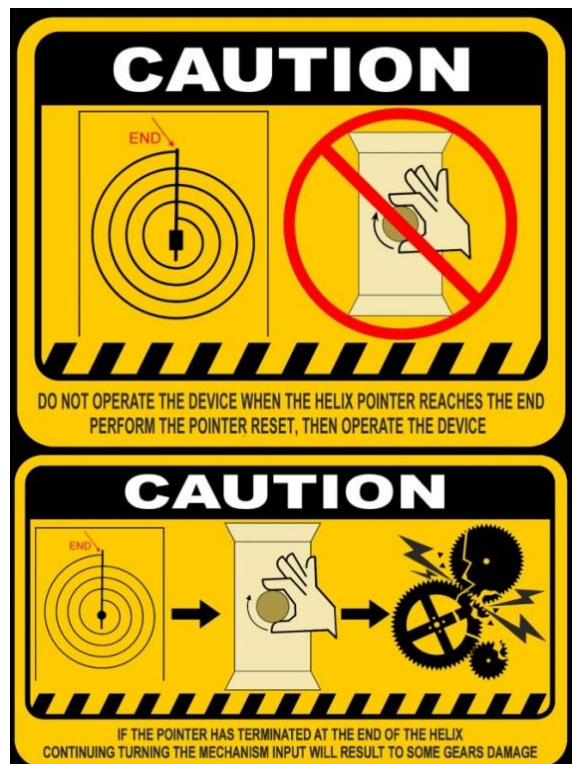

**Figure 9:** Two of the *Caution* signs on the Antikythera Mechanism Instruction Manual, in modern terms. The signs of ⚠**CAUTION** warn the User of the Antikythera Mechanism to take appropriate action: Do not operate the Mechanism if the helix pointer reaches the end, in order avoid the destruction of a number of gears (destruction of their teeth). Idea, design and scheme by the authors.

The reset position of the two helices' pointers in current terms is a *Caution*. The *Caution* which is accompanied with its corresponding sign indicates a situation which, if not avoided, may result in damage of the device or a part of it, **Figure 9**: If the Periodicos pointer reaches the end of helix (end of the furrow), the pointer will be immobilized and if the user continues to rotate the Input of the Mechanism (the Lunar Cylinder, Voulgaris et al., 2018b and 2022), several gears will be destroyed! since the torque of the Input is relatively high (Roumeliotis 2018; Voulgaris et al., 2018b and 2022).

 (A similar "Caution" exists in the Instruction Manual Part-1 reconstruction, Authors' submitted BCI Part-1 *Lines 10-11*: Ο ΤΗΣ ΣΕΛΗΝΗΣ ΚΥΛΙΝΔΡΟΣ ΦΕΡΟΜΕΝΟΣ ΕΣΤΙΝ ΚΑΙ ΕΙΣ ΤΟ ΔΙΗΝΕΚΕΣ ΔΕΞΙΟΣΤΡΟΦ**ΩΣ ΜΕΝΩΝ,** *the Lunar cylinder is rotated continuously in CW direction*).

- **Lines 15-16**: After the *Periodicos Pointer Reset Position Procedure* the same text should be given for the Metonic *Pointer Reset Position Procedure*. Although a satisfactory space for the text repetition exists, the specific position of the phrase **ΤΗΣ ΠΡΩΤΗΣ ΧΩΡΑΣ-M**[……. and the following letter Μ makes the fitting of a similar text quite difficult. It seems that after ΠΡΩΤΗΣ ΧΩΡΑΣ, a new sentence begins with letter Μ. In the previous sentence the word ΟΜΟΙΩΣ (similarly) was introduced in order to describe briefly the same procedure for the Metonic pointer reset position (the word ΟΜΟΙΩΣ is also preserved in Line 8). By introducing the word ΟΜΟΙΩΣ, the repetition of the full same text for the second pointer reset position is avoided.

| |
|---|
| 15) ΤΗΝ **ΠΕΡΟΝΗΝ ΟΘΕΝ ΕΞΗΛΚΥΣ**ΘΑΙ ΚΑΙ ΕΠΙ ΤΗΝ ΤΗΣ ΠΡΩΤΗΣ ΧΩΡΑΣ ΑΡΧΗΝ ΤΙΘΗΝΑΙ. ΟΜΟΙΩΣ ΚΑΙ ΕΠΙ ΤΗΝ ΤΟΥ ΜΕΤΩΝΟΣ ΑΡ-<br>16) ΧΗΝ **ΤΗΣ ΠΡΩΤΗΣ ΧΩΡΑΣ. –Μ**……………………………………………………………………………………………………………..<br>*15) pull out the pointer's pin end set it at the beginning of the $1^{st}$ Area. (Apply) the same (procedure for setting the pointer) at the Metonic*<br>*16) $1^{st}$ Area beginning. ……………………………………..* |

- **Line 16**: Right after the *Pointer's Reset Position Procedure*, a clarification/comment regarding "*what it means to reset the position of the pointer*" would be useful for the *User*: When the pointer is relocated on the beginning of the helix's $1^{st}$ Area, this means that "*A new cycle starts*" (ΕΝΙΑΥΤΟΣ ΑΠΟΚΑΘΙΣΤΑΤΑΙ or ΑΡΧΕΤΑΙ), **Figure 10**. The word ΑΡΧΕΤΑΙ means "*begins/starts*" (Geminus in Manitius 1898, p. 182 *ἀρχήν ἐνιαυτοῦ; p. 184 τας ἀρχάς τῶν ἐνιαυτῶν*) and the word ΑΠΟΚΑΘΙΣΤΑΤΑΙ (ΑΠΟΚΑΤΑΣΤΑΣΙΣ *return to origin*, *restitution*) means "*return to the initial position-beginning-reset position, in repeated times*", Voulgaris et al., 2021). Geminus in Chap. XVIII (Manitius 1898 and Spandagos 2004) and Ptolemy (Heiberg 1898) mention many times the word ΑΠΟΚΑΤΑΣΤΑΣΙΣ (Chap. III.1; IV.2; IV.3; V.2; VI.9) in order to describe the restitution of the cycles of Sun and Moon. It is also mentioned seven times in the Inscription of the Front Cover plate (Anastasiou et al., 2016a).

During the reset position of the Metonic pointer, three ΑΠΟΚΑΤΑΣΤΑΣΕΙΣ take place:
i) The 19 year Metonic cycle,
ii) The solar tropical year,
iii) The Lunar Synodic cycle (new month),

As a result of the above, the Sun and the Moon are located at the same solstice (or equinox point), since each Metonic cycle starts at a specific season change point (equinoxes and solstices) (Danezis and Theodosiou 1995).

During the reset position of Periodicos pointer two ΑΠΟΚΑΤΑΣΤΑΣΕΙΣ take place:

i) The Periodicos cycle,

ii) The Lunar Synodic cycle (new month),

Also, the rest two lunar cycles, the Anomalistic and the Draconic, are also located at their reset positon, as inferred from the definition of Exeligmos/(Saros) by Geminus (Chap. XVIII in Spandagos 2002). As result of the above, a large duration annular solar eclipse occurs during the beginning each of Periodicos/Exeligmos cycle.

The phrase ΟΙ ΕΝΙΑΥΤΟΙ ΑΠΟΚΑΘΙΣΤΑΝΤΑΙ is proper in order to describe the restitution of the cycles and is preferred to the word ΚΥΚΛΟΣ (as ΚΥΚΛΟΙ-cycles is the probable word for the two circular scales of Pentaeteris and Exeligmos, see Line 17).

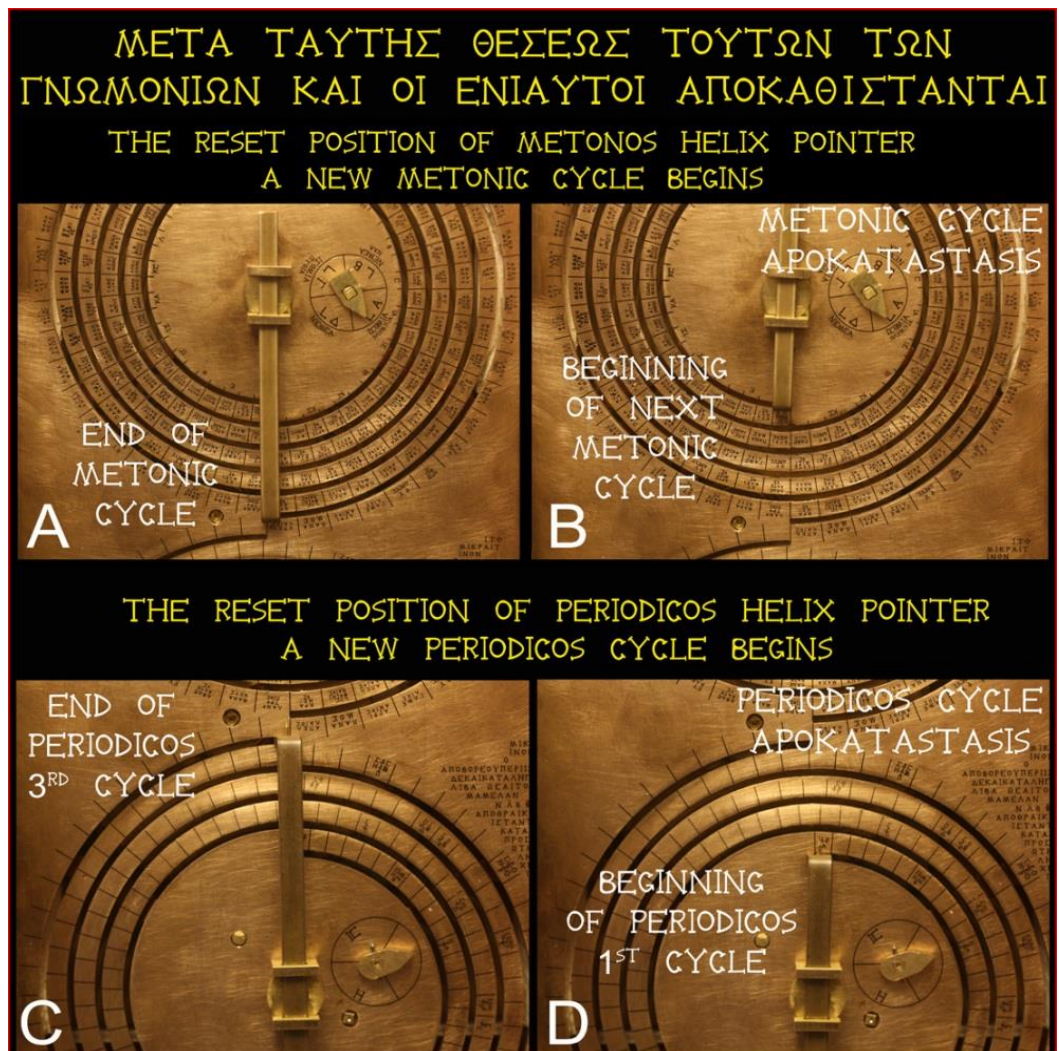

**Figure 10:** The reset position of the pointers, apart from the mechanical necessity, offers the reset position of the astronomical/calendrical cycles. The mechanical reset position of the pointers (A-B and C-D) signifies the restart (ΑΠΟΚΑΤΑΣΤΑΣΙΣ-return to origin) of the time measuring based on the Metonic and Periodicos cycles. Close up image of the functional model of Antikythera Mechanism designed and constructed by the authors.

---

16) ΧΗΝ **ΤΗΣ ΠΡΩΤΗΣ ΧΩΡΑΣ. -Μ**ΕΤΑ ΤΑΥΤΗΣ ΘΕΣΕΩΣ ΤΟΥΤΩΝ ΤΩΝ ΓΝΩΜΟΝΙΩΝ ΚΑΙ ΟΙ ΕΝΙΑΥΤΟΙ ΑΠΟΚΑΘΙΣΤΑΝΤΑΙ (/ΑΡΧΟΝΤΑΙ). ...................

*16) 1ˢᵗ Area beginning. After the position reset of the pointers, the eniautoi (cycles) return on their beginning (origin). ............*

---

    **- Lines 17-20**: In this part of the text, the *Description* and the *Position* for the two small circular scales and their pointers is presented.

    **- Line 17:** ΓΝΩ**ΜΟΝΙΑ ΔΥΟ-ΩΝ ΤΑ ΑΚΡΑ ΦΕ**ΡΟΝΤΑΙ (*two pointers, the edges of which they travel on...*) is the *Definition* of two pointers and their *Operation* is partially preserved (*travel on ...*).

    The two pointers travel around their corresponding circles ΚΥΚΛΟΙ. The two circular scales are in much smaller diameter than the helices dimension, so the word ΕΛΑΣΣΟΝΩΝ (small) seems to be necessary:

    **- Line 17:** ΓΝΩ**ΜΟΝΙΑ ΔΥΟ-ΩΝ ΤΑ ΑΚΡΑ ΦΕ**ΡΟΝΤΑΙ (*there are two pointers, the edges of which travel along...)* is the *Definition* of the two pointers and their operation is partially preserved *(travel along...).*

    The two pointers travel along their corresponding circles ΚΥΚΛΟΙ. The two circular scales have much smaller diameters than the helices, so the word ΕΛΑΣΣΟΝΩΝ (smalls) seems to be necessary:

    **- Line 17:** ΓΝΩ**ΜΟΝΙΑ ΔΥΟ-ΩΝ ΤΑ ΑΚΡΑ ΦΕ**ΡΟΝΤΑΙ ΕΠΙ ΕΛΑΣΣΟΝΩΝ ΚΥΚΛΩΝ (*there are two pointers and their edges travel along two small circles*). After that, the *Position* of the two small circles should follow (see further below).

    **- Line 18:** The phrase **ΔΗΛΟΙ Δ´ Ο ΜΕΝ** (*one of them indicates*) well cooperated to the phrase ΕΛΑΣΣΟΝΩΝ ΚΥΚΛΩΝ of Line 17... The word **Ο ΜΕΝ** (*the one-* in Masculine noun) implies the existence of a second scale (...Ο ΔΕ), i.e. Ο ΜΕΝ ΚΥΚΛΟΣ ...... ΚΑΙ Ο ΔΕ ΚΥΚΛΟΣ ...... (*the one of the circular scale .... and the other one.....*).

    In Line 19: [ . ]**Σ ΤΗΝ ΤΗΣ -_OCL_IΘL_ΤΟΥ** the two numbers and the symbol of year *L* are preserved: OCL = 76 years, IΘL= 19 years. These two numbers are related to the Callippic period of the 76 years as a result of 4X19 Metonic years (minus one day). Obviously the numbers/letters **OCL_IΘL** are an abbreviation (abridgment) of a more extended phrase, but the precise translation of this abbreviation it is unclear, as the previous words are not preserved. The turn of the two numbers is 76 years_19 years. Freeth et al., 2008; Bitsakis and Jones 2016 based on the preserved words of (...*the 19-year period of the 76-year period i.e. the 19 year/Metonic cycle of the Callippic period*) suggest that a second measuring scale should be located within the internal area of the Metonic helix (the first is the Stephanites Games Dial scale). This second circle should be the Callippic scale divided in four quadrants (4 Metonic cycles of 19 years =76 years), and the pointer aimed at each of the corresponding Metonic cycles (no mechanical evidence exists for this suggestion).

The author's attempts to reconstruct the text assuming that two circular scales are located on the Metonic helix (instead of one), by applying the *DDPO* method (which is fully operational for all the rest parts located on Front and Back plate), was not was not successful (bellow the unsuccessful *Option 1* and *2* are discussed and analyzed). Additionally, the hypothesis of the two small circular scales on the Metonic helix creates syntax and text positioning problems for the second and third circular scales (hypothetical Callippic scale and the preserved Exeligmos scale).

In the beginning of Line 20, the letters **ΜΟΣ** are preserved. These letters perfectly fit for the word ΕΞΕΛΙΓ**ΜΟΣ** (the authors' attempt using the data base from *Thesaurus Linguae Graecae*-TLG, to detect a different but proper/relative word …ΜΟΣ was not successful). Then follows the phrase **ΕΙΣ ΙΣΑ_ΣΚΓ-ΣΥΝ ΤΕΣ**[ and therefore the word ΕΧΕΛΙΓ|**ΜΟΣ** is the last word of the sentence in Line 19. This means that the *Description* of Exeligmos circular scale is located in Line 19. Therefore, before the *Description* of Exeligmos circular scale, the Pentaeteris circular scale should be mentioned.

Below we present the reconstructed text for Lines 17-20, assuming that there exist only two small circular scales on the Back Dial plate: one scale (the Stephanites/Sacred Games dial scale) is located within the internal area of the Metonic helix and the other (Exeligmos scale) is located in the internal area of Periodicos helix. Further below the unsuccessful attempt to reconstruct text assuming there are three scales on the Back Dial plate.

Line 17 should cntain the *Definition* and the *Position* of the two circular scales: ΔΥΟ ΕΛΑΣΣΟΝΕΣ ΚΥΚΛΟΙ ΚΕΙΝΤΑΙ ΠΑΡΑ ΤΑΣ ΤΩΝ ΕΛΙΚΩΝ ΑΡΧΑΣ (two *small circles are located close to the helices origin*).

The Stephanites Games were repeated every four/(two) years. This time span is called ΠΕΝΤΑΕΤΗΡΙΣ-Pentaeteris (Pindar Olympia X, 57 and Nemea XI,27 in Mommsen 1864) or ΠΕΝΤΕΤΗΡΙΣ-Penteteris or ΠΕΡΙΟΔΟΣ (*Periodos*, *circuit*). The Pentaeteris (from ΠΕΝΤΕ ΕΤΗ) means *5 years* and corresponds to a 4 year period. The "inconsistency" between the name and the actual time period is due to the specific way of this period measurement: in order to calculate the next Olympic Games the measurement started from the year of the current Games (Robertson 2010).

The number **ΤΕΣΣΑΡΑ** (four) is related to the Stephanites Games scale, which is divided in ΤΕΣΣΑΡΑ ΜΕΡΗ (4 sectors). The Exeligmos scale is divided in ΜΕΡΗ ΤΡΙΑ (three sectors), **Figure 11**.

The Line 19 can be well correlated to a reference for the time duration of the Metonic helix: ΕΙ**Σ ΤΗΝ ΤΗΣ -_OCL_IΘL_ΤΟΥ** ΜΕΤΩΝΟΣ ΕΛΙΚΑΝ (…on *the 19 year period of the Metonic helix, out of the 76 year of the Callippic period*).

- **The Circle/Scale of Pentataeris**:
*Definition/Description*: ΕΛΑΣΣΩΝ ΚΥΚΛΟΣ ΕΙΣ ΜΕΡΗ ΤΕΣΣΑΡΑ (*small circular scale divided in four sectors*) with ΓΝΩΜΟΝΙΟΝ (*pointer*) and ΜΕΤ' ΑΝΑΓΡΑΦΟΜΕΝΩΝ ΤΩΝ ΣΤΕΦΑΝΙΤΩΝ (or ΙΕΡΩΝ) ΑΓΩΝΩΝ (*having engraved the Stephanitic or Sacred Games-* these are the "*units*" of the scale).
*Position*: ΚΕΙΤΑΙ ΠΑΡΑ ΤΗΣ (ΤΟΥ ΜΕΤΩΝΟΣ) ΕΛΙΚΟΣ ΑΡΧΗΝ (*is located close to/off the Metonic Helix origin*)
*Operation*: **ΔΗΛΟΙ (Δ' Ο ΜΕΝ) ΤΗ**Ν ΠΕΝΤΑΕΤΗΡΙΔΑΝ (*indicates the Pentaeteris*). Instead of the word **ΤΗ**Ν ΠΕΝΤΑΕΤΗΡΙΔΑΝ one might suggest **ΤΗ**Ν ΠΕΡΙΟΔΟΝ, but this word can be confused with the word ΠΕΡΙΟΔΙΚΟΣ which is the word for the Saros helix and cycle. Moreover, the word ΠΕΡΙΟΔΟΣ has been linked to many different durations of time.

- **The Circle/Scale of Exeligmos:**
*Definition/Description*: ΕΛΑΣΣΩΝ ΚΥΚΛΟΣ ΕΙΣ ΜΕΡΗ ΤΡΙΑ (*small circular scale divided in three sectors*) with ΓΝΩΜΟΝΙΟΝ (*pointer*)

*Position*: ΚΕΙΤΑΙ ΠΑΡΑ ΤΗΣ (ΤΟΥ ΠΕΡΙΟΔΙΚΟΥ) ΕΛΙΚΟΣ ΑΡΧΗΝ (*is located close to/off the Periodicos Helix origin*) and is called ΕΞΕΛΙΓΜΟΣ.

*Operation*: The Operation and the units of Exeligmos (hours) is directly related to the Periodicos helix operation and is extensively described in Lines 23, 24, 25.

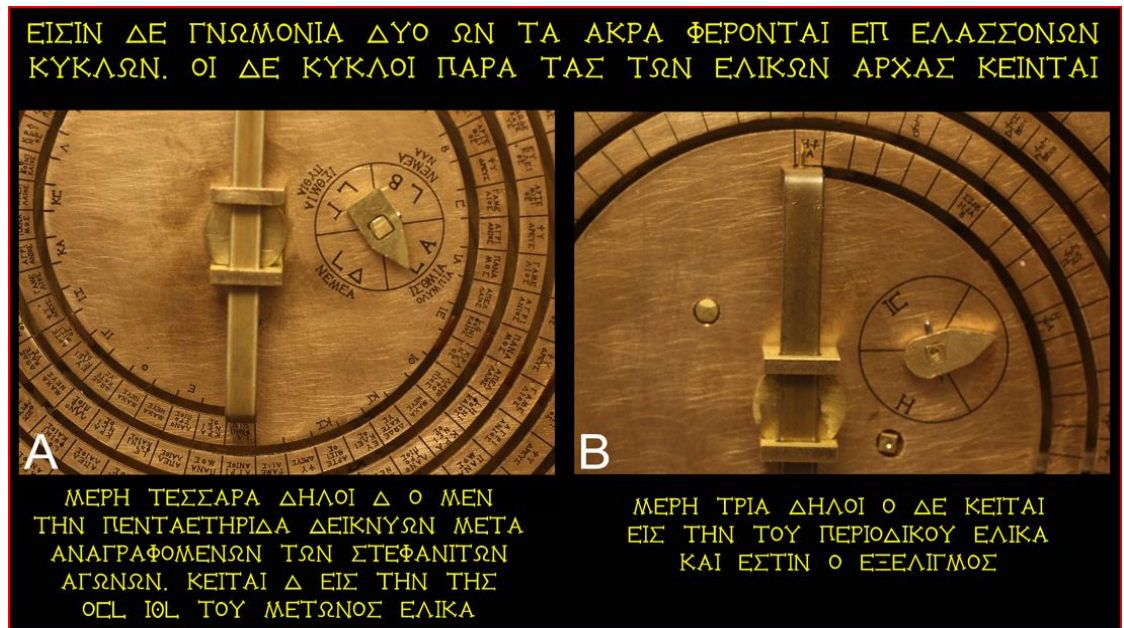

**Figure 11:** A) The small circular scale of Pentaeteris (ΕΛΑΣΣΩΝ ΚΥΚΛΟΣ ΤΗΣ ΠΕΝΤΑΕΤΗΡΙΔΟΣ) is divided in four sectors and is preserved on the right of the Metonic pointer. B) The small circular scale of Exeligmos is divided in three sectors and is preserved on the right of the Periodicos pointer. Close up image of the functional model of Antikythera Mechanism designed/constructed by the authors.

16) ΧΗΝ **ΤΗΣ ΠΡΩΤΗΣ ΧΩΡΑΣ. -Μ**ΕΤΑ ΤΑΥΤΗΣ ΘΕΣΕΩΣ ΤΟΥΤΩΝ ΤΩΝ ΓΝΩΜΟΝΙΩΝ ΚΑΙ ΟΙ ΕΝΙΑΥΤΟΙ ΑΠΟΚΑΘΙΣΤΑΝΤΑΙ. ΕΙΣΙΝ ΔΕ

17) ΓΝΩ**ΜΟΝΙΑ ΔΥΟ, -ΩΝ ΤΑ ΑΚΡΑ Φ**ΕΡΟΝΤΑΙ ΕΠ' ΕΛΑΣΣΟΝΩΝ ΚΥΚΛΩΝ. ΟΙ ΔΕ ΚΥΚΛΟΙ ΠΑΡΑ ΤΑΣ ΤΩΝ ΕΛΙΚΩΝ ΑΡΧΑΣ ΚΕΙΝΤΑΙ. ΜΕ-

18) ΡΗ **ΤΕΣΣΑΡΑ ΔΗΛΟΙ Δ' Ο ΜΕΝ, ΤΗ**Ν ΠΕΝΤΑΕΤΗΡΙΔΑ ΔΕΙΚΝΥΩΝ ΜΕΤ' ΑΝΑΓΡΑΦΟΜΕΝΩΝ ΤΩΝ ΣΤΕΦΑΝΙΤΩΝ ΑΓΩΝΩΝ. ΚΕΙΤΑΙ Δ' Ε-

19) ΙΣ **ΤΗΝ ΤΗΣ_-OCL_IΘL_ΤΟΥ** ΜΕΤΩΝΟΣ ΕΛΙΚΑ. ΜΕΡΗ ΤΡΙΑ ΔΗΛΟΙ Ο ΔΕ, ΚΕΙΤΑΙ ΕΙΣ ΤΗΝ ΤΟΥ ΠΕΡΙΟΔΙΚΟΥ (ΕΛΙΚΑ) ΚΑΙ ΕΣΤΙΝ Ο ΕΞΕΛΙΓ-

20) **Μ**ΟΣ.

16) *1st Area beginning. After the position reset of the pointers, the eniautoi (cycles) return on their beginning (origin). And there are*

17) *two pointers, which their edges are rotating on two small circular scales, which are located next to the helices' origin.*

18) *Four sectors is divided the one (circle) of them, indicates the Pentaeteris period and has the Stephanites Games engraved. It is located*

19) *in the 19 year-out of 76 year period, of the Metonic Helix. In three sectors is divided the other (circle), which is located on the Periodicos (helix) and it is the Exelig-*

20) *mos. ................................*

In this way, the *pattern* design of the two small scales presentation is equal to the two helices *pattern* presentation in Lines I, ii and iii.

| Helices presentation (Lines I, ii, iii) | The Back plate is divided in two helices…. *At top* there is the Metonic cycle helix and *at bottom* there is the Helix of Periodicos time. |
|---|---|
| **Helix pointers presentation** | At the beginning of the Metonic helix there is a Base. At the beginning of the Periodicos helix there is a similar Base. |

| Small circular scales presentation (Lines 17-20) | There are two small circular scales located close to/off the helices origin. One is located on the Metonic helix, is divided in four sectors and is the scale of the Stephanitic Games. The second is located on the Periodicos helix, is divided in three sectors and is the scale of Exeligmos. |
|---|---|

- If the suggestion by Freeth et al., 2008; Bitsakis and Jones 2016 that two small circular scales divided in four sectors existed on the Metonic helix is real, then a probable text for the *Description/Definition* (the kind of scales and their corresponding measuring units: Stephanites Games and four Metonic cycles)*, Position, Operation* of the two scales, should be written, in accordance to the preserved letters and their position.

*Option 1*: Firstly, Pentaeteris and then the Callippic period are mentioned.

*Option 2*: Firstly, the Callippic period and then the Pentaeteris is mentioned.

*Option 1 (Pentaeteris and then Callippic):*
17) ΓΝΩ**ΜΟΝΙΑ ΔΥΟ-ΩΝ ΤΑ ΑΚΡΑ ΦΕ**ΡΟΝΤΑΙ ΕΠΙ (ΕΛΑΣΣΟΝΩΝ) ΚΥΚΛΩΝ, ΟΙ ΕΣΤΙΝ ΕΝΘΕΝ ΚΑΙ ΕΝΘΕΝ ΤΗΣ ΤΟΥ ΜΕΤΩΝΟΣ ΕΛΙΚΑΣ ΑΡΧΗΣ. ΕΚΑΣΤΟΣ (ΚΥΚΛΟΣ) ΜΕ-          (letters 89.5-104.5)
18) ΡΗ **ΤΕΣΣΑΡΑ ΔΗΛΟΙ Δ' Ο ΜΕΝ ΤΗ**Ν ΠΕΝΤΑΕΤΗΡΙΔΑΝ ΜΕΤ' ΑΝΑΓΡΑΦΟΜΕΝΩΝ ΤΩΝ ΣΤΕΦΑΝΙΤΩΝ ΑΓΩΝΩΝ ΚΑΙ Ο ΔΕ (ΔΗΛΟΙ) ΤΗΝ ΤΟΥ ΚΑΛΛΙΠΠΟΥ ΠΕΡΙΟΔΟΝ ΤΩΝ ΤΕΣΣΑΡΩΝ ΕΝΝΕΑΚΑΙΔΕΚΑΕΤΗΡΙΔΩΝ (letters *130+*)
19) **. Σ ΤΗΝ ΤΗΣ_-OCL_IΘL_ΤΟΥ** ……………………………………………………………………………………… […… -
20) **ΜΟΣ. ΕΙΣ ΙΣΑ_ΣΚΓ-ΣΥΝ ΤΕΣ**…………………………………………………………………………………………………
*17) There are two pointers rotating on their circular scales, which are located left and right of the Metonic helix origin. Each (cycle) is divided in pa-*
*18) –rts four, one of them indicates the Pentaeteris with the Stephanites Games engraved and the second (circle indicates) the Callippic period (consisting) of four Enneakaidekateris ………………………*
*19) the 19 year period of the 76 year period ………………………………………………………………………*

1) The phrase ΜΕΤ' ΑΝΑΓΡΑΦΟΜΕΝΩΝ ΤΩΝ ΣΤΕΦΑΝΙΤΩΝ ΑΓΩΝΩΝ (*with the Stephanites Games engraved*) is difficult to be removed because it is the *Description* of the Penteteris "*units*". Also, as the circular scales have much smaller diameters than the helices dimension, the word ΕΛΑΣΣΟΝΩΝ (*small*) seems to be necessary (101.5 letters). Line 18 has 130+ letters (the word ΕΝΝΕΑΚΑΙΔΕΚΑΕΤΗΡΙΔΩΝ should be there as it is the *Definition* of the Callippic cycle and the units of the hypothetical scale).

2) The connection of the end of Line 18/begin of Line 19 is problematic, because the letter Σ (begin of Line 19) could not be the last letter of an article. The word ΕΙ**Σ ΤΗΝ ΤΗΣ** fits well.

3) In Line 19 there should be an additional phrase with max 67 letters and the last letters of the line must be the first letters of a word ending with the three last letters located at the beginning of Line 20.

4) There is only one word option in the text: […]|**ΜΟΣ** as the following phrase is **ΕΙΣ ΙΣΑ**-in equal) in the beginning of Line 20. The authors could find no Greek word that could be well related to the text except the word ΕΞΕΛΙΓ]|**ΜΟΣ**, which fits well.

*Option 2 (Callippic and then Pentaeteris):*
17) ΓΝΩ**ΜΟΝΙΑ ΔΥΟ-ΩΝ ΤΑ ΑΚΡΑ ΦΕ**ΡΟΝΤΑΙ ΕΠΙ (ΕΛΑΣΣΟΝΩΝ) ΚΥΚΛΩΝ, ΟΙ ΕΣΤΙΝ ΕΝΘΕΝ ΚΑΙ ΕΝΘΕΝ ΤΗΣ ΤΟΥ ΜΕΤΩΝΟΣ ΕΛΙΚΑΣ ΑΡΧΗΣ. ΜΕ-                    (82.5- 91.5)
18) ΡΗ **ΤΕΣΣΑΡΑ ΔΗΛΟΙ Δ' Ο ΜΕΝ ΤΗ**Ν ΤΟΥ ΚΑΛΛΙΠΠΟΥ ΠΕΡΙΟΔΟΝ ΤΩΝ ΤΕΣΣΑΡΩΝ ΕΝΝΕΑΚΑΙΔΕΚΑΕΤΗΡΙΔΩΝ ……                                                (73+)
19) =**Σ ΤΗΝ ΤΗΣ_-OCL_IΘL_ΤΟΥ** …… ΚΑΙ Ο ΔΕ ΤΗΝ ΠΕΝΤΑΕΤΗΡΙΔΑΝ ΜΕΤ' ΑΝΑΓΡΑΦΟΜΕΝΩΝ ΤΩΝ ΣΤΕΦΑΝΙΤΩΝ ΑΓΩΝΩΝ …………………………………… […-                          (77.5)
20) **ΜΟΣ. ΕΙΣ ΙΣΑ_ΣΚΓ-ΣΥΝ ΤΕΣ**………………………………………………………………………………

*17) There are two pointers rotating on their circular scales, which are located left and right of the Metonic helix origin. Each (cycle) is divided in pa-*
*18) –rts four, one of them indicates the Callippic period (consisting) of four Enneakaidekateris …….*
*19) the 19 year period of the 76 year period and the second (circle indicates) the Pentaeteris, with the Stephanites Games engraved ……………………………………………………………………………*

1) As the circular scales are in much smaller diameter than the helices dimension, the word ΕΛΑΣΣΟΝΩΝ (small) seems to be necessary (94.5 letters). Line 18 is hard to be filled out, especially in connection to the next line.

2) Line 19 is also hard to be filled out, especially in connection to the next line. The letters [……]**ΜΟΣ** are definitely the last letters of a word which starts at the end of the previous line.

By the previous analysis it seems that only two small circular scales existed on the Back Dial plate, one on the Metonic and one on the Periodicos helix {the authors' opinion is based on their (unsuccessful) attempt to reconstruct the text assuming the existence of three small scales on the Back Dial plate of the Mechanism and taking into account the word [……]|**ΜΟΣ**)}.

- **Line 20**: In this Line, the ΕΛΙΞ ΤΟΥ ΠΕΡΙΟΔΙΚΟΥ (Periodicos helix) information (*Description*) is presented. Periodicos helix is divided in 223 ΤΜΗΜΑΤΑ (parts) in ΤΕΣΣΑΡΑΙΣ ΠΕΡΙΦΟΡΑΙΣ (4 turns) covering the full path of the helix (**Figures 2, 10**). The same pattern was repeated for the Metonic helix, see Line 3.

| |
|---|
| 20) **ΜΟΣ. ΕΙΣ ΙΣΑ_ΣΚΓ-ΣΥΝ ΤΕΣ**ΣΑΡΣΙ ΠΕΡΙΦΟΡΑΙΣ, ΤΜΗΜΑΤΑ ΕΙΣΙΝ, ΕΝ ΟΛΗΙ ΤΗΙ ΕΛΙΚΙ, ΕΠΙ ΤΗΝ ΧΩΡΑΝ ΤΟΥ ΠΕΡΙΟΔΙΚΟΥ, ……. |
| *20) mos. Equal 223, in four turns, divisions were made along the full path of the helix on the area of Periodicos ………..* |

- **Lines 20-21:** The preserved letters in the beginning of Line 21 are indistinct (Bitsakis and Jones 2016) and don't allow for safe reconstruction beyond any doubts.

By making two hypotheses a probable text can be reconstructed:

1) The doubtful first letter of Line 21 (Τ) is the letter Π, which has a similar design to the letter Τ and

2) Instead of the word **ΔΙΑΙΡΕΘΗ**<I> there is the word **ΔΙΗΙΡΕΘΗ** i.e. by a spelling error (the word ΔΙΑΙΡΕΘΗ was detected in a text by C. Galenus, (Kühn 1830, p. 139, 1), then a probable text follows:

| |
|---|
| 20) **ΜΟΣ. ΕΙΣ ΙΣΑ_ΣΚΓ-ΣΥΝ ΤΕΣΣ**ΑΡΣΙ ΠΕΡΙΦΟΡΑΙΣ, ΤΜΗΜΑΤΑ ΕΙΣΙΝ, ΕΝ ΟΛΗΙ ΤΗΙ ΕΛΙΚΙ, ΕΠΙ ΤΗΝ ΧΩΡΑΝ ΤΟΥ ΠΕΡΙΟΔΙΚΟΥ, ΩΣ- |
| 21) ΠΕΡ ΑΥΤΟΣ ΔΙΑ{Η}ΙΡΕΘΗ. -Η-ΟΛΗ …………………………………………………………………………………………. |
| *20) mos. Equal 223, in four turns, divisions were made along the full path of the helix on the area of Periodicos as* |
| *21) indeed (Periodicos cycle) was divided. The full ………* |

By dividing Periodicos cycle in four parts, useful astronomical information regarding the Lunar cycles appears: at the half cycle of Periodicos the Moon is located at Perigee and a Lunar eclipse occurs (Voulgaris et al., 2021), while at each Periodicos quarter, the Moon is located at a Node (alternately at Node 1 and Node 2). Also in the 1[st] quarter (cycle beginning) the Moon travels in the sky with its minimum angular velocity (it is at Apogee), in the 2nd quarter travels with its mean velocity, in the 3rd quarter with its fastest velocity (it is at Perigee) and on 4th quarter with its mean velocity.

   **- Line 21:** The word **Η ΟΛΗ** (all, the full …) can be related to the eclipse sequence of all of the events, which is engraved in some of the cells, **Figure 12**. A similar phrase relating to the eclipses was mentioned (reconstructed) in the BCI Part-1, (Authors' submitted BCI Part-1) Line 13: …*ΤΑ ΜΕΓΙΣΤΑ ΤΩΝ **ΟΛΩΝ ΓΕΓΟΝΟΤΩΝ** ΚΑΙ Η ΣΕΛΗΝΗΣ ΚΑΙ Η ΗΛΙΟΥ ΕΓΛΕΙΨΙΣ…* (*The most important events are the Lunar and Solar eclipses…*).

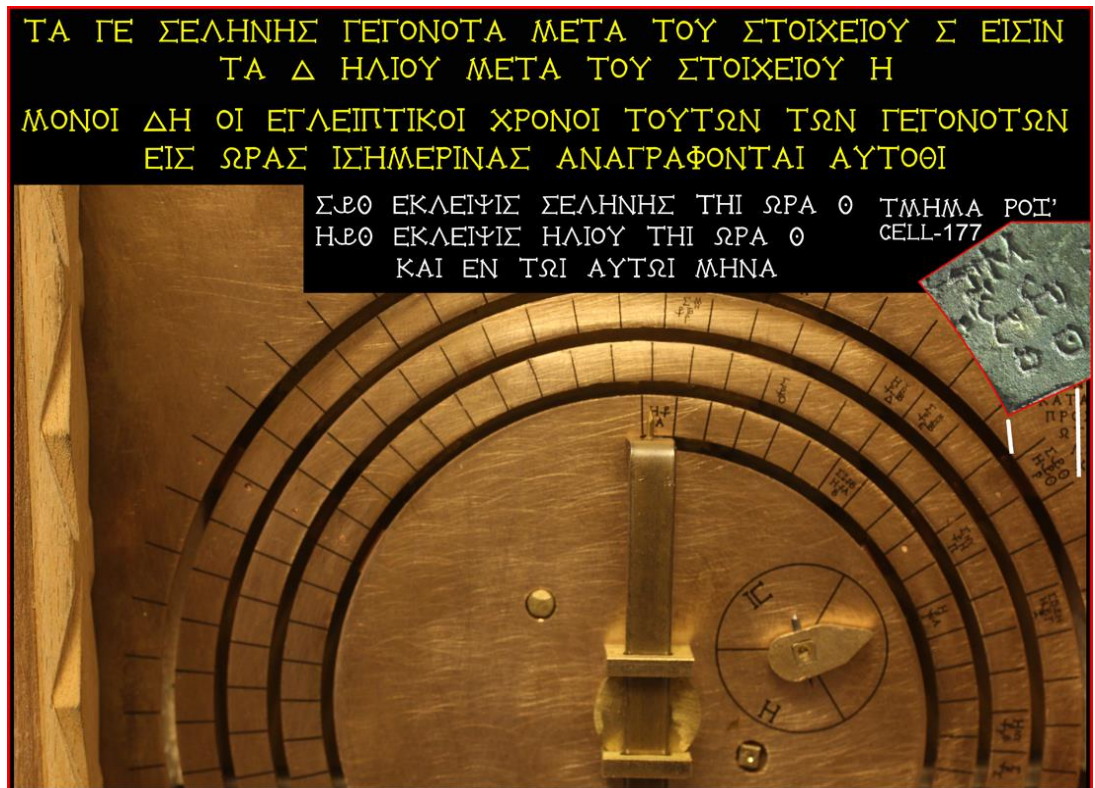

**Figure 12:** The Periodicos pointer is located at the 1st area beginning/Cell-1, indicating the first month of the Periodicos eniautos. After the cell re-numbering by Voulgaris et al., 2021, the first month of the Periodicos cycle includes a solar eclipse occurred on its last day, which is an annular solar eclipse of a very large duration (the largest duration of all the Saros period solar eclipses). At the same time, the Exeligmos pointer aims at the 1st sector (without any number symbol), which it means the 1st Periodicos cycle and there is no need for a time correction of the eclipse events hours. The characteristic cell-177, see Voulgaris et al., 2021 (old numbering 178, Freeth et al., 2008; Freeth, 2014; Anastasiou et al., 2016; Freeth, 2019; Iversen and Jones, 2019; Jones, 2020) presents two eclipses occurring in the same month and at the same hour, but half synodic month (≈15 days) apart. Close up image of the functional model of Antikythera Mechanism designed/constructed by the authors.

20) ΜΟΣ. **ΕΙΣ ΙΣΑ_ΣΚΓ-ΣΥΝ ΤΕΣ**ΣΑΡΣΙ ΠΕΡΙΦΟΡΑΙΣ, ΤΜΗΜΑΤΑ ΕΙΣΙΝ, ΕΝ ΟΛΗΙ ΤΗΙ ΕΛΙΚΙ, ΕΠΙ ΤΗΝ ΧΩΡΑΝ ΤΟΥ ΠΕΡΙΟΔΙΚΟΥ, ΩΣ-
21) ΠΕΡ **ΑΥΤΟΣ ΔΙΑ**{Η}**ΙΡΕΘΗ. -Η-ΟΛΗ** ΤΩΝ ΓΕΓΟΝΟΤΩΝ ΑΚΟΛΟΥΘΙΑ ΕΠΙ ΤΙΝΩΝ ΤΩΝ ΤΗΣ ΕΛΙΚΑΣ ΤΜΗΜΑΤΩΝ ΠΑΡΑΤΙΘΕΤΑΙ.

*20) mos. Equal 223, in four turns, divisions were made along the full path of the helix on the area of Periodicos as*
*21) indeed (Periodicos cycle) was divided. The full sequence of the (eclipse) events is presented on some of the helix divisions (cells).*

- **Line 22:** On the cells where the eclipse events sequence is engraved, the times of the eclipse events are also engraved. On some of the cells containing the times of the events, *The Engraver* has noted the letters ΝΥ (ΝΥΧΤΑ-*Nighttime*) or ΗΜ (ΗΜΕΡΑ-*Daytime*), i.e. 12 daytime hours and 12 nighttime hours (Freeth et al., 2006 and 2008; Anastasiou et al., 2016). The time of an eclipse event and the correction of +8 hours/Saros must be in the same unit and at constant hour duration, i.e. equinoctial hours (ΙΣΗΜΕΡΙΝΑΙ ΩΡΑΙ) and not seasonal (ΚΑΙΡΙΚΑΙ ΩΡΑΙ) the duration of which vary along the year (Anastasiou et al., 2016).

The phrase **ΕΓΛΕΙΠΤΙΚΟΙ ΧΡ**ΟΝΟΙ well relates to the definition of the time during of which the eclipse event occurs (times of the eclipses, see Bitsakis and Jones 2016).

| |
|---|
| 22) **ΜΟΝ**ΟΙ ΔΗ **ΟΙ ΕΓΛΕΙΠΤΙΚΟΙ ΧΡ**ΟΝΟΙ ΤΟΥΤΩΝ ΤΩΝ ΓΕΓΟΝΟΤΩΝ ΕΙΣ ΩΡΑΣ ΙΣΗΜΕΡΙΝΑΣ ΑΝΑΓΡΑΦΟΝΤΑΙ ΑΥΤΟΘΙ. ΚΑΙ |
| *22) Only the times of these eclipse events, are engraved (on the cells) in Equinoctial hours. …* |

The phrase **ΜΟΝ**ΟΙ ΔΗ **ΟΙ ΕΓΛΕΙΠΤΙΚΟΙ ΧΡ**ΟΝΟΙ ΑΝΑΓΡΑΦΟΝΤΑΙ (*only the times of the eclipse events are engraved*) clarifies to the User that only the eclipse times are mentioned, not the places at which the eclipse event will be visible.

- **Line 23**: The word **ΟΜΟ**ΙΩΣ (similarly) can be well related to the times engraved on the Exeligmos scale sectors (ΜΕΡΗ) as these times are engraved on some of the Periodicos Helix ΤΜΗΜΑΤΑ-subdivisions as well (**Figures 11, 12**).

| |
|---|
| 22) **ΜΟΝ**ΟΙ ΔΗ **ΟΙ ΕΓΛΕΙΠΤΙΚΟΙ ΧΡ**ΟΝΟΙ ΤΟΥΤΩΝ ΤΩΝ ΓΕΓΟΝΟΤΩΝ ΕΙΣ ΩΡΑΣ ΙΣΗΜΕΡΙΝΑΣ ΑΝΑΓΡΑΦΟΝΤΑΙ ΑΥΤΟΘΙ. ΚΑΙ<br>23) **ΟΜΟ**ΙΩΣ **ΤΟΙΣ ΕΠΙ ΤΗΣ Ε**ΛΙΚΟΣ ΤΜΗΜΑΣΙ, ΩΡΑΙ ΑΝΑΓΡΑΦΟΝΤΑΙ ΚΑΙ ΤΟΙΣ ΕΠΙ ΤΟΥ ΕΞΕΛΙΓΜΟΥ ΜΕΡΕΣΙ. ………………… |
| *22) …………………………………………………………………………………………………… And*<br>*23) similarly to the helix parts (cells), the hours are also engraved on the Exeligmos' sectors. …………..* |

- **Lines 23-24**: In this part of the text, the operation of the Exeligmos pointer and what it depicts on the Exeligmos scale are presented. The Exeligmos cycle consists of three Saros cycles/Periodicos eniautoi, the 1[st], the 2[nd] and the 3[rd] Periodicos eniautos. The Exeligmos scale is divided in three sectors and each sector corresponds to one of the three Saros cycles, **Figure 13**.

| |
|---|
| 23) **ΟΜΟ**ΙΩΣ **ΤΟΙΣ ΕΠΙ ΤΗΣ Ε**ΛΙΚΟΣ ΤΜΗΜΑΣΙ, ΩΡΑΙ ΑΝΑΓΡΑΦΟΝΤΑΙ ΚΑΙ ΤΟΙΣ ΕΠΙ ΤΟΥ ΕΞΕΛΙΓΜΟΥ ΜΕΡΕΣΙ. ΤΟ ΓΝΩΜΟΝΙΟΝ, ΟΥ ΤΟ<br>24) **ΑΚΡΟΝ ΦΕΡΕΤΑΙ Κ**ΑΘ ΕΚΑΣΤΟΝ ΤΩΝ ΤΟΥ ΕΞΕΛΙΓΜΟΥ ΜΕΡΩΝ, ΕΚΑΣΤΟΝ ΤΩΝ ΤΡΙΩΝ ΕΝΙΑΥΤΩΝ ΤΟΥ ΠΕΡΙΟΔΙΚΟΥ ΔΗΛΟΙ. …………… |
| *23) …………………………………………………………………………………………………… The pointer's*<br>*24) edge travels through each of Exeligmos' sector and indicates each of the three Periodic cycles. ……* |

- **Lines 24-25**: In the last preserved sentences of the BCI Part-2 text, the relation between the position (sector) of each Periodicos cycle and the hour events correction should be explained, i.e. the time correction procedure for the eclipse event(s) hour *versus* the number of Periodicos eniautos: After one Saros cycle, the same eclipse event is repeated (the pointer aims to the same cell), but with a delay of 8 hours. Therefore, for the 2[nd] Saros cycle events, the user must add +8 hours to the event hour which is engraved in the corresponding cell. For the 3[rd] Saros events, the user must add +16 hours to the event hour which is engraved on the corresponding cell, **Figure 13**.

24) **ΑΚΡΟΝ ΦΕΡΕΤΑΙ Κ**ΑΘ ΕΚΑΣΤΟΝ ΤΩΝ ΤΟΥ ΕΞΕΛΙΓΜΟΥ ΜΕΡΩΝ, ΕΚΑΣΤΟΝ ΤΩΝ ΤΡΙΩΝ ΕΝΙΑΥΤΩΝ ΤΟΥ ΠΕΡΙΟΔΙΚΟΥ ΔΗΛΟΙ. ΔΕΥΤΕ-
25) ΡΟΥ **ΜΕΝ ΤΟΥ Π**ΕΡΙΟΔΙΚΟΥ ΟΝΤΟΣ, ΔΕΙ ΤΟΥΣ ΕΓΛΕΙΠΤΙΚΟΥΣ ΧΡΟΝΟΥΣ ΑΥΞΕΣΘΑΙ ΚΑΤΑ ΩΡΑΣ -Η-, ΤΡΙΤΟΥ ΔΕ ΚΑΤΑ ΩΡΑΣ -ΙϚ-.

*24) ………………………………………………………… (When the pointer aims) at the Second (/Middle)*
*25) Periodicos (cycle), add to the time of the eclipse events 8 hours, and (when the pointer aims) at the Third (/Last) (Periodicos cycle), add (to the time of the eclipse events) 16 hours.*

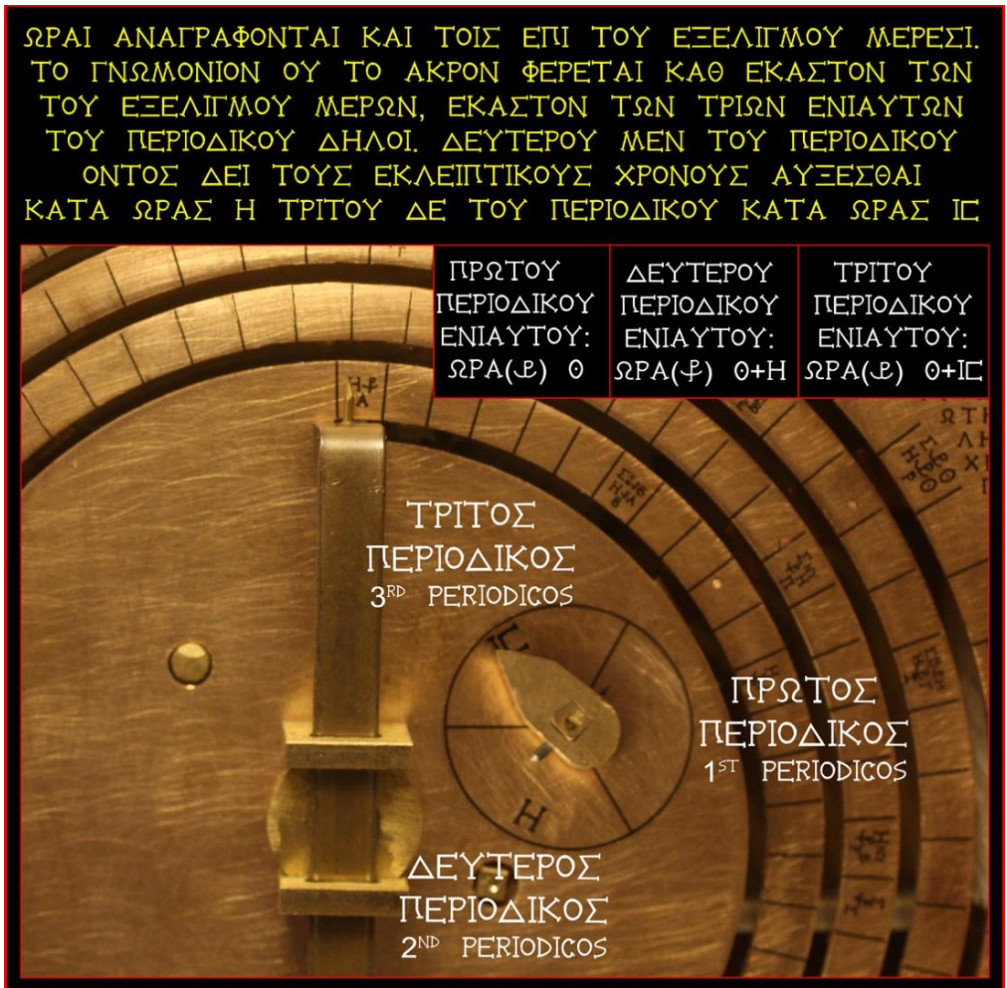

**Figure 13:** The Exeligmos pointer aims at the 3[rd] sector (with the number 16), meaning that this Periodicos eniautos is the 3[rd] Periodicos cycle and a time correction for the eclipse events hours of +16h is needed. The Periodicos pointer is located at the 1[st] area beginning/Cell-1, indicating the first month of the 3[rd] Periodicos eniautos. Close up image of the functional model of Antikythera Mechanism designed/constructed by the authors.

After Line 25, it is very probable that the text of the Instruction Manual of the Antikythera Mechanism was continued. In Authors' submitted BCI Part-1 reconstruction, the final number of the full text lines was estimated at about 91-87 lines. Today, the total number of (partially/poorly preserved) lines of Part-1 and Part-2 is 30+25= 55 lines. The total number of the Back cover letters is estimated at about 7350.

## 9. Symmetry in the BCI Part-2 text reconstruction

The Symmetry is a feature of the Antikythera Mechanism design and construction (Voulgaris et al., 2021). Many years before the Hellenistic era, Proto-Geometric, Geometric, Archaic and Classical period, the ancient Greeks used the idea of Symmetry (Horup, 2000;

Cold-stream, 1977; Roes, 1933; Schweitzer, 1971; Wide, 1899). Even today the use of Symmetry in many applications, designs, and constructions is necessary.

From the preserved words on the BCI Part-2, as well as on Part-1, Symmetry also exists in the design of the text. The idea of Symmetry helped us to reconstruct of the Instruction Manual for the operational parts located on the Antikythera Mechanism Back plate. In **Table III** the existence of this Symmetry is evident.

**Table III**: The two columns depicting the Symmetry existing in the preserved and reconstructed text of BCI Part-2.

| TEXT DESCRIPTION FOR THE PARTS POSITIONED ON THE BACK PLATE UPPER AREA | TEXT DESCRIPTION FOR THE PARTS POSITIONED ON THE BACK PLATE LOWER AREA |
|---|---|
| Η ΑΝΩΤΕΡΑ ΕΣΤΙΝ Η ΤΟΥ ΜΕΤΩΝΟΣ ΕΝΙΑΥΤΟΥ ΕΛΙΞ | ΚΑΤΩΤΕΡΑ ΕΣΤΙΝ Η ΕΛΙΞ ΤΟΥ ΠΕΡΙΟΔΙΚΟΥ ΕΝΙΑΥΤΟΥ |
| ΕΠΙ ΤΗΣ ΤΟΥ ΜΕΤΩΝΟΣ ΧΩΡΑΣ, ΕΙΣΙΝ Ε**Ν ΟΛΗ<I> ΤΗΙ ΕΛΙΚΙ, ΤΜΗΜΑΤΑ_Σ Λ̅ Ε** ΣΥΝ ΠΕΝΤΕ ΠΕΡΙΦΟΡΑΙΣ | **ΕΙΣ ΙΣΑ_ΣΚΓ-ΣΥΝ ΤΕΣΣ**ΑΡΣΙ ΠΕΡΙΦΟΡΑΙΣ, ΤΜΗΜΑΤΑ ΕΙΣΙΝ, ΕΝ ΟΛΗΙ ΤΗΙ ΕΛΙΚΙ, ΕΠΙ ΤΗΝ ΧΩΡΑΝ ΤΟΥ ΠΕΡΙΟΔΙΚΟΥ |
| - ΤΗΙ ΕΠΙ ΤΗΣ ΕΛΙΚΟΣ ΑΡΧΗΙ ΤΥΜΠΑΝΙΟΝ ΚΕΙΤΑΙ Ε**ΧΟΝ ΣΤΗΜΑΤΙΑ ΔΥΟ**.<br>- **ΠΕΡΙ ΤΥΜΠΑΝΙ**ΟΥ ΤΑ ΣΤΗΜΑΤΙΑ ΚΑΙ ΟΜΟΙΑ ΚΑΙ ΠΑΡΑΛΛΗΛΑ ΕΙΣΙΝ ΚΑΙ ΟΡΘΑ ΙΣΤΑΝΤΑΙ ΕΠ' ΑΥΤΟΥ.<br>- ΤΑ **ΠΡΟΕΙΡΗΜΕΝΑ ΣΤΗΜΑΤΙΑ ΤΡΗΜΑ**ΤΑ ΟΜΟΙΑ ΕΧΟΥΣΙΝ Α ΚΑΙ ΕΤΕΡΟΜΗΚΗ ΤΩΙ ΣΧΗΜΑΤΙ ΚΑΙ ΕΝΑΝΤΙΑ ΚΑΙ ΙΣΟΫΨΗ ΕΙΣΙΝ.<br>- ΔΙΑ **ΤΩΝ ΤΡΗΜΑΤΩΝ ΔΙΕΛΚΕΣΘΑΙ** ΓΝΩΜΟΝΙΟΝ ΜΕΤΑ ΠΕΡΟΝΗΣ, ΔΕΞΙΟΣΤΡΟΦΩΣ ΦΕΡΕΤΑΙ ΚΑΙ ΤΟΥΣ ΜΗΝΑΣ ΔΕΙΚΝΥΣΙ. | - ΤΗΙ ΕΠΙ ΤΗΣ ΤΟΥ ΠΕΡΙΟΔΙΚΟΥ ΕΛΙΚΟΣ ΑΡΧΗΙ ΣΥΜ **ΦΥΕΣ ΠΟΙΗ**ΣΑΙ **ΤΥΜΠ**ΑΝΙΟΝ ΕΧΟΝ ΣΤΗΜΑΤΙΑ ΔΥΟ Α ΚΑΙ ΟΜΟΙΑ ΚΑΙ ΠΑΡΑΛΛΗΛΑ ΕΙΣΙΝ, ΚΑΙ ΟΡΘΑ ΙΣΤΑΝΤΑΙ ΕΠ ΑΥΤΟΥ.<br>- ΤΡΗΜΑΤΑ ΟΜΟΙΑ ΚΑΙ ΕΤΕΡΟΜΗΚΗ ΚΑΙ ΕΝΑΝΤΙΑ ΚΑΙ ΙΣΟΫΨΗ ΕΧΟΥΣΙΝ ΤΑ **ΣΤΗΜΑΤΙΑ**.<br>- ΔΙΑ ΤΩΝ ΤΡΗΜΑΤΩΝ ΔΙΕΛΚΕΤΑΙ ΤΟ ΓΝΩΜΟΝΙΟΝ, ΔΕΞΙΟΣΤΡΟΦΩΣ ΦΕΡΕΤΑΙ ΚΑΙ ΤΩΝ ΕΓΛΕΙΨΕΩΝ ΤΑ ΓΕΓΟΝΟΤΑ ΔΕΙΚΝΥΣΙ. |
| ΚΑΤΑ **ΤΗΝ ΕΝΑΝΤΙΑΝ_ΕΙΣ** ΤΗΝ ΤΟΥ ΠΕΡΙΟΔΙΚΟΥ ΤΗΣ ΠΡΩΤΗΣ ΧΩΡΑΣ ΑΡΧΗΝ ΠΡΟΣΙΟΝΤΟΣ ΤΟΥ ΓΝΩΜΟΝΙΟΥ, ΤΟΤΕ ΔΕΙ ΓΕ ΤΗΝ **ΠΕΡΟΝΗΝ ΟΘΕΝ ΕΞΗΛΚΥΣ**ΘΑΙ ΚΑΙ ΕΠΙ ΤΗΝ ΤΗΣ ΠΡΩΤΗΣ ΧΩΡΑΣ ΑΡΧΗΝ ΤΙΘΗΝΑΙ. | ΟΜΟΙΩΣ ΚΑΙ (ΤΗΝ ΠΕΡΟΝΗΝ ΟΘΕΝ ΕΞΗΛΚΥΣΘΑΙ ΚΑΙ ΤΙΘΗΝΑΙ) ΕΠΙ ΤΗΝ ΤΟΥ ΜΕΤΩΝΟΣ ΑΡΧΗΝ **ΤΗΣ ΠΡΩΤΗΣ ΧΩΡΑΣ.** |
| ΜΕΡΗ **ΤΕΣΣΑΡΑ ΔΗΛΟΙ Δ' Ο ΜΕΝ, ΤΗ**Ν ΠΕΝΤΑΕΤΗΡΙΔΑ ΔΕΙΚΝΥΩΝ ΜΕΤ' ΑΝΑΓΡΑΦΟΜΕΝΩΝ ΤΩΝ ΣΤΕΦΑΝΙΤΩΝ ΑΓΩΝΩΝ. ΚΕΙΤΑΙ Δ' ΕΙΣ **ΤΗΝ ΤΗΣ_-OCL_IΘL_ΤΟΥ** ΜΕΤΩΝΟΣ ΕΛΙΚΑ. | ΜΕΡΗ ΤΡΙΑ ΔΗΛΟΙ Ο ΔΕ, ΚΕΙΤΑΙ ΕΙΣ ΤΗΝ ΤΟΥ ΠΕΡΙΟΔΙΚΟΥ ΚΑΙ ΕΣΤΙΝ Ο ΕΞΕΛΙΓ**ΜΟΣ** ….<br>… ΩΡΑΙ ΑΝΑΓΡΑΦΟΝΤΑΙ ΚΑΙ ΤΟΙΣ ΕΠΙ ΤΟΥ ΕΞΕΛΙΓΜΟΥ ΜΕΡΕΣΙ. |
| ΤΗΝ ΕΚΑΣΤΗΝ ΤΗΣ ΕΛΙΚΟΣ ΑΥΛΑΚΑ ΑΓΕΣΘΑΙ ΚΑΙ ΔΙΑΠΟΡΕΥΕΣΘΑΙ ΑΠΟ ΤΗΝ ΤΟΥ ΓΝΩΜΟΝΙΟΥ ΠΕΡΟΝΗΝ. ||
| **Μ**ΕΤΑ ΤΑΥΤΗΣ ΘΕΣΕΩΣ ΤΟΥΤΩΝ ΤΩΝ ΓΝΩΜΟΝΙΩΝ ΚΑΙ ΟΙ ΕΝΙΑΥΤΟΙ ΑΠΟΚΑΘΙΣΤΑΝΤΑΙ. ||

**Translation in English of Table III**

| TEXT DESCRIPTION FOR THE PARTS POSITIONED ON THE BACK PLATE UPPER AREA | TEXT DESCRIPTION FOR THE PARTS POSITIONED ON THE BACK PLATE LOWER AREA |
|---|---|
| At the top there is the Helix of Metonic eniautos | At the bottom there is the Helix of Periodicos Eniautos |
| The Metonic helix area is made (divided) along the full path of the helix, in 235 parts in five turns. | Equal 223, in four turns, divisions were made along the full path of the helix on the area of Periodicos |
| At the beginning (origin) of the (Metonic) helix there is a base which has two bearings. The bearings of the base are similar, parallel to each other and perpendicular to the base and fixed | On the beginning (origin) of the Periodicos helix a same kind of base is made (/there is). On the base there are two bearings, which are similar and parallel to each other, perpendicular and fixed on |
| on it. The aforementioned bearings have holes, which are similar, in oblong shape, opposite to each other and of the same height. Through the holes slides a pointer which has a perpendicular pin and is rotated CW indicating the (lunar synodic) months. | the base. And a same kind of pointer is made, which has a perpendicular pin. Similar holes, , in oblong shape, opposite each other and of the same height, have the bearings. Through the holes the pointer slides and is rotated in CW direction, and the eclipse events indicates. |
| When the pointer reaches the opposite position to the beginning of the Periodicos' 1st Area, pull out the pointer's pin end set it at the beginning of the 1st Area. | (Apply) the same (procedure for setting the pointer) at the Metonic 1st Area beginning. |
| Four sectors is divided the one (circle) of them indicates the Pentaeteris period and has the Stephanites Games engraved. It is located in the 19 year-out of 76 year period, of the Metonic Helix. | In three sectors is divided the other (circle), which is located on the Periodicos (helix) and it is the Exeligmos. |
| Each of the pointers is driven and travels along the helix's furrow. | |
| After the position reset of the pointers, the eniautoi (cycles) return on their beginning (origin). | |

In **Table IV** the outer operational parts visible on the Back plate as well as their *Definition/Description*/Position and *Operation* are presented.

**Table IV**: Synoptic Catalogue of the Antikythera Mechanism Back plate operational parts and their *Definition/Description/Position/Operation* as presented on the BCI Part-2 text reconstruction.

| | Parts of the Back plate with Ancient Greek name | OPERATIONAL PART/ DEFINITION | DESCRIPTION | POSITION | OPERATION |
|---|---|---|---|---|---|
| 1 | ΕΛΙΞ ΤΟΥ ΜΕΤΩΝΟΣ ΕΝΙΑΥΤΟΥ | Metonic Helix | The helix has a furrow, is divided in 235 parts | On the upper part of the Back plate | The calendar (scale) of the device |
| 2 | ΕΛΙΞ ΤΟΥ ΠΕΡΙΟΔΙΚΟΥ ΧΡΟΝΟΥ | Helix of Periodicos (Saros helix) | The helix has a furrow, is divided in 223 parts | Located at the lower part of the Back plate | The eclipse information scale |
| 3 | ΕΛΙΚΩΝ ΤΜΗΜΑΤΑ (on Metonic and Periodicos helix) | Parts (subdivisions) of the 2 helices | The parts of the 2 helices are divided in equal epicenter angles) | Along the full length of the area of helices | Scale subdivisions |
| 4 | ΤΥΜΠΑΝΙΟΝ (on Metonic and Periodicos helix) | Base | Is the base of the Metonic/Saros pointer | Located at the helix origin | Via the base, the pointer travels through the full length of the helix |
| 5 | ΣΤΗΜΑΤΙΑ (on Metonic and Periodicos helix) | Bearings | Bearings perpendicular on their base, opposite between them | Are stabilized on the base | On the bearings is adapted the Metonic/Saros pointer |
| 6 | ΤΡΗΜΑΤΑ (on Metonic and Periodicos helix) | Holes | The holes have oblong shape, are equal, opposite and in same height between them | The holes are located on the bearings | Via the holes, the Metonic/Saros pointer slides |
| 7 | ΓΝΩΜΟΝΙΟΝ ΜΕΤΑ ΠΕΡΟΝΗΣ (on Metonic and Periodicos helix) | Metonic/Saros pointer | Metonic/Saros pointer has a (perpendicular) pin | Perforates the holes of the bearings | Metonic/Saros pointer is rotated in CW direction and depicts the months/eclipse events |

| 8 | ΕΛΑΣΣΩΝ ΚΥΚΛΟΣ +ΓΝΩΜΟΝΙΟΝ (on Metonic helix) | Small circular scale Pentateris Dial | Is divided in 4 sectors | Located close to the Metonic helix origin | Scale units: Stephanites Athletic Games |
|---|---|---|---|---|---|
| 9 | ΕΛΑΣΣΩΝ ΚΥΚΛΟΣ +ΓΝΩΜΟΝΙΟΝ (on Helix of Periodicos) | Small circular scale Exeligmos Dial | Is divided in 3 sectors | Located close to the Saros helix origin | Scale units: Periodicos cycles/ Hours (correction of eclipse events times) |

## 8. The final arrangement/refinition of the BCI Part-2 reconstructed text

Based on the previous analysis and the text reconstruction line by line, the reconstructed text parts were firstly connected in an intergraded text, keeping the specific position of the preserved words and the maximum limit of 88 letters per line. The final reconstructed text is presented on **Table V** (also on **Table VI**: Comments and clarifications in the reconstructed text BCI Part-2, **Table VII**: BCI PART-2 text translation in the English language, **Table VIII**: BCI PART-2 text translation in the Modern Greek language, **Table IX**: Positional analysis of the Back dial operational parts).

**Table V**: Reconstructed text of the Back Cover Inscription Part-2 based on the analysis of Section 8.

```
..............................................................................................................................................................
..............................................................................................................................................................
..............................................................................................................................................................
 I) ΕΙΣ ΤΗΝ ΟΠΙΣΘΙΑΝ ΟΨΙΝ, ΤΟ ΗΜΕΡΟΛΟΓΙΩΝ ΚΑΙ ΤΩΝ ΕΓΛΕΙΨΕΩΝ ΤΑ ΓΕΓΟΝΟΤΑ ΚΑΙ ΣΕΛΗΝΗΣ ΚΑΙ ΗΛΙΟΥ, ΑΝΑΓΡΑΦΟΝΤΑΙ
 II) ΑΥΤΟΘΙ. ΔΙΗΡΗΜΕΝΗ ΕΣΤΙΝ Η ΠΛΑΞ ΕΙΣ ΕΛΙΚΑΣ ΔΥΟ ΚΑΙ ΜΕΤ' ΑΥΛΑΚΟΣ ΕΚΑΣΤΗ ΕΣΤΙΝ. ΑΝΩΘΕΝ ΚΕΙΤΑΙ Η ΤΟΥ ΜΕΤΩΝΟΣ
 III) ΕΝΙΑΥΤΟΥ ΕΛΙΞ ΚΑΙ ΚΑΤΩΘΕΝ Η ΕΛΙΞ ΤΟΥ ΠΕΡΙΟΔΙΚΟΥ ΕΝΙΑΥΤΟΥ.  .........................................
01) . . . . . ΛΟΣ.                                  ΤΑ ΤΩΝ ΕΛΙΚΩΝ ΤΜΗΜΑΤΑ, Α ΜΗΝΕΣ ΕΙΣΙΝ, ΠΕΠΟΙΗ-
02) ΝΤΑΙ ΑΠΟ ΤΩΝ ΔΙΑΙΡΕΣΕΩΝ ΤΗΣ ΤΟΥ ΜΕΤΩΝΟΣ ΚΑΙ ΤΗΣ ΤΟΥ ΠΕΡΙΟΔΙΚΟΥ ΧΩΡΑΣ. ΕΠΙ ΤΗΝ ΤΟΥ ΜΕΤΩΝΟΣ ΧΩΡΑΝ ΕΙΣ-
03) ΙΝ ΕΝ ΟΛΗ<I> ΤΗΙ ΕΛΙΚΙ, ΤΜΗΜΑΤΑ_ΣΛΕ_ΣΥΝ ΠΕΝΤΕ ΠΕΡΙΦΟΡΑΙΣ. ΚΑΤΑ ΤΗΝ ΤΗΣ ΕΛΙΚΟΣ ΠΡΩΤΗΝ ΠΕΡΙΦΟΡΑΝ ΑΝΑΓΡΑΦΟΝ-
04) ΤΑΙ ΔΕ ΚΑΙ ΑΙ ΕΞΑΙΡΕΣΙΜΟΙ ΗΜΕΡΑΙ ΚΑΤΑ ΤΩΝ ΜΕΤΩΝΑΝ ΛΕΛΟΓΙΣΜΕΝΑΙ. ΤΗΙ ΕΠΙ ΤΗΣ ΕΛΙΚΟΣ ΑΡΧΗΙ ΤΥΜΠΑΝΙΟΝ ΚΕΙΤΑΙ
05) ΕΧΟΝ ΣΤΗΜΑΤΙΑ ΔΥΟ. -ΠΕΡΙ ΤΥΜΠΑΝΙΟΥ ΤΑ ΣΤΗΜΑΤΙΑ ΚΑΙ ΟΜΟΙΑ ΚΑΙ ΠΑΡΑΛΛΗΛΑ ΕΙΣΙΝ ΚΑΙ ΟΡΘΑ ΙΣΤΑΝΤΑΙ ΕΠ ΑΥΤΟΥ.
06) ΤΑ ΠΡΟΕΙΡΗΜΕΝΑ ΣΤΗΜΑΤΙΑ ΤΡΗΜΑΤΑ ΟΜΟΙΑ ΕΧΟΥΣΙΝ Α ΚΑΙ ΕΤΕΡΟΜΗΚΗ ΤΩΙ ΣΧΗΜΑΤΙ ΚΑΙ ΕΝΑΝΤΙΑ ΚΑΙ ΙΣΟΫΨΗ ΕΙΣΙΝ.
07) ΔΙΑ ΤΩΝ ΤΡΗΜΑΤΩΝ ΔΙΕΛΚΕΣΘΑΙ ΓΝΩΜΟΝΙΟΝ ΜΕΤΑ ΠΕΡΟΝΗΣ, ΔΕΞΙΟΣΤΡΟΦΩΣ ΦΕΡΕΤΑΙ ΚΑΙ ΤΟΥΣ ΜΗΝΑΣ ΔΕΙΚΝΥΣΙ. ΚΑΙ
08) ΟΜΟΙΩΣ ΤΟΙΣ ΠΡΩΤΟΙΣ ΛΟΓΟΙΣ ΚΑΙ ΤΟΙΟΥΤΟΙ ΛΟΓΟΙ ΕΠΟΝΤΑΙ. ΤΗΙ ΕΠΙ ΤΗΣ ΤΟΥ ΠΕΡΙΟΔΙΚΟΥ ΕΛΙΚΟΣ ΑΡΧΗΙ ΣΥΜ-
09) ΦΥΕΣ ΠΟΙΗΣΑΙ ΤΥΜΠΑΝΙΟΝ ΕΧΟΝ ΣΤΗΜΑΤΙΑ ΔΥΟ Α ΚΑΙ ΟΜΟΙΑ ΚΑΙ ΠΑΡΑΛΛΗΛΑ ΕΙΣΙΝ, ΚΑΙ ΟΡΘΑ ΙΣΤΑΝΤΑΙ ΕΠ ΑΥΤΟΥ.
10) ΚΑΙ ΣΥΜΦΥΕΣ ΠΟΙΗΣΑΙ ΓΝΩΜΟΝΙΟΝ ΜΕΤΑ ΠΕΡΟΝΗΣ. ΤΡΗΜΑΤΑ ΟΜΟΙΑ ΚΑΙ ΕΤΕΡΟΜΗΚΗ ΚΑΙ ΕΝΑΝΤΙΑ ΚΑΙ ΙΣΟΫΨΗ ΕΧΟΥΣΙΝ
11) ΤΑ ΣΤΗΜΑΤΙΑ.  ΔΙΑ ΤΩΝ ΤΡΗΜΑΤΩΝ ΔΙΕΛΚΕΤΑΙ ΤΟ ΓΝΩΜΟΝΙΟΝ,  ΔΕΞΙΟΣΤΡΟΦΩΣ ΦΕΡΕΤΑΙ ΚΑΙ ΤΩΝ ΕΓΛΕΙΨΕΩΝ ΤΑ ΓΕΓΟΝΟΤ-
12) Α ΔΕΙΚΝΥΣΙ. ΤΑ ΓΕ ΣΕΛΗΝΗΣ ΓΕΓΟΝΟΤΑ ΜΕΤΑ ΤΟΥ ΣΤΟΙΧΕΙΟΥ -Σ:- ΕΙΣΙΝ, ΤΑ Δ' ΗΛΙΟΥ ΜΕΤΑ ΤΟΥ ΣΤΟΙΧΕΙΟΥ -Η-. ....................
13) [. . .]ΡΟΥ . . ΦΟΔΟΥ ΤΗΝ ΕΚΑΣΤΗΣ ΤΗΣ ΕΛΙΚΟΣ ΑΥΛΑΚΑ ΑΓΕΣΘΑΙ ΚΑΙ ΔΙΑΠΟΡΕΥΕΣΘΑΙ ΑΠΟ ΤΗΝ ΤΟΥ ΓΝΩΜΟΝΙΟΥ ΠΕΡΟΝΗΝ.
14) ΚΑΤΑ ΤΗΝ ΕΝΑΝΤΙΑΝ_ΕΙΣ ΤΗΝ ΤΟΥ ΠΕΡΙΟΔΙΚΟΥ ΤΗΣ ΠΡΩΤΗΣ ΧΩΡΑΣ ΑΡΧΗΝ ΠΡΟΣΙΟΝΤΟΣ ΤΟΥ ΓΝΩΜΟΝΙΟΥ, ΤΟΤΕ ΔΕΙ ΓΕ
15) ΤΗΝ ΠΕΡΟΝΗΝ ΟΘΕΝ ΕΞΗΛΚΥΣΘΑΙ ΚΑΙ ΕΠΙ ΤΗΝ ΤΗΣ ΠΡΩΤΗΣ ΧΩΡΑΣ ΑΡΧΗΝ ΤΙΘΗΝΑΙ. ΟΜΟΙΩΣ ΚΑΙ ΕΠΙ ΤΗΝ ΤΟΥ ΜΕΤΩΝΟΣ ΑΡ-
16) ΧΗΝ ΤΗΣ ΠΡΩΤΗΣ ΧΩΡΑΣ. -ΜΕΤΑ ΤΑΥΤΗΣ ΘΕΣΕΩΣ ΤΟΥΤΩΝ ΤΩΝ ΓΝΩΜΟΝΙΩΝ ΚΑΙ ΟΙ ΕΝΙΑΥΤΟΙ ΑΠΟΚΑΘΙΣΤΑΝΤΑΙ. ΕΙΣΙΝ ΔΕ
17) ΓΝΩΜΟΝΙΑ ΔΥΟ, -ΩΝ ΤΑ ΑΚΡΑ ΦΕΡΟΝΤΑΙ ΕΠ' ΕΛΑΣΣΟΝΩΝ ΚΥΚΛΩΝ. ΟΙ ΔΕ ΚΥΚΛΟΙ ΠΑΡΑ ΤΑΣ ΤΩΝ ΕΛΙΚΩΝ ΑΡΧΑΣ ΚΕΙΝΤΑΙ. ΜΕ-
18) ΡΗ ΤΕΣΣΑΡΑ ΔΗΛΟΙ Δ' Ο ΜΕΝ, ΤΗΝ ΠΕΝΤΑΕΤΗΡΙΔΑ ΔΕΙΚΝΥΩΝ ΜΕΤ' ΑΝΑΓΡΑΦΟΜΕΝΩΝ ΤΩΝ ΣΤΕΦΑΝΙΤΩΝ ΑΓΩΝΩΝ. ΚΕΙΤΑΙ Δ' Ε-
19) ΙΣ ΤΗΝ ΤΗΣ_-ΟCL_ΙΟL_ΤΟΥ ΜΕΤΩΝΟΣ ΕΛΙΚΑ. ΜΕΡΗ ΤΡΙΑ ΔΗΛΟΙ Ο ΔΕ, ΚΕΙΤΑΙ ΕΙΣ ΤΗΝ ΤΟΥ ΠΕΡΙΟΔΙΚΟΥ ΚΑΙ ΕΣΤΙΝ Ο ΕΞΕΛΙΓ-
20) ΜΟΣ, ΕΙΣ ΙΣΑ _ΣΚΓ-ΣΥΝ ΤΕΣΣΑΡΣΙ ΠΕΡΙΦΟΡΑΙΣ, ΤΜΗΜΑΤΑ ΕΙΣΙΝ, ΕΝ ΟΛΗΙ ΤΗΙ ΕΛΙΚΙ, ΕΠΙ ΤΗΝ ΧΩΡΑΝ ΤΟΥ ΠΕΡΙΟΔΙΚΟΥ, ΩΣ-
21) ΠΕΡ ΑΥΤΟΣ ΔΙΑ(Η)ΙΡΕΘΗ. -Η-ΟΛΗ ΤΩΝ ΓΕΓΟΝΟΤΩΝ ΑΚΟΛΟΥΘΙΑ ΕΠΙ ΤΙΝΩΝ ΤΩΝ ΤΗΣ ΕΛΙΚΑΣ ΤΜΗΜΑΤΩΝ ΠΑΡΑΤΙΘΕΤΑΙ.
22) ΜΟΝΟΙ ΔΗ ΟΙ ΕΓΛΕΙΠΤΙΚΟΙ ΧΡΟΝΟΙ ΤΟΥΤΩΝ ΤΩΝ ΓΕΓΟΝΟΤΩΝ ΕΙΣ ΩΡΑΣ ΙΣΗΜΕΡΙΝΑΣ ΑΝΑΓΡΑΦΟΝΤΑΙ ΑΥΤΟΘΙ. ΚΑΙ
23) ΟΜΟΙΩΣ ΤΟΙΣ ΕΠΙ ΤΗΣ ΕΛΙΚΟΣ ΤΜΗΜΑΣΙ, ΩΡΑΙ ΑΝΑΓΡΑΦΟΝΤΑΙ ΚΑΙ ΤΟΙΣ ΕΠΙ ΤΟΥ ΕΞΕΛΙΓΜΟΥ ΜΕΡΕΣΙ. ΤΟ ΓΝΩΜΟΝΙΟΝ, ΟΥ ΤΟ
24) ΑΚΡΟΝ ΦΕΡΕΤΑΙ ΚΑΘ ΕΚΑΣΤΟΝ ΤΩΝ ΤΟΥ ΕΞΕΛΙΓΜΟΥ ΜΕΡΩΝ, ΕΚΑΣΤΟΝ ΤΩΝ ΤΡΙΩΝ ΕΝΙΑΥΤΩΝ ΤΟΥ ΠΕΡΙΟΔΙΚΟΥ ΔΗΛΟΙ, ΔΕΥΤΕ-
25) ΡΟΥ ΜΕΝ ΤΟΥ ΠΕΡΙΟΔΙΚΟΥ ΟΝΤΟΣ, ΔΕΙ ΤΟΥΣ ΕΓΛΕΙΠΤΙΚΟΥΣ ΧΡΟΝΟΥΣ ΑΥΞΕΣΘΑΙ ΚΑΤΑ ΩΡΑΣ -Η-, ΤΡΙΤΟΥ ΔΕ ΚΑΤΑ ΩΡΑΣ -ΙΣ-.
..............................................................................................................................................................
..............................................................................................................................................................
..............................................................................................................................................................
```

**Table VI**: Comments/corrections/synonyms in the reconstructed text BCI Part-2

- **Line III:** For the phrase ΚΑΤΩΘΕΝ * Η ΕΛΙΞ ΤΟΥ ΠΕΡΙΟΔΙΚΟΥ ΕΝΙΑΥΤΟΥ, the word *ΚΕΙΤΑΙ (is located) is implied as is referred on Line II.
- **Line 01:** The ΜΗΝΕΣ corresponds into the lunar synodic months (ΣΥΝΟΔΙΚΟΙ). When Geminus and Ptolemy use the word ΜΗΝΑΙ they mean the "mean lunar synodic month".
- **Line 05:** Instead of ΟΜΟΙΑ (*equal*) could be ΙΣΑ (ΤΕ).
- **Line 07:** Instead of **ΔΙΕΛΚΕΣΘΑΙ** (*perforates/slides*) it should be written ΔΙΕΛΚΕΤΑΙ.
- **Line 09:** Instead of **ΠΟΙΗΣ**ΑΙ (infinitive) it should be written ΠΕΠΟΙΗΤΑΙ (verb).
- **Line 09:** Instead of ΟΜΟΙΑ could be ΙΣΑ (as on Line 05).
- **Line 15:** For the phrase ΟΜΟΙΩΣ (*) ΚΑΙ ΕΠΙ ΤΗΝ ΤΟΥ ΜΕΤΩΝΟΣ, the word * ΠΡΑΞΑΙ (*make all the aforementioned procedure*) or the phrase (* ΤΙΘΗΝΑΙ ΤΟ ΓΝΩΜΟΝΙΟΝ) is implied.
- **Line 16:** Instead of ΑΠΟΚΑΘΙΣΤΑΝΤΑΙ (they *return to their beginning*) it could be ΑΡΧΟΝΤΑΙ (they *begin*).
- **Line 17:** ΩΝ ΤΑ ΑΚΡΑ **ΦΕ**ΡΟΝΤΑΙ or **ΩΝ ΤΑ ΑΚΡΑ ΦΕ**ΡΕΤΑΙ (in a syntax of Attic Greek).
- **Line 18:** Instead of **ΤΗ**Ν ΠΕΝΤΑΕΤΗΡΙΔΑ (*Pentateris*) could be **ΤΗ**Ν ΠΕΡΙΟΔΟΝ (*Periodos/Circuit*), but less probable because it could be confused to the word ΠΕΡΙΟΔΙΚΟΣ (*Periodicos/Saros cycle*).
- **Line 18:** For the phrase **ΤΕΣΣΑΡΑ ΔΗΛΟΙ Δ' Ο ΜΕΝ** (*), the phrase * ΕΛΑΣΣΩΝ ΚΥΚΛΟΣ is implied, as is presented before.
- **Line 18:** Instead of ΣΤΕΦΑΝΙΤΩΝ (*Stephanites/Crown*) it could be ΙΕΡΩΝ (*Sacred*).
- **Line 19:** For the phrase ΜΕΡΗ ΤΡΙΑ ΔΗΛΟΙ Ο ΔΕ (*), the phrase ΕΛΑΣΣΩΝ ΚΥΚΛΟΣ is implied, as is presented before.
- **Line 19:** For the phrase ΚΕΙΤΑΙ ΕΙΣ ΤΗΝ ΤΟΥ ΠΕΡΙΟΔΙΚΟΥ (*), the word *ΕΛΙΚΑ is implied, as is mentioned before.
- **Line 21:** ΤΕ . Δ . . ΟΣΔΙΑΙΡΕΘΗ<I> Η ΟΛΗ, the first letter is considered as the letter ΠΕ . Δ . . ΟΣ ΔΙΑΙΡΕΘΗ Η ΟΛΗ
- **Line 21:** Considering the first letter of Line 21 as Π, then instead of ΔΙΑΙΡΕΘΗ it should be as ΔΙΗΙΡΕΘΗ (ΩΣ|ΠΕΡ **Α**ΥΤΟΣ ΔΙΗΙΡΕΘΗ. Η ΟΛΗ……)
- **Line 21:** Instead of ΑΚΟΛΟΥΘΙΑ (*sequence/succession*) it could be ΔΙΑΔΟΧΗ.
- **Line 24 (end)/Line 25 (beginning):** Before the phrase *ΔΕΥΤΕ|ΡΟΥ ΜΕΝ ΤΟΥ ΠΕΡΙΟΔΙΚΟΥ, the phrase *ΓΝΩΜΟΝΙΟΥ ΔΗΛΟΥΝΤΟΣ ΤΟΥ (*as the pointer aims to the 2$^{nd}$ Periodicos*), is implied as the word ΔΗΛΟΙ is mentioned before.
- **Line 24 (end)/Line 25 (beginning):** Instead of ΔΕΥΤΕ|ΡΟΥ it could be the word ΜΕ|ΣΟΥ (*middle/medial*).
- **Line 25:** Correspondingly instead of ΤΡΙΤΟΥ (*3$^{rd}$*) it could be ΕΣΧΑΤΟΥ (*last/ultimate*).
- **Line 25:** For the phrase ΔΕΥΤΕ|ΡΟΥ **ΜΕΝ ΤΟΥ** ΠΕΡΙΟΔΙΚΟΥ (*) ΟΝΤΟΣ, the word *ΕΝΙΑΥΤΟΥ is implied, as is presented before.
- **Line 25:** For the phrase ΤΡΙΤΟΥ ΔΕ (*) ΚΑΤΑ ΩΡΑΣ the phrase ΤΟΥ ΠΕΡΙΟΔΙΚΟΥ ΕΝΙΑΥΤΟΥ is implied, as is presented before.

**Table VII**: BCI PART-2 text translation in the English language

i) On the Back face, the calendar and the lunar and solar eclipse events are engraved. The Back plate is divided
ii) in two helices and each of them has a furrow. At the top there is the Helix of Metonic eniautos and
iii) at the bottom there is the Helix of Periodicos Eniautos. …………………………………………………………………………
01) ………………………………………………………… The parts of helix, which are synodic months, have been
02) created by the division of the Metonic and Periodicos helix area. The Metonic helix area is made (divided)
03) along the full path of the helix, in 235 parts in five turns. Along the first turn there are also engraved
04) the Exeresimoi (omitted) days, calculated by Meton. At the beginning (origin) of the (Metonic) helix there is a base
05) which has two bearings. The bearings of the base are similar, parallel to each other and perpendicular to the base and fixed on it.
06) The aforementioned bearings have holes, which are similar, in oblong shape, opposite to each other and of the same height.
07) Through the holes slides a pointer which has a perpendicular pin and is rotated CW indicating the (lunar synodic) months.
08) Similar to the previous sentences (/description) are the sentences (/description) that follow: On the beginning (origin) of the Periodicos helix
09) a same kind of base is made (/there is). On the base there are two bearings, which are similar and parallel to each other, perpendicular and fixed on the base.
10) And a same kind of pointer is made, which has a perpendicular pin. Similar holes, in oblong shape, opposite each other and of the same height, have
11) the bearings. Through the holes the pointer slides and is rotated in CW direction, and the eclipse events
12) indicates. The lunar (eclipse) events are marked with the letter (symbol) *Σ* and the solar (eclipse) events are marked with the letter *H*. …………
13) ……………………………. Each of the pointers is driven and travels along the helix's furrow.
14) When the pointer reaches the opposite position to the beginning of the Periodicos' 1$^{st}$ Area,

15) pull out the pointer's pin end set it at the beginning of the 1$^{st}$ Area. (Apply) the same (procedure for setting the pointer) at the Metonic

16) 1$^{st}$ Area beginning. After the position reset of the pointers, the eniautoi (cycles) return on their beginning (origin). And there are

17) two pointers, which their edges are rotating on two small circular scales, which are located next to the helices' origin.

18) Four sectors is divided the one (circle) of them indicates the Pentaeteris period and has the Stephanites Games engraved. It is located

19) in the 19 year-out of 76 year period, of the Metonic Helix. In three sectors is divided the other (circle), which is located on the Periodicos (helix) and it is the Exelig-

20) mos. Equal 223, in four turns, divisions were made along the full path of the helix on the area of Periodicos as

21) indeed (Periodicos cycle) was divided. The full sequence of the (eclipse) events is presented on some of the helix divisions (cells).

22) Only the times of these eclipse events, are engraved (on the cells) in Equinoctial hours. And

23) similarly to the helix parts (cells), the hours are also engraved on the Exeligmos' sectors. The pointer's

24) edge travels through each of Exeligmos' sector and indicates each of the three Periodic cycles. (When the pointer aims) at the Second (/Middle)

25) Periodicos (cycle), add to the time of the eclipse events 8 hours, and (when the pointer aims) at the Third (/Last) (Periodicos cycle), add (to the time of the eclipse events) 16 hours.

**Table VIII:** BCI PART-2 text translation in the Modern Greek language

..............................................................................................................................................................................
..............................................................................................................................................................................
Ι) ΣΤΗΝ ΠΙΣΩ ΟΨΗ ΑΝΑΓΡΑΦΕΤΑΙ ΤΟ ΗΜΕΡΟΛΟΓΙΟ ΚΑΙ ΤΑ ΓΕΓΟΝΟΤΑ ΤΩΝ ΕΚΛΕΙΨΕΩΝ ΤΗΣ ΣΕΛΗΝΗΣ ΚΑΙ ΤΟΥ ΗΛΙΟΥ.

ΙΙ) Η ΠΛΑΚΑ ΕΙΝΑΙ ΔΙΑΙΡΕΜΕΝΗ ΣΕ ΔΥΟ ΕΛΙΚΕΣ ΚΑΙ ΚΑΘΕ ΜΙΑ ΕΧΕΙ ΕΝΑ ΑΥΛΑΚΙ. ΣΤΟ ΕΠΑΝΩ ΤΜΗΜΑ ΤΗΣ ΠΛΑΚΑΣ ΒΡΙΣΚΕΤΑΙ Η ΕΛΙΚΑ ΤΟΥ ΜΕΤΩΝΟΣ

ΙΙΙ) ΕΝΙΑΥΤΟΥ ΚΑΙ ΣΤΟ ΚΑΤΩ (ΒΡΙΣΚΕΤΑΙ) Η ΕΛΙΚΑ ΤΟΥ ΠΕΡΙΟΔΙΚΟΥ ΧΡΟΝΟΥ. ......................................................

01) . . . . . **ΛΟΣ**........................................................................ ΟΙ ΥΠΟΔΙΑΙΡΕΣΕΙΣ ΤΩΝ ΕΛΙΚΩΝ ΟΙ ΟΠΟΙΕΣ ΕΙΝΑΙ ΣΥΝΟΔΙΚΟΙ ΜΗΝΕΣ, ΕΧΟΥΝ ΓΙΝ-

02) ΝΕΙ ΔΙΑΙΡΩΝΤΑΣ ΤΗΝ ΠΕΡΙΟΧΗ ΤΗΣ ΕΛΙΚΑΣ ΤΟΥ ΜΕΤΩΝΑ ΚΑΙ ΤΗΣ ΕΛΙΚΑΣ ΤΟΥ ΠΕΡΙΟΔΙΚΟΥ. Η ΠΕΡΙΟΧΗ ΤΗΣ ΕΛΙΚΑΣ ΤΟΥ ΜΕΤΩΝΑ ΔΙΑΙΡΕΙΤΑΙ

03) ΚΑΤΑ ΜΗΚΟΣ ΟΛΗΣ ΤΗΣ ΕΛΙΚΑΣ ΣΕ 223 ΤΜΗΜΑΤΑ ΕΝΤΟΣ 5 ΕΛΙΚΟΣΤΡΟΦΩΝ. ΚΑΤΑ ΤΗΝ ΠΡΩΤΗ ΕΛΙΚΟΣΤΡΟΦΗ , ΑΝΑΓΡΑΦΟΝ-

04) ΤΑΙ ΚΑΙ ΟΙ ΕΞΑΙΡΕΣΙΜΕΣ ΗΜΕΡΕΣ ΠΟΥ ΕΧΟΥΝ ΥΠΟΛΟΓΙΣΤΕΙ ΑΠΟ ΤΟΝ ΜΕΤΩΝΑ. ΣΤΗΝ ΑΡΧΗ ΤΗΣ ΕΛΙΚΑΣ ΒΡΙΣΚΕΤΑΙ ΜΙΑ ΒΑΣΗ

05) ΠΟΥ ΕΧΕΙ ΔΥΟ ΣΤΗΡΙΓΜΑΤΑ. ΤΑ ΣΤΗΡΙΓΜΑΤΑ ΤΗΣ ΒΑΣΗΣ ΕΙΝΑΙ ΟΜΟΙΑ, ΚΑΙ ΠΑΡΑΛΛΗΛΑ ΚΑΙ ΕΙΝΑΙ ΚΑΘΕΤΑ ΣΤΕΡΕΩΜΕΝΑ ΣΤΗ ΒΑΣΗ.

06) ΤΑ ΠΡΟΑΝΑΦΕΡΘΕΝΤΑ ΣΤΗΡΙΓΜΑΤΑ ΕΧΟΥΝ ΟΠΕΣ ΠΟΥ ΕΙΝΑΙ ΟΜΟΙΕΣ ΚΑΙ ΟΙ ΟΠΟΙΕΣ ΕΧΟΥΝ ΠΑΡΑΛΛΗΛΟΓΡΑΜΜΟ ΣΧΗΜΑ, ΕΙΝΑΙ ΑΠΕΝΑΝΤΙ ΜΕΤΑΞΥ ΤΟΥΣ ΚΑΙ ΣΤΟ ΙΔΙΟ ΥΨΟΣ.

07) ΔΙΑ ΜΕΣΟΥ ΤΩΝ ΟΠΩΝ ΔΙΕΡΧΕΤΑΙ ΕΝΑΣ ΔΕΙΚΤΗΣ ΜΕ ΠΕΡΟΝΗ, Ο ΟΠΟΙΟΣ ΠΕΡΙΣΤΡΕΦΕΤΑΙ ΔΕΞΙΟΣΤΡΟΦΑ ΚΑΙ ΔΕΙΧΝΕΙ ΤΟΥΣ ΜΗΝΕΣ. ΚΑΙ

08) ΟΜΟΙΩΣ ΜΕ ΤΙΣ ΠΡΟΗΓΟΥΜΕΝΕΣ ΠΕΡΙΓΡΑΦΕΣ, ΤΕΤΟΙΕΣ ΠΕΡΙΓΡΑΦΕΣ ΑΚΟΛΟΥΘΟΥΝ. ΣΤΗΝ ΑΡΧΗ ΤΗΣ ΕΛΙΚΑΣ ΤΟΥ ΠΕΡΙΟΔΙΚΟΥ ΥΠΑΡΧΕΙ ΜΙΑ ΙΔΙΑ

09) ΒΑΣΗ ΠΟΥ ΕΧΕΙ ΔΥΟ ΣΤΗΡΙΓΜΑΤΑ, ΤΑ ΟΠΟΙΑ ΕΙΝΑΙ ΟΜΟΙΑ, ΠΑΡΑΛΛΗΛΑ ΜΕΤΑΞΥ ΤΟΥΣ ΚΑΙ ΚΑΘΕΤΑ ΣΤΕΡΕΩΜΕΝΑ ΣΤΗ ΒΑΣΗ.

10) ΚΑΙ ΥΠΑΡΧΕΙ ΕΝΑΣ ΟΜΟΙΟΣ ΔΕΙΚΤΗΣ ΜΕ ΠΕΡΟΝΗ. ΟΠΕΣ ΟΜΟΙΕΣ ΚΑΙ ΜΕ ΠΑΡΑΛΛΗΛΟΓΡΑΜΜΟ ΣΧΗΜΑ ΚΑΙ ΑΠΕΝΑΝΤΙ ΜΕΤΑΞΥ ΤΟΥΣ ΚΑΙ ΣΤΟ ΙΔΙΟ ΥΨΟΣ ΕΧΟΥΝ

11) ΤΑ ΣΤΗΡΙΓΜΑΤΑ. ΔΙΑ ΜΕΣΟΥ ΤΩΝ ΟΠΩΝ Ο ΔΕΙΚΤΗΣ ΔΙΕΡΧΕΤΑΙ, ΠΕΡΙΣΤΡΕΦΕΤΑΙ ΔΕΞΙΟΣΤΡΟΦΑ ΚΑΙ ΤΑ ΓΕΓΟΝΟΤΑ ΤΩΝ ΕΚΛΕΙΨΕΩΝ

12) ΔΕΙΧΝΕΙ. ΤΑ ΓΕΓΟΝΟΤΑ ΤΩΝ ΕΚΛΕΙΨΕΩΝ ΤΗΣ ΣΕΛΗΝΗΣ ΣΗΜΕΙΩΝΟΝΤΑΙ ΜΕ ΤΟ ΓΡΑΜΜΑ *Σ* ΚΑΙ (ΤΑ ΓΕΓΟΝΟΤΑ ) ΤΩΝ ΕΚΛΕΙΨΕΩΝ ΤΟΥ ΗΛΙΟΥ ΜΕ ΤΟ ΓΡΑΜΜΑ *Η*. .......................

13) ........................ ΜΕΣΑ ΣΤΟ ΑΥΛΑΚΙ ΤΗΣ ΚΑΘΕ ΕΛΙΚΑΣ, ΟΔΗΓΕΙΤΑΙ ΚΑΙ ΚΙΝΕΙΤΑΙ Η ΠΕΡΟΝΗ ΤΟΥ ΔΕΙΚΤΗ.

14) ΣΤΟ ΑΠΕΝΑΝΤΙ ΣΗΜΕΙΟ ΑΠΟ ΤΗΝ ΑΡΧΗ ΤΗΣ ΠΡΩΤΗΣ ΠΕΡΙΟΧΗΣ ΤΗΣ ΕΛΙΚΑΣ ΤΟΥ ΠΕΡΙΟΔΙΚΟΥ, ΦΤΑΝΟΝΤΑΣ Ο ΔΕΙΚΤΗΣ, ΤΟΤΕ ΠΡΕΠΕΙ ΝΑ

15) ΤΡΑΒΗΞΤΕ ΕΞΩ (ΑΠΟ ΤΟ ΑΥΛΑΚΙ) ΤΗΝ ΠΕΡΟΝΗ ΚΑΙ ΝΑ ΤΗΝ ΤΟΠΟΘΕΤΗΣΕΤΕ ΣΤΗΝ ΑΡΧΗ ΤΗΣ ΠΡΩΤΗΣ ΠΕΡΙΟΧΗΣ ΤΗΣ ΕΛΙΚΑΣ. ΟΜΟΙΩΣ ΠΡΑΞΤΕ ΤΟΠΟΘΕΤΩΝΤΑΣ (ΤΗΝ ΠΕΡΟΝΗ) ΚΑΙ ΣΤΗΝ ΑΡΧΗ

16) ΤΗΣ ΠΡΩΤΗΣ ΠΕΡΙΟΧΗΣ ΤΗΣ ΕΛΙΚΑΣ ΤΟΥ ΜΕΤΩΝΑ. ΜΕ ΤΗΝ ΕΠΑΝΑΘΕΣΗ ΤΩΝ ΔΕΙΚΤΩΝ ΚΑΙ ΟΙ ΕΝΙΑΥΤΟΙ

ΞΑΝΑΞΕΚΙΝΟΥΝ ΑΠΟ ΤΗΝ ΑΡΧΗ (ΑΠΟΚΑΘΙΣΤΑΝΤΑΙ). ΚΑΙ ΕΠΙΣΗΣ ΥΠΑΡΧΟΥΝ
17) ΔΥΟ ΔΕΙΚΤΕΣ, ΤΩΝ ΟΠΟΙΩΝ ΤΑ ΑΚΡΑ ΠΕΡΙΣΤΡΕΦΟΝΤΑΙ ΣΕ ΔΥΟ ΜΙΚΡΟΥΣ ΚΥΚΛΟΥΣ. ΟΙ ΚΥΚΛΟΙ ΒΡΙΣΚΟΝΤΑΙ ΚΟΝΤΑ ΣΤΙΣ ΑΡΧΕΣ ΤΩΝ ΕΛΙΚΩΝ. ΣΕ ΜΕ-
18) ΡΗ ΤΕΣΣΕΡΑ (ΕΙΝΑΙ ΔΙΑΙΡΕΜΕΝΟΣ) Ο ΕΝΑΣ, ΔΕΙΧΝΟΝΤΑΣ ΤΗΝ ΠΕΝΤΑΕΤΗΡΙΔΑ (ΕΧΟΝΤΑΣ) ΧΑΡΑΓΜΕΝΟΥΣ ΤΟΥΣ ΣΤΕΦΑΝΙΤΕΣ (/ΙΕΡΟΥΣ) ΑΓΩΝΕΣ, ΚΑΙ ΒΡΙΣΚΕΤΑΙ
19) ΣΤΗΝ 19ΕΤΗ, ΑΠΟ ΤΗΝ 76ΕΤΗ ΠΕΡΙΟΔΟ, ΤΗΣ ΕΛΙΚΑΣ ΤΟΥ ΜΕΤΩΝΑ ΚΑΙ ΣΕ ΤΡΙΑ ΜΕΡΗ (ΕΙΝΑΙ ΔΙΑΙΡΕΜΕΝΟΣ) Ο ΑΛΛΟΣ ΚΥΚΛΟΣ, ΟΠΟΙΟΣ ΒΡΙΣΚΕΤΑΙ ΣΤΗΝ ΕΛΙΚΑ ΤΟΥ ΠΕΡΙΟΔΙΚΟΥ ΚΑΙ ΕΙΝΑΙ Ο ΕΞΕΛΙΓ-
20) ΜΟΣ. ΣΕ ΙΣΑ 223, ΕΝΤΟΣ ΤΕΣΣΑΡΩΝ ΕΛΙΚΟΣΤΡΟΦΩΝ ΚΑΙ ΚΑΤΑ ΜΗΚΟΣ ΟΛΗΣ ΤΗΣ ΕΛΙΚΑΣ, ΤΜΗΜΑΤΑ ΔΙΑΙΡΕΘΗΚΕ Η ΠΕΡΙΟΧΗ ΤΟΥ ΠΕΡΙΟΔΙΚΟΥ, ΟΠΩΣ ΑΚΡΙΒΩΣ
21) ΚΑΙ ΑΥΤΟΣ ΕΙΝΑΙ ΔΙΑΙΡΕΜΕΝΟΣ. Η ΠΛΗΡΗΣ ΑΚΟΛΟΥΘΙΑ ΤΩΝ ΕΚΛΕΙΨΕΩΝ ΕΙΝΑΙ ΓΡΑΜΜΕΝΗ ΣΕ ΚΑΠΟΙΑ ΑΠΟ ΤΑ ΤΜΗΜΑΤΑ.
22) ΜΟΝΟ ΟΙ ΧΡΟΝΟΙ ΑΥΤΩΝ ΤΩΝ ΓΕΓΟΝΟΤΩΝ (ΕΚΛΕΙΨΕΩΝ), ΕΙΝΑΙ ΑΝΑΓΡΑΜΜΕΝΟΙ ΣΕ ΙΣΗΜΕΡΙΝΕΣ ΩΡΕΣ.
23) ΟΠΩΣ ΣΤΑ ΤΜΗΜΑΤΑ ΤΗΣ ΕΛΙΚΑΣ (ΤΟΥ ΠΕΡΙΟΔΙΚΟΥ), ΕΤΣΙ ΚΑΙ ΣΤΑ ΜΕΡΗ ΤΟΥ ΕΞΕΛΙΓΜΟΥ ΑΝΑΓΡΑΦΟΝΤΑΙ ΩΡΕΣ. Ο ΔΕΙΚΤΗΣ, ΤΟΥ ΟΠΟΙΟΥ ΤΟ
24) ΑΚΡΟ ΠΕΡΙΦΕΡΕΤΑΙ ΣΕ ΚΑΘΕ ΥΠΟΔΙΑΙΡΕΣΗ ΤΟΥ ΕΞΕΛΙΓΜΟΥ, ΔΕΙΧΝΕΙ ΤΟΝ ΚΑΘΕ ΕΝΑΝ ΑΠΟ ΤΟΥΣ ΤΡΕΙΣ ΠΕΡΙΟΔΙΚΟΥΣ ΚΥΚΛΟΥΣ. (ΔΕΙΧΝΟΝΤΑΣ) ΤΟΝ ΔΕΥΤΕ-
25) ΡΟ ΠΕΡΙΟΔΙΚΟ (ΚΥΚΛΟ), ΠΡΕΠΕΙ ΟΙ ΧΡΟΝΟΙ ΤΩΝ ΕΚΛΕΙΨΕΩΝ ΝΑ ΠΡΟΣΑΥΞΗΘΟΥΝ ΚΑΤΑ 8 ΩΡΕΣ, ΚΑΙ (ΔΕΙΧΝΟΝΤΑΣ) ΤΟΝ ΤΡΙΤΟ (ΠΕΡΙΟΔΙΚΟ ΚΥΚΛΟ, ΟΙ ΧΡΟΝΟΙ ΤΩΝ ΕΚΛΕΙΨΕΩΝ ΠΡΕΠΕΙ ΝΑ ΠΡΟΣΑΥΞΗΘΟΥΝ) ΚΑΤΑ 16 ΩΡΕΣ.
.................................................................................................................................................................
.................................................................................................................................................................

**Table IX:** Positional analysis of the Back dial operational parts, based on the reconstructed BCI Part-2 text.

| Line | Definition/Description-Position-Operation of the Back Dial plate operational parts |
|------|-----------------------------------------------------------------------------------|
| i-iii and 1-2 | A general presentation of the Back Dial plate and its function |
| 2-4 | Presentation of the Metonic helix |
| 4-7 | Presentation of the Metonic pointer |
| 8-12 | Presentation of the Periodicos pointer |
| 12 | General Description of the Eclipse events |
| 13 | Operation of the traveling pointers along the helices' furrow |
| 14-16 | Reset procedure for the two pointers |
| 16-20 | Presentation of the two small circular scales of Pentateris and Exeligmos |
| 20-21 | Presentation of the Periodicos helix |
| 22 | Clarification for the Periodicos units |
| 23 | Clarification for the Exeligmos units |
| 24 | Clarification for the Exeligmos pointer operation |
| 25 | Clarification for the Periodicos results calculation |

## 10. Epilogue

In this work, the reconstruction of the BCI Part-2 text was presented following the analysis of the preserved text. Since a significant part of the original text was missing (as was also the case with Part-1) the reconstruction of the BCI Part-2 text was based on a) the preserved phrases/words which correspond to specific mechanical parts, b) the visual photographs, c) the AMRP tomographies and their digital reconstructions, d) taking into consideration the common way of Instruction Manual writing for any device. The Symmetry existed in the preserved phrases played a significant role in the text reconstruction. In **Figure 14** a realistic view of the reconstructed BCI Part-2 text is presented.

The construction and the assembly of the authors' Antikythera Mechanism functional models as well as the great number of hours working with their models helped a lot in the reconstruction of the Back Cover Inscription Part-2 text, since the use of the Mechanism created the conditions and the necessity for the Antikythera Mechanism Instruction Manual reconstruction.

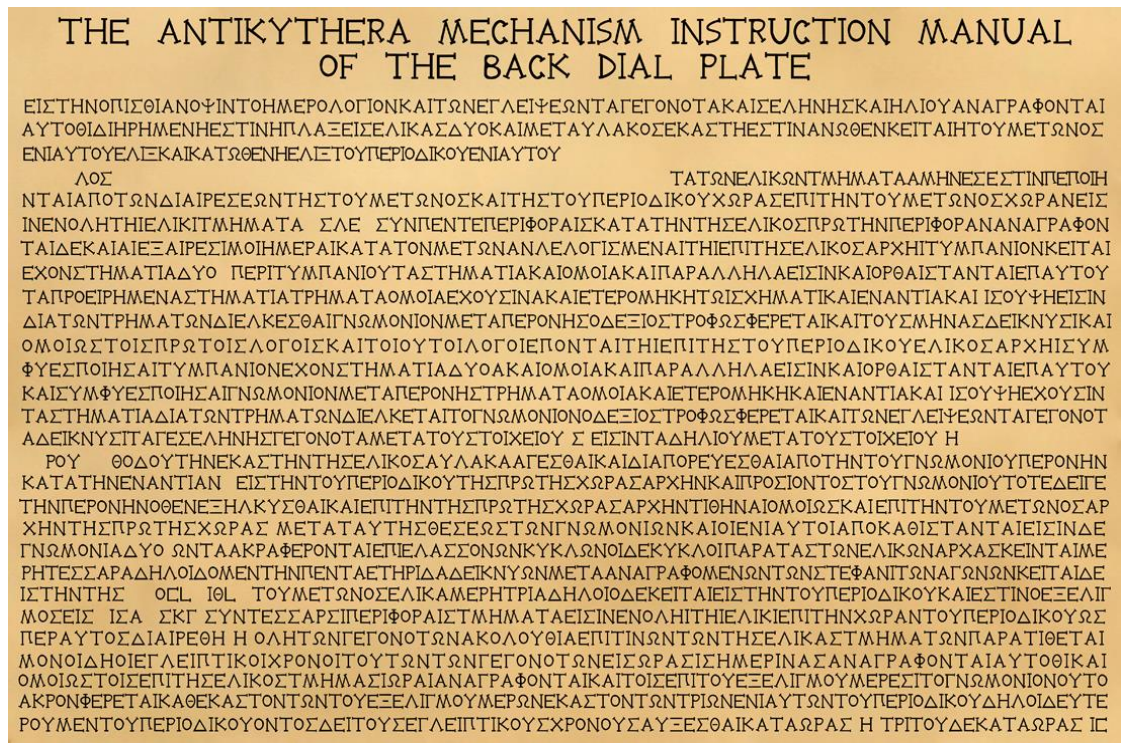

**Figure 14:** A digital representation of the reconstructed BCI Part-2 text on a bronze plate. The text is presented in the style of the original preserved text (words without gaps, except the original gaps). The Fragment 19 corresponds to the bottom left area. Image by the authors.

## Acknowledgments

We are very grateful to Professor Xenophon Moussas (National and Kapodistrian University of Athens, Greece) who provided us with the full AMRP X-ray Volume Raw data of Antikythera Mechanism. John Hugh Seiradakis (Aristotle University of Thessaloniki, Greece) provided us with AMRP X-ray CTs of some of the large fragments, before he passed away in May 2020. Thanks are due to the National Archaeological Museum of Athens, Greece, for granting us permission to photograph and study the Antikythera Mechanism fragments. The Instruction Manual for the Antikythera Mechanism Back Dial plate operational parts was reconstructed in the ancient Greek language by the authors. We thank Dr. C. Boussiou for her help on the final syntax of the ancient Greek text.